\documentclass{article}

\usepackage{arxiv}

\usepackage[utf8]{inputenc} 
\usepackage[T1]{fontenc}    
\usepackage{hyperref}       
\usepackage{url}            
\usepackage{booktabs}       
\usepackage{amsmath}        
\usepackage{amssymb}        
\usepackage{nicefrac}       
\usepackage{microtype}      
\usepackage{graphicx}       
\usepackage{natbib}         
\usepackage{doi}            
\usepackage{xcolor}         
\usepackage{CJK}            
\usepackage{enumerate}      
\usepackage{bm}             
\usepackage{indentfirst} 

\hypersetup{
  colorlinks=true,
  linkcolor=red,
  citecolor=orange,
  urlcolor=blue,
}

\graphicspath{{../figure}}

\title{Deformation and breakup of the liquid ligament \\with various disturbances on the interface in shear flow
}

\author{\href{https://orcid.org/0000-0002-4875-8174}{\includegraphics[scale=0.06]{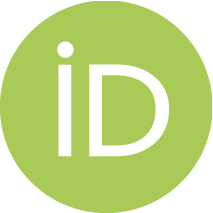}\hspace{1mm}Hideki Yanaoka (柳岡英樹)}
\thanks{Email address for correspondence: yanaoka@iwate-u.ac.jp} \\
	Department of Systems Innovation Engineering, \\
    Faculty of Science and Engineering, Iwate University, \\
    4-3-5 Ueda, Morioka, Iwate 020-8551, Japan \\
	\And
	Wataru Sakamoto (坂本亘) \\
    Department of Science and Engineering, \\
    Graduate School of Integrated Arts and Sciences, Iwate University, \\
    4-3-5 Ueda, Morioka, Iwate 020-8551, Japan \\
	\texttt{} \\
}

\renewcommand{\shorttitle}{Deformation and breakup of liquid ligament with disturbances on interface in shear flow}

\makeatletter
\hypersetup{
pdfusetitle,
bookmarksnumbered,
pdftitle={\@title},
pdfsubject={},
pdfauthor={Hideki Yanaoka, Wataru Sakamoto},
pdfkeywords={Ligament, Droplet, Deformation, Breakup, Shear flow, Vortex, Wavenumber, Disturbance, Numerical simulation}
}
\makeatother

\begin{document}

\begin{CJK*}{UTF8}{ipxg} 
\maketitle
\end{CJK*}

\begin{abstract}
This study performed a numerical analysis of the deformation and breakup 
of a liquid ligament with various disturbances on the interface in shear flow. 
The shear flow generates a three-dimensional flow and vortices 
around the liquid ligament. 
These vortices promote the movement of the liquid inside the liquid ligament. 
When the velocity difference of shear flow increases, 
a nonlinear effect becomes strong, 
and turbulence with higher wavenumber components than the initial disturbance occurs 
at the interface. 
This turbulence accelerates the ligament splitting 
and increases the number of breakup droplets. 
Then, the droplet diameters become uniform, 
and the atomization quality improves. 
As the wavenumber of the disturbance applied to the interface increases, 
the liquid moving velocity along the central axis of the liquid ligament increases. 
Furthermore, the breakup time of the liquid ligament becomes short. 
In addition to the initial reference disturbance with a low wavenumber, 
when turbulence with twice the wavenumber of the reference disturbance is applied to the interface, 
the interface deformation and the splitting of the liquid ligament 
are similar to those with the single reference disturbance. 
When turbulence with four times the wavenumber is added to the initial disturbance 
with the low wavenumber at the interface, 
the deformation of the liquid ligament is accelerated, 
and the liquid ligament splits faster. 
Additionally, the number of breakup droplets increases; 
hence, the total surface area of the liquid increases. 
The droplet diameter becomes uniform; 
therefore, the atomization quality of the liquid ligament is improved.
\end{abstract}


\section{INTRODUCTION}

Liquid atomization technology is used in various fields, 
including industry, agriculture, and medicine. 
In the industrial field, 
atomization equipment is applied in internal combustion engines, 
spray coating, spray drying, and powder production. 
For automobile engines, atomization technology is utilized in fuel injectors. 
Therefore, further clarification of atomization characteristics is required 
to improve engine thermal efficiency and achieve cleaner exhaust gas.

In the general atomization process, 
when the ejected liquid in the form of a liquid sheet or column is destabilized 
by the shearing of gas and liquid, 
a fine-scale liquid column, called a ligament, is generated, 
and the liquid eventually splits into small-scale droplets \citep{Hashimoto_1995}. 
Numerous studies have been reported on the destabilization of liquid films, 
liquid columns, and droplets 
\citep{Squire_1953, O'Rourke&Amsden_1987, Li&Tankin_1991, Li_et_al_2000, Kourmatzis&Masri_2006, Fernandez_et_al_2009, Shinjo&Umemura_2010, Li_et_al_2011, Odier_et_al_2015, Gong_et_al_2016}. 
As spray consists of many small-scale droplets, 
it is considered that the behavior and splitting of liquid ligaments 
that generate droplets significantly affect the spray characteristics. 
However, the destabilization and breakup of liquid ligaments have not been elucidated in detail; 
therefore, further research focusing on the behavior of liquid ligaments is necessary.

\citet{Marmottant&Villermaux_2004a} 
experimentally investigated the splitting behavior of liquid ligaments 
formed by pulling up a circular tube immersed in liquid and stretching the liquid. 
It has been clarified from this investigation 
that the ligament breakup time depends on the diameter of the tube, 
not on the speed with which the liquid is pulled up. 
\citet{Bizjan_et_al_2014} conducted experiments on the breakup of liquid ligaments 
formed by centrifugal force using a rotating disk. 
Consequently, it has been clarified that the Weber number, Ohnesorge number, 
and liquid volumetric flow rate greatly influence the behavior and splitting length of the liquid ligament. 
In addition to the above studies, 
experimental \citep{Henderson_et_al_2000} 
and numerical \citep{Dumouchel_et_al_2015,Zhang&Yuan_2022} studies on liquid ligaments have been reported. 
These studies did not focus on the deformation of liquid caused by gas--liquid shear velocity 
and did not investigate the effect of surrounding airflows on the behavior of liquid ligaments. 
\citet{Constante-Amores_et_al_2021} 
captured a ligament around a liquid jet interface with direct numerical simulation. 
The ligament detached from the interface and formed droplets. 
In this manner, fine-scale liquids such as ligaments split in atomization processes. 
Therefore, the breakup characteristics of ligaments should be investigated 
to clarify spray characteristics. 
The authors \citep{Yanaoka&Nakayama_2022} performed a three-dimensional analysis 
of the destabilization and splitting of a liquid ligament in shear flow. 
When the wavenumber of initial disturbance increases, 
the breakup time of the ligament in the shear flow becomes short, 
and the diameter of the formed droplet becomes small. 
Furthermore, as the velocity difference of the shear flow increases, 
the dispersion of the droplet diameter decreases, 
and the droplet size becomes uniform.

Airflows exist around liquids injected from injectors. 
As liquid ligaments are destabilized by gas--liquid shear velocity, 
it is necessary to investigate the effect of shear flow on the deformation 
and splitting of liquid ligaments. 
Additionally, both air and liquid flows are turbulent in high-speed jets. 
In such a turbulent flow field, 
the velocity turbulence intensity increases around the gas--liquid interface. 
At this time, disturbances with various wavenumber components occur at the interface, 
and it is considered that these turbulences promote the deformation and splitting of liquid ligaments. 
However, it is difficult to experimentally investigate the effects of various turbulences 
on interface deformation, 
focusing on fine-scale liquid ligaments generated in a turbulent flow. 
Hence, the influences of wide-ranging wavenumber disturbances 
on interface deformation have not been fully elucidated. 
Thus, it is necessary to investigate the effects of various disturbances 
on the behavior of liquid ligaments.

From the above viewpoints, 
to clarify the behavior of a liquid ligament in turbulent flow, 
we perform a numerical analysis of a ligament in shear flow 
and investigate the effects of the airflow velocity and disturbances 
with various wavenumbers at the interface 
on the deformation and breakup of the liquid ligament.

\section{NUMERICAL PROCEDURES}

This study considers incompressible viscous flow. 
We apply the level set method \citep{Sussman_et_al_1994} to track the interface. 
Additionally, the continuum surface force method \citep{Brackbill_et_al_1992} is used 
to evaluate surface tension, which is treated as a body force. 
The governing equations for incompressible two-phase flow are 
the continuity equation, the Navier-Stokes equation, 
the advection equation for the level set function, 
and the reinitialization equation, which are given as follows:
\begin{equation}
   \nabla \cdot {\bf u} = 0,
   \label{continuity}
\end{equation}
\begin{equation}
   \frac{\partial {\bf u}}{\partial t} + \nabla \cdot ({\bf u} {\bf u}) 
   = \frac{1}{\rho} \left[ - \nabla p 
   + \frac{1}{Re} \nabla \cdot (2 \mu {\bf D}) 
   + \frac{1}{We} \kappa \hat{\bf n} \delta_s \right],
   \label{navier-stokes}
\end{equation}
\begin{equation}
   \frac{\partial \phi}{\partial t} + {\bf u} \cdot \nabla \phi = 0,
   \label{level-set}
\end{equation}
\begin{equation}
   \frac{\partial \phi}{\partial \tau} + {\bf V} \cdot \nabla \phi = S(\phi_0),
   \label{reinitialization}
\end{equation}
\begin{equation}
   {\bf V} = S(\phi_0) \frac{\nabla \phi}{|\nabla \phi|},
\end{equation}
\begin{equation}
   S(\phi_0) = \frac{\phi_0}{\sqrt{\phi_0^2 + \epsilon^2}},
\end{equation}
where $t$ is the time, ${\bf u} = (u, v, w)$ is the velocity vector 
at the coordinates ${\bf x} = (x, y, z)$, $\rho$ is the density, 
$p$ is the pressure, $\mu$ is the viscosity coefficient, 
and ${\bf D}$ is the strain rate tensor. 
$\kappa$ is the interface curvature, 
$\hat{\bf n}$ is the unit normal vector at the interface, 
$\delta_s$ is the delta function, 
and $\phi$ is the level set function. 
$\tau$ is a pseudo time for reinitialization, 
and $\phi_0$ is the level set function before reinitialization. 
The variables in the fundamental equations are nondimensionalized 
as follows using the reference values of length $L_\mathrm{ref}$, 
velocity $U_\mathrm{ref}$, density $\rho_\mathrm{ref}$, 
and viscosity coefficient $\mu_\mathrm{ref}$:
\[
   t^{*} = \frac{U_\mathrm{ref} t}{L_\mathrm{ref}}, \quad 
   {\bf x}^{*} = \frac{\bf x}{L_\mathrm{ref}}, \quad 
   {\bf u}^{*} = \frac{\bf u}{U_\mathrm{ref}}, \quad 
   p^{*} = \frac{p}{\rho U_\mathrm{ref}^{2}}, \quad 
   \phi^{*} = \frac{\phi}{L_\mathrm{ref}}, \quad 
\]
\begin{equation}
   \rho^{*} = \frac{\rho}{\rho_\mathrm{ref}}, \quad 
   \mu^{*} = \frac{\mu}{\mu_\mathrm{ref}},
\end{equation}
where the superscript $*$ represents a dimensionless variable 
and is omitted in the above governing equations. 
As dimensionless parameters in these equations, 
$Re = U_\mathrm{ref} L_\mathrm{ref}/\nu$ is the Reynolds number, 
$We = \rho_\mathrm{ref} L_\mathrm{ref} U_\mathrm{ref}^2/\sigma$ 
is the Weber number, and $\sigma$ is the surface tension coefficient. 
The strain rate tensor is defined as
\begin{equation}
   {\bf D} = \frac{1}{2} \left[ \nabla {\bf u} + (\nabla {\bf u})^T \right].
\end{equation}
The density and viscosity coefficients for the gas--liquid two-phase flow 
are expressed as
\begin{equation}
   \rho = \rho_g + (\rho_l - \rho_g) H(\phi), \quad 
   \mu = \mu_g + (\mu_l - \mu_g) H(\phi),
\end{equation}
where the subscripts $l$ and $g$ represent liquid and gas, respectively. 
$H(\phi)$ is the Heaviside function and is given as
\begin{equation}
   H(\phi) = \left\{
   \begin{array}{ll}
      1 & \quad \mbox{if} \quad \phi > \epsilon, \\
      0 & \quad \mbox{if} \quad \phi < -\epsilon, \\
      \displaystyle
      \frac{\phi + \epsilon}{2 \epsilon} 
      + \frac{1}{2 \pi} \sin \left( \frac{\pi \phi}{\epsilon} \right) & 
      \quad \mbox{if} \quad |\phi| \le \epsilon.
   \end{array}
   \right.
\end{equation}
The gas--liquid interface is represented as a transition region 
with a thickness of $2\epsilon$, where 
$\epsilon$ is set to $1.0-2.0$ times the grid width. 
The interface curvature and the unit normal vector are expressed as
\begin{equation}
   \kappa = - \nabla \cdot \hat{\bf n},
\end{equation}
\begin{equation}
   \hat{\bf n} = \frac{\nabla \phi}{|\nabla \phi|}.
\end{equation}
The delta function at the interface is obtained as the gradient 
of the Heaviside function as follows:
\begin{equation}
   \delta_\mathrm{s} = \nabla_{\phi} H(\phi) = \left\{
   \begin{array}{ll}
      0 & \quad \mbox{if} \quad |\phi| > \epsilon, \\
      \displaystyle
      \frac{1}{2 \epsilon} \left[ 
      1 + \cos \left( \frac{\pi \phi}{\epsilon} \right) \right] & 
      \quad \mbox{if} \quad |\phi| \le \epsilon.
   \end{array}
   \right.
\end{equation}
The numerical method is almost the same as the previous research 
\citep{Yanaoka&Nakayama_2022}, but the time discretization is different. 
The simplified marker and cell method \citep{Amsden&Harlow_1970} is used to solve 
Eqs. (\ref{continuity}) and (\ref{navier-stokes}). 
The implicit midpoint rule is used to discretize the time derivatives, 
and time marching is performed. 
The second-order central difference scheme is used 
for the discretization of the space derivatives. 
Similarly to existing studies \citep{Yanaoka&Inafune_2023,Yanaoka_2023}, 
we used a simultaneous relaxation method for velocity and pressure. 
For the discretization of Eqs. (\ref{level-set}) and (\ref{reinitialization}), 
the implicit midpoint rule is applied for the time derivatives, 
and the fifth-order weighted essentially nonoscillatory (WENO) scheme \citep{Jiang&Peng_2000} is applied 
for the space derivatives. 
The existing research \citep{Yanaoka&Nakayama_2022} used the explicit method 
for the time derivatives in Eqs. (\ref{level-set}) and (\ref{reinitialization}), 
but considering the calculation efficiency, 
the time discretization was changed to the implicit method. 
In preparation for large-scale calculations to be performed in the future, 
we have improved the numerical method.

\section{CALCULATION CONDITIONS}

\subsection{Instability analysis of a liquid ligament in a stationary fluid}
\label{theory}

\begin{figure}[!t]
\centering
\begin{minipage}{0.48\linewidth}
\begin{center}
\includegraphics[trim=5mm 20mm 0mm 0mm, clip, width=70mm]{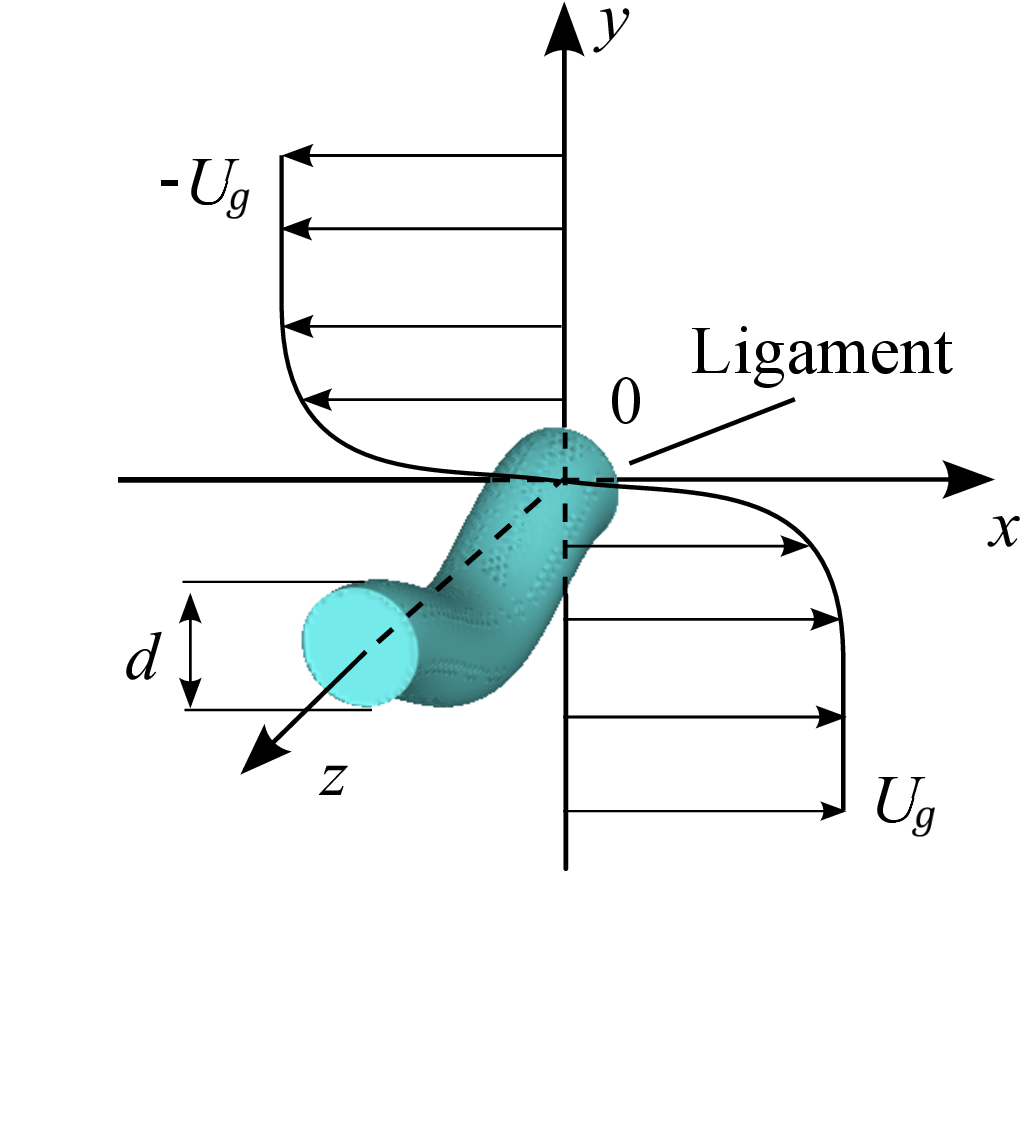}
\end{center}
\end{minipage}
\begin{minipage}{0.48\linewidth}
\begin{center}
\vspace{-13mm}
\includegraphics[trim=10mm 5mm 0mm 0mm, clip, width=72mm]{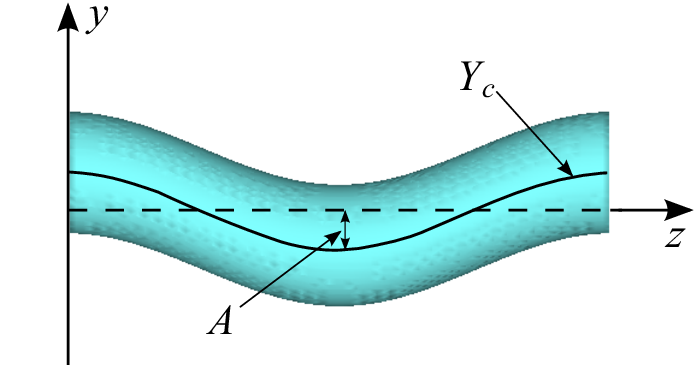}
\end{center}
\end{minipage}
\vspace*{-1.0\baselineskip}
\caption{Flow configuration and coordinate system for ligament in shear flow.}
\label{flow_model3d_b}
\end{figure}

Figure \ref{flow_model3d_b} shows the flow configuration and coordinate system. 
We first analyze the instability of the interface of a liquid ligament 
in stationary fluid to verify the validity of the present numerical method. 
The validity of the computational method has been verified in existing research \citep{Yanaoka&Nakayama_2022}, 
but as the time discretization has been changed, 
the calculation accuracy is confirmed again. 
In this analysis, we deal with a liquid ligament without distortion. 
The origin is placed on the central axis of a ligament with diameter $d$. 
The horizontal direction is the $x$-axis, 
the vertical direction is the $y$-axis, 
and the central axis of the ligament is the $z$-axis. 

The calculation region is $10d$ in the $x$-direction, 
$8d$ in the $y$-direction, and $\lambda$ in the $z$-direction. 
The setting of this computational domain was adapted to the analysis of liquid ligaments 
in shear flow, which will be described later. 
Herein, $\lambda$ is the wavenumber of an initial disturbance on the ligament interface. 
To confirm the dependence of the grid on calculation results, 
we use three grids of $103 \times 87 \times 41$ (grid1), 
$159 \times 141 \times 81$ (grid2), and $209 \times 197 \times 101$ (grid3). 
The minimum grid widths of three grids are $0.08d$, $0.04d$, and $0.02d$, respectively. 
As will be described later, 
when we compared the growth rate of the liquid ligament interface using three grids, 
we found that reasonable results could be obtained with the grid resolution of grid2.

As an initial condition in this calculation, 
similarly to the previous study \citep{Yanaoka&Nakayama_2022}, 
a disturbance with wavelength $\lambda$ and initial amplitude $\eta_{0}$ is given 
to the interface as follows:
\begin{equation}
   \eta = \eta_0 \cos (2 \pi z/\lambda) = \eta_0 \cos (k z), 
   \label{eta}
\end{equation}
where $k = 2\pi/\lambda$ is the wavenumber. 
The interface of the stationary ligament interface grows 
owing to the influence of this disturbance. 
For the boundary conditions of the velocity and level set function, 
periodic boundary conditions are given at all boundaries.

To examine the effect of the initial amplitude $\eta_0$ on the growth of the interface, 
the results obtained using three amplitudes ($\eta_0/d = 0.01$, 0.001, and 0.0001) were compared. 
There was no difference in the results for $\eta_0/d \leq 0.01$. 
It was found that $\eta_0/d = 0.01$ was sufficiently small. 
Therefore, the results obtained using $\eta_0/d = 0.01$ are shown below.

The liquid is a light oil and the surrounding gas is air. 
The densities of the liquid and gas are $\rho_l = 851$ kg/m$^3$ 
and $\rho_g = 1.20$ kg/m$^3$ 
and the viscosity coefficients are $\mu_l = 2.225 \times 10^{-3}$ Pa s 
and $\mu_g = 1.827 \times 10^{-5}$ Pa s, respectively. 
The surface tension coefficient is $\sigma = 2.44 \times 10^{-2}$ N/m. 
The reference values used in the calculation are $L_\mathrm{ref}=d$, 
$U_\mathrm{ref} = \sqrt{\sigma/(\rho_l d)}$, $\rho_\mathrm{ref} = \rho_{l}$, and $\mu_\mathrm{ref} = \mu_{l}$. 
The time intervals used in the calculation are 
$\Delta t/(d/U_\mathrm{ref}) = 1.0 \times 10^{-3}$, $5.0 \times 10^{-4}$, and $2.5 \times 10^{-4}$ for grid1, grid2, and grid3, respectively. 
The time interval is set such that the Courant number is smaller than one.

To verify the validity of this calculation result, 
we varied the wavenumber $k$ of the initial disturbance given to the interface 
and compared the growth rate of the interface with the theoretical value \citep{Weber_1931} 
of a liquid column at each wavenumber. 
The interface growth rate $\omega$ for the liquid column 
in a stationary viscous fluid is defined by linear theory as follows:
\begin{equation}
   \omega^2 + \frac{3 \mu_{l} (ka)^2}{\rho_{l} a^2} \omega 
            = \frac{{\sigma}(ka)^2}{2 \rho_{l} a^3}\left[ 1 - (ka)^2 \right],
\end{equation}
where $a$ is the radius of the liquid column, 
and $ka$ is the dimensionless wavenumber. 
The growth rate $\omega$ is made dimensionless 
as $\Omega$ = $\omega/\sqrt{\sigma/(\rho_{l} a^3)}$. 
In this calculation, we set $ka = 0.3$, 0.5, 0.7, and 0.9.

\subsection{Analysis of the deformation of liquid ligament interface in shear flow}
\label{shear_curve}

Subsequently, we investigate the effect of shear flow on the deformation of a liquid ligament. 
As shown in Fig. \ref{flow_model3d_b}, 
the gas flows in the positive and negative directions of the $x$-axis 
at the velocity $U_g$ on the lower and upper sides of the liquid ligament, respectively. 
The velocity difference between the upper and lower boundaries is ${\Delta} U = 2U_g$. 
In a liquid jet ejected from a nozzle, 
the direction of the shear velocity between gas and liquid is parallel 
to the central axis of a liquid column. 
Conversely, in the model targeted by this study, 
the two directions are orthogonal; 
thus, it is considered that the interface is likely to be deformed. 
We assumed a situation in which a ligament breaks up from the interface 
of a liquid jet or a liquid sheet, and the ligament is exposed to airflow. 
Such ligaments are present in nonuniform shear flows rather than uniform flows. 
Hence, this study analyzes the behavior of a ligament in a shear flow.

As for the calculation region, 
as the liquid ligament is elongated in the $x$-direction owing to the shear flow, 
the wide area in the $x$-direction is set. 
The region is $10d$, $8d$, and $\lambda$ in the $x$-, $y$-, and $z$-directions, respectively. 
The grid used for the calculations is the same as in Subsection \ref{theory}.

As the initial condition in this calculation, 
the following velocity is applied to the gas, 
as in the previous study \citep{Yanaoka&Nakayama_2022}:
\begin{equation}
   u = - U_g \tanh (y/\delta),
\end{equation}
where $\delta$ is a parameter related to the velocity boundary layer thickness. 
In this study, we set $\delta/d = 0.5$ 
and set the velocity boundary layer thickness to about the radius of the ligament 
so that the entire ligament is in an airflow with a velocity gradient. 
Additionally, as the liquid flow in the high-speed flow field is turbulent, 
the formed liquid ligaments are not straight 
but curved \citep{Marmottant&Villermaux_2004b, Wahono_et_al_2008}. 
To model such a deformed liquid ligament, in this calculation, 
the position $Y_c$ of the central axis of the liquid ligament is changed 
in the $y$-direction as shown in the following equation:
\begin{equation}
   Y_c = A \cos (2 \pi z/\lambda) = A \cos (k z),
   \label{centerline}
\end{equation}
where $A$ is a fluctuation amplitude. 
We set $A/d = 0.1$ so that the liquid ligament exists approximately 
in the velocity boundary layer. 
On the interface of this curved liquid ligament, 
we give the initial disturbance of Eq. (\ref{eta}).

Regarding the boundary conditions for the velocity and level set function, 
periodic boundary conditions are given in the $x$- and $z$-directions. 
At the upper and lower boundaries in the $y$-direction, 
uniform flow velocities $-U_g$ and $U_g$ are given, respectively, 
and the level set function is extrapolated.

The physical properties and the reference values are the same as those described 
in Subsection \ref{theory}. 
In our calculation, as in the previous study \citep{Yanaoka&Nakayama_2022}, 
the ligament diameter is determined 
by back calculation from the droplet diameter using linear theory \citep{Rayleigh_1878}. 
The diameter of a droplet formed by an atomizer depends on several factors, 
such as the nozzle diameter, injection pressure, and nozzle shape, 
but is generally about $1.0 \times 10^{-5}-2.0 \times 10^{-4}$ m 
\citep{Tamaki_et_al_2001b, Lee&Park_2002, Liu_et_al_2006, Suh_et_al_2007, Suh&Lee_2008, Kim&Lee_2008}. 
In the linear theory \citep{Rayleigh_1878}, 
the relationship between the diameter $d$ of the liquid column 
and the diameter $d_\mathrm{drop}$ of the droplet is expressed as 
$d_\mathrm{drop} = 1.89d$.
From this, the ligament diameter is 
$d = 5.3 \times 10^{-6}-1.0 \times 10^{-4}$ m. 
In this calculation, the ligament diameter is set to $d = 1.0 \times 10^{-5}$ m. 

The airflow velocity difference in this calculation is $\Delta U = 10$, 20, and 30m/s, 
referring to a study \citep{Zha_et_al_2015} on the airflow behavior inside the cylinder of a diesel engine. 
The dimensionless velocity difference $\Delta U/U_\mathrm{ref}$ at this time 
is $\Delta U/U_\mathrm{ref}$ = 5.9, 11.9, and 17.8. 
The time intervals used in the calculation are 
$\Delta t/(d/U_\mathrm{ref}) = 1.0 \times 10^{-3}$, $5.0 \times 10^{- 4}$, 
and $2.5 \times 10^{-4}$ for grid1, grid2, and grid3, respectively. 
In this analysis, we show the results for the condition of wavenumber $ka = 0.7$, 
which gives the fastest growth rate for a ligament in a stationary fluid.

\subsection{Analysis of the effects of disturbance with various wavenumber components 
on the deformation and breakup of liquid ligament}
\label{wave_number}

In a turbulent flow field, 
disturbance with various wavenumber components occurs at the interface of a liquid ligament. 
In this analysis, 
we investigate the effects of varied disturbances generated 
at the interface of a liquid ligament in shear flow on the deformation and splitting of the liquid ligament. 
The flow field and coordinate system are the same as Subsection \ref{shear_curve}.

The computational domain is the same as the analysis in Subsection \ref{shear_curve}. 
We confirmed the grid dependence on calculation results 
using the same three grids as in Subsection \ref{theory}. 
Consequently, we found that valid results could be obtained with the grid resolution of grid2.

To model disturbance with various wavenumber components, 
we combine higher wavenumber disturbances with a reference fluctuation 
defined by Eq. (\ref{eta}). 
The initial disturbance is given to the interface of the liquid ligament curved using Eq.(\ref{centerline}) 
as follows:
\begin{equation}
   \eta = \eta_h \cos (2 \pi z/\lambda_h) + \eta_0 \cos (2 \pi z/\lambda) 
   = \eta_h \cos (k_h z) + \eta_0 \cos (k z),
   \label{eta_synthetic}
\end{equation}
where $\eta_h$, $\lambda_h$, and $k_h$ represent the amplitude, wavelength, 
and wavenumber of the high wavenumber component disturbance, respectively. 
The amplitude and wavenumber ratios of the high wavenumber component 
to the reference disturbance are defined as $\eta_r = \eta_h/\eta_0$ 
and $k_r = k_h/k$, respectively. 
When we give a single reference disturbance component 
to the liquid ligament interface, we set $\eta_r = 0$ and $k_r =0$. 
Under this condition, we vary the wavenumber as $ka = $ 0.3, 0.5, 0.7, and 0.9. 
We investigated the effects of $\eta_r$ and $k_r$ on the deformation of the interface 
and found that $\eta_r = 0.9$ had the greatest effect. 
Additionally, when the wavenumbers of the high wavenumber disturbance were 
two and four times the wavenumbers of the low wavenumber disturbance, 
i.e., $k_r = $2.0 and 4.0, the effect of the high wavenumber disturbance appeared. 
In this calculation, we vary the wavenumber ratio as $k_r = $ 2.0 and 4.0 under $\eta_r = 0.9$.

Calculation conditions for initial conditions, boundary conditions, physical property values, 
reference values, and time steps are the same as in Subsection \ref{shear_curve}. 
In this calculation, 
we use the velocity difference $\Delta U/U_\mathrm{ref} = 17.8$ 
where the effect of shear velocity is the strongest. 
In the results shown below, the dimensionless time when the sampling starts is $T = 0$.

\section{RESULTS AND DISCUSSION}

\subsection{Instability analysis of liquid ligament in a stationary fluid}

\begin{figure}[!t]
\centering
\begin{minipage}{0.48\linewidth}
\begin{center}
\includegraphics[trim=0mm 0mm 0mm 0mm, clip, width=70mm]{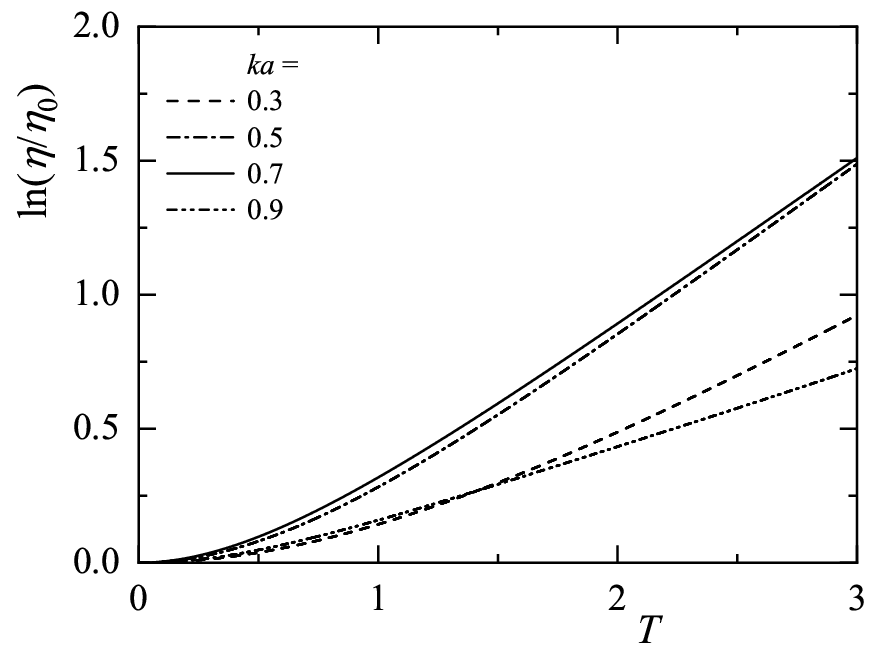} \\
\vspace*{-0.5\baselineskip}
(a)
\end{center}
\end{minipage}
\begin{minipage}{0.48\linewidth}
\begin{center}
\includegraphics[trim=0mm 0mm 0mm 0mm, clip, width=75mm]{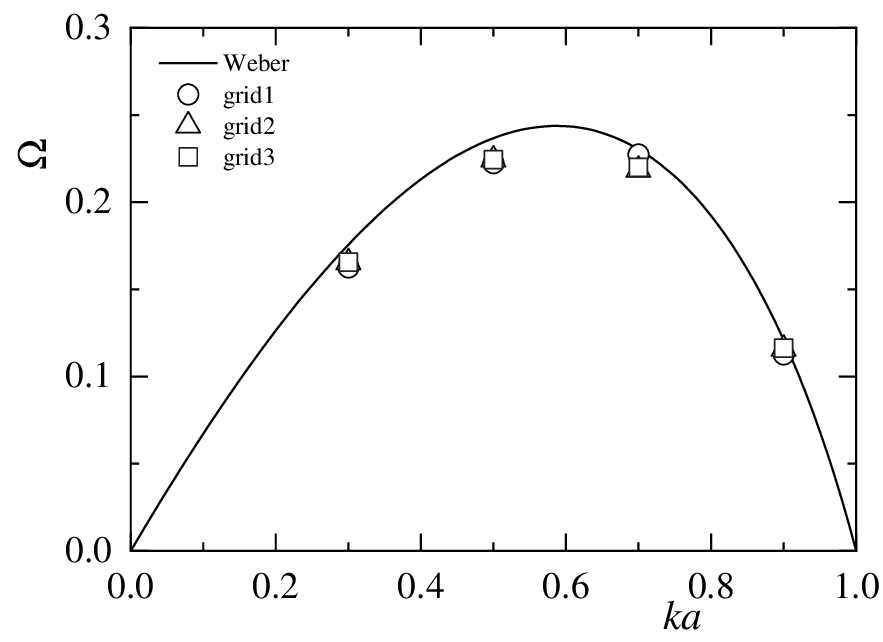} \\
\vspace*{-0.5\baselineskip}
(b)
\end{center}
\end{minipage}
\caption{Time variations of amplitudes and growth rates for various wave numbers: 
(a) amplitude (grid2) and (b) growth rate.}
\label{growth_rate}
\end{figure}

First, we verify the validity of this numerical method 
by comparing the growth rate of the liquid ligament interface with the theoretical value 
\citep{Weber_1931}. 
Figure \ref{growth_rate}(a) shows the time variation of the amplitude 
of the liquid ligament interface at each dimensionless wavenumber $ka$. 
The amplitude is the mean value of the amplitudes at $z/\lambda = 0.5$ and 1.0. 
This result was obtained using grid2, 
and the same result was obtained with other grids. 
The amplitude of the interface increases exponentially with time. 
The interface amplitude grows fastest at $ka = 0.7$ and slowest at $ka = 0.9$.

Figure \ref{growth_rate}(b) shows the dimensionless growth rate $\Omega$ 
in each grid compared with the theoretical value \citep{Weber_1931}. 
At all wavenumbers, the calculated results support the theoretical values.

\subsection{Analysis of the deformation of liquid ligament interface in shear flow}

\begin{figure}[!t]
\centering
\begin{minipage}{0.32\linewidth}
\begin{center}
\includegraphics[trim=0mm 0mm 0mm 0mm, clip, width=55mm]{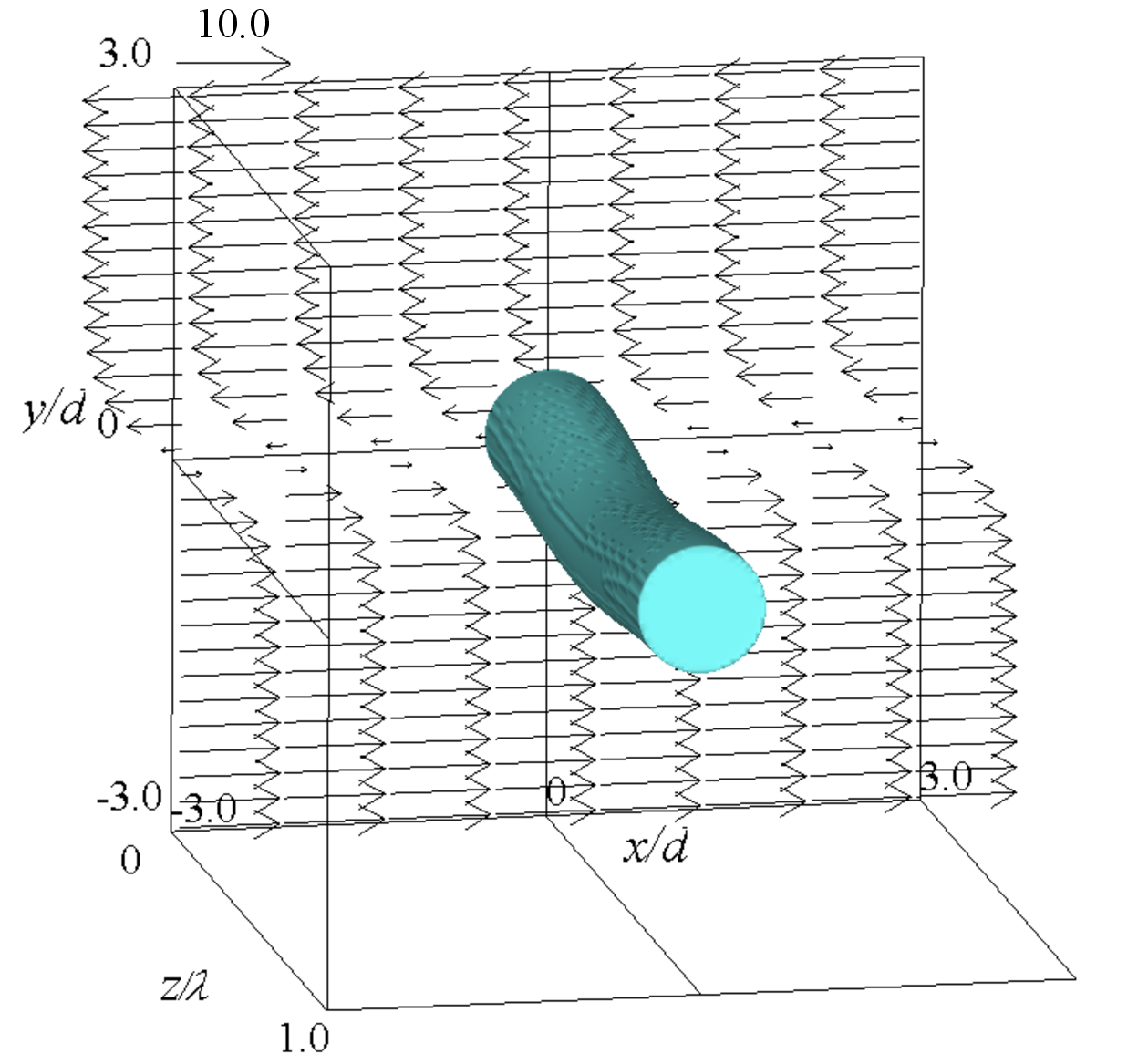} \\
\vspace*{-0.5\baselineskip}
(a)
\end{center}
\end{minipage}
\begin{minipage}{0.32\linewidth}
\begin{center}
\includegraphics[trim=0mm 0mm 0mm 0mm, clip, width=55mm]{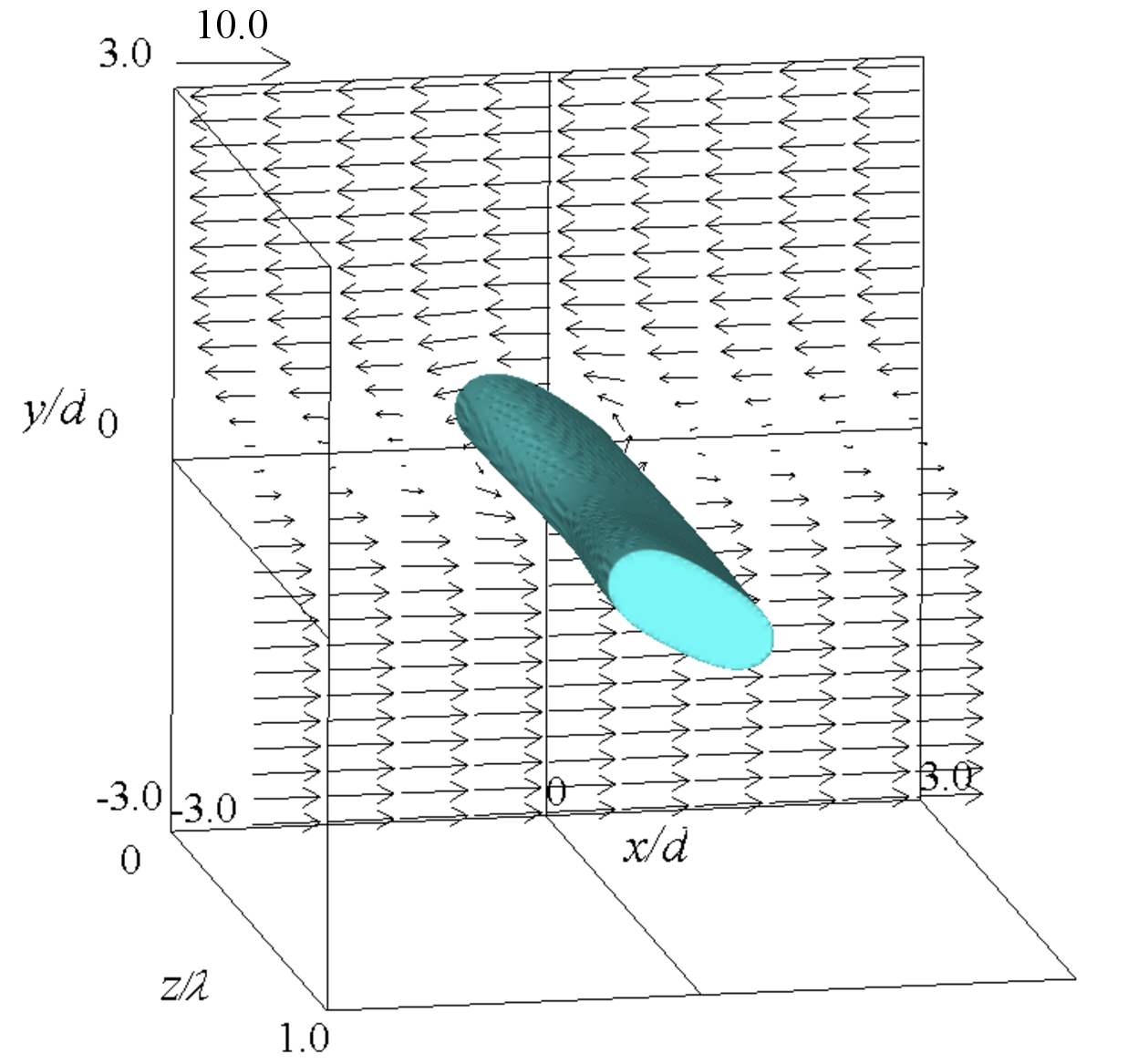} \\
\vspace*{-0.5\baselineskip}
(b)
\end{center}
\end{minipage}
\begin{minipage}{0.32\linewidth}
\begin{center}
\includegraphics[trim=0mm 0mm 0mm 0mm, clip, width=55mm]{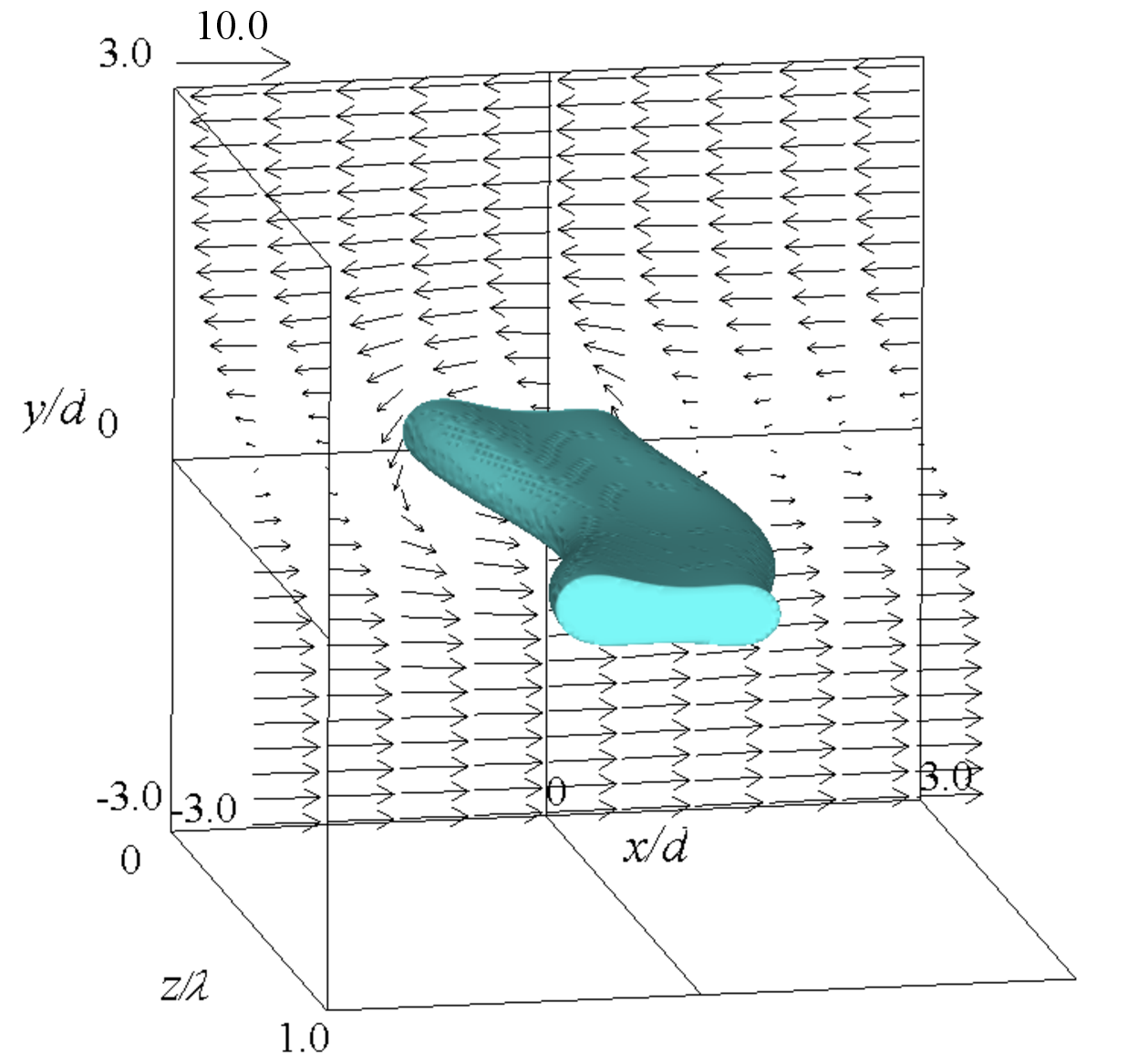} \\
\vspace*{-0.5\baselineskip}
(c)
\end{center}
\end{minipage}

\vspace*{0.5\baselineskip}
\begin{minipage}{0.32\linewidth}
\begin{center}
\includegraphics[trim=0mm 0mm 0mm 0mm, clip, width=55mm]{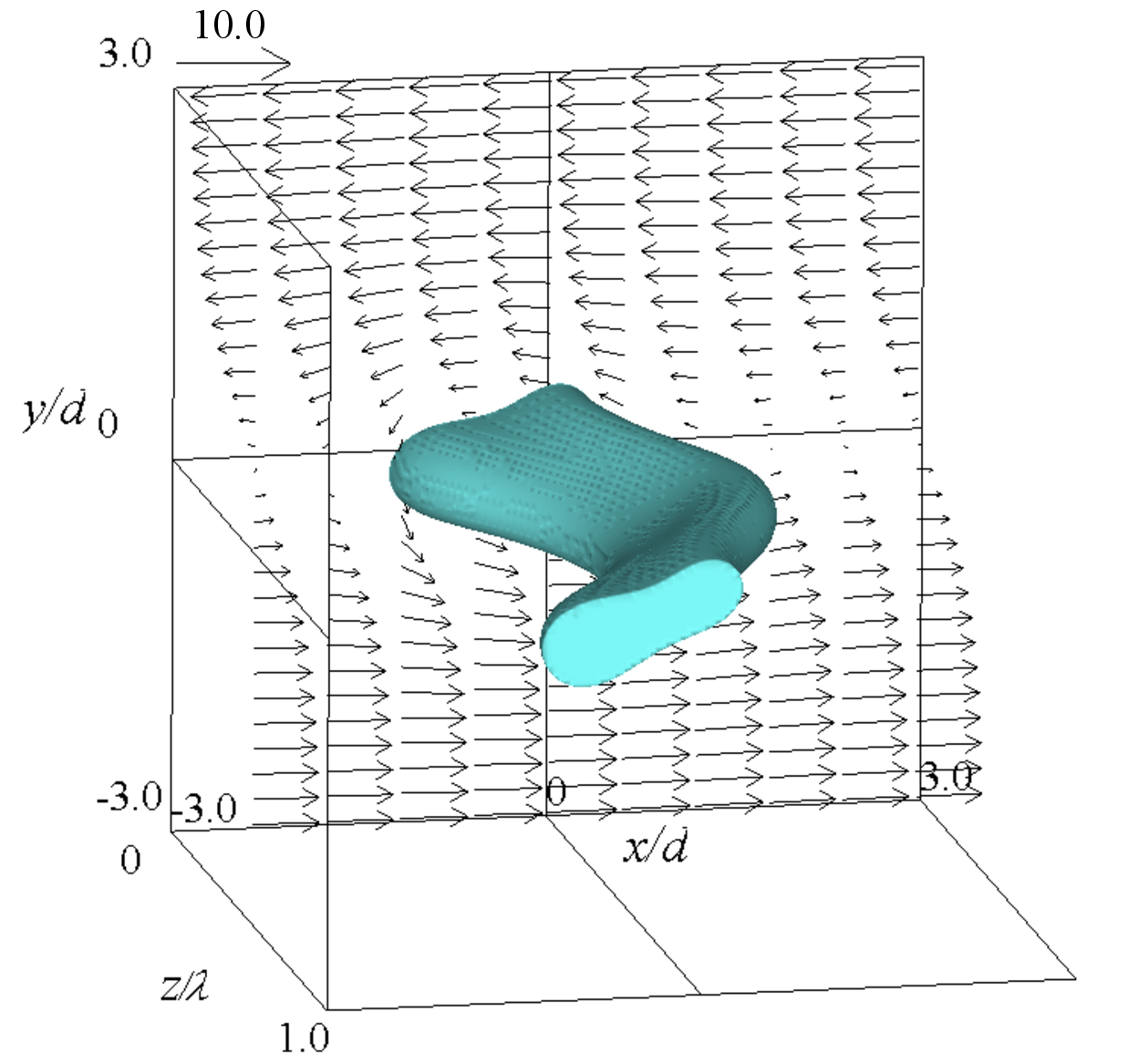} \\
\vspace*{-0.5\baselineskip}
(d)
\end{center}
\end{minipage}
\begin{minipage}{0.32\linewidth}
\begin{center}
\includegraphics[trim=0mm 0mm 0mm 0mm, clip, width=55mm]{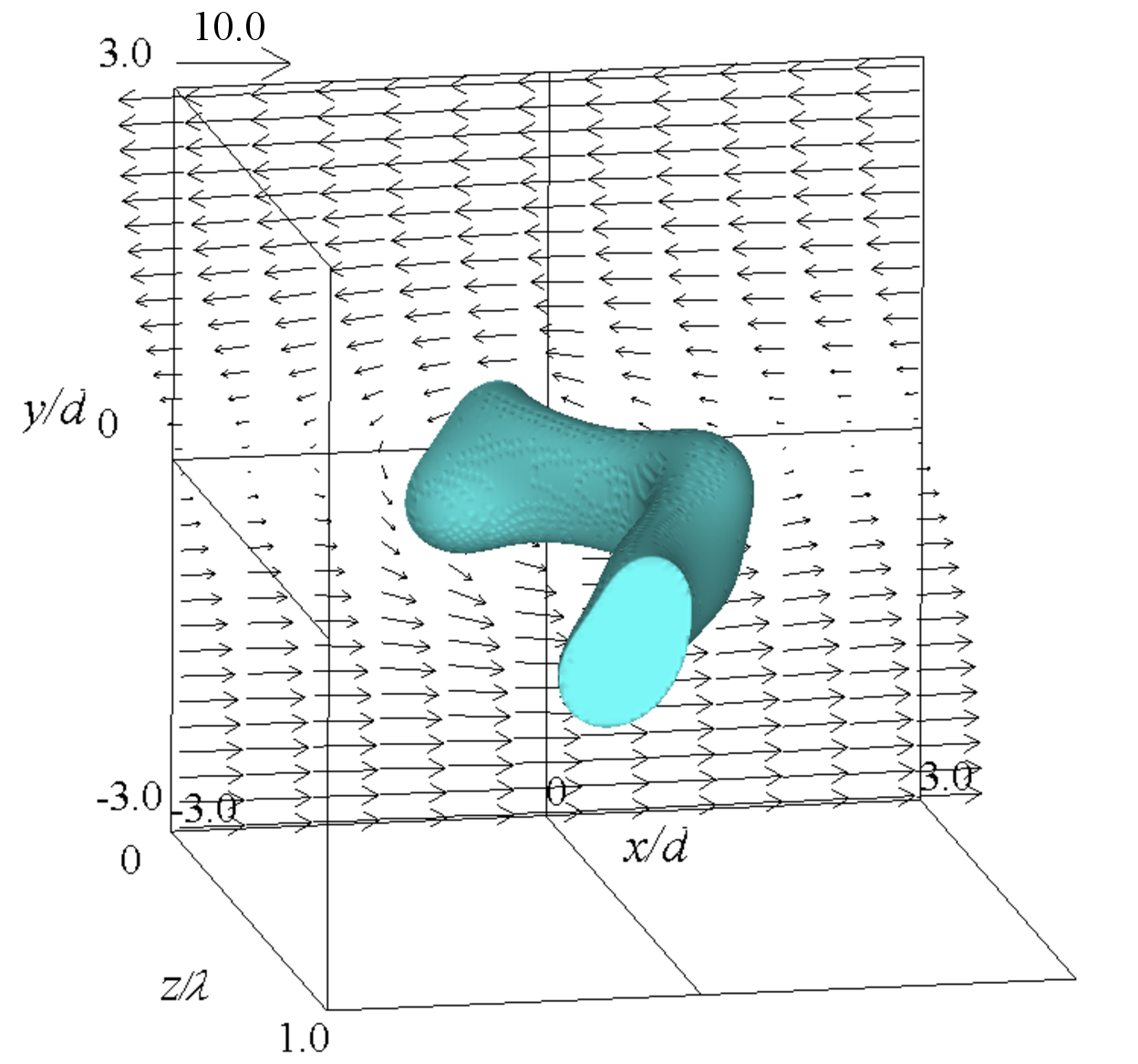} \\
\vspace*{-0.5\baselineskip}
(e)
\end{center}
\end{minipage}
\begin{minipage}{0.32\linewidth}
\begin{center}
\includegraphics[trim=0mm 0mm 0mm 0mm, clip, width=55mm]{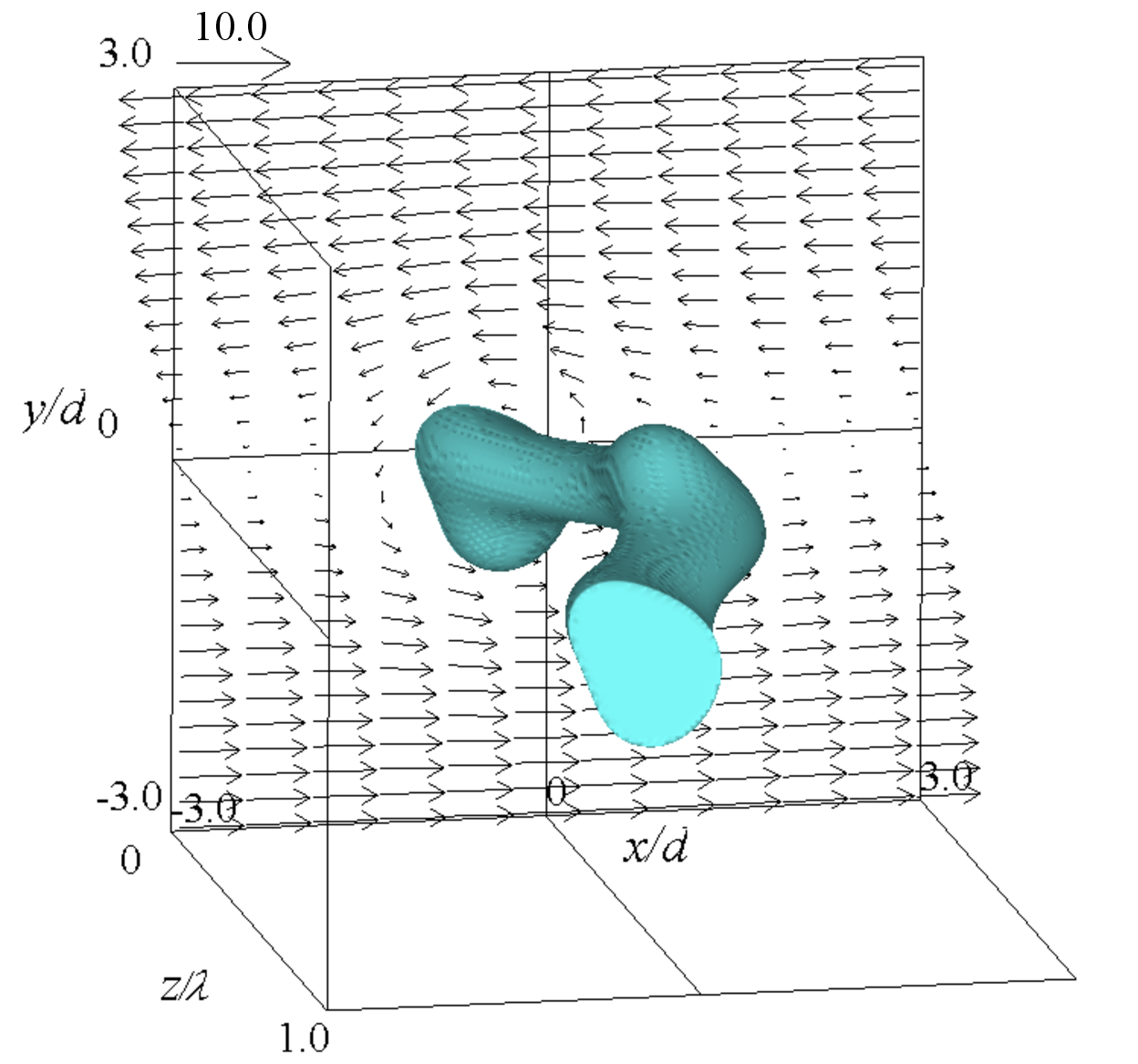} \\
\vspace*{-0.5\baselineskip}
(f)
\end{center}
\end{minipage}
\caption{Time variations of velocity vectors at $z/\lambda = 0.01$ 
and ligament interface: 
(a) $T = 0$, (b) $T = 0.1$, (c) $T = 0.3$, (d) $T = 0.5$, (e) $T = 0.7$, and (f) $T = 0.9$: 
$\Delta U/U_\mathrm{ref} = 11.9$, $ka = 0.7$.}
\label{u20_t03-09}
\end{figure}

Subsequently, we investigate the effect of shear flow on the deformation of the liquid ligament interface. 
Figure \ref{u20_t03-09} shows the time variation of the flow field 
and liquid ligament for $\Delta U/U_\mathrm{ref} = 11.9$ and $ka = 0.7$. 
The velocity vector is the distribution in the $x$--$y$ cross-section 
at $z/\lambda = 0.01$. 
At $T = 0-3$, the parts of $z/\lambda = 0.5$ and $z/\lambda = $ 0 
in the ligament are stretched by the shear flow 
in the positive and negative directions of the $x$-axis, respectively. 
Around the interface of the liquid ligament, 
the airflow flows along the interface, 
so the liquid ligament rotates counterclockwise with time. 
At $T = 0.9$, we can see that the cross-sections at $z/\lambda = 0$ and 0.5 
in the liquid ligament become large.

\begin{figure}[!t]
\centering
\begin{minipage}{0.48\linewidth}
\begin{center}
\includegraphics[trim=0mm 0mm 0mm 0mm, clip, width=70mm]{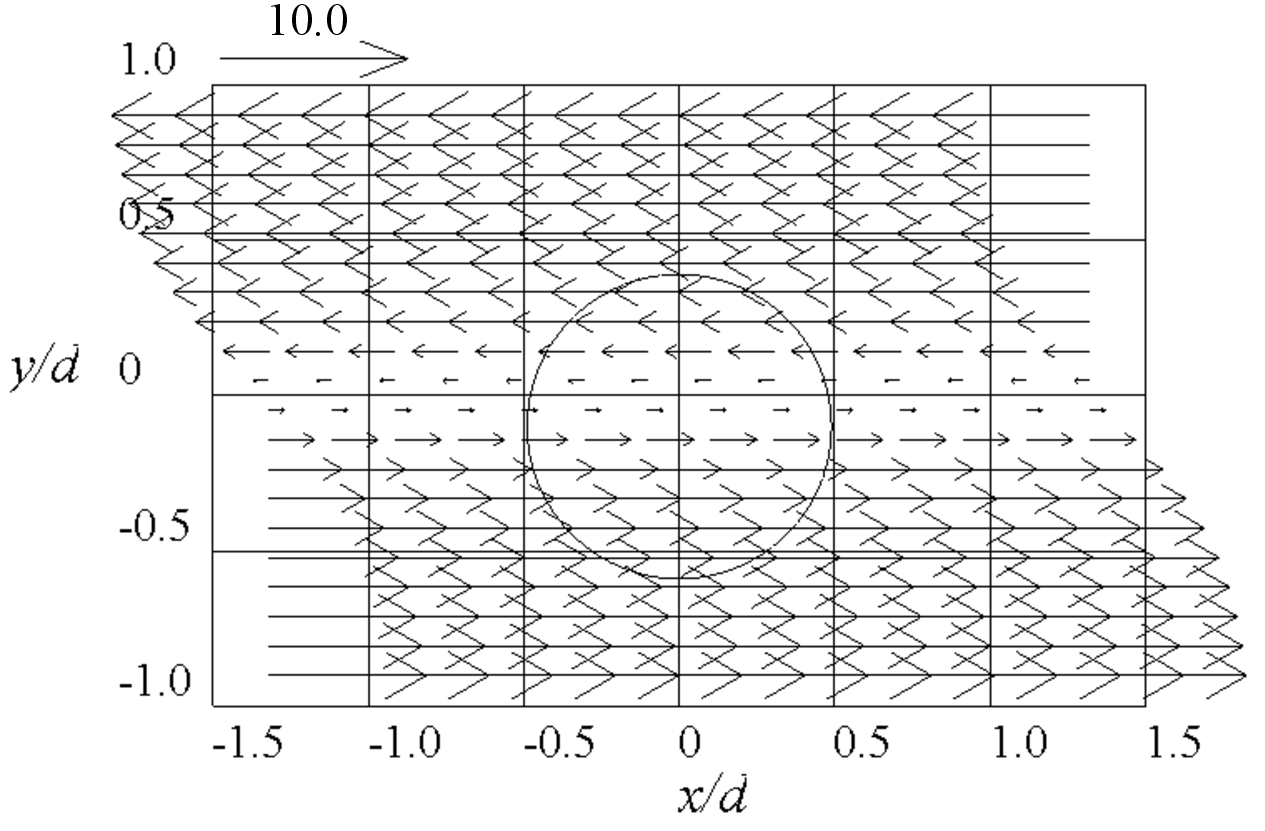} \\
\end{center}
\end{minipage}
\begin{minipage}{0.48\linewidth}
\begin{center}
\includegraphics[trim=0mm 0mm 0mm 0mm, clip, width=70mm]{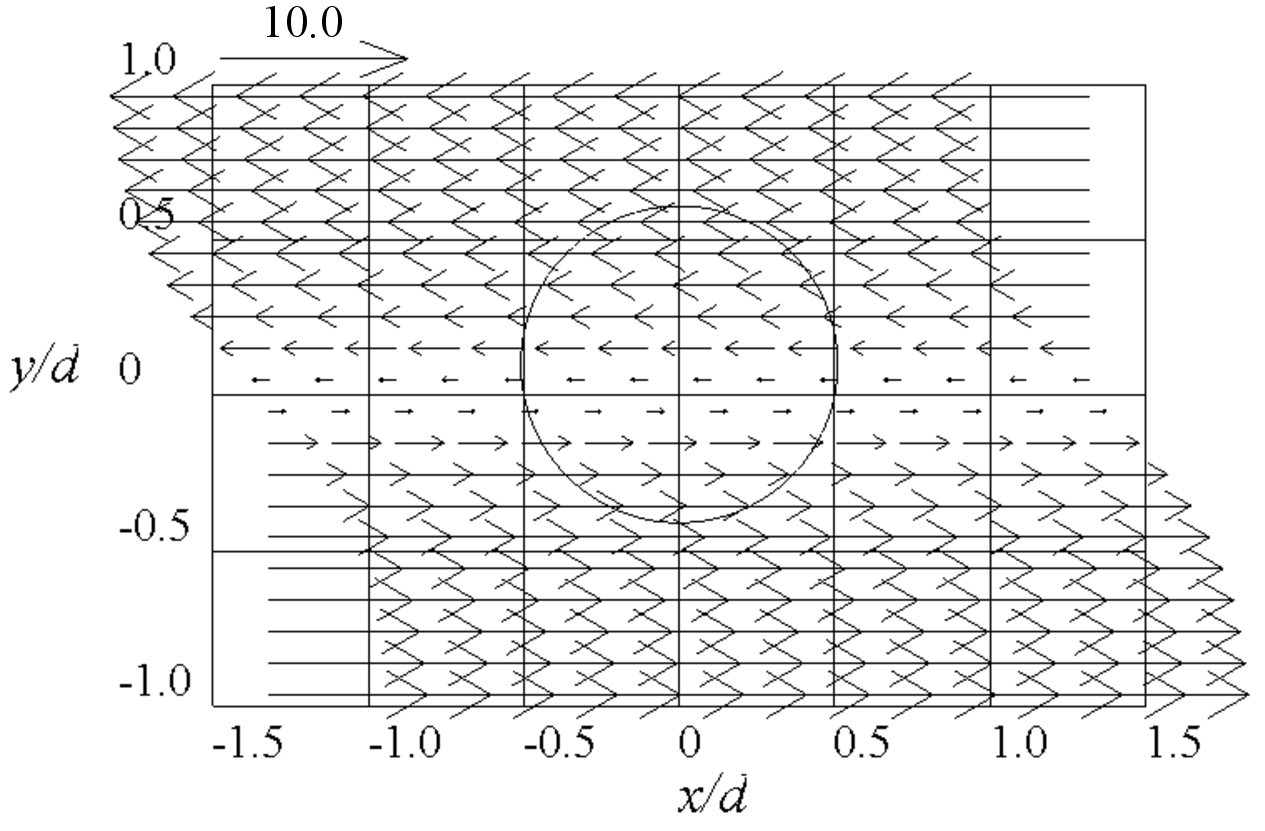} \\
\end{center}
\end{minipage}

(a)

\begin{minipage}{0.48\linewidth}
\begin{center}
\includegraphics[trim=0mm 0mm 0mm 0mm, clip, width=70mm]{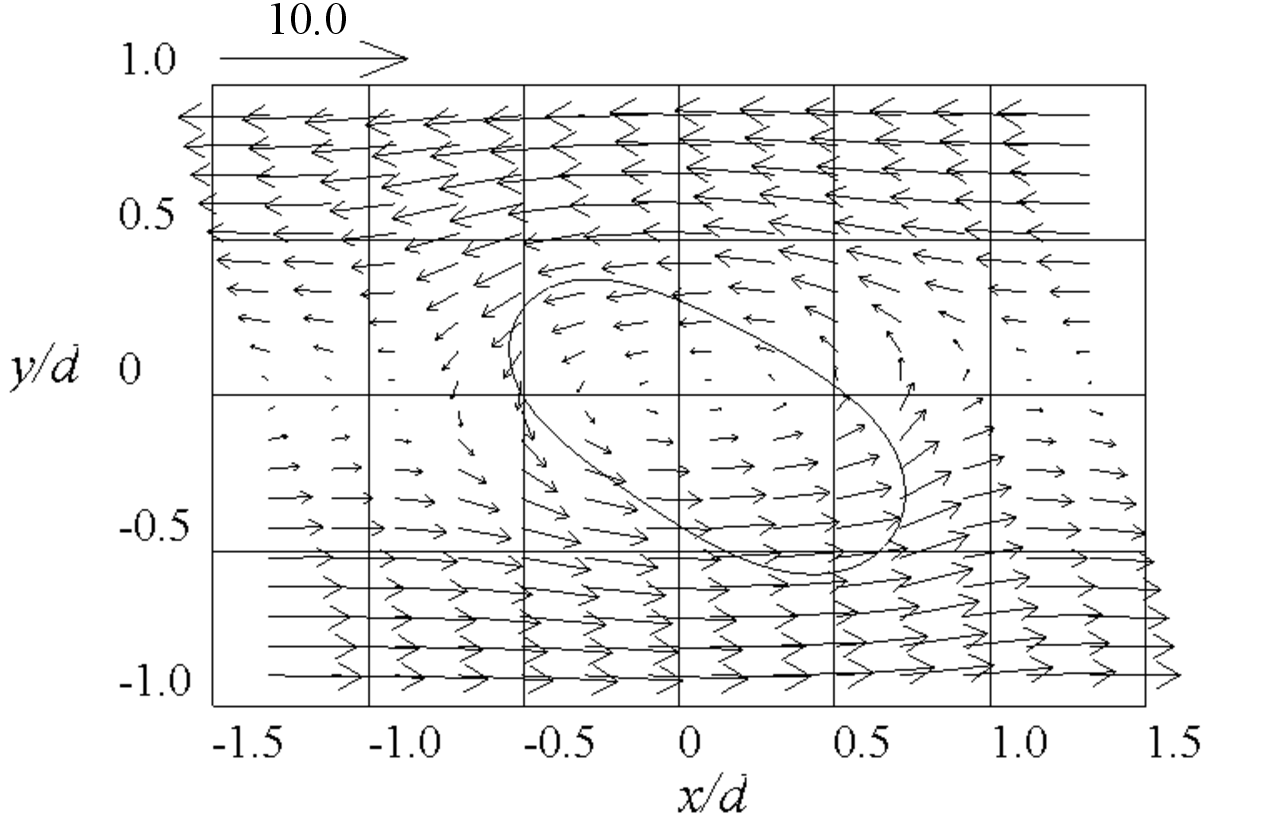} \\
\end{center}
\end{minipage}
\centering
\begin{minipage}{0.48\linewidth}
\begin{center}
\includegraphics[trim=0mm 0mm 0mm 0mm, clip, width=70mm]{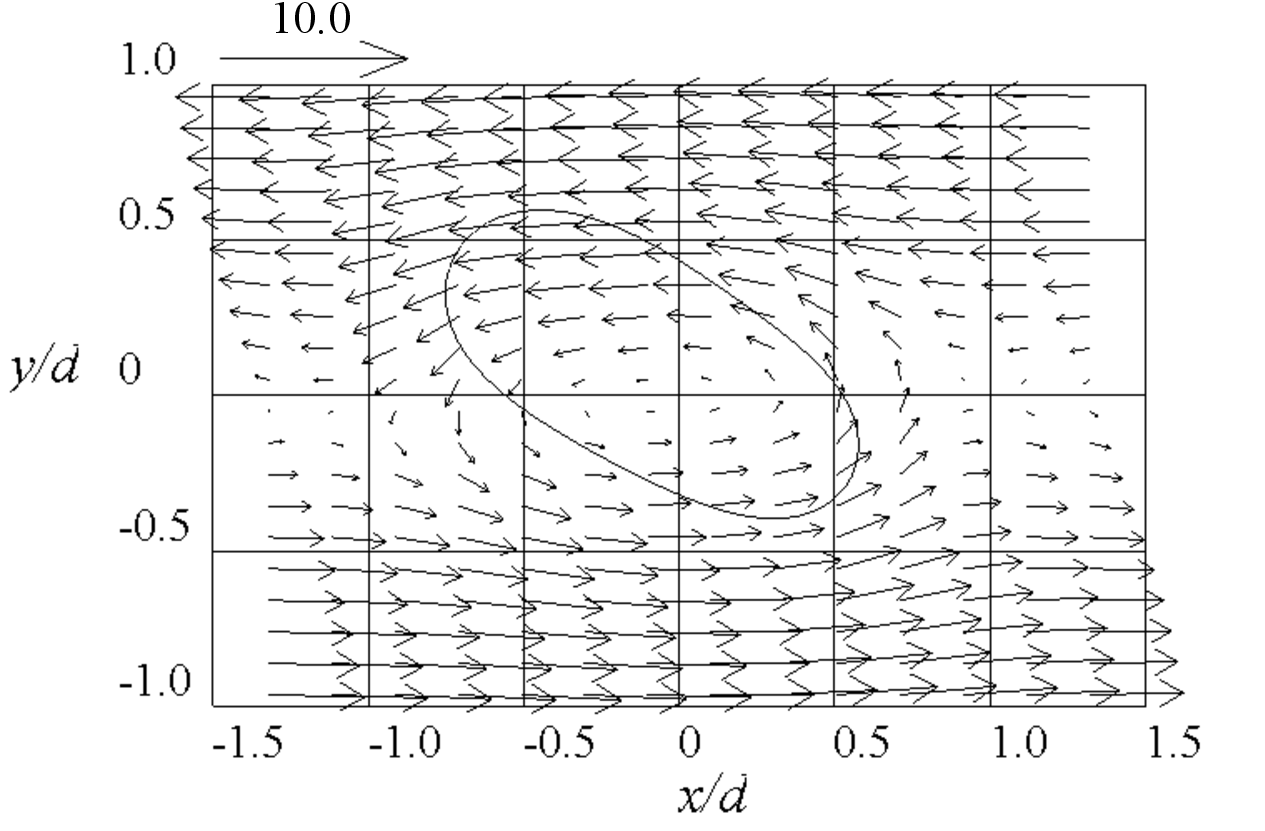} \\
\end{center}
\end{minipage}

(b)

\begin{minipage}{0.48\linewidth}
\begin{center}
\includegraphics[trim=0mm 0mm 0mm 0mm, clip, width=70mm]{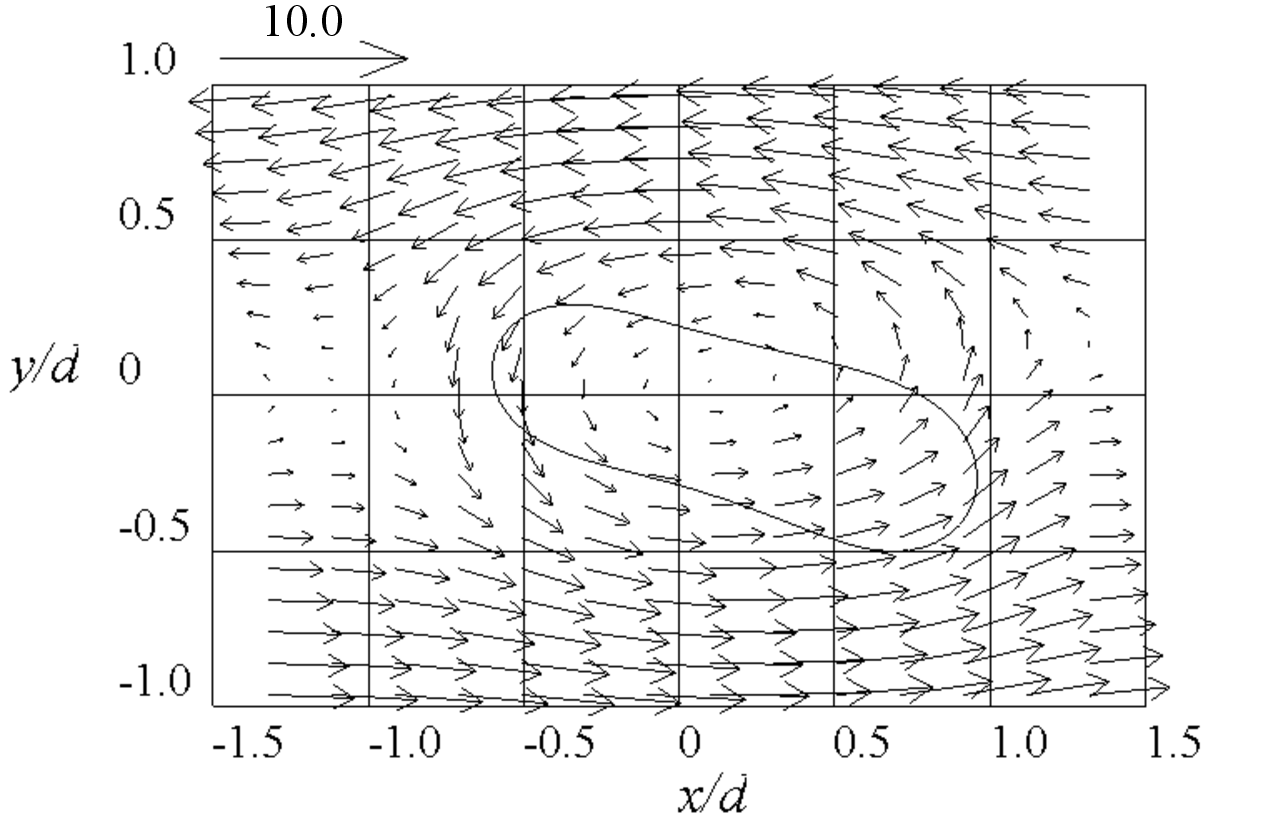} \\
\end{center}
\end{minipage}
\centering
\begin{minipage}{0.48\linewidth}
\begin{center}
\includegraphics[trim=0mm 0mm 0mm 0mm, clip, width=70mm]{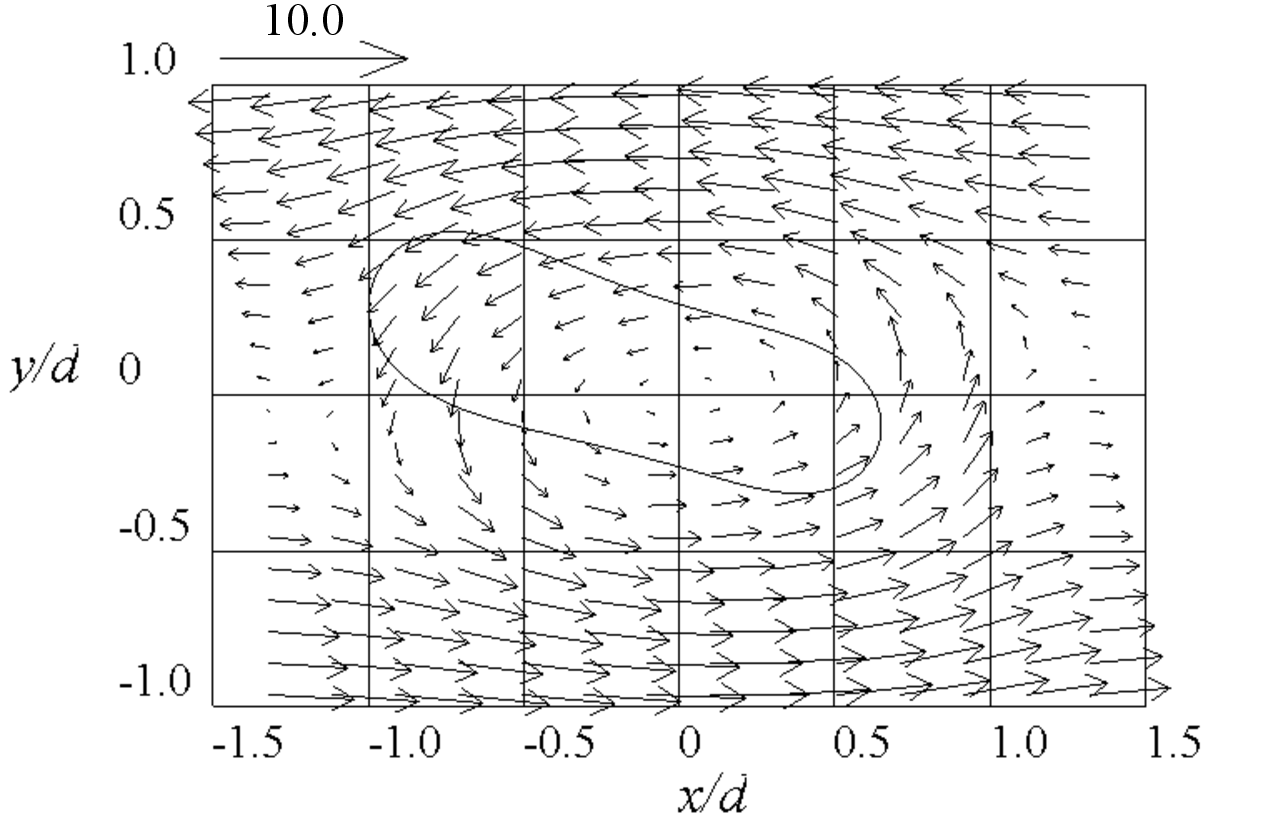} \\
\end{center}
\end{minipage}

(c)

\begin{minipage}{0.48\linewidth}
\begin{center}
\includegraphics[trim=0mm 0mm 0mm 0mm, clip, width=70mm]{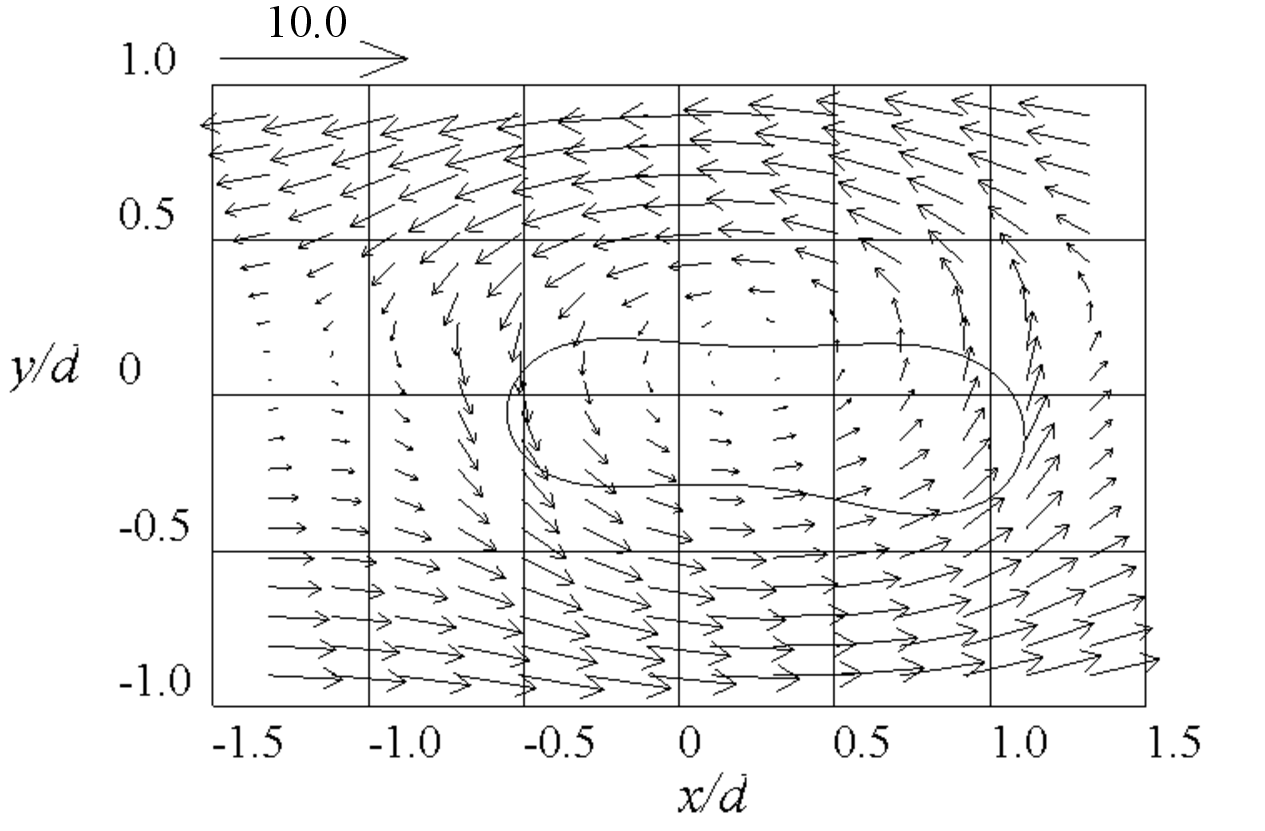} \\
\end{center}
\end{minipage}
\begin{minipage}{0.48\linewidth}
\begin{center}
\includegraphics[trim=0mm 0mm 0mm 0mm, clip, width=70mm]{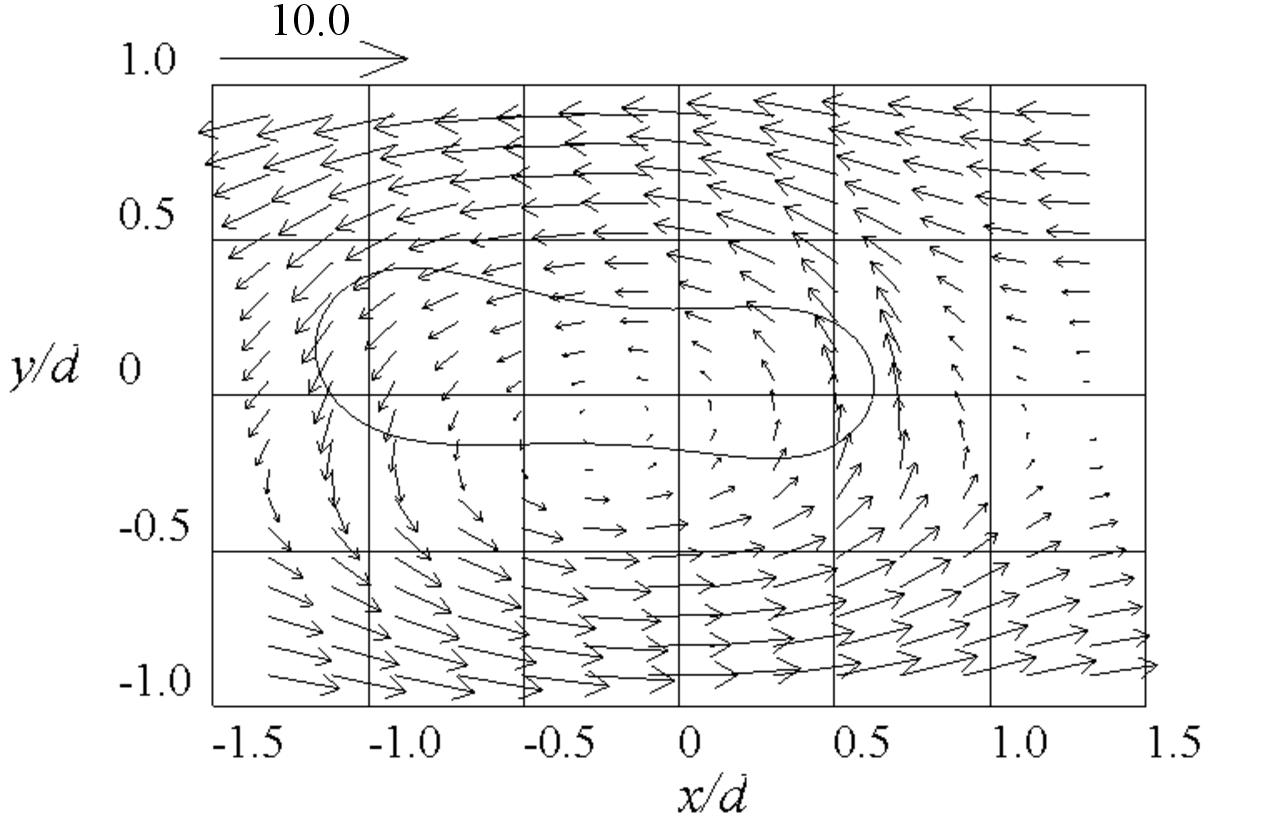} \\
\end{center}
\end{minipage}

(d)

\caption{Time variations of ligament interface and velocity vectors 
at $z/\lambda = 0.5$ (left) and $z/\lambda = 1.0$ (right): 
(a) $T = 0$, (b) $T = 0.1$, (c) $T = 0.2$, and (d) $T = 0.3$: 
$\Delta U/U_\mathrm{ref} = 11.9$, $ka = 0.7$.}
\label{u20k7timevcxy}
\end{figure}

Figure \ref{u20k7timevcxy} shows the interface and velocity vectors 
in the $x$--$y$ cross-section at $T = 0-0.3$. 
A solid black line indicates the interface. 
The left and right figures are the results at $z/\lambda = 0.5$ and 1.0, respectively. 
At $T = 0$, the liquid ligament is curved like a cosine function in the $z$-direction; 
thus, the lower interface ($y/d < 0$) of $z/\lambda = 0.5$ is exposed to faster airflow 
than the velocity near the upper interface ($y/d > 0$). 
Consequently, the liquid inside the liquid ligament is further accelerated, 
and at $T = 0.1$, the liquid velocity around $(x/d, y/d) = (0.5, -0.5)$ 
is faster than that around $(x/d, y/d) = (-0.5, 0.25)$. 
Similarly, in $z/\lambda = 1.0$ at $T = 0.1$, 
the liquid velocity around $(x/d, y/d) = (-0.5, 0.5)$ is faster than 
that around $(x/d, y/d) = (0.5, -0.2)$. 
At the two cross-sections at $T = 0.1$, 
vortices centered at $(x/d, y/d) = (0, 0)$ are formed by the airflow. 
The airflow flows along the interface; 
thus, the vortex centers at the cross-sections of $z/\lambda = 0.5$ and 1.0 move with time 
to the upper and lower sides of the interface, respectively. 
Consequently, at $T = 0.3$, the vortex centers exist 
near $(x/d, y/d) = (0.25, 0.2)$ for $z/\lambda = 0.5$ 
and near $(x/d, y/d) = (-0.25, -0.2)$ for $z/\lambda = 1.0$.

\begin{figure}[!t]
\centering
\begin{minipage}{0.48\linewidth}
\begin{center}
\includegraphics[trim=0mm 0mm 0mm 0mm, clip, width=72mm]{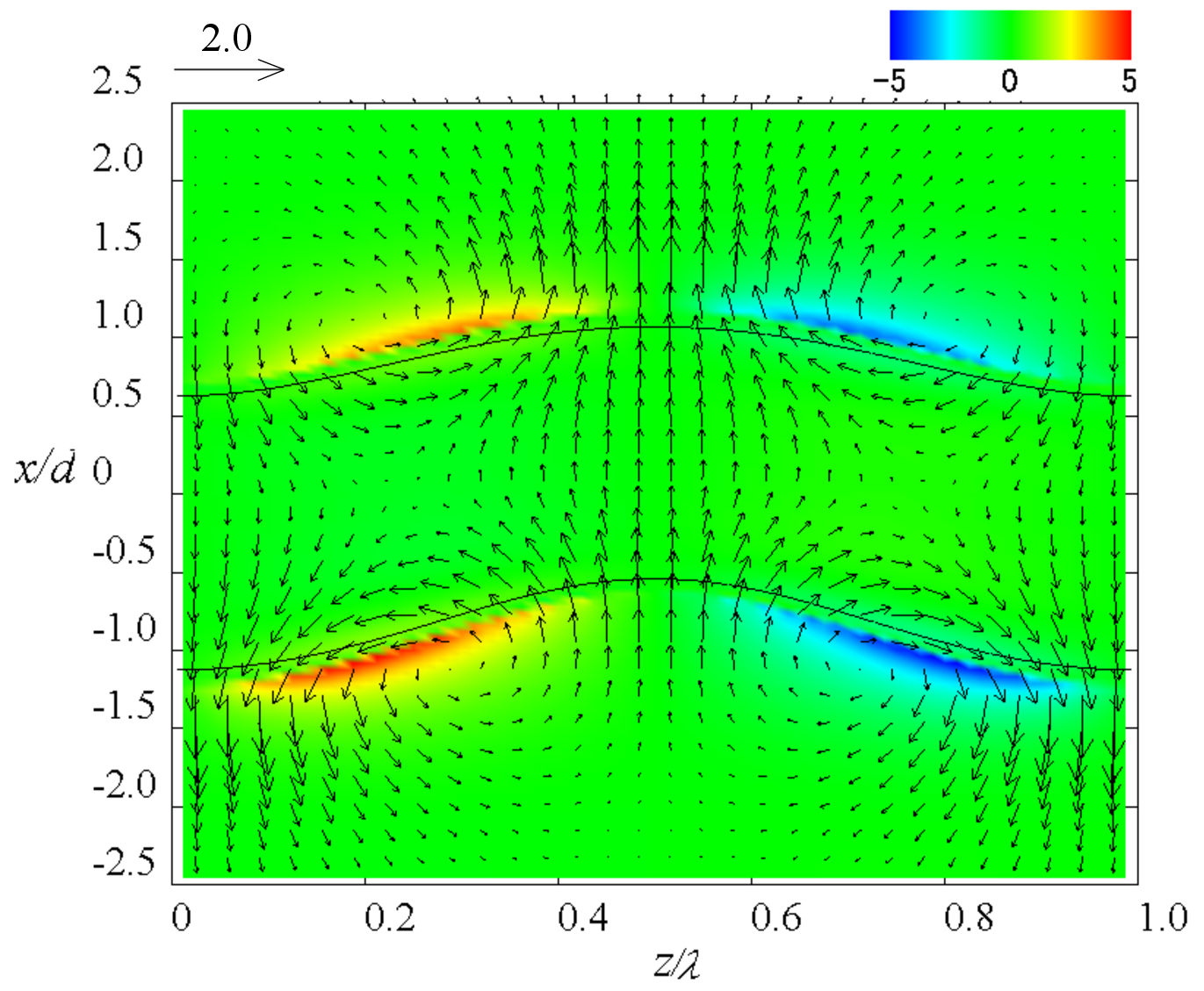} \\
\vspace*{-0.5\baselineskip}
(a)
\end{center}
\end{minipage}
\begin{minipage}{0.48\linewidth}
\begin{center}
\includegraphics[trim=0mm 0mm 0mm 0mm, clip, width=80mm]{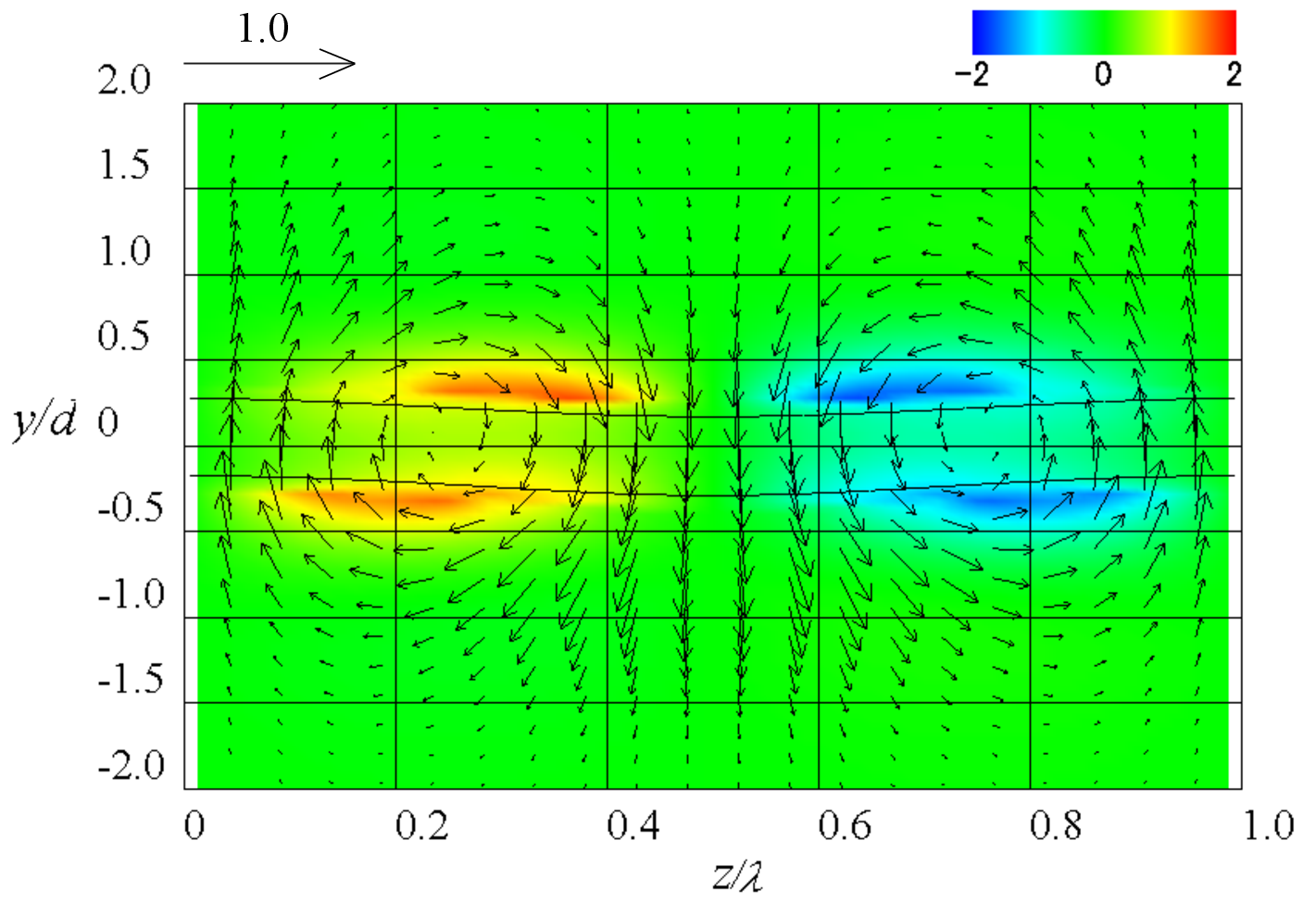} \\
\vspace*{-0.5\baselineskip}
(b)
\end{center}
\end{minipage}
\caption{Ligament interface, velocity vectors, and vorticity contours at $T = 0.3$: 
(a) $y/d = 0$ and (b) $x/d = 0$:
$\Delta U/U_\mathrm{ref} = 11.9$, $ka = 0.7$.}
\label{u20k7t03vcxzyz}
\end{figure}

Figure \ref{u20k7t03vcxzyz} shows the interface, velocity vectors, 
and vorticity distribution in the $x$--$z$ and $y$--$z$ cross-sections 
of $y/d = 0$ and $x/d = 0$, respectively, at $T = 0.3$. 
The vorticity distributions in the $y$- and $x$-directions are shown 
in Figs. \ref{u20k7t03vcxzyz}(a) and (b), respectively. 
In Fig. \ref{u20k7timevcxy}(d), 
the vortex centers in the cross-section of $z/\lambda = 0.5$ and 1.0 exist 
at $y/d > 0$ and $y/d < 0$, respectively. 
Thus, at $z/\lambda = 0.5$ and $z/\lambda = 0$, 1.0 in Fig. \ref{u20k7t03vcxzyz}(a), 
the airflows flow in the positive and negative directions of the $x$-axis at $y/d = 0$. 
Around $(x/d, z/\lambda) = (1.0, 0.2)$, $(-1.0, 0.3)$ 
and $(x/d, z/\lambda) = (1.0, 0.8)$, $(-1.0, 0.7)$, 
counterclockwise and clockwise vortices are formed by the difference 
in the direction of movement of the airflow, respectively. 
Similarly, around $(y/d, z/\lambda) = (0, 0.25)$ and $(y/d, z/\lambda) = (0, 0.75)$ 
in Fig. \ref{u20k7t03vcxzyz}(b), 
clockwise and counterclockwise vortices are formed 
by the deviation of the center positions of the vortices 
at $z/\lambda = 0.5$ and 1.0 in Fig. \ref{u20k7timevcxy}(d). 
Therefore, it is considered that the vortex caused by the airflow 
near the interface of the liquid ligament significantly influences the behaviors 
at the initial stage of the deformation of the liquid ligament.

\begin{figure}[!t]
\centering
\begin{minipage}{0.48\linewidth}
\begin{center}
\includegraphics[trim=0mm 0mm 0mm 0mm, clip, width=75mm]{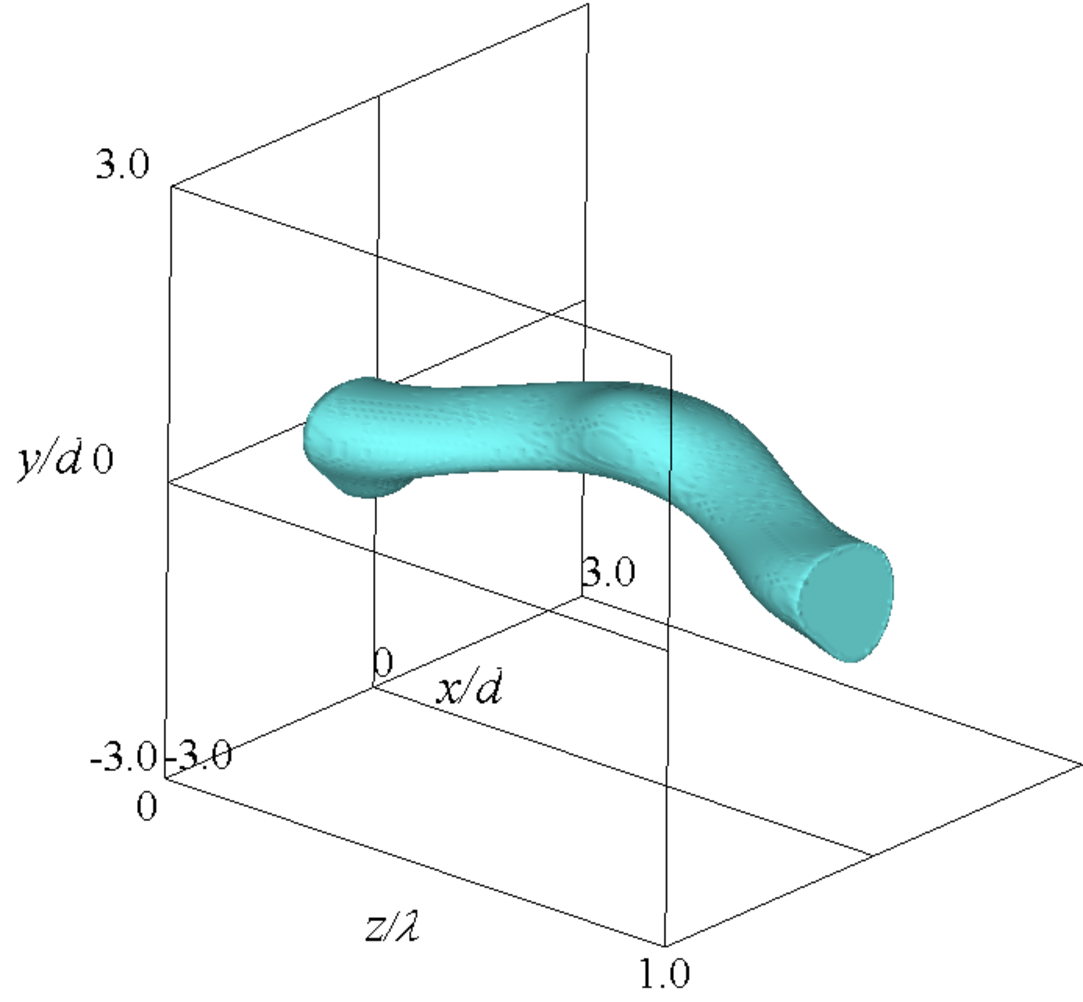} \\
\vspace*{-0.5\baselineskip}
(a)
\end{center}
\end{minipage}
\begin{minipage}{0.48\linewidth}
\begin{center}
\includegraphics[trim=0mm 0mm 0mm 0mm, clip, width=75mm]{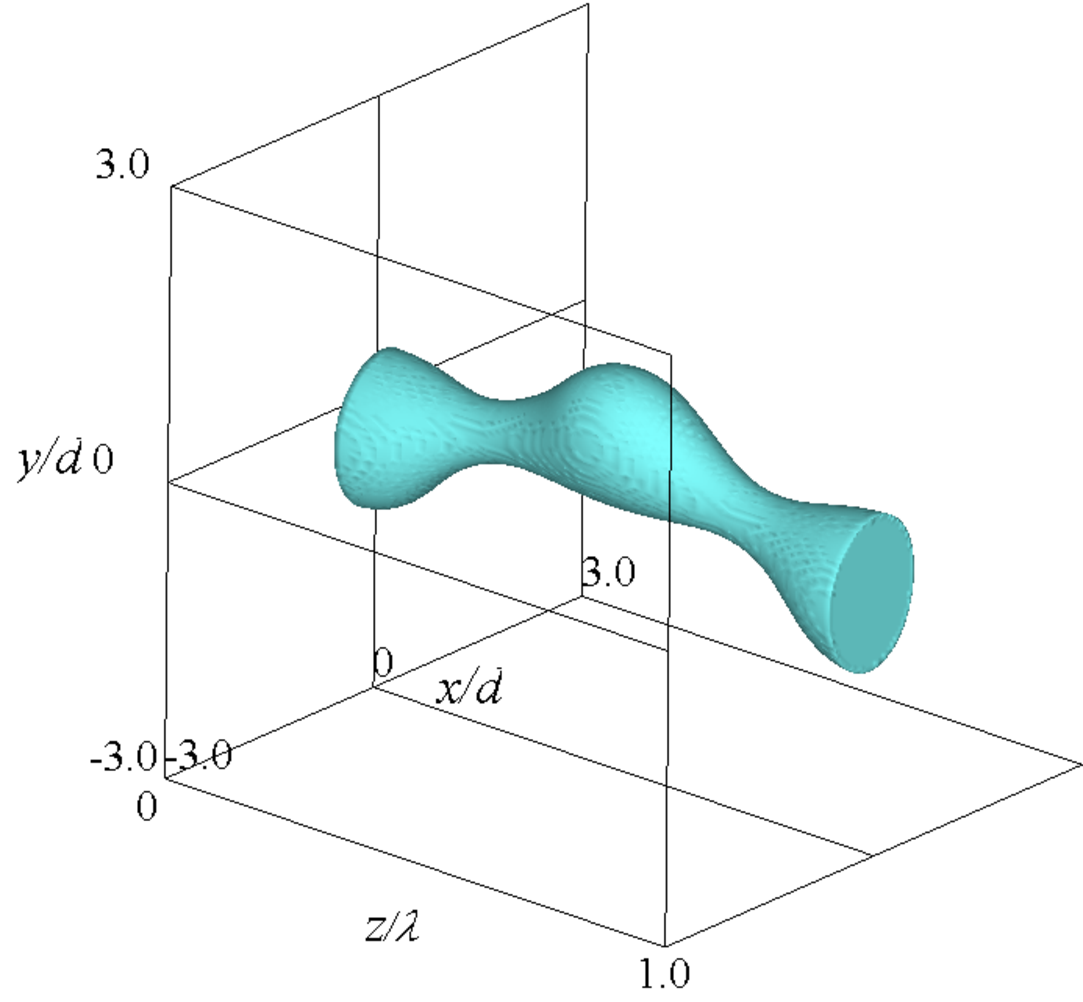} \\
\vspace*{-0.5\baselineskip}
(b)
\end{center}
\end{minipage}

\vspace*{0.5\baselineskip}
\begin{minipage}{0.48\linewidth}
\begin{center}
\includegraphics[trim=0mm 0mm 0mm 0mm, clip, width=75mm]{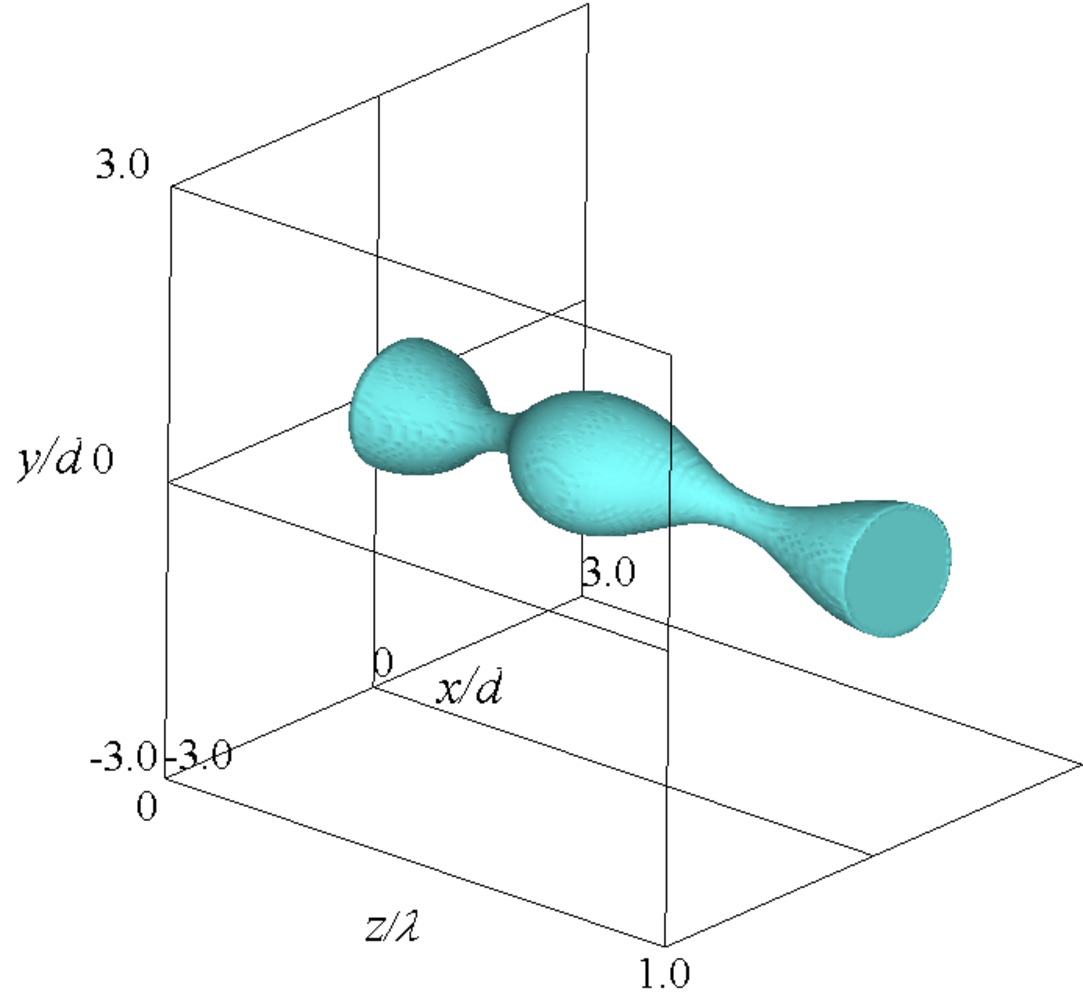} \\
\vspace*{-0.5\baselineskip}
(c)
\end{center}
\end{minipage}
\begin{minipage}{0.48\linewidth}
\begin{center}
\includegraphics[trim=0mm 0mm 0mm 0mm, clip, width=75mm]{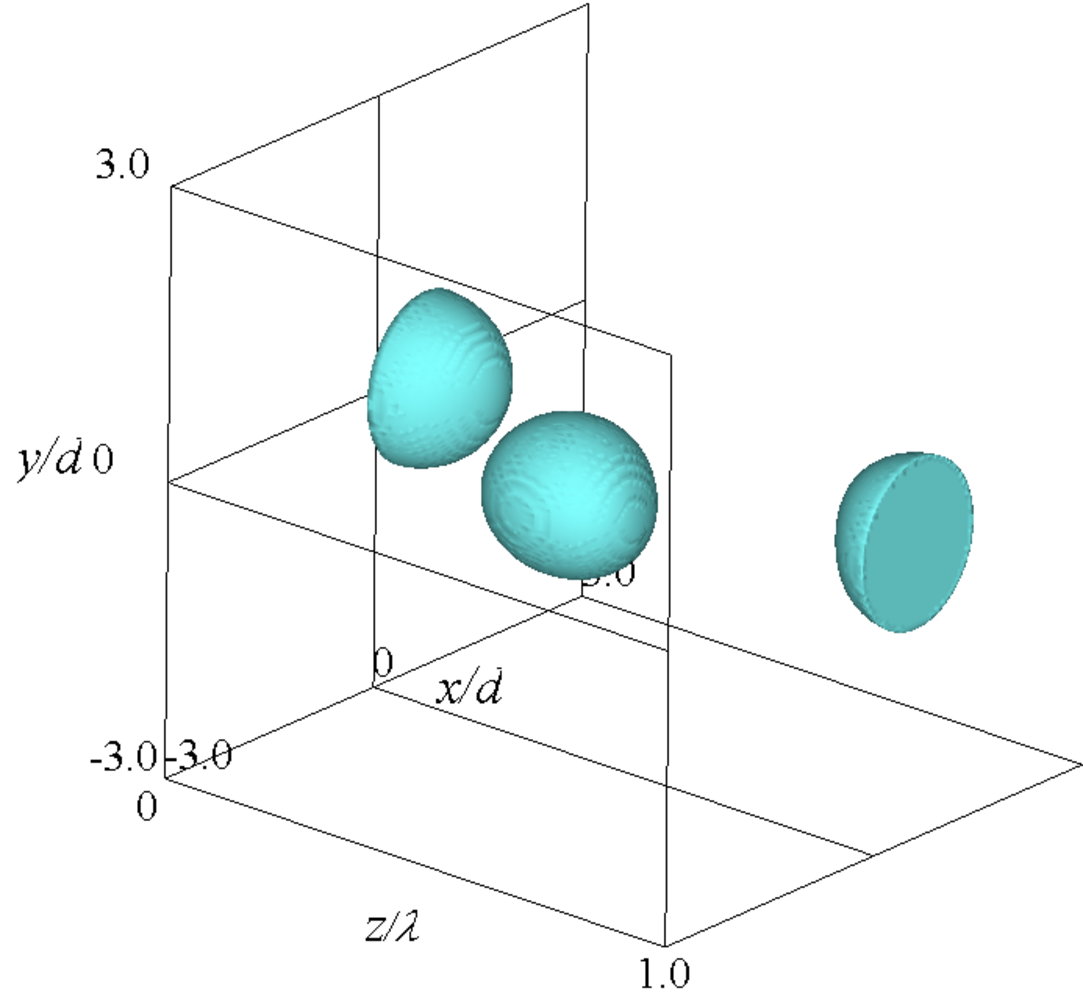} \\
\vspace*{-0.5\baselineskip}
(d)
\end{center}
\end{minipage}
\caption{Time variations of ligament interface: 
(a) $T = 1.0$, (b) $T = 1.5$, (c) $T = 2.0$, and (d) $T = 2.7$: 
$\Delta U/U_\mathrm{ref} = 11.9$, $ka = 0.7$.}
\label{u20_time}
\end{figure}

Figure \ref{u20_time} shows the temporal variation of the interface of the liquid ligament after $T = 0.9$. 
The $z/\lambda = 0.5$ and $z/\lambda = 1.0$ portions of the liquid ligament 
move away from each other along the $x$-axis over time. 
Constrictions occur around $z/\lambda = 0.25$ and 0.75 of the liquid ligament. 
Finally, the ligament splits at these constrictions 
and two droplets are formed at $T = 2.7$.

\begin{figure}[!t]
\centering
\begin{minipage}{0.32\linewidth}
\begin{center}
\includegraphics[trim=0mm 0mm 0mm 0mm, clip, width=55mm]{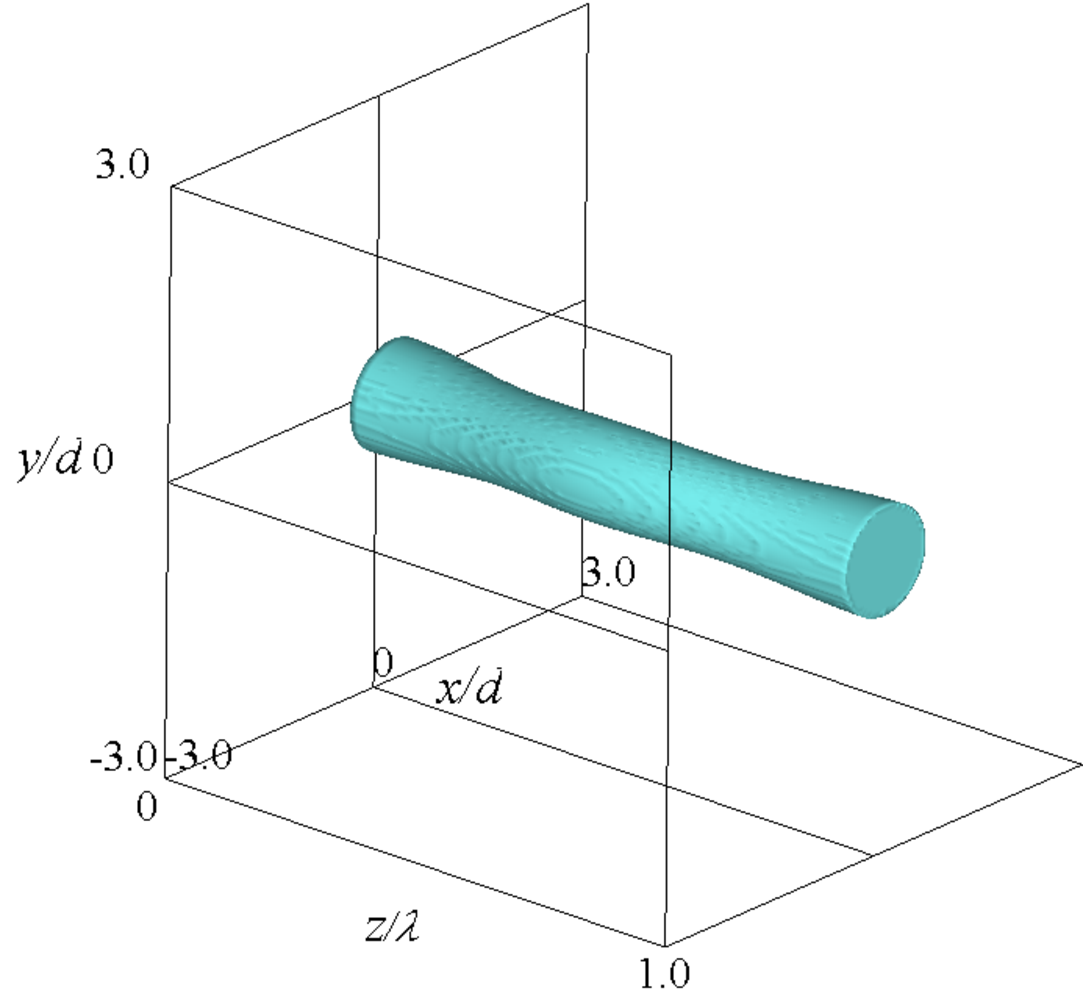} \\
(a)
\end{center}
\end{minipage}
\begin{minipage}{0.32\linewidth}
\begin{center}
\includegraphics[trim=0mm 0mm 0mm 0mm, clip, width=55mm]{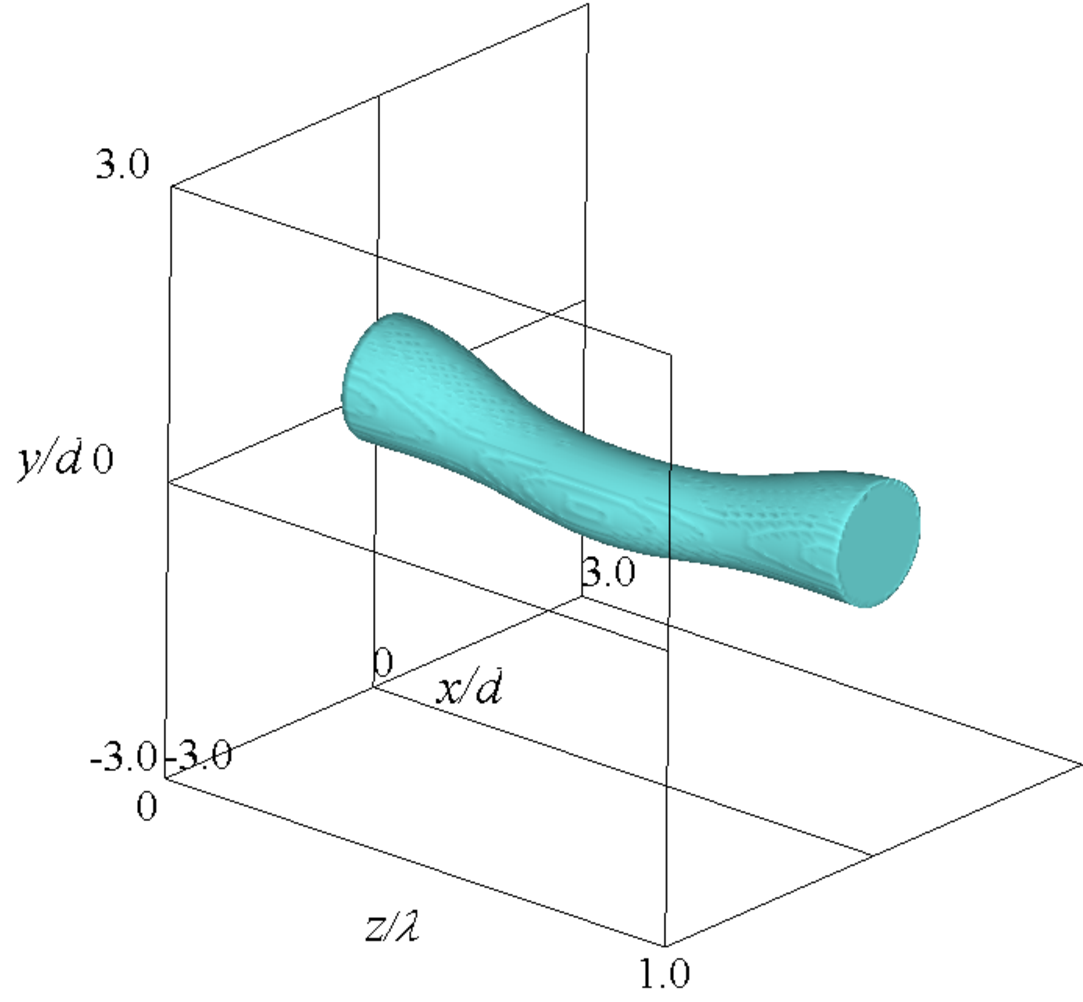} \\
(b)
\end{center}
\end{minipage}
\begin{minipage}{0.32\linewidth}
\begin{center}
\includegraphics[trim=0mm 0mm 0mm 0mm, clip, width=55mm]{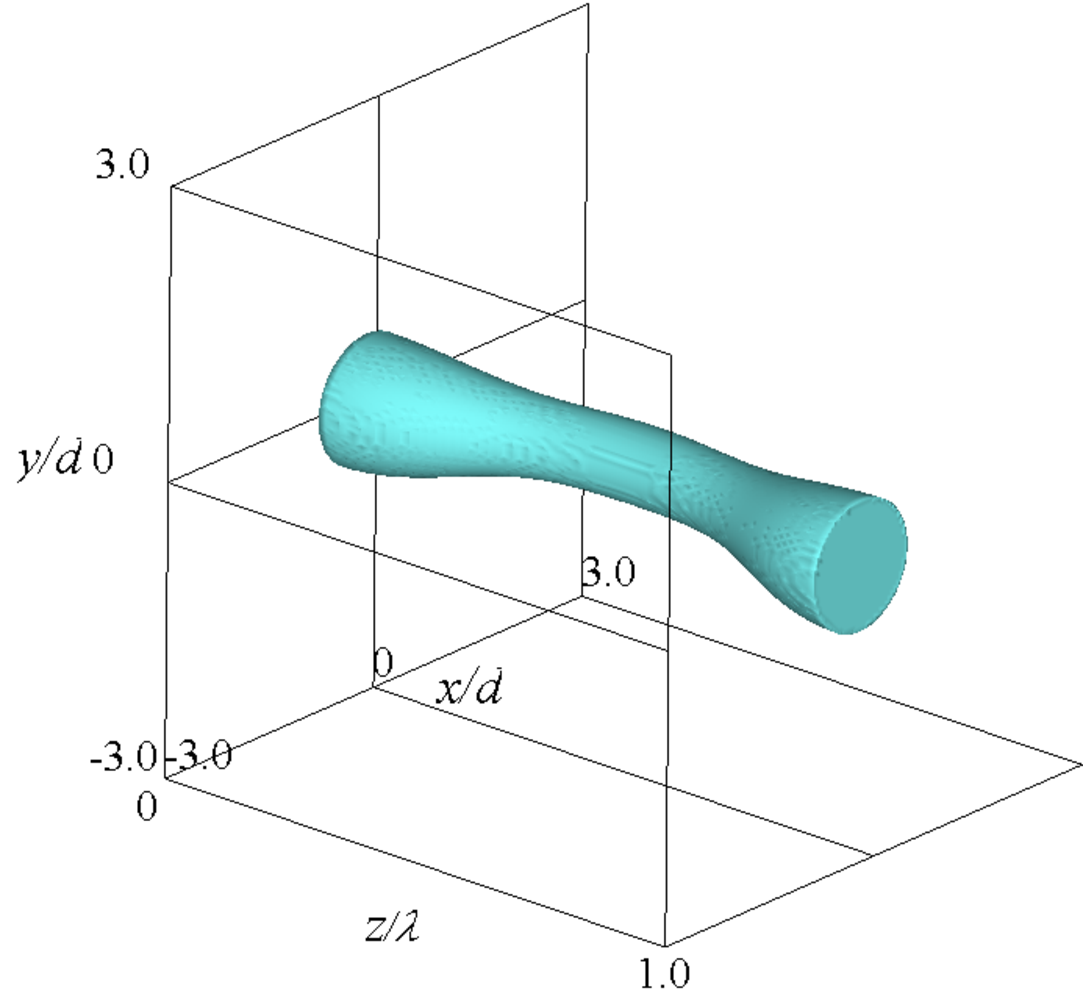} \\
(c)
\end{center}
\end{minipage}

\vspace*{0.5\baselineskip}
\begin{minipage}{0.32\linewidth}
\begin{center}
\includegraphics[trim=0mm 0mm 0mm 0mm, clip, width=55mm]{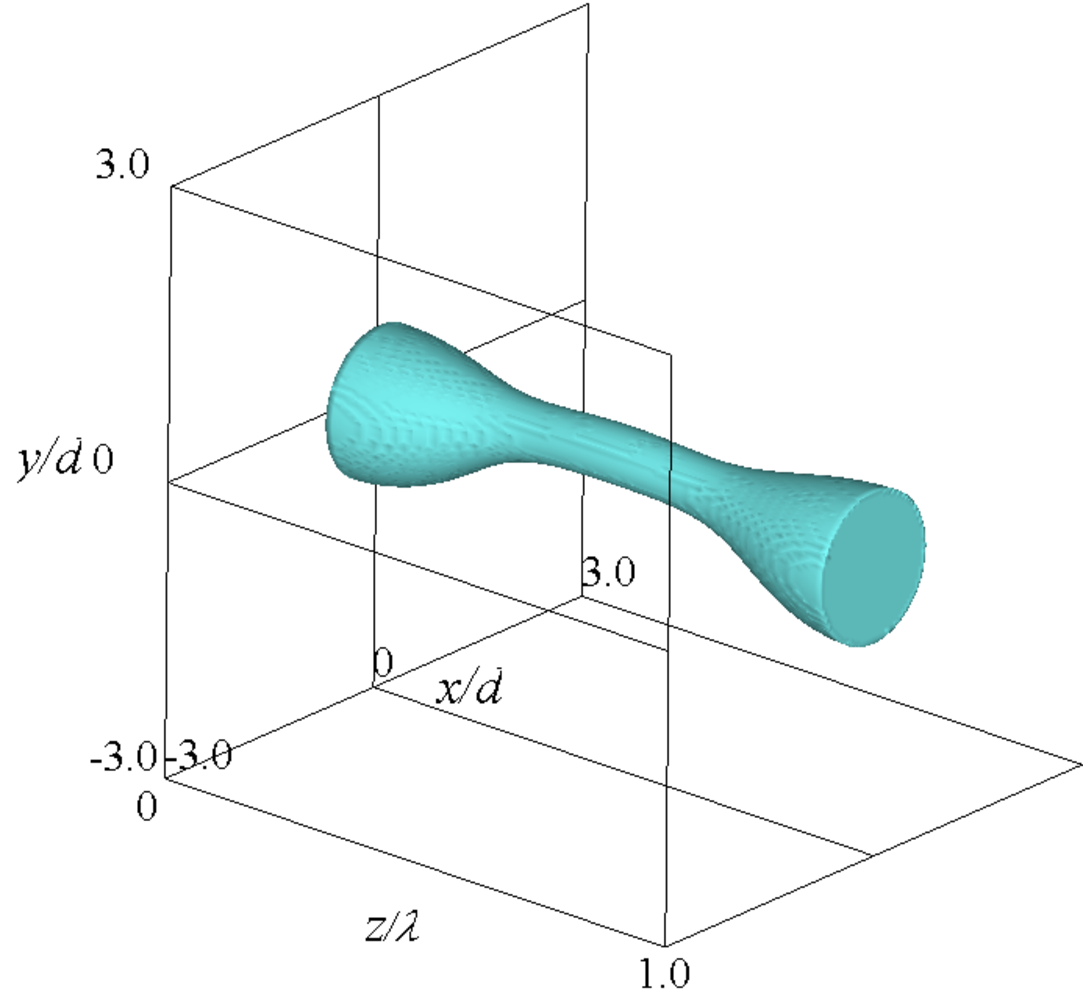} \\
(d)
\end{center}
\end{minipage}
\begin{minipage}{0.32\linewidth}
\begin{center}
\includegraphics[trim=0mm 0mm 0mm 0mm, clip, width=55mm]{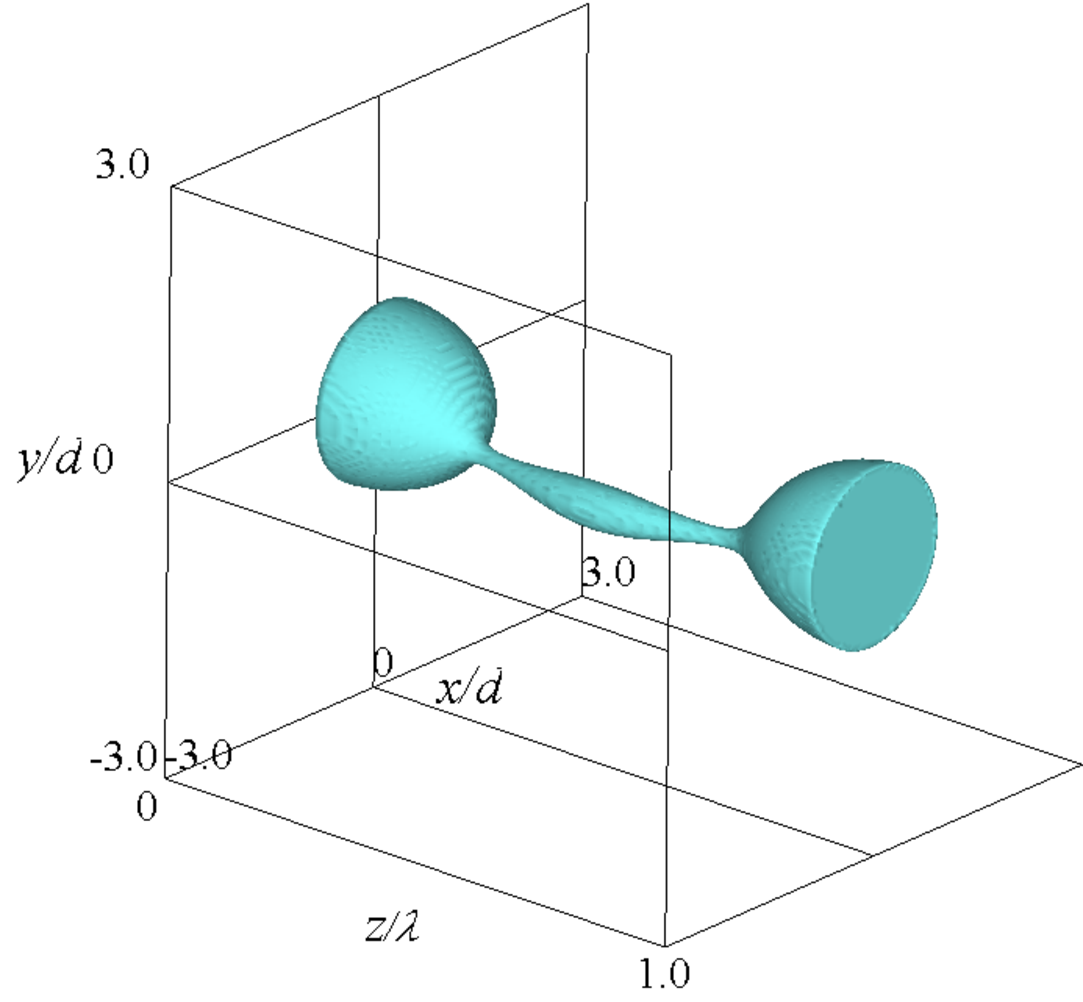} \\
(e)
\end{center}
\end{minipage}
\hspace{0.5\baselineskip}
\begin{minipage}{0.32\linewidth}
\begin{center}
\includegraphics[trim=0mm 0mm 0mm 0mm, clip, width=55mm]{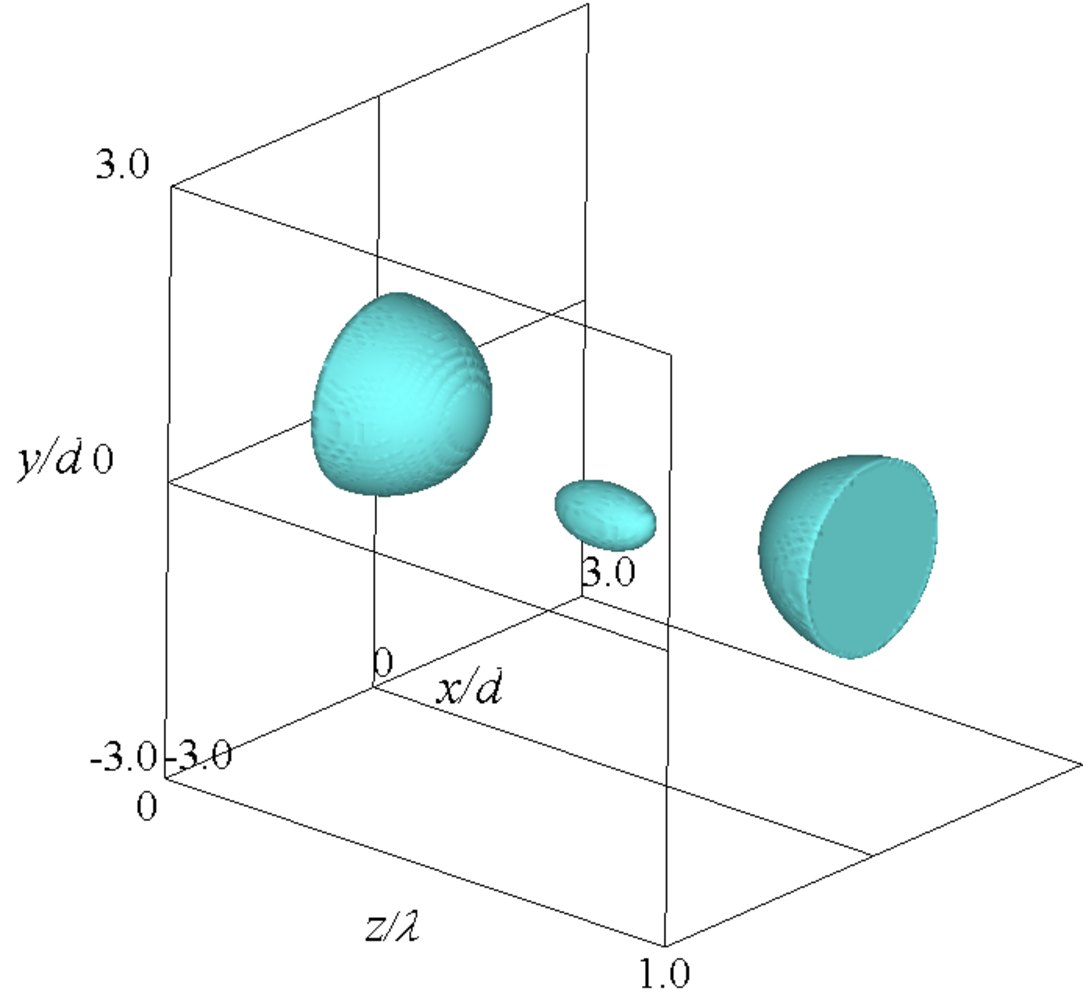} \\
(f)
\end{center}
\end{minipage}
\caption{Time variations of ligament interface: 
(a) $T = 2.0$, (b) $T = 3.0$, (c) $T = 4.0$, (d) $T = 5.0$, (e) $T = 6.0$, and (f) $T = 6.3$: 
$\Delta U/U_\mathrm{ref} = 5.9$, $ka = 0.7$.}
\label{u10_time}
\end{figure}

\begin{figure}[!t]
\centering
\begin{minipage}{0.32\linewidth}
\begin{center}
\includegraphics[trim=0mm 0mm 0mm 0mm, clip, width=55mm]{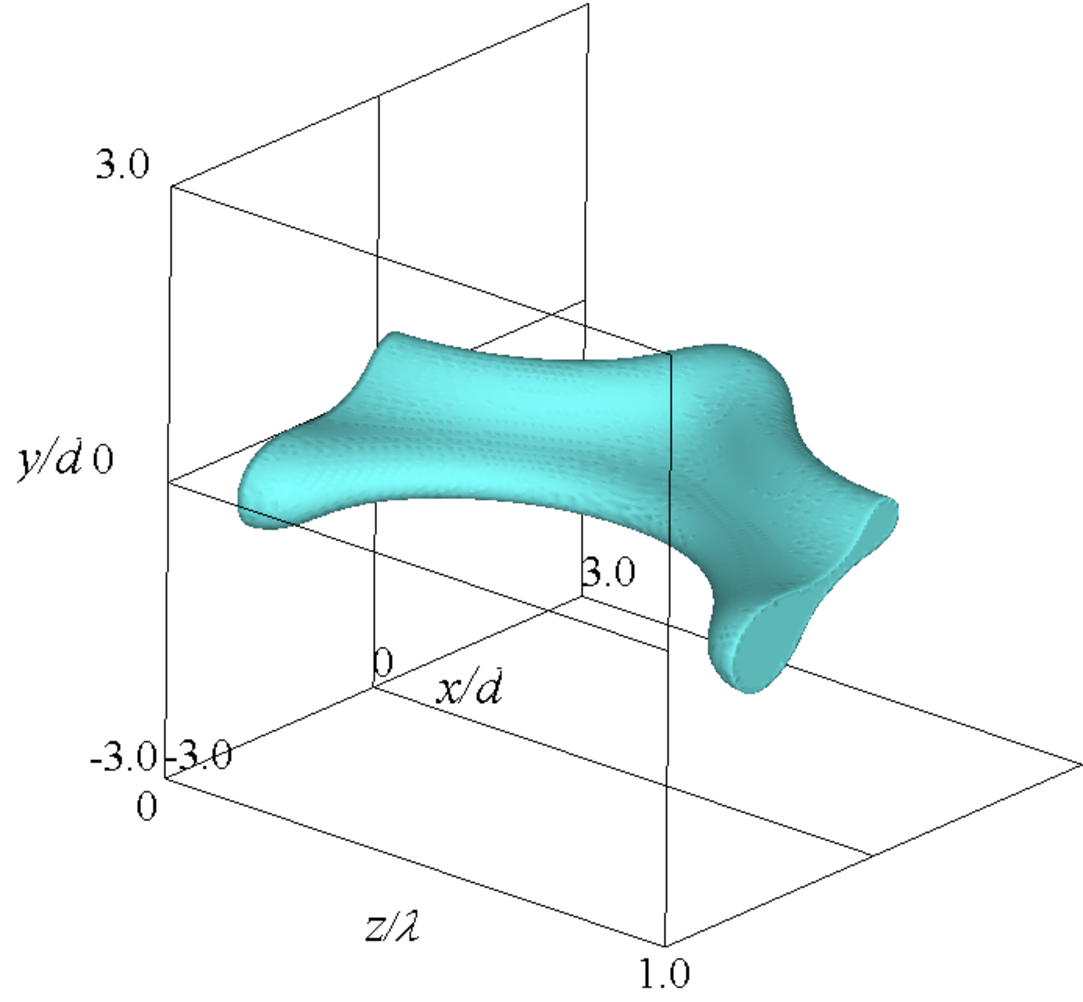} \\
(a)
\end{center}
\end{minipage}
\begin{minipage}{0.32\linewidth}
\begin{center}
\includegraphics[trim=0mm 0mm 0mm 0mm, clip, width=55mm]{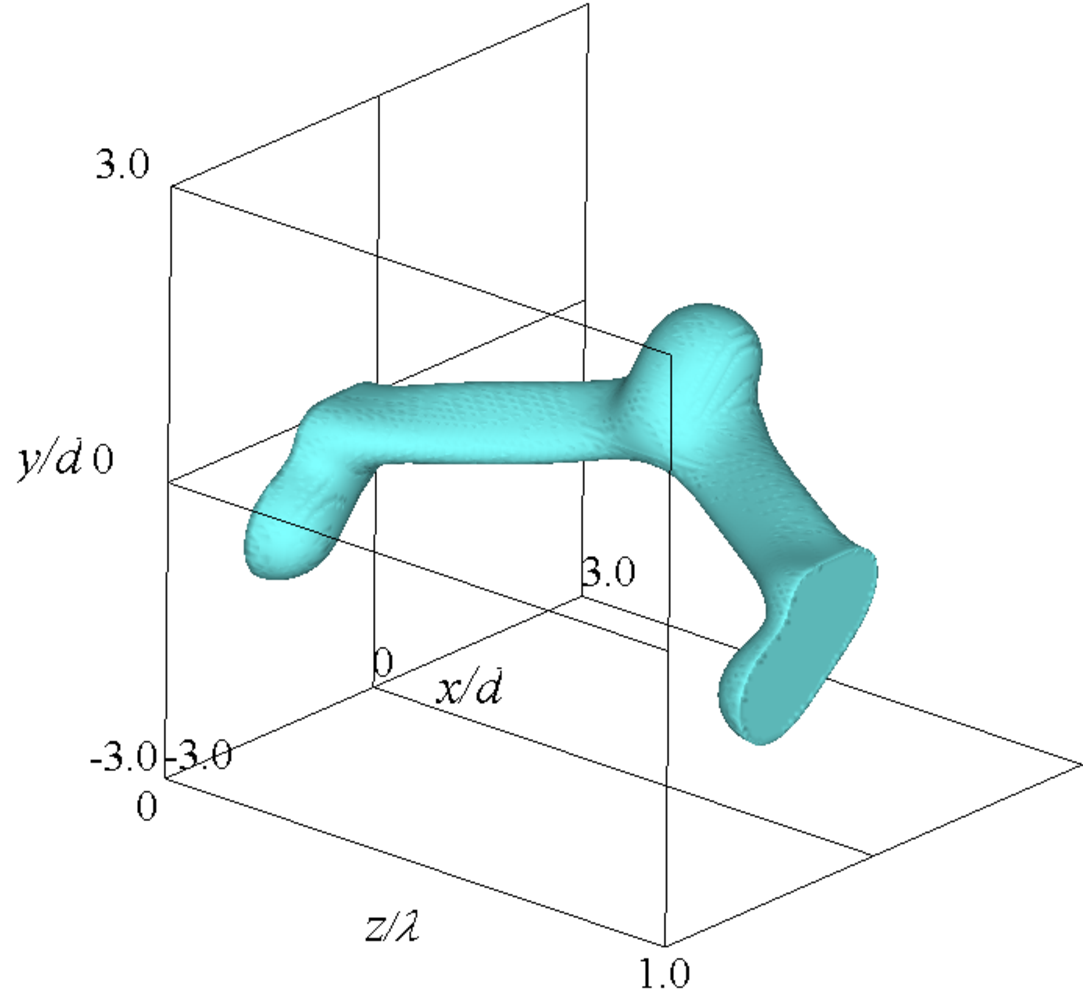} \\
(b)
\end{center}
\end{minipage}
\begin{minipage}{0.32\linewidth}
\begin{center}
\includegraphics[trim=0mm 0mm 0mm 0mm, clip, width=55mm]{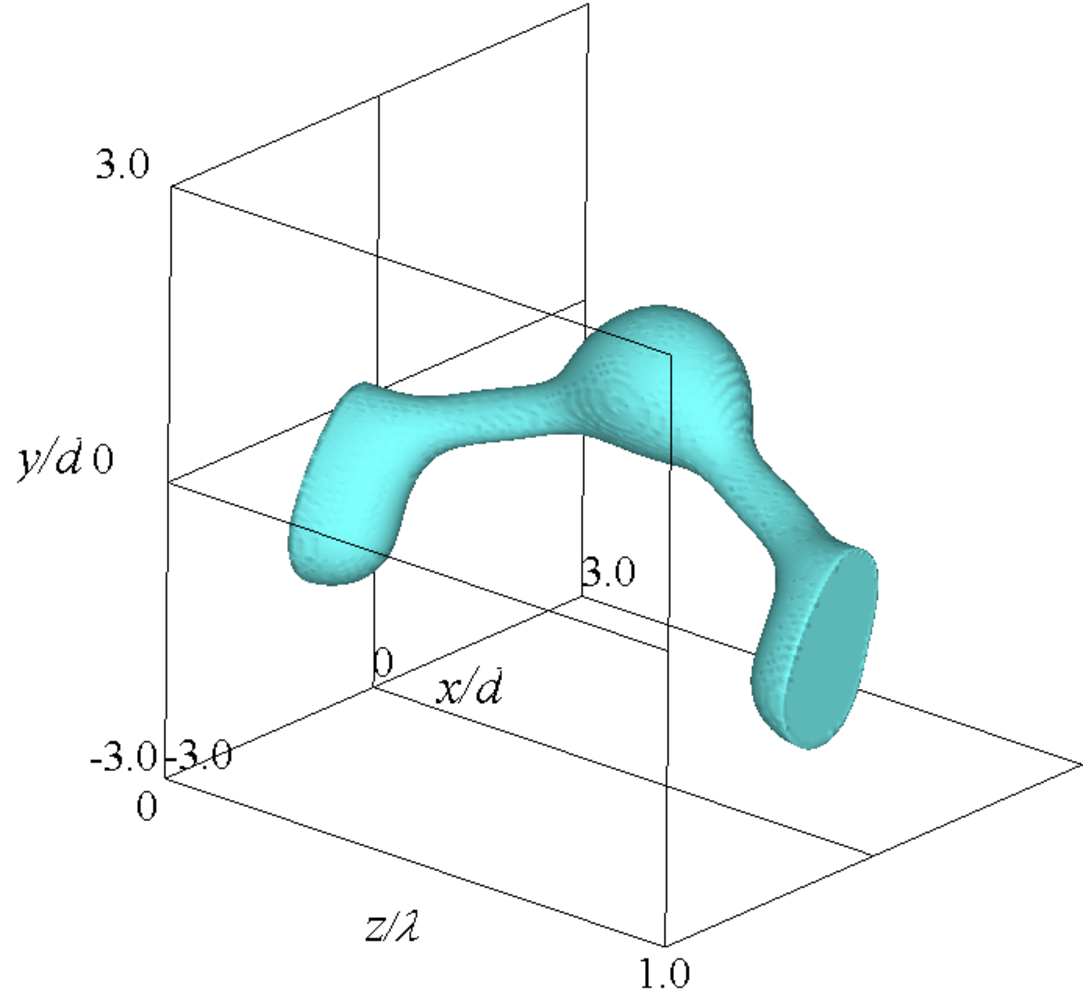} \\
(c)
\end{center}
\end{minipage}

\vspace*{0.5\baselineskip}
\begin{minipage}{0.32\linewidth}
\begin{center}
\includegraphics[trim=0mm 0mm 0mm 0mm, clip, width=55mm]{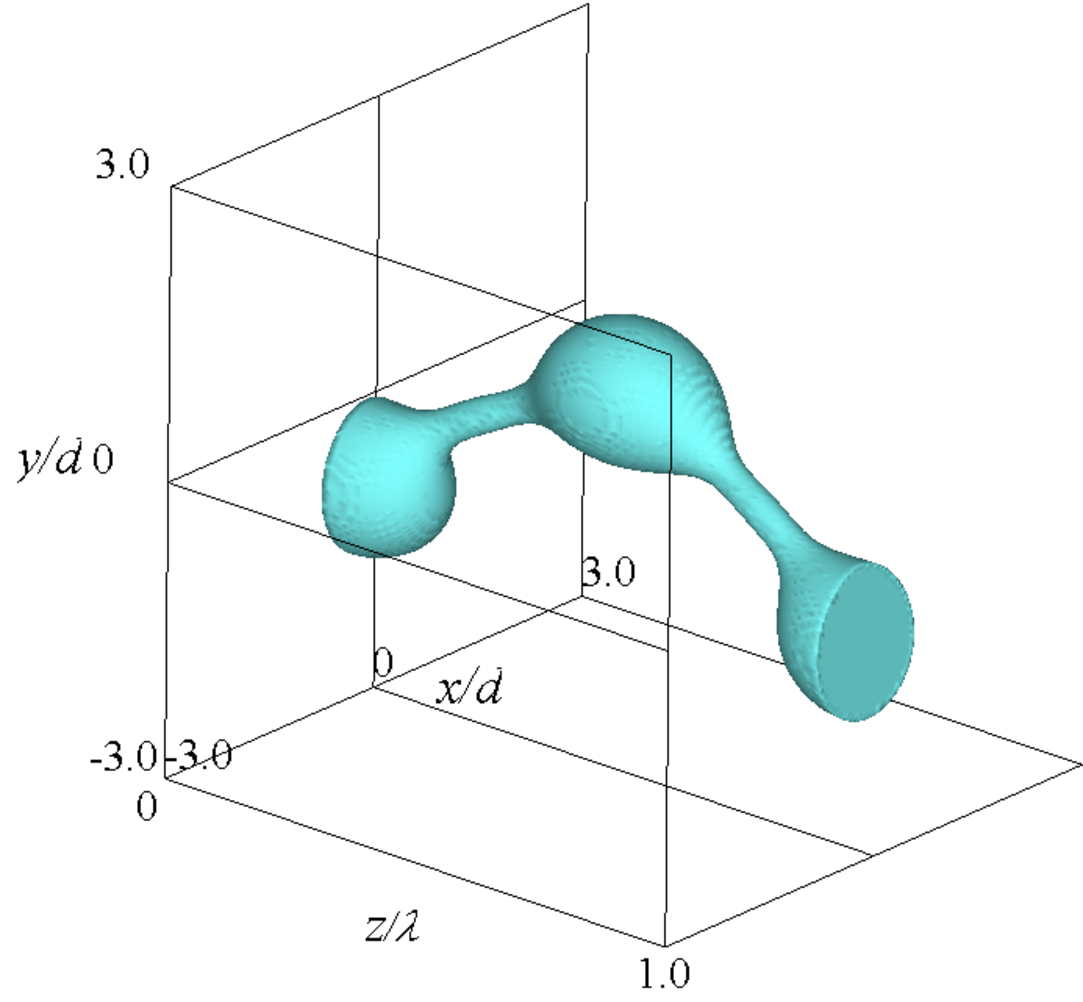} \\
(d)
\end{center}
\end{minipage}
\begin{minipage}{0.32\linewidth}
\begin{center}
\includegraphics[trim=0mm 0mm 0mm 0mm, clip, width=55mm]{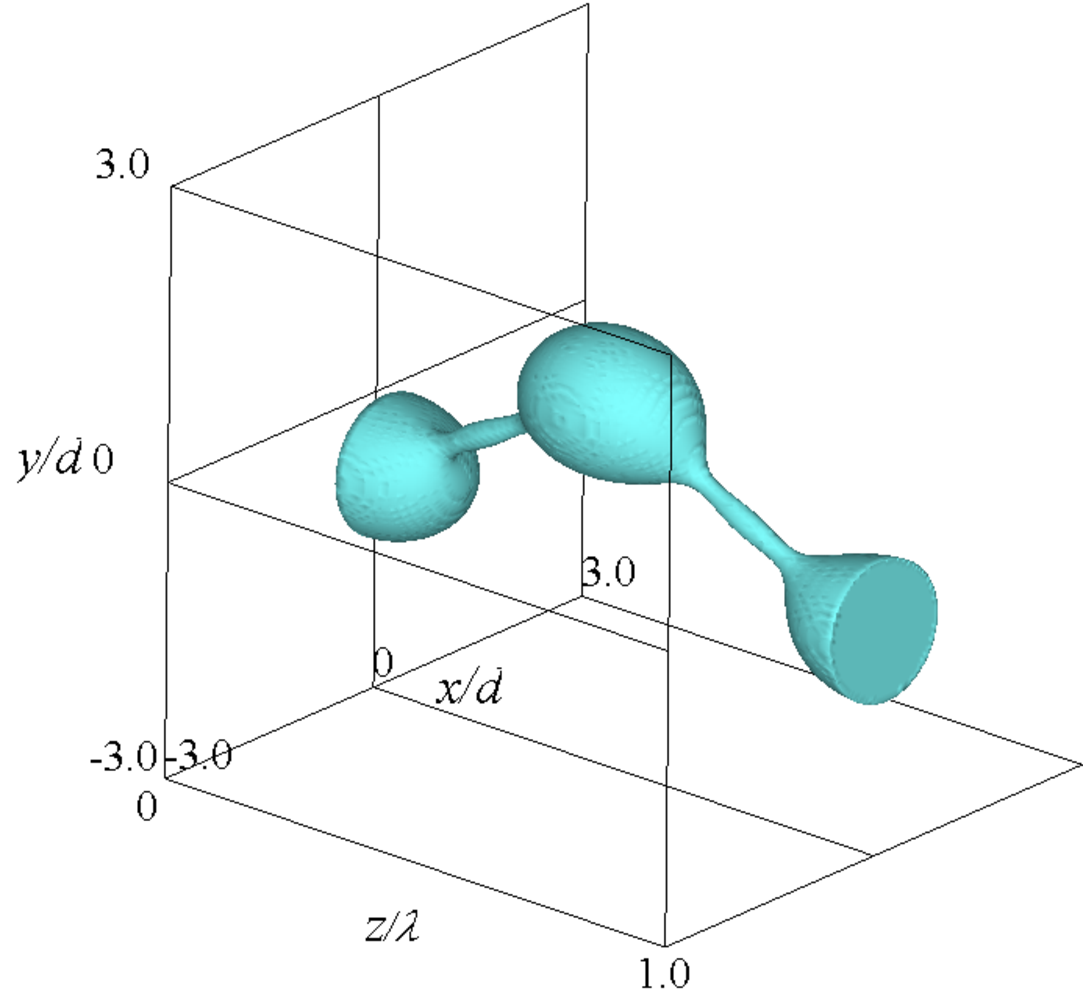} \\
(e)
\end{center}
\end{minipage}
\begin{minipage}{0.32\linewidth}
\begin{center}
\includegraphics[trim=0mm 0mm 0mm 0mm, clip, width=55mm]{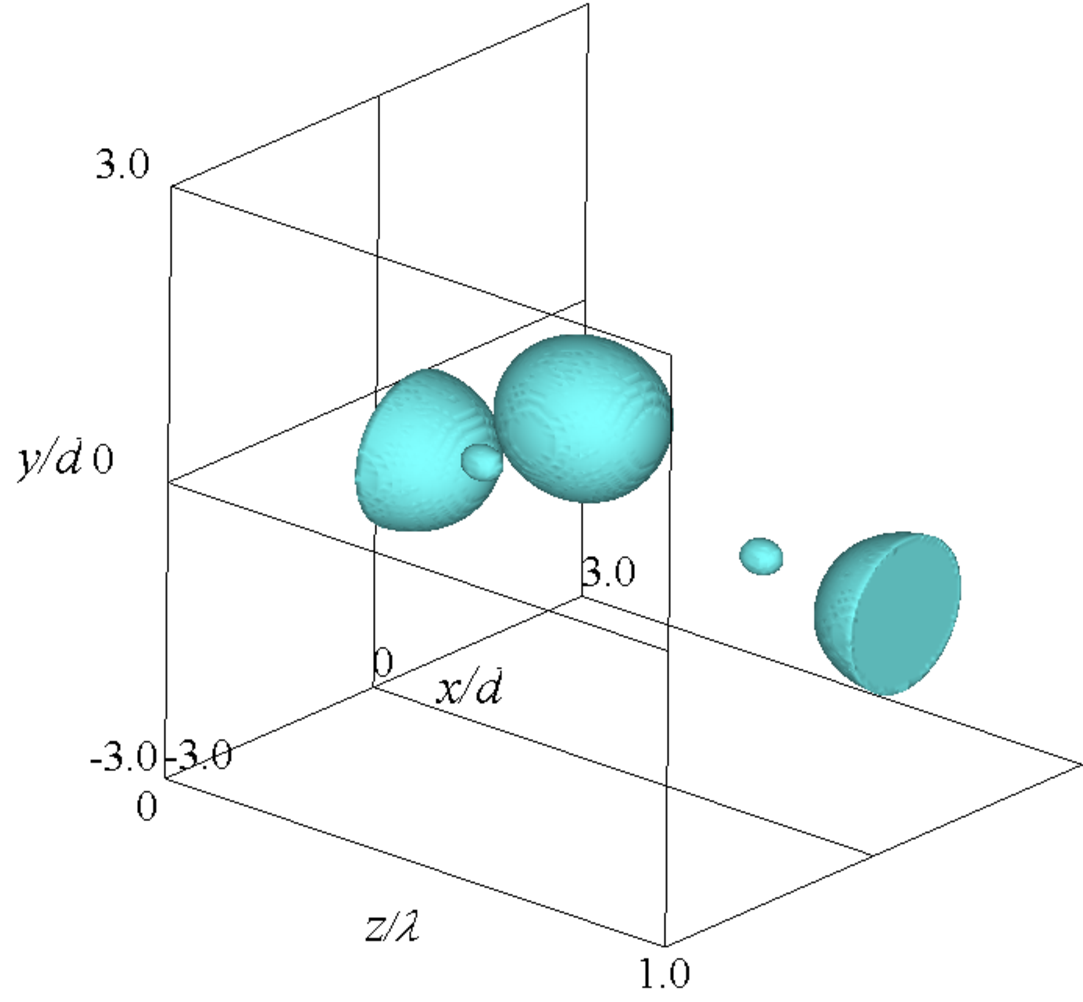} \\
(f)
\end{center}
\end{minipage}
\caption{Time variations of ligament interface: 
(a) $T = 0.6$, (b) $T = 1.0$, (c) $T = 1.5$, (d) $T = 2.0$, (e) $T = 2.3$, and (f) $T = 2.7$: 
$\Delta U/U_\mathrm{ref} = 17.8$, $ka = 0.7$.}
\label{u30k7_time}
\end{figure}

Figures \ref{u10_time} and \ref{u30k7_time} show the time evolution of the liquid ligament interface 
for $\Delta U/U_\mathrm{ref} = 5.9$ and 17.8, respectively. 
The result for $\Delta U/U_\mathrm{ref} = 11.9$ was shown in Fig. \ref{u20_time}. 
For $\Delta U/U_\mathrm{ref} = 5.9$, 
ligament breakup is confirmed at $T = 6.3$, 
and two droplets are formed. 
For $\Delta U/U_\mathrm{ref} = 11.9$, 
the ligament splits into two droplets at $T = 2.7$, 
which is earlier than the breakup time for $\Delta U/U_\mathrm{ref} = 5.9$. 
It can be seen that an increase in $\Delta U/U_\mathrm{ref}$ promotes 
the deformation and splitting of the liquid ligament. 
For $\Delta U/U_\mathrm{ref} = $ 17.8, 
in addition to two large droplets, small droplets are formed 
near $z/\lambda = $ 0.25 and 0.75. 
Comparing the ligaments for $\Delta U/U_\mathrm{ref} = 11.9$ 
and $\Delta U/U_\mathrm{ref} = 17.8$, 
the ligaments break up at almost the same time, 
and the number of droplets increases with an increase of $\Delta U/U_\mathrm{ref}$.

\begin{figure}[!t]
\centering
\begin{minipage}{0.48\linewidth}
\begin{center}
\includegraphics[trim=0mm 0mm 0mm 0mm, clip, width=70mm]{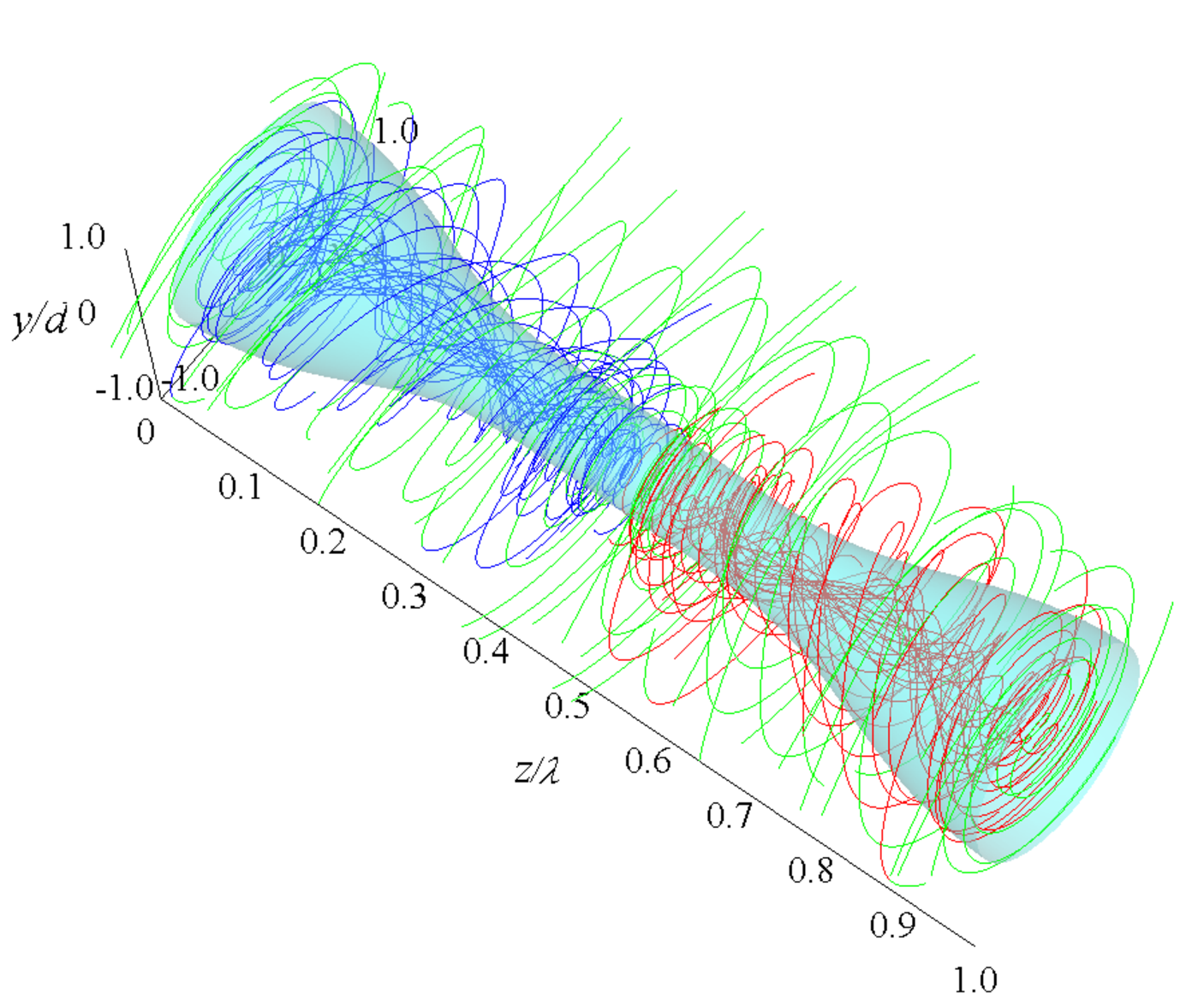} \\
\end{center}
\end{minipage}
\begin{minipage}{0.48\linewidth}
\begin{center}
\includegraphics[trim=0mm 0mm 0mm 0mm, clip, width=78mm]{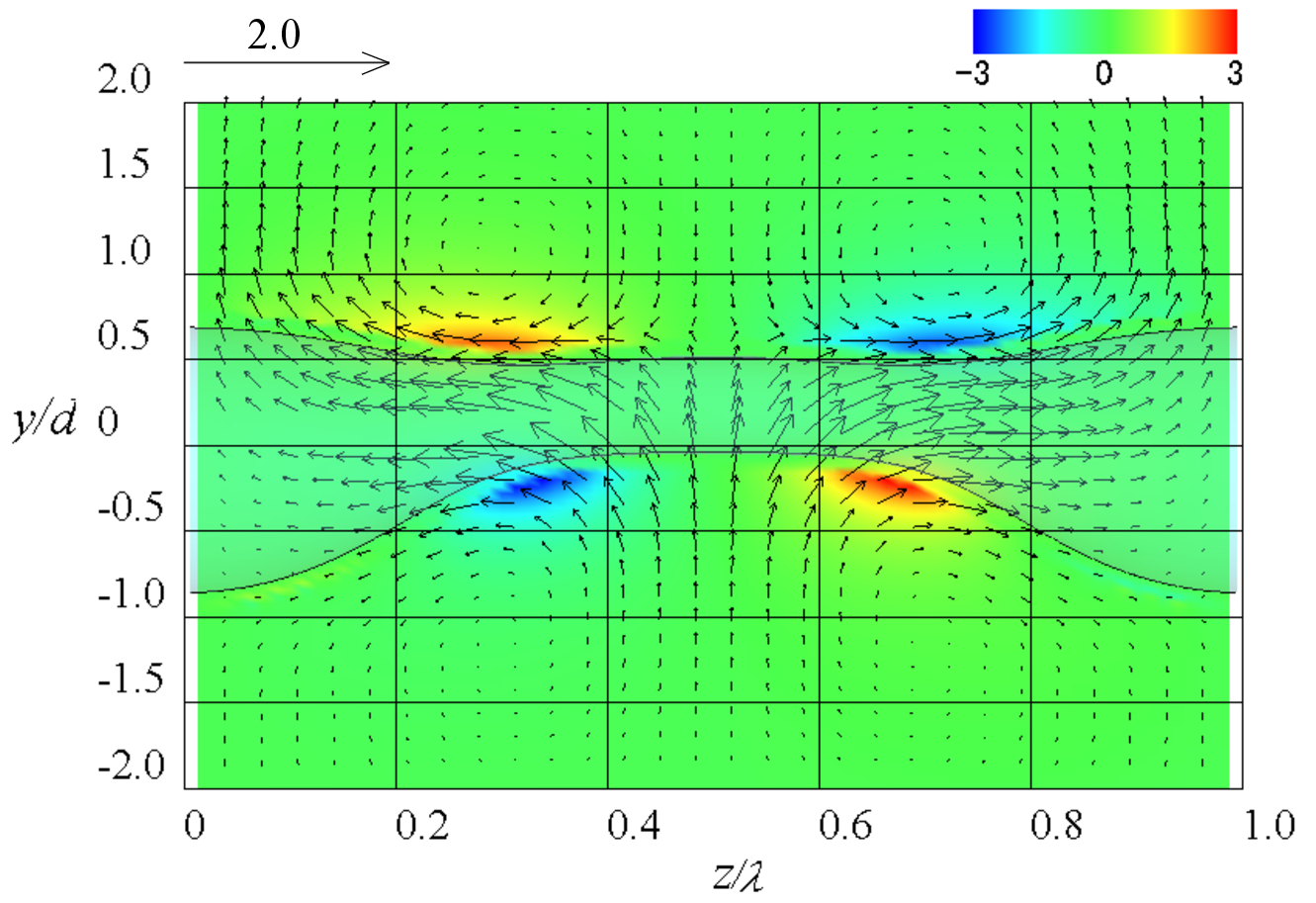} \\
\end{center}
\end{minipage}

(a)

\begin{minipage}{0.48\linewidth}
\begin{center}
\includegraphics[trim=0mm 0mm 0mm 0mm, clip, width=70mm]{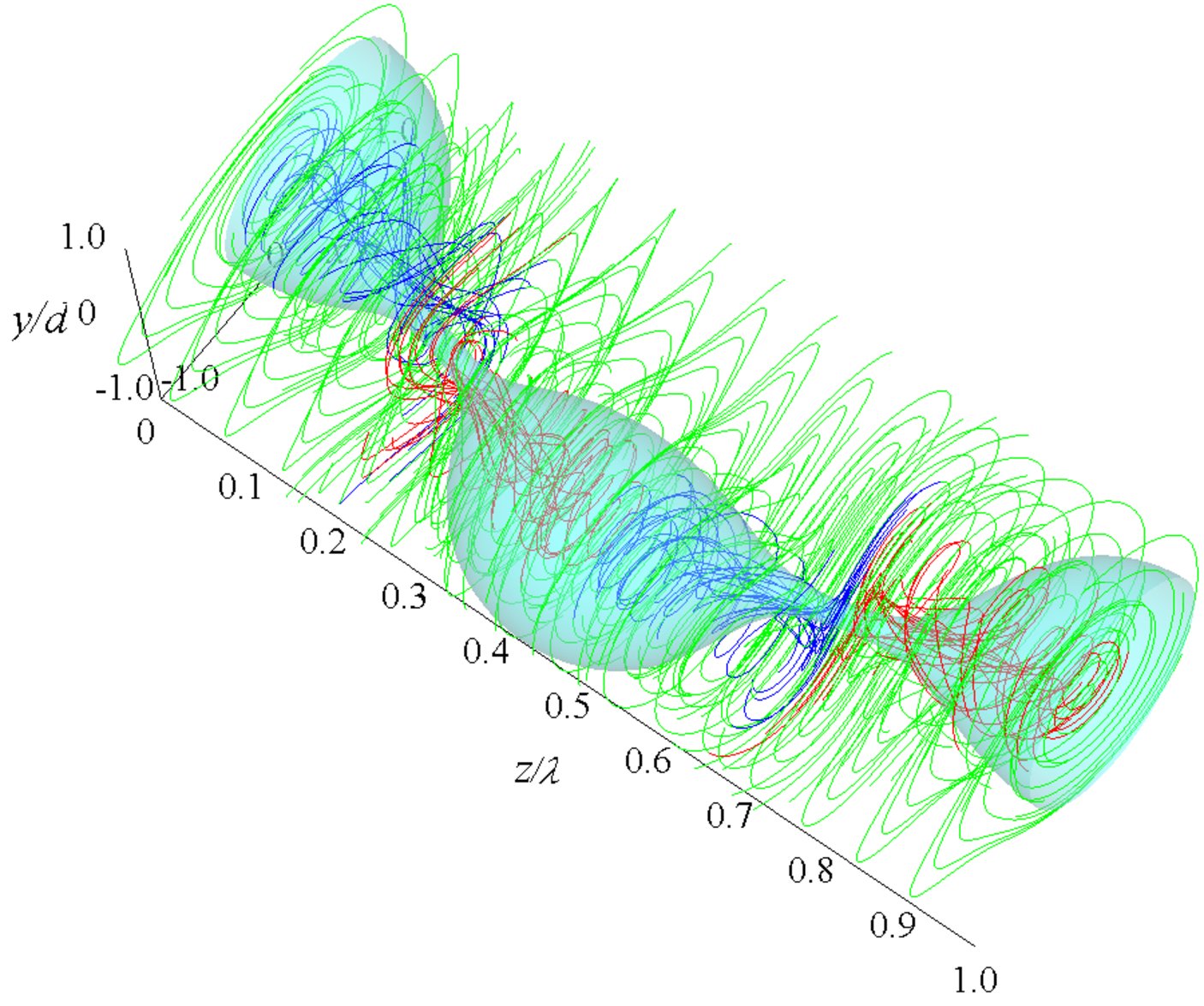} \\
\end{center}
\end{minipage}
\begin{minipage}{0.48\linewidth}
\begin{center}
\includegraphics[trim=0mm 0mm 0mm 0mm, clip, width=78mm]{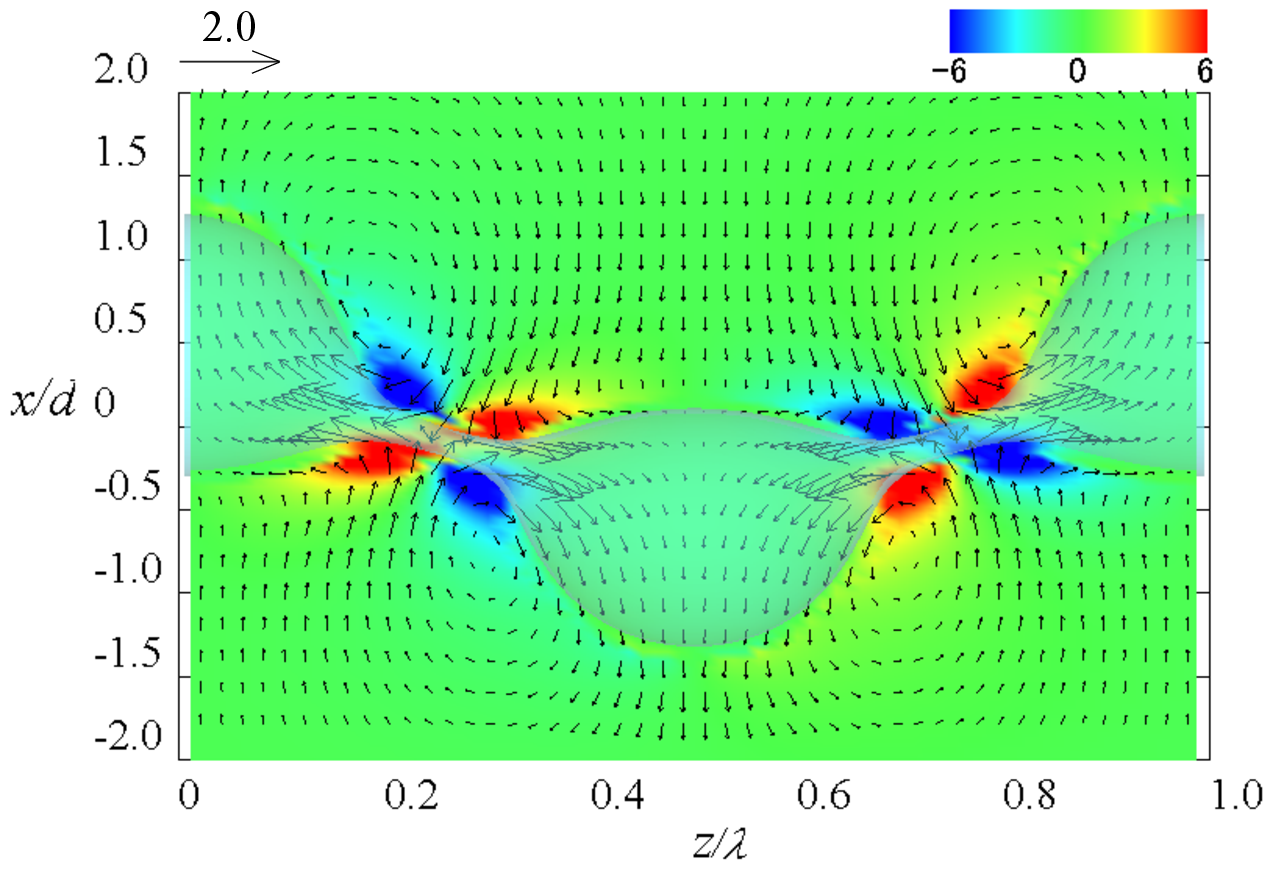} \\
\end{center}
\end{minipage}

(b)

\begin{minipage}{0.48\linewidth}
\begin{center}
\includegraphics[trim=0mm 0mm 0mm 0mm, clip, width=70mm]{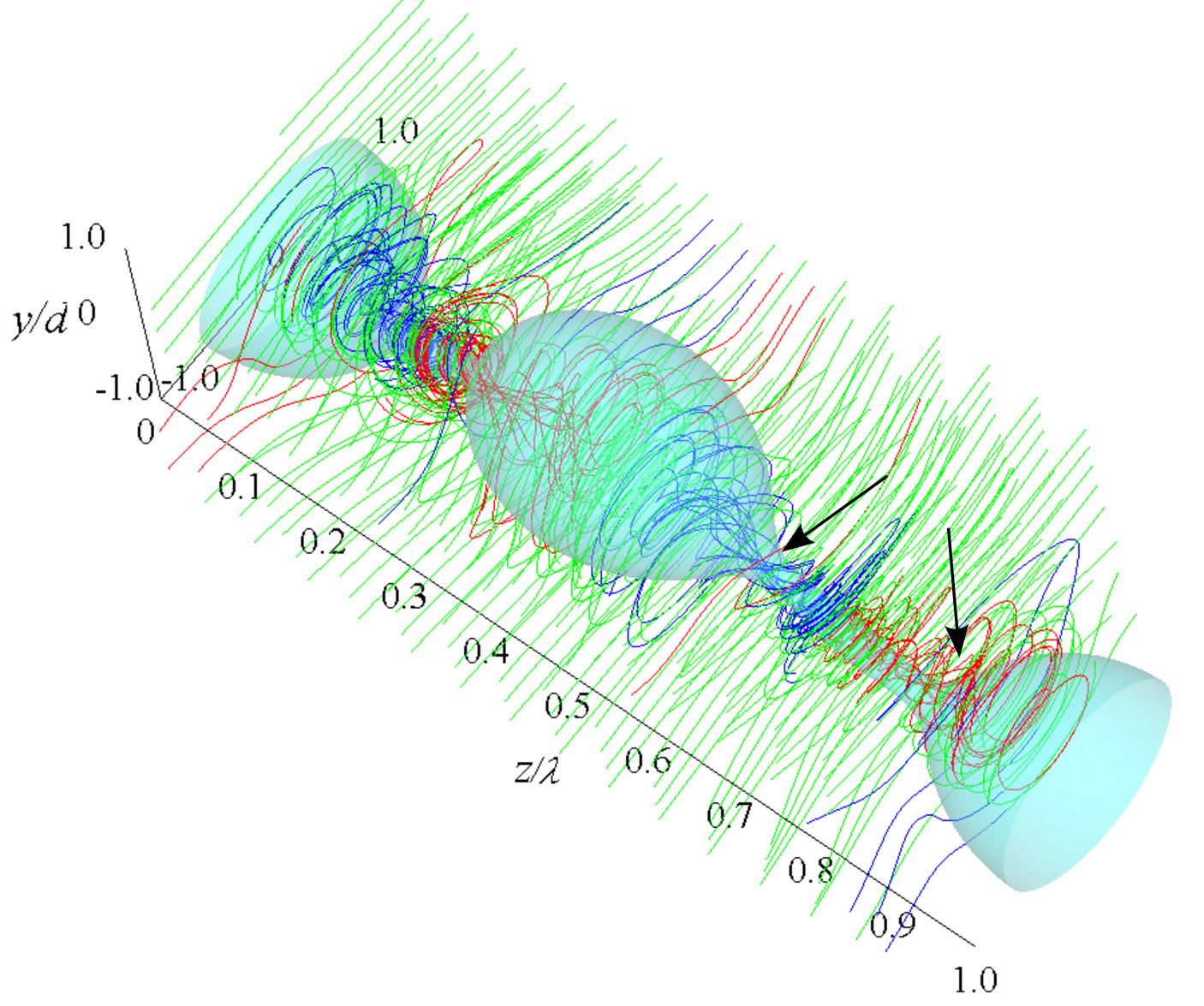} \\
\end{center}
\end{minipage}
\begin{minipage}{0.48\linewidth}
\begin{center}
\includegraphics[trim=0mm 0mm 0mm 0mm, clip, width=78mm]{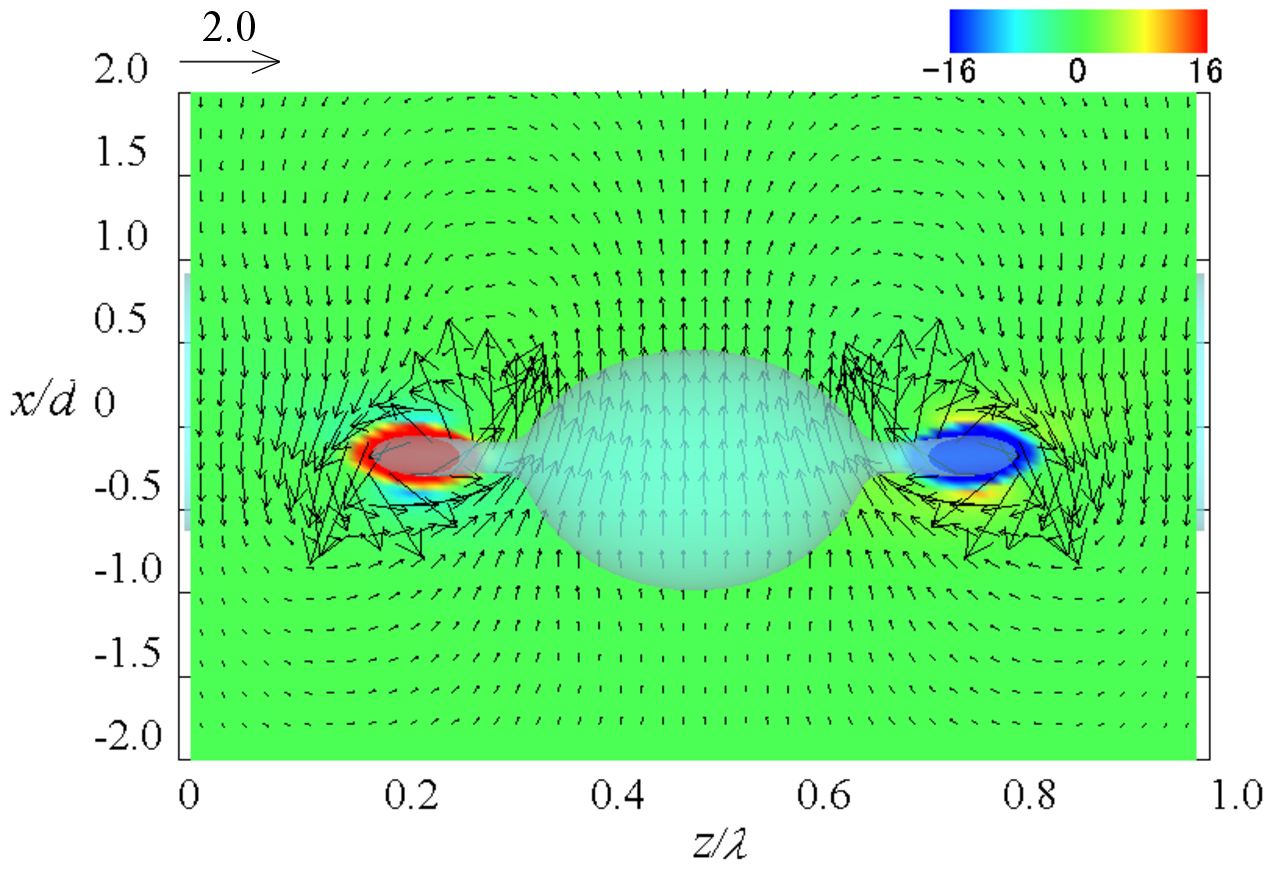} \\
\end{center}
\end{minipage}

(c)

\caption{Ligament interface, streamlines, velocity vectors, and vorticity contours for $ka = 0.7$: 
(a) $\Delta U/U_\mathrm{ref} = 5.9$, $T = 5.0$, 
(b) $\Delta U/U_\mathrm{ref} = 11.9$, $T = 2.3$, and 
(c) $\Delta U/U_\mathrm{ref} = 17.8$, $T = 2.3$.}
\label{u102030vcvor}
\end{figure}

We investigate the changes in the flow field and liquid movement 
inside the liquid ligament owing to the velocity difference 
at the time just before the liquid ligament breaks up. 
Figure \ref{u102030vcvor} shows the interface, velocity vectors, 
vorticity contours, and streamlines at each $\Delta U/U_\mathrm{ref}$. 
Figure \ref{u102030vcvor}(a) is the result at $T = 5.0$, 
and Figs. \ref{u102030vcvor}(b) and (c) are the results at $T = 2.3$. 
In the left figure, 
the red solid line shows the streamline with a clockwise swirling component 
to the flow direction, and the blue solid line shows the reverse rotation. 
The velocity vector and $x$-direction vorticity distribution 
in Fig. \ref{u102030vcvor}(a) are the results of the $y$--$z$ cross-section at $x/d = 0$. 
The velocity vector and $y$-direction vorticity distribution 
in Figs. \ref{u102030vcvor}(b) and (c) are the results of the $x$--$z$ cross-section at $y/d = 0$. 
From the streamlines in Fig. \ref{u102030vcvor}(a), 
it can be seen that the liquid around $z/\lambda = 0.5$ flows spirally 
toward $z/\lambda = 0$ and 1.0. 
Because of such a three-dimensional flow, 
counter-rotating vortices are formed symmetrically to $y/d = 0$ 
around $z/\lambda = 0.3$ and 0.7 in the airflow. 
As these vortices induce the liquid movement along the central axis of the liquid ligament, 
the cross-sectional area of the liquid ligament decreases at $z/\lambda = 0.5$ 
and increases at $z/\lambda = 0$. 
Therefore, in Fig. \ref{u10_time}, 
the droplet at $z/\lambda = 0.5$ is smaller than that at $z/\lambda = 0$.

In Fig. \ref{u102030vcvor}(b), 
the liquid existing near $z/\lambda = 0.25$ and $z/\lambda = 0.75$ flows 
toward $z/\lambda = 0$, 0.5 and $z/\lambda = 0.5$, 1.0, respectively. 
Additionally, vortices near the interface develop. 
These vortices induce liquid movement along the central axis of the liquid ligament.

In Fig. \ref{u102030vcvor}(c), 
streamlines become dense near $z/\lambda = 0.25$ and 0.75. 
Near the large bulge indicated by the arrow, 
the liquid flows to the bulge while rotating. 
At that time, around $z/\lambda = 0.25$ and 0.75, 
a vortex centered inside the liquid ligament is formed by the airflow, 
and the magnitude of the vorticity increases near the interface. 
Compared with the results for $\Delta U/U_\mathrm{ref} = 5.9$ and 11.9, 
the largest magnitude of the vorticity occurs for $\Delta U/U_\mathrm{ref} = 17.8$.

\begin{figure}[!t]
\centering
\includegraphics[trim=0mm 0mm 0mm 0mm, clip, width=80mm]{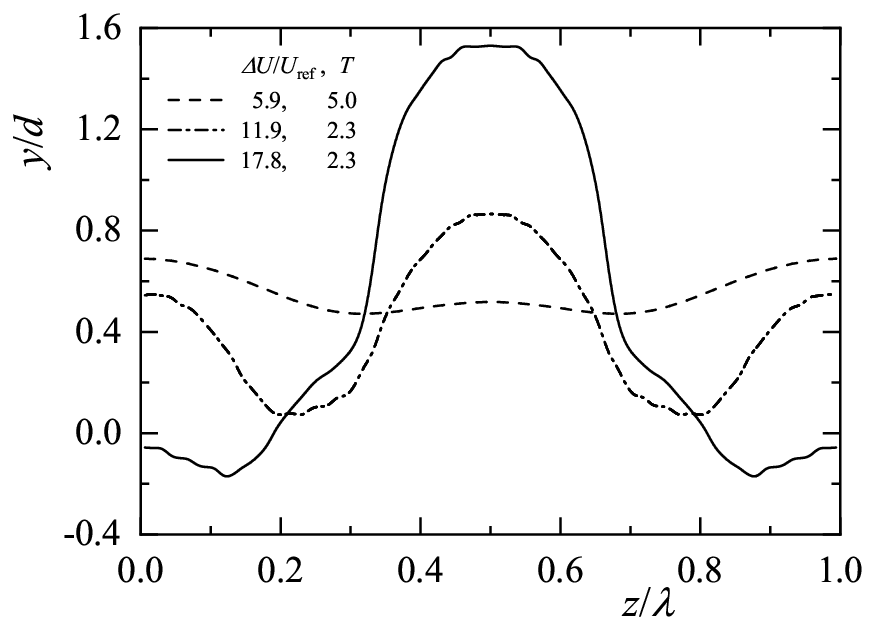}
\vspace*{-0.5\baselineskip}
\caption{Position of ligament interface for $ka = 0.7$.}
\label{u102030int}
\end{figure}

Figure \ref{u102030int} shows the position $y/d$ 
where the interface at each $\Delta U/U_\mathrm{ref}$ is the highest 
at the time just before the splitting of the liquid ligament. 
For $\Delta U/U_\mathrm{ref} = 5.9$ and 11.9, 
the interface around $z/\lambda = 0$, 0.5, and 1.0 bulges toward the gas side. 
This result is owing to a nonlinear effect, 
and the turbulence with twice the wavenumber of the initial disturbance occurs at the interface. 
For $\Delta U/U_\mathrm{ref} = 17.8$, in addition to $z/\lambda = 0$, 0.5, and 1.0, 
the interface also expands to the gas side around $z/\lambda = 0.25$ and 0.75. 
Compared with the results of $\Delta U/U_\mathrm{ref} = 5.9$ and 11.9, 
higher wavenumber turbulence occurs at the interface. 
This high wavenumber turbulence grows, and the liquid ligament in the vicinity 
indicated by the arrows in Fig. \ref{u102030vcvor}(c) are constricted and split, 
resulting in an increase in the number of breakup droplets 
for $\Delta U/U_\mathrm{ref} = 17.8$.

\begin{figure}[!t]
\centering
\includegraphics[trim=0mm 0mm 0mm 0mm, clip, width=80mm]{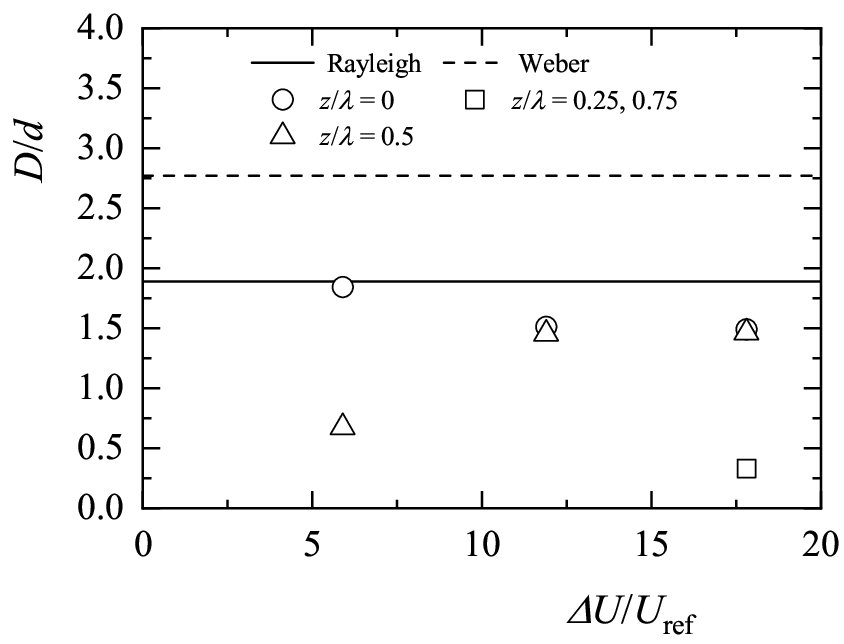}
\vspace*{-0.5\baselineskip}
\caption{Variations of droplet diameter with $\Delta U/U_\mathrm{ref}$ for $ka = 0.7$.}
\label{dsize_u10u20u30}
\end{figure}

Figure \ref{dsize_u10u20u30} shows the diameters $D$ of the droplets 
that broke up for $\Delta U/U_{\rm ref} = 5.9$, 11.9, and 17.8. 
The coordinate $z/\lambda$ represents the location of the droplet. 
In this study, as in the existing report \cite{Yanaoka&Nakayama_2022}, 
the droplet diameter is defined as the equivalent diameter obtained 
from the droplet volume. 
The solid and dashed lines in the figure are the theoretical values 
for an inviscid liquid column \citep{Rayleigh_1878} 
and for a viscous liquid column \citep{Weber_1931}, respectively. 
Weber's theoretical value is expressed as follows:
\begin{equation}
   D = 1.89 \left[ 1+\frac{3\mu_{l}} {\sqrt{d \rho_{l} \sigma}} \right]d.
\end{equation}
The droplet size in this calculation is smaller than the two theoretical values 
\citep{Rayleigh_1878, Weber_1931}. 
The theoretical values are based on linear theory 
and do not consider the nonlinear effects of flow. 
In this calculation, as shown in Fig. \ref{u102030int}, 
turbulence with higher wavenumber components than the initial disturbance occurs 
at the interface owing to the nonlinear effect. 
It is considered that as the liquid ligament with a short wavelength breaks up 
owing to this high wavenumber turbulence, 
the droplet diameter is smaller than the theoretical value \citep{Rayleigh_1878, Weber_1931}. 
When $\Delta U/U_\mathrm{ref}$ increases, 
the droplet diameters at $z/\lambda = 0$ and 0.5 decrease and increase, respectively. 
The ratio of the droplet diameters of $z/\lambda = 0.5$ and $z/\lambda = 0$, 
$D_{z/\lambda = 0.5}$/$D_{z/\lambda = 0}$, is 0.37, 0.96, and 0.98 
for $\Delta U/U_\mathrm{ref} = 5.9$, 11.9 and 17.8, respectively. 
It can be seen that the droplet diameter becomes uniform 
as $\Delta U/U_\mathrm{ref}$ increases. 
Additionally, for $\Delta U/U_\mathrm{ref} = 17.8$, 
the number of breakup droplets increases; 
hence, the total surface area of the liquid increases. 
Therefore, it can be said that increasing $\Delta U/U_\mathrm{ref}$ improved 
the atomization quality of the liquid ligament.

\subsection{Analysis of the effects of disturbance with various wavenumber components 
on deformation and splitting of liquid ligament}

\begin{figure}[!t]
\centering
\begin{minipage}{0.32\linewidth}
\begin{center}
\includegraphics[trim=0mm 0mm 0mm 0mm, clip, width=55mm]{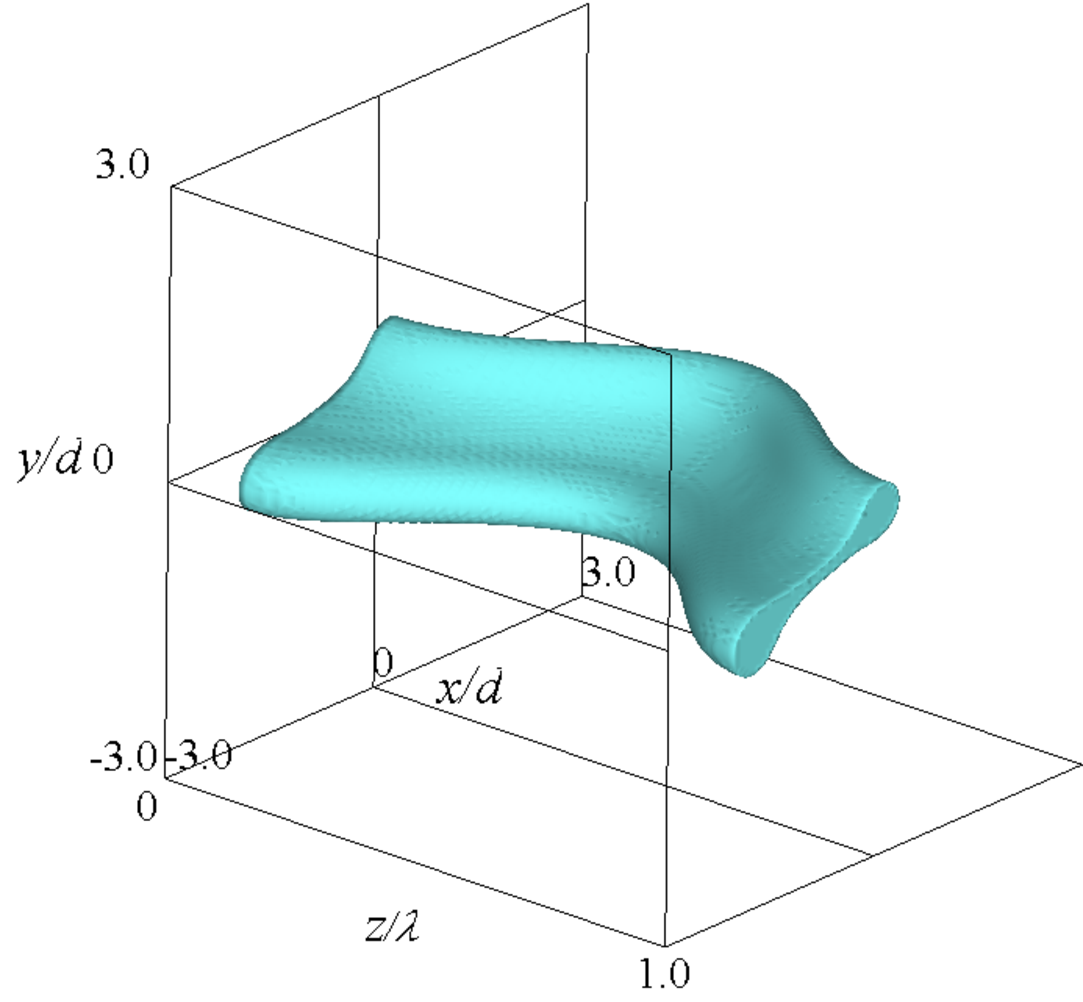} \\
\vspace*{-0.5\baselineskip}
(a)
\end{center}
\end{minipage}
\begin{minipage}{0.32\linewidth}
\begin{center}
\includegraphics[trim=0mm 0mm 0mm 0mm, clip, width=55mm]{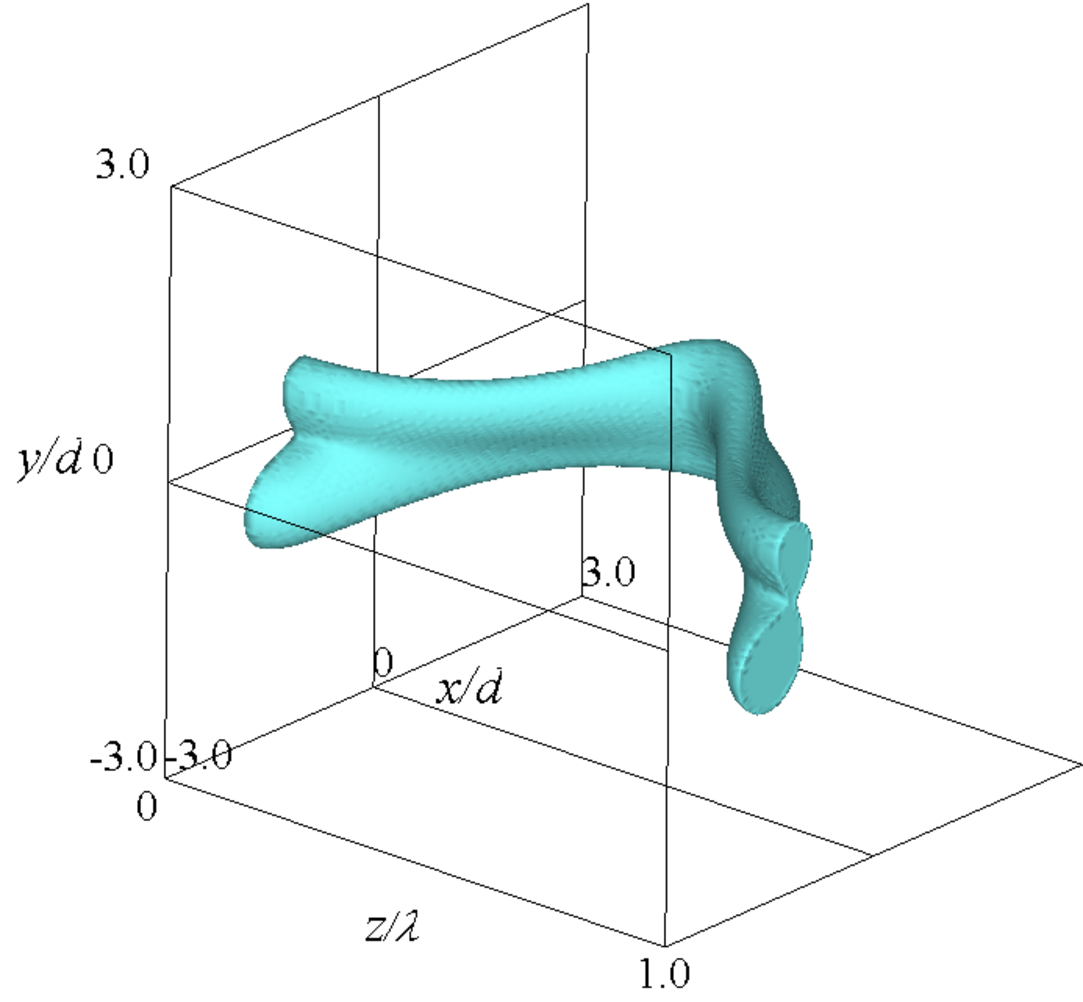} \\
\vspace*{-0.5\baselineskip}
(b)
\end{center}
\end{minipage}
\begin{minipage}{0.32\linewidth}
\begin{center}
\includegraphics[trim=0mm 0mm 0mm 0mm, clip, width=55mm]{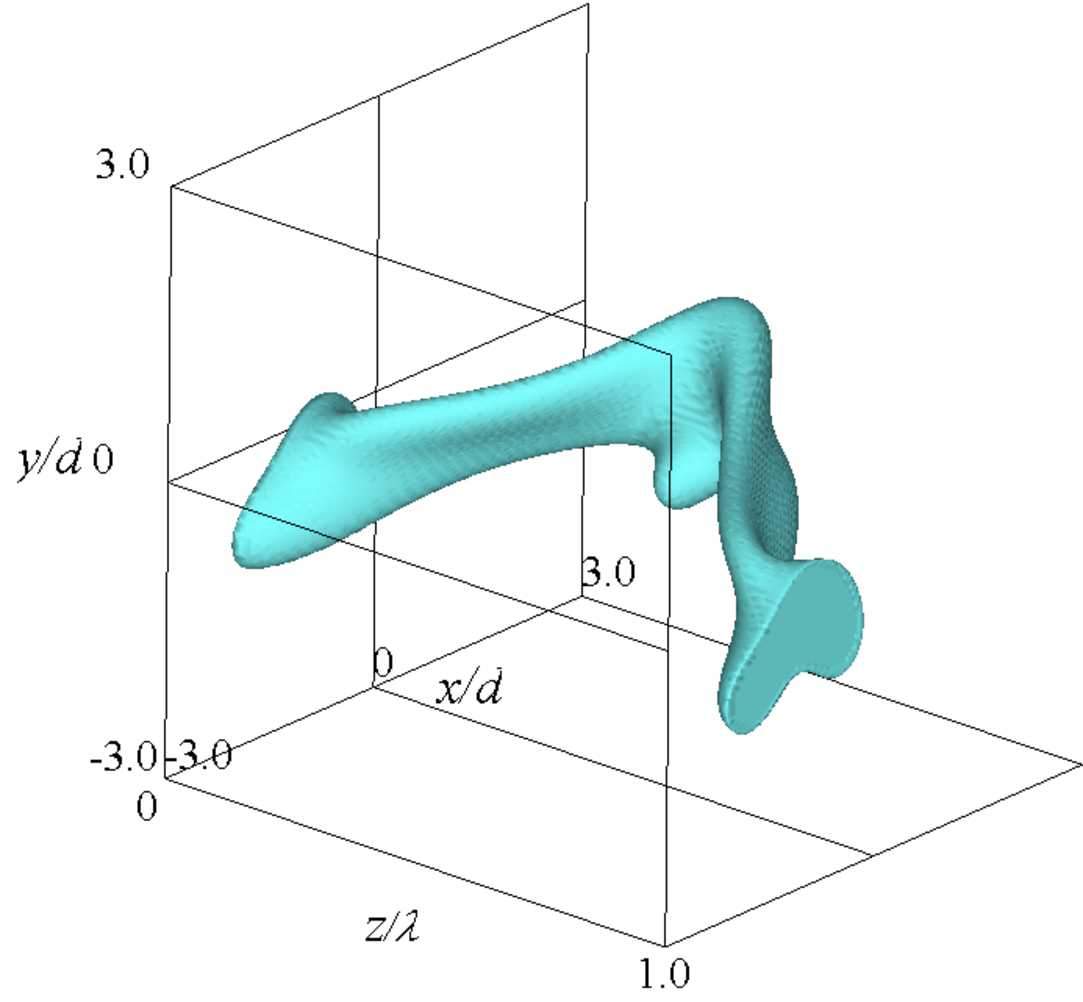} \\
\vspace*{-0.5\baselineskip}
(c)
\end{center}
\end{minipage}

\vspace*{0.5\baselineskip}
\begin{minipage}{0.32\linewidth}
\begin{center}
\includegraphics[trim=0mm 0mm 0mm 0mm, clip, width=55mm]{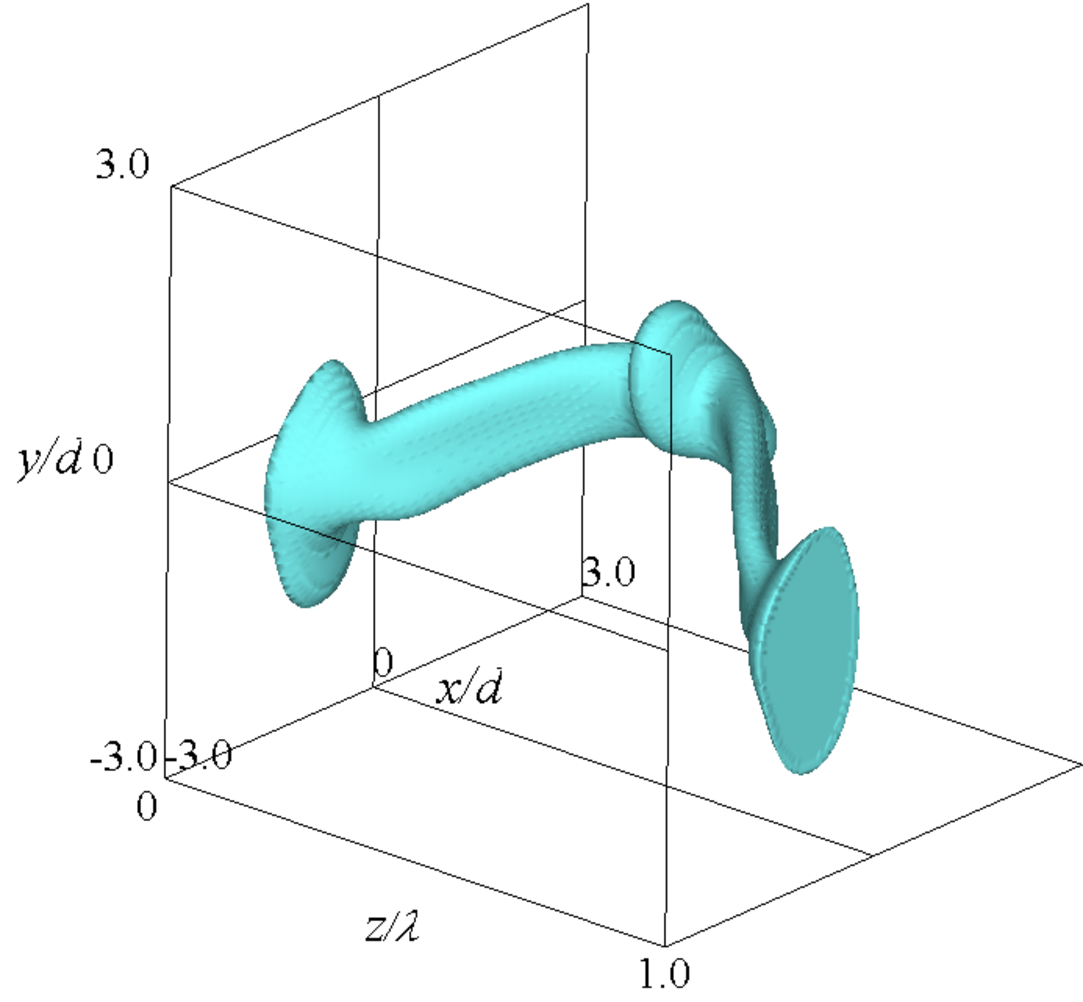} \\
\vspace*{-0.5\baselineskip}
(d)
\end{center}
\end{minipage}
\begin{minipage}{0.32\linewidth}
\begin{center}
\includegraphics[trim=0mm 0mm 0mm 0mm, clip, width=55mm]{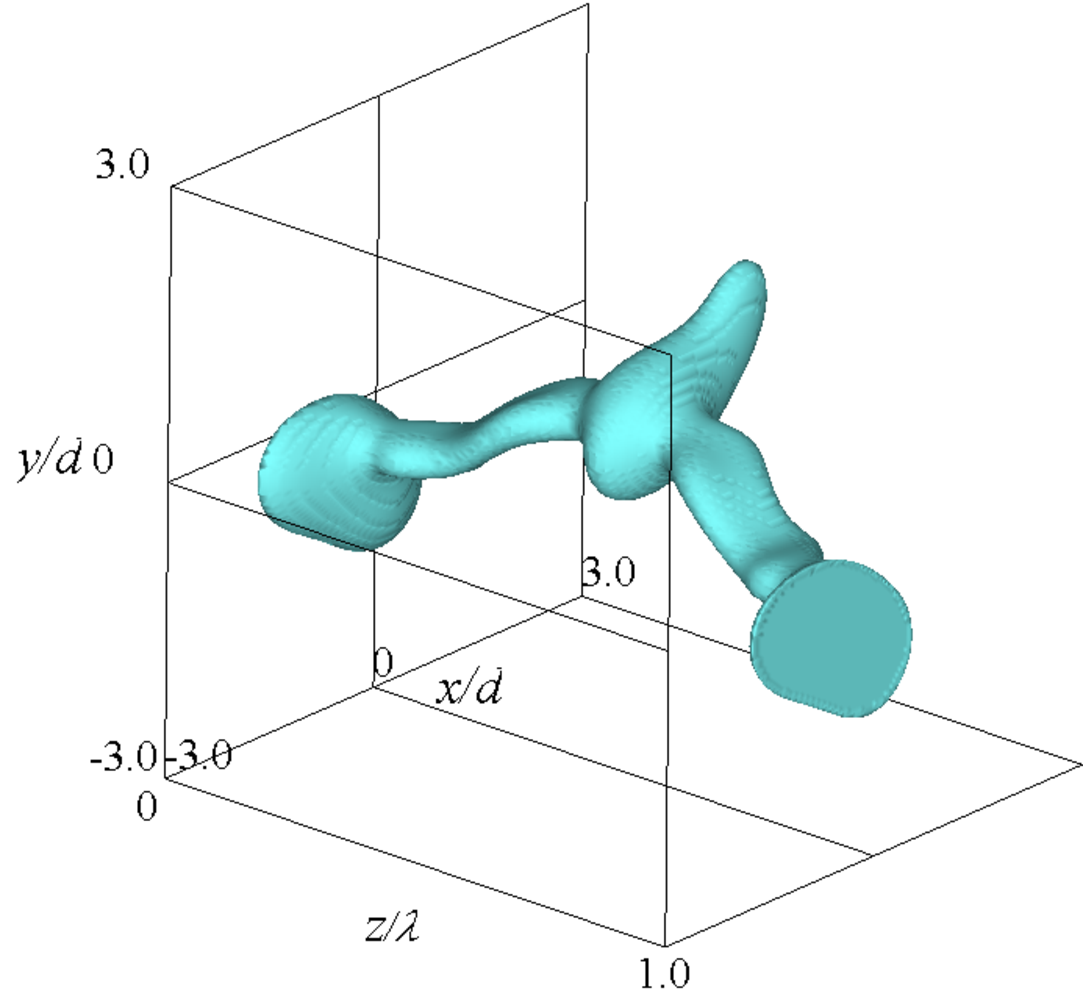} \\
\vspace*{-0.5\baselineskip}
(e)
\end{center}
\end{minipage}
\begin{minipage}{0.32\linewidth}
\begin{center}
\includegraphics[trim=0mm 0mm 0mm 0mm, clip, width=55mm]{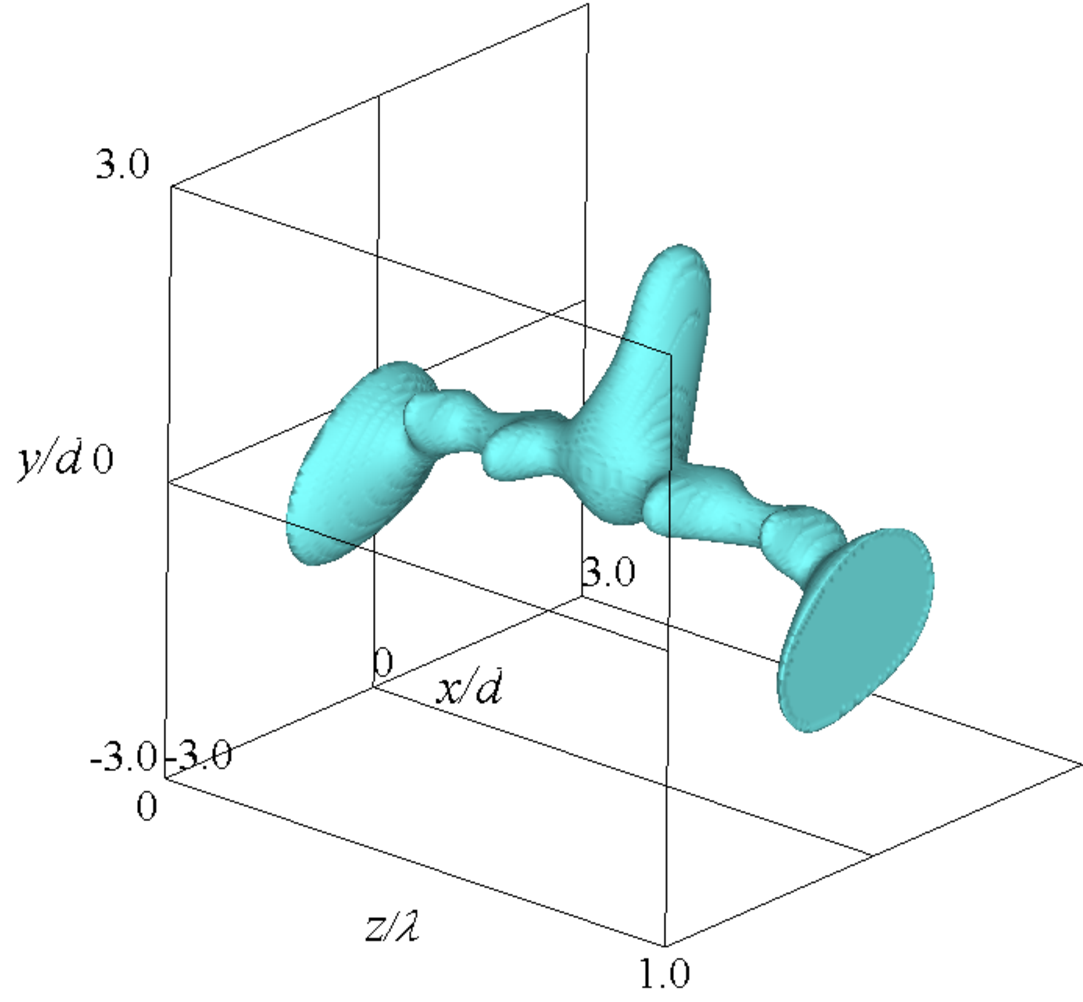} \\
\vspace*{-0.5\baselineskip}
(f)
\end{center}
\end{minipage}

\vspace*{0.5\baselineskip}
\begin{minipage}{0.32\linewidth}
\begin{center}
\includegraphics[trim=0mm 0mm 0mm 0mm, clip, width=55mm]{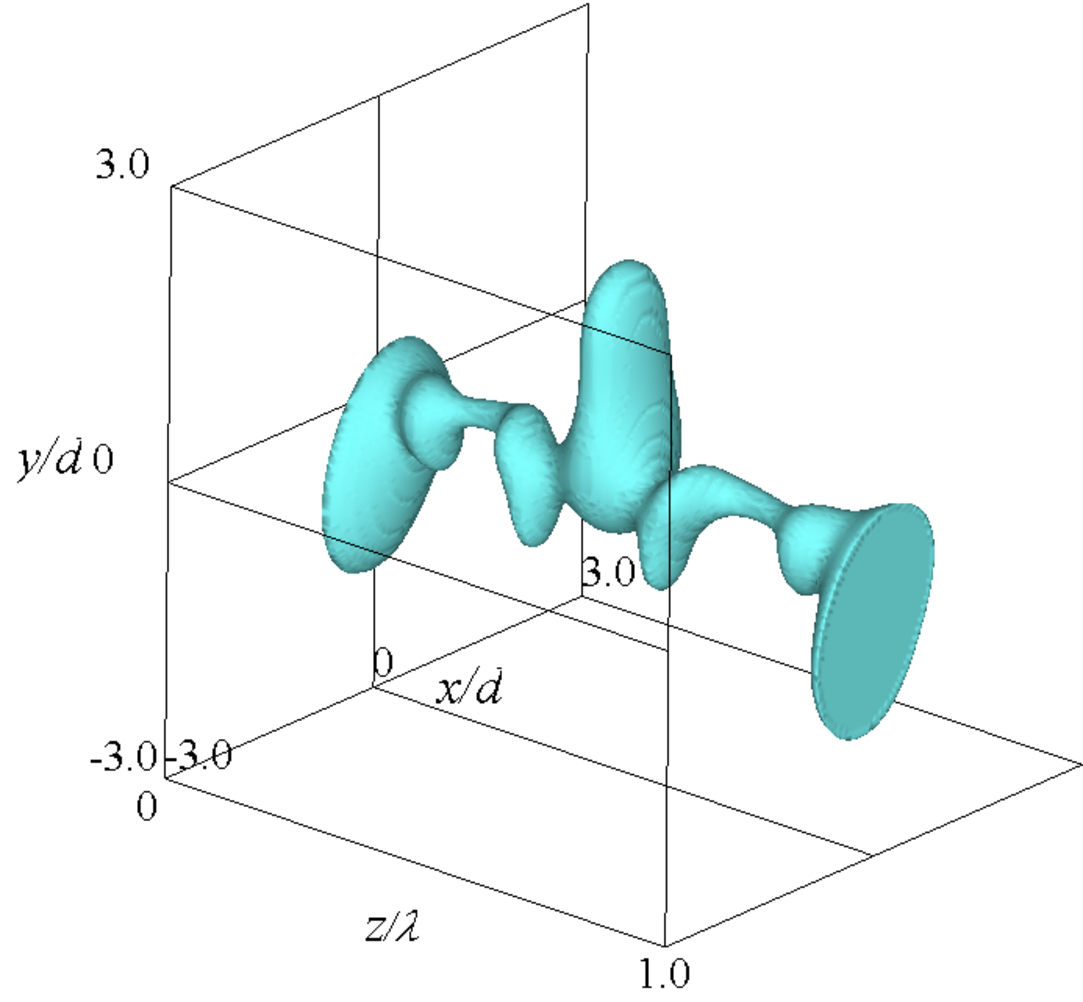} \\
\vspace*{-0.5\baselineskip}
(g)
\end{center}
\end{minipage}
\begin{minipage}{0.32\linewidth}
\begin{center}
\includegraphics[trim=0mm 0mm 0mm 0mm, clip, width=55mm]{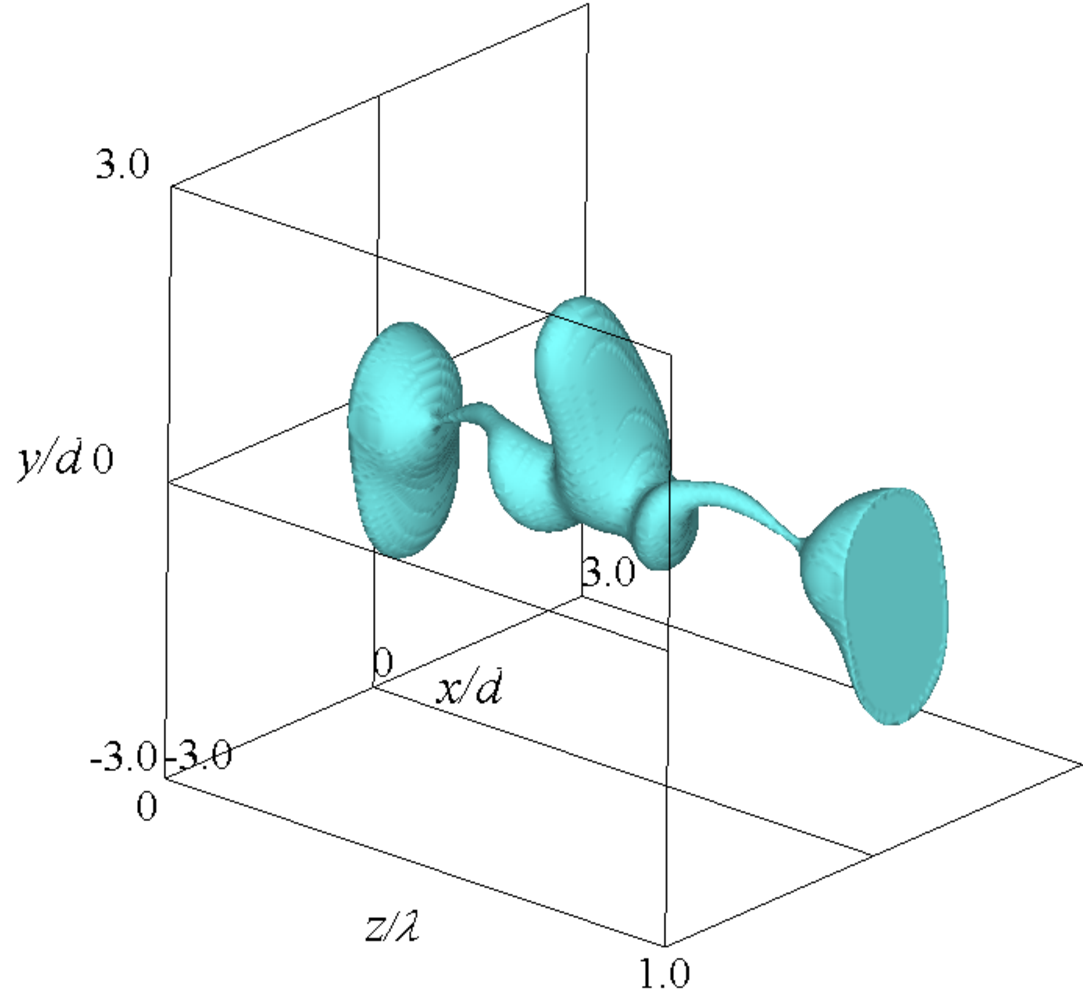} \\
\vspace*{-0.5\baselineskip}
(h)
\end{center}
\end{minipage}
\begin{minipage}{0.32\linewidth}
\begin{center}
\includegraphics[trim=0mm 0mm 0mm 0mm, clip, width=55mm]{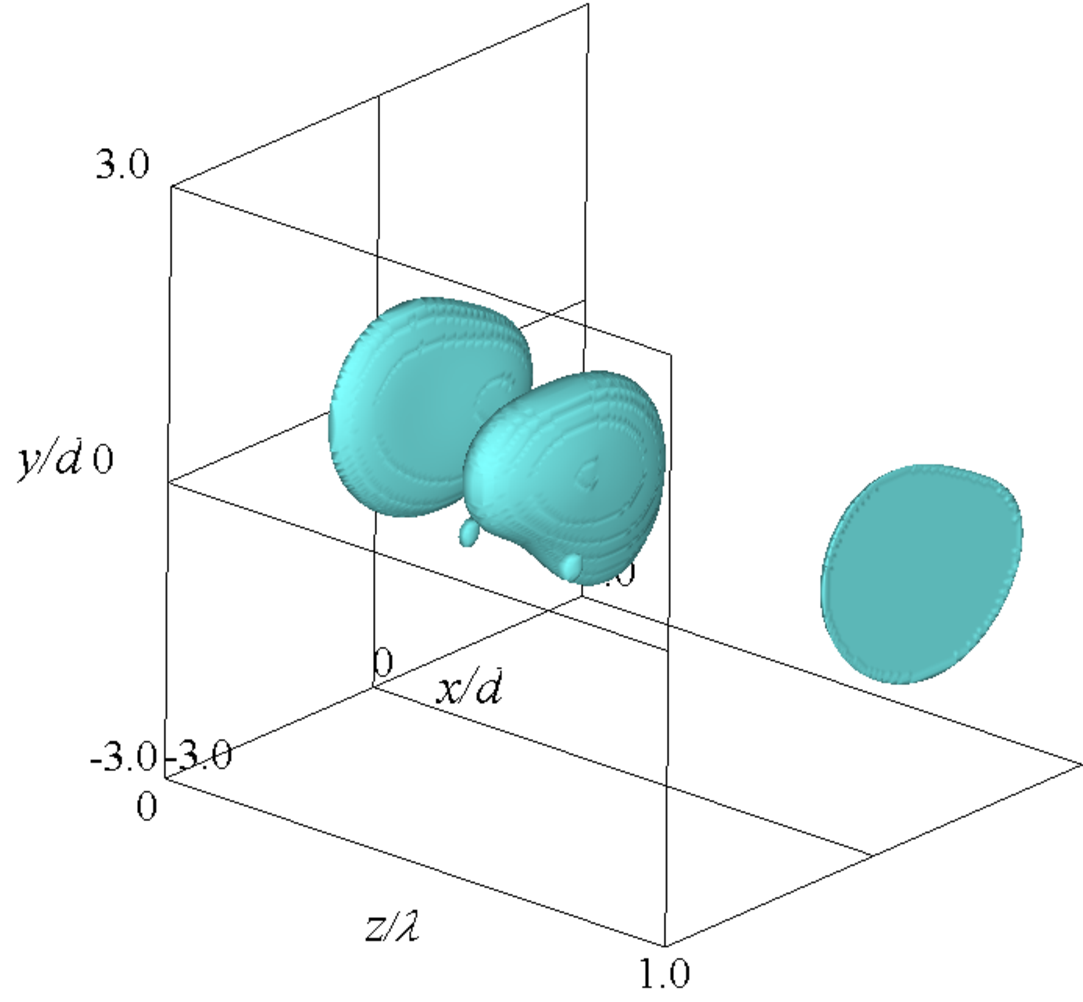} \\
\vspace*{-0.5\baselineskip}
(i)
\end{center}
\end{minipage}
\caption{Time variations of ligament interface: 
(a) $T = 0.6$, (b) $T = 1.0$, (c) $T = 1.6$, (d) $T = 2.0$, (e) $T = 2.6$, 
(f) $T = 3.0$, (g) $T = 3.3$, (h) $T = 3.6$, and (i) $T = 4.3$: 
$\Delta U/U_\mathrm{ref} = 17.8$, $ka = 0.3$.}
\label{u30k3_time}
\end{figure}

\begin{figure}[!t]
\centering
\begin{minipage}{0.32\linewidth}
\begin{center}
\includegraphics[trim=0mm 0mm 0mm 0mm, clip, width=55mm]{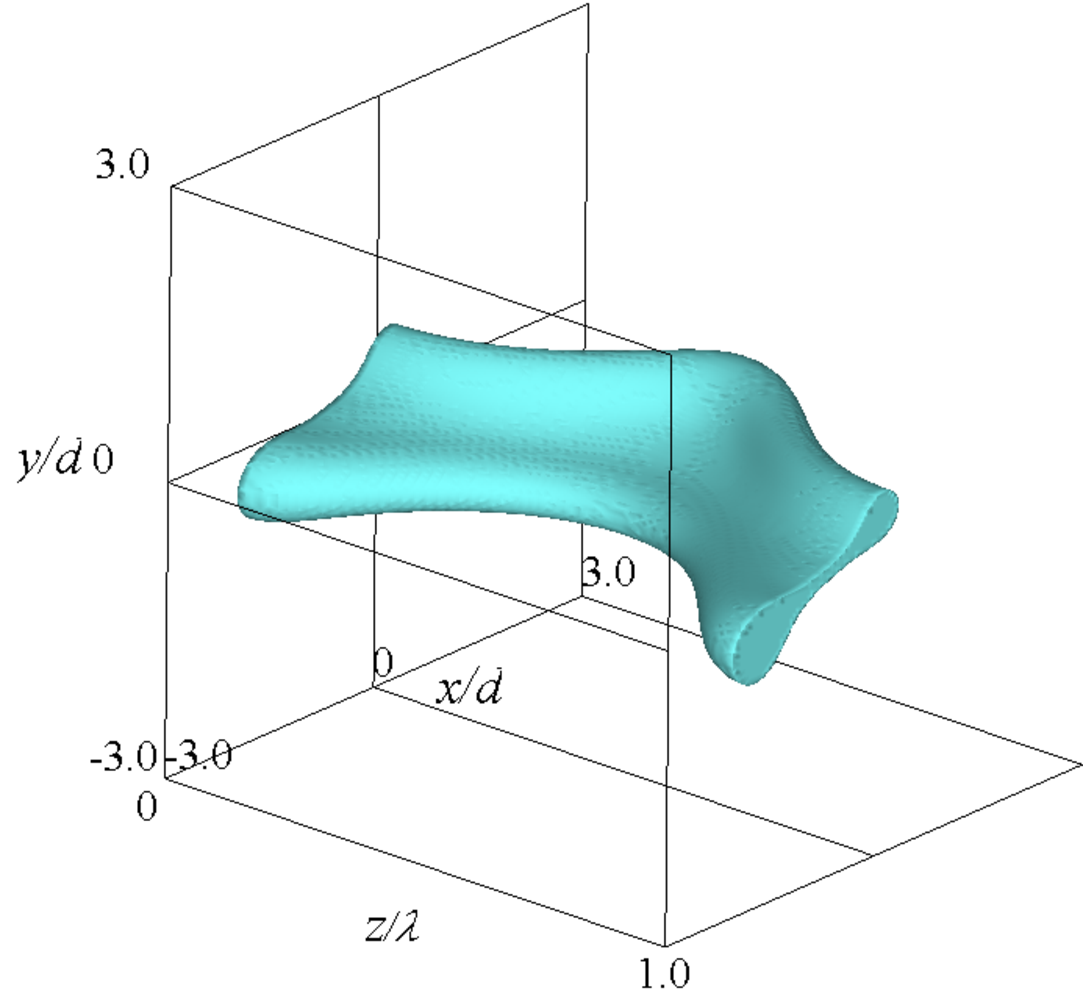} \\
\vspace*{-0.5\baselineskip}
(a)
\end{center}
\end{minipage}
\begin{minipage}{0.32\linewidth}
\begin{center}
\includegraphics[trim=0mm 0mm 0mm 0mm, clip, width=55mm]{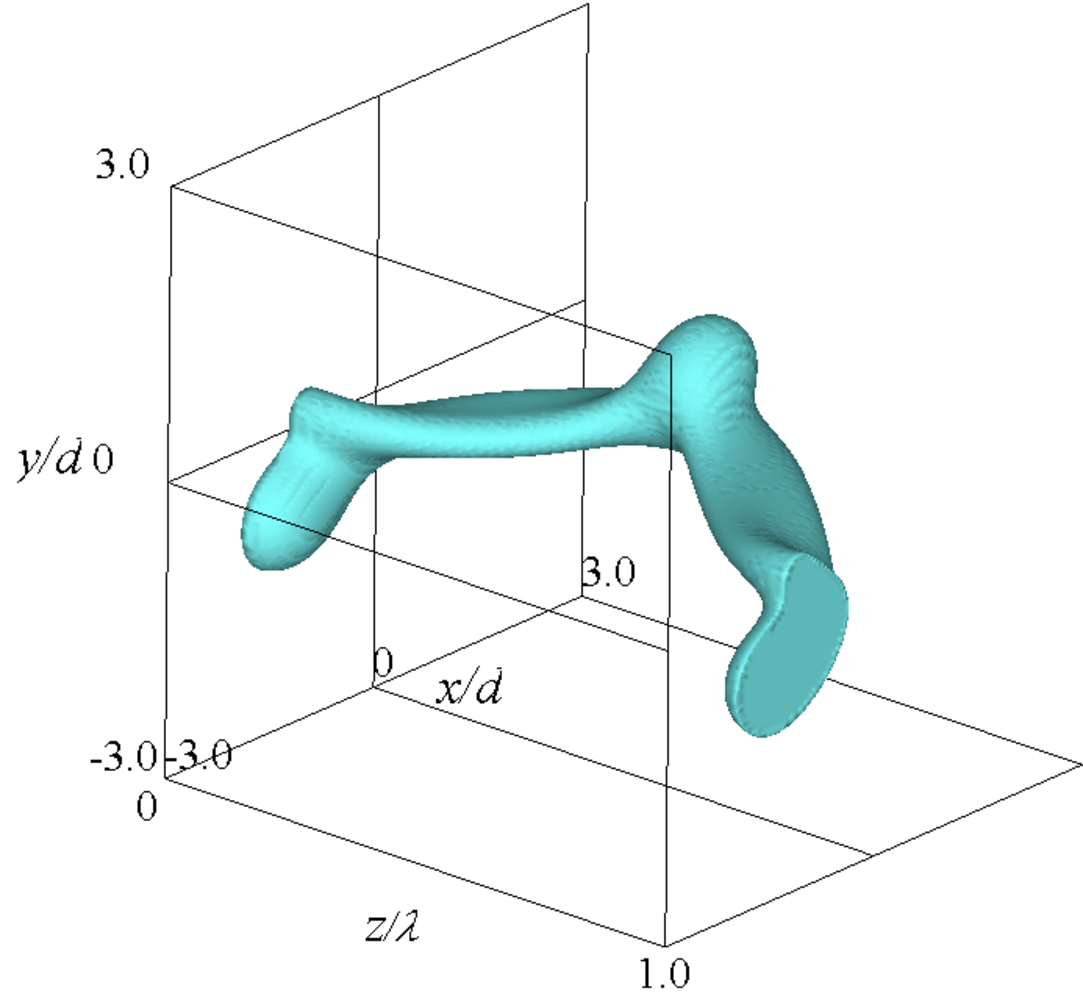} \\
\vspace*{-0.5\baselineskip}
(b)
\end{center}
\end{minipage}
\begin{minipage}{0.32\linewidth}
\begin{center}
\includegraphics[trim=0mm 0mm 0mm 0mm, clip, width=55mm]{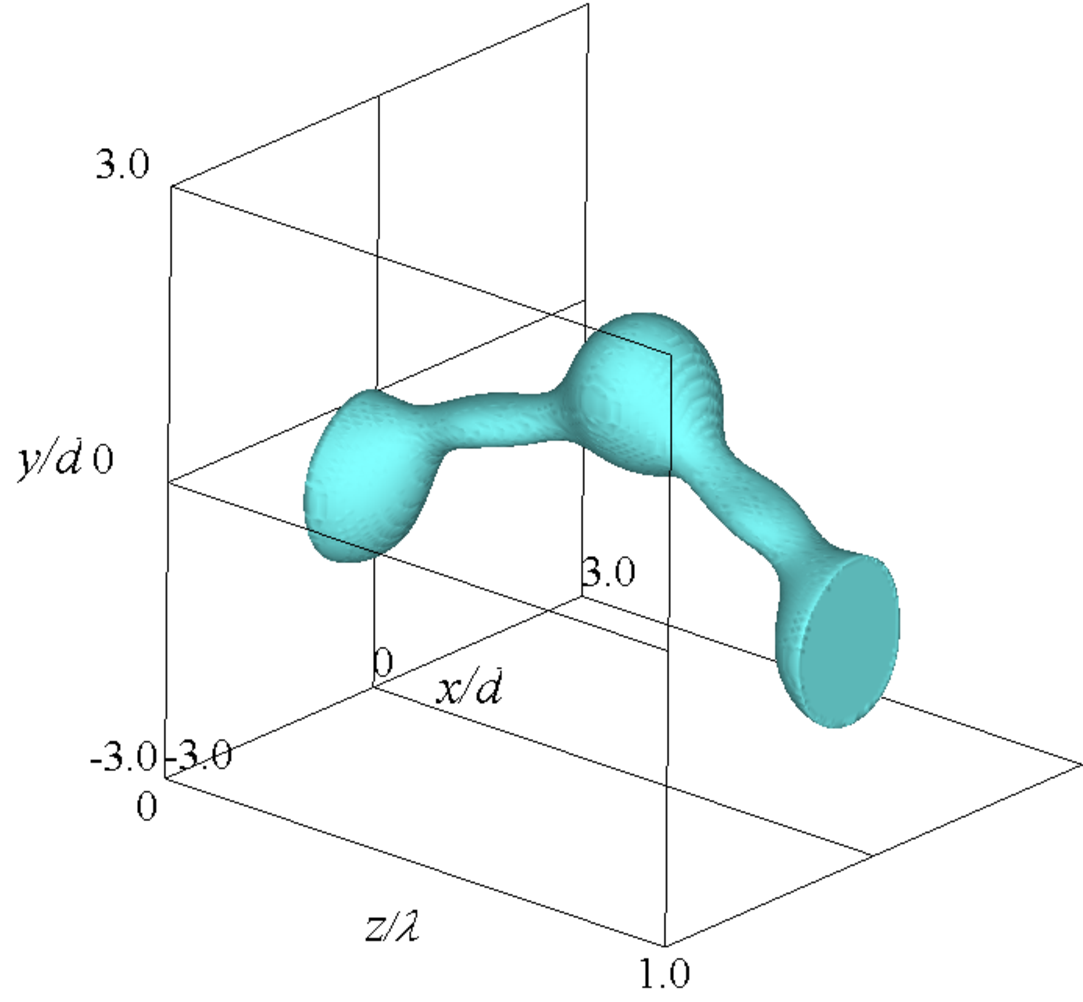} \\
\vspace*{-0.5\baselineskip}
(c)
\end{center}
\end{minipage}

\vspace*{0.5\baselineskip}
\begin{minipage}{0.32\linewidth}
\begin{center}
\includegraphics[trim=0mm 0mm 0mm 0mm, clip, width=55mm]{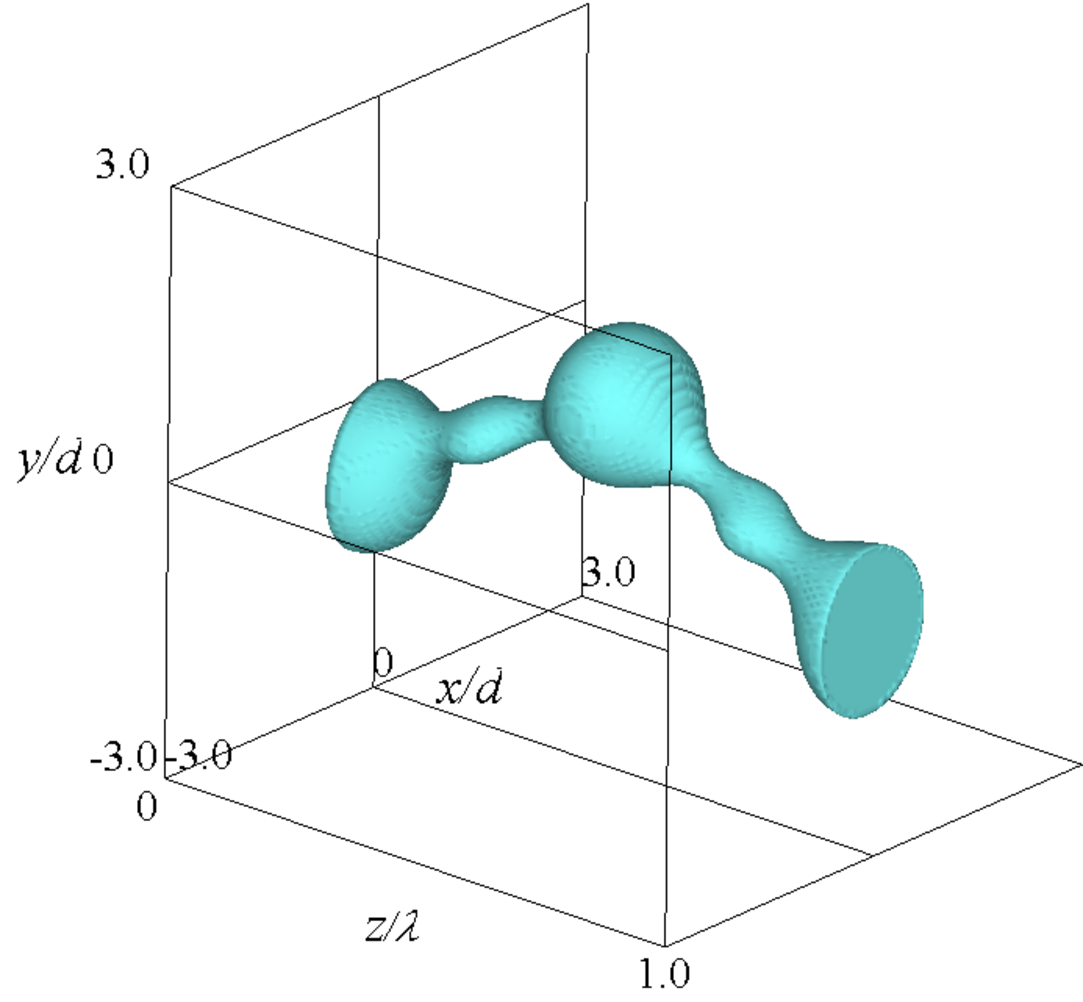} \\
\vspace*{-0.5\baselineskip}
(d)
\end{center}
\end{minipage}
\begin{minipage}{0.32\linewidth}
\begin{center}
\includegraphics[trim=0mm 0mm 0mm 0mm, clip, width=55mm]{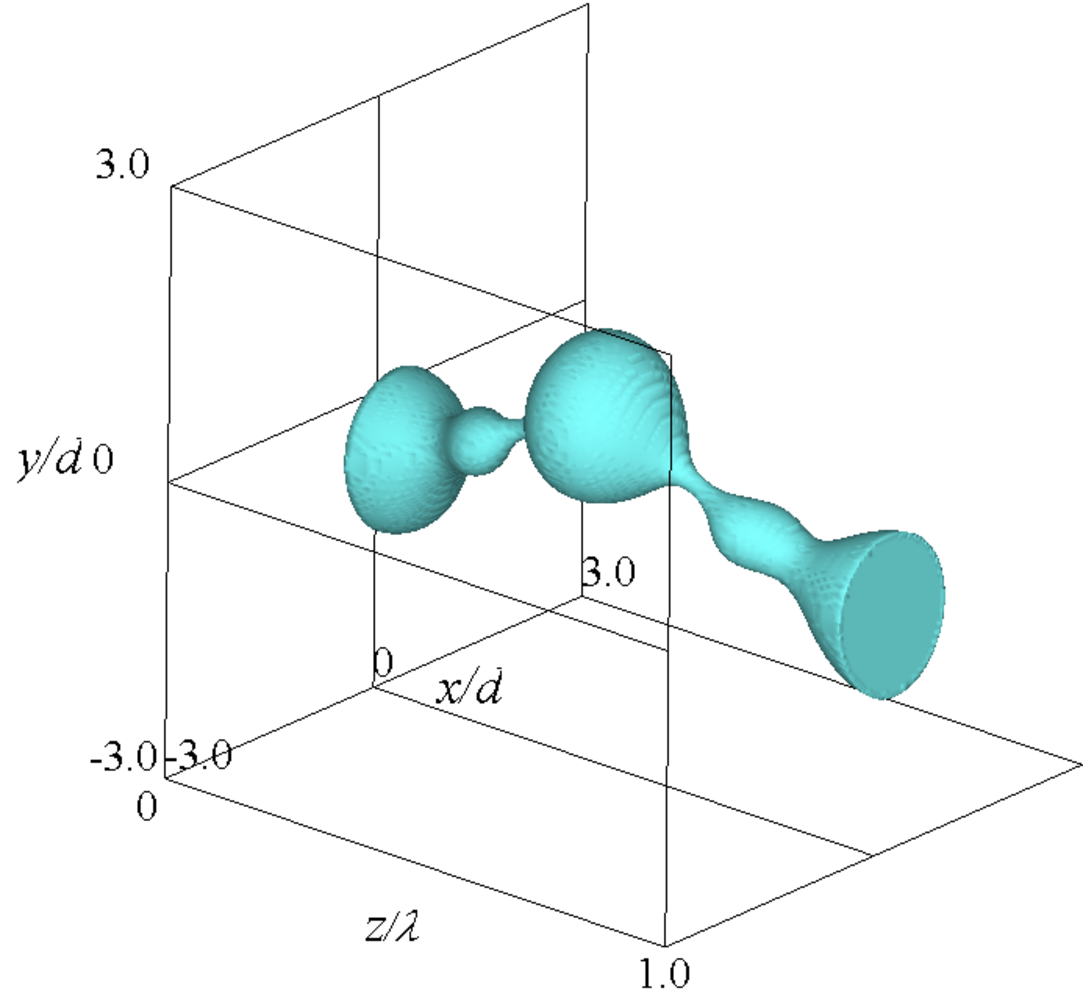} \\
\vspace*{-0.5\baselineskip}
(e)
\end{center}
\end{minipage}
\begin{minipage}{0.32\linewidth}
\begin{center}
\includegraphics[trim=0mm 0mm 0mm 0mm, clip, width=55mm]{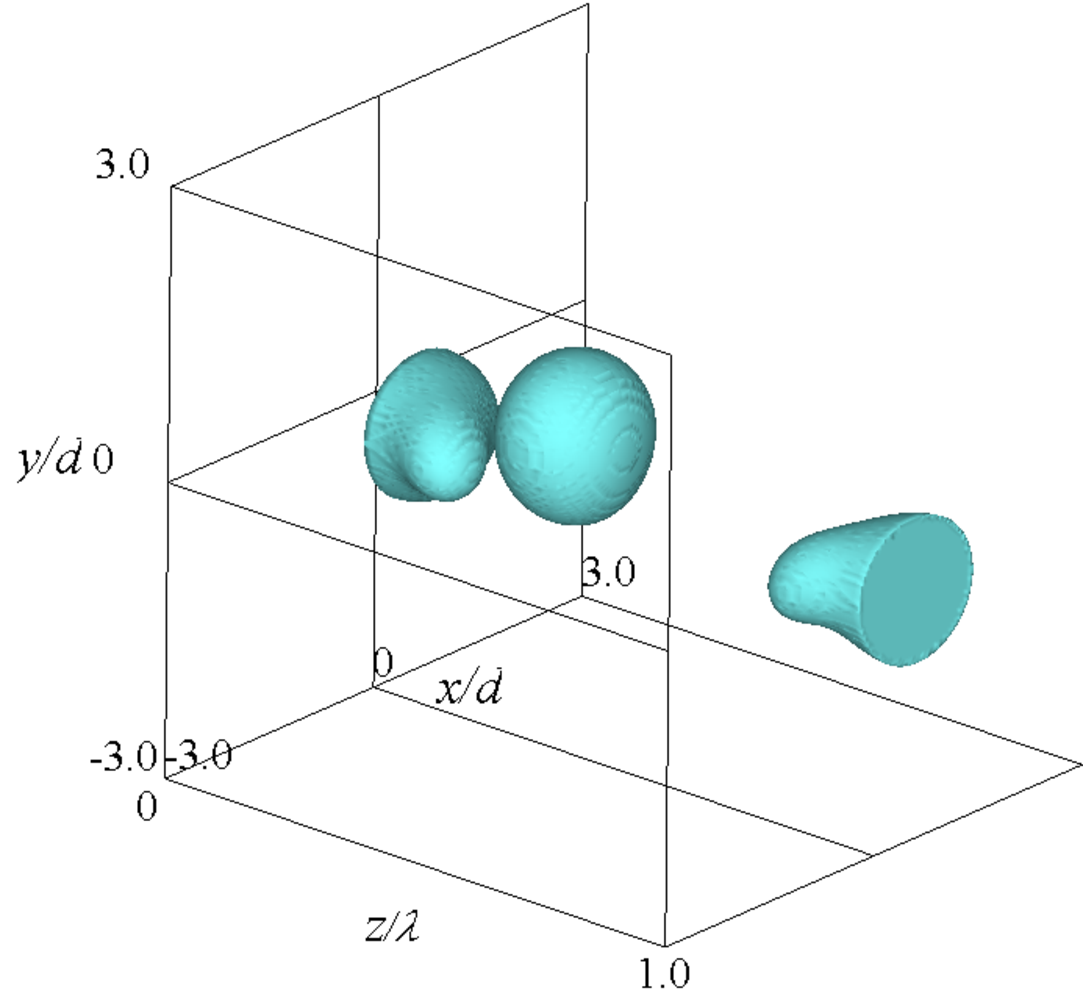} \\
\vspace*{-0.5\baselineskip}
(f)
\end{center}
\end{minipage}
\caption{Time variations of ligament interface: 
(a) $T = 0.6$, (b) $T = 1.0$, (c) $T = 2.0$, (d) $T = 2.3$, (e) $T = 2.6$, and (f) $T = 3.0$: 
$\Delta U/U_\mathrm{ref} = 17.8$, $ka = 0.5$.}
\label{u30k5_time}
\end{figure}

\begin{figure}[!t]
\centering
\begin{minipage}{0.48\linewidth}
\begin{center}
\includegraphics[trim=0mm 0mm 0mm 0mm, clip, width=75mm]{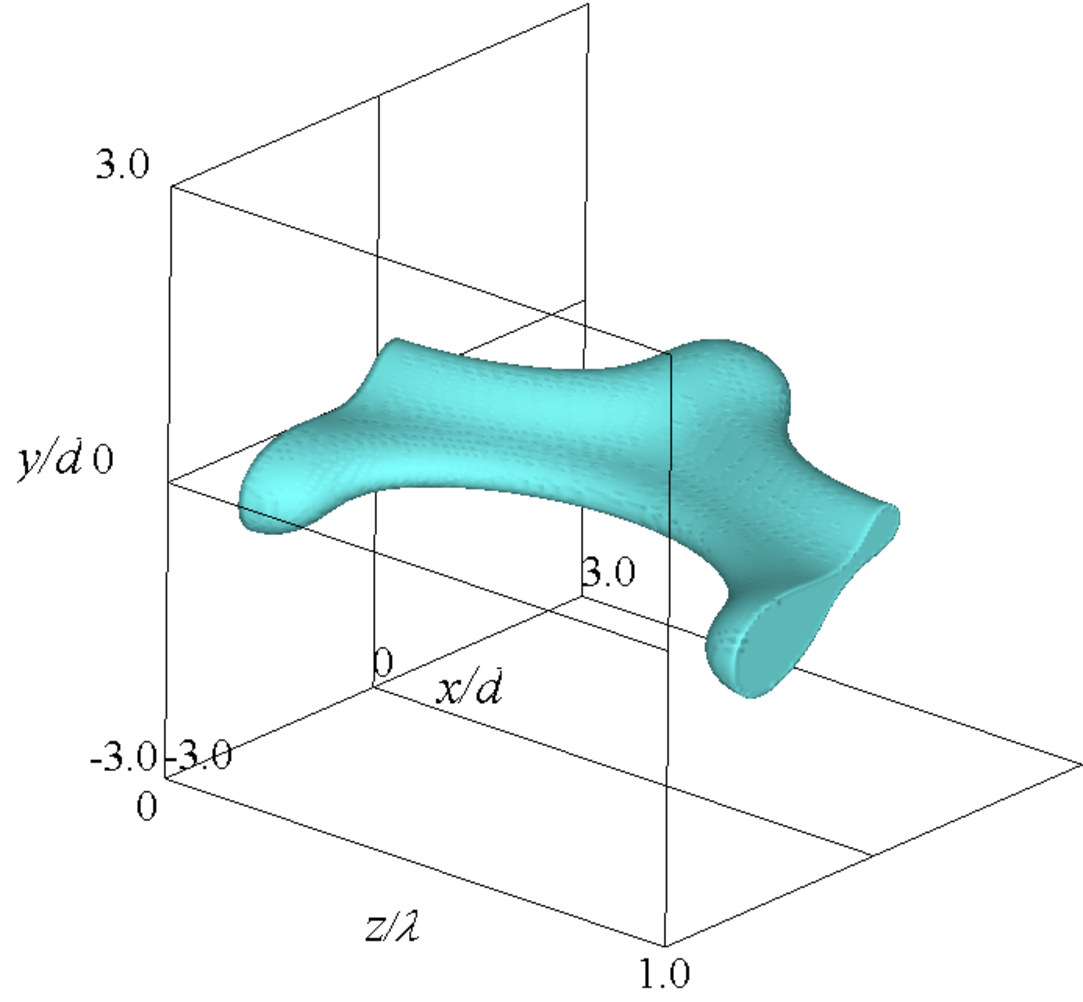} \\
\vspace*{-0.5\baselineskip}
(a)
\end{center}
\end{minipage}
\begin{minipage}{0.48\linewidth}
\begin{center}
\includegraphics[trim=0mm 0mm 0mm 0mm, clip, width=75mm]{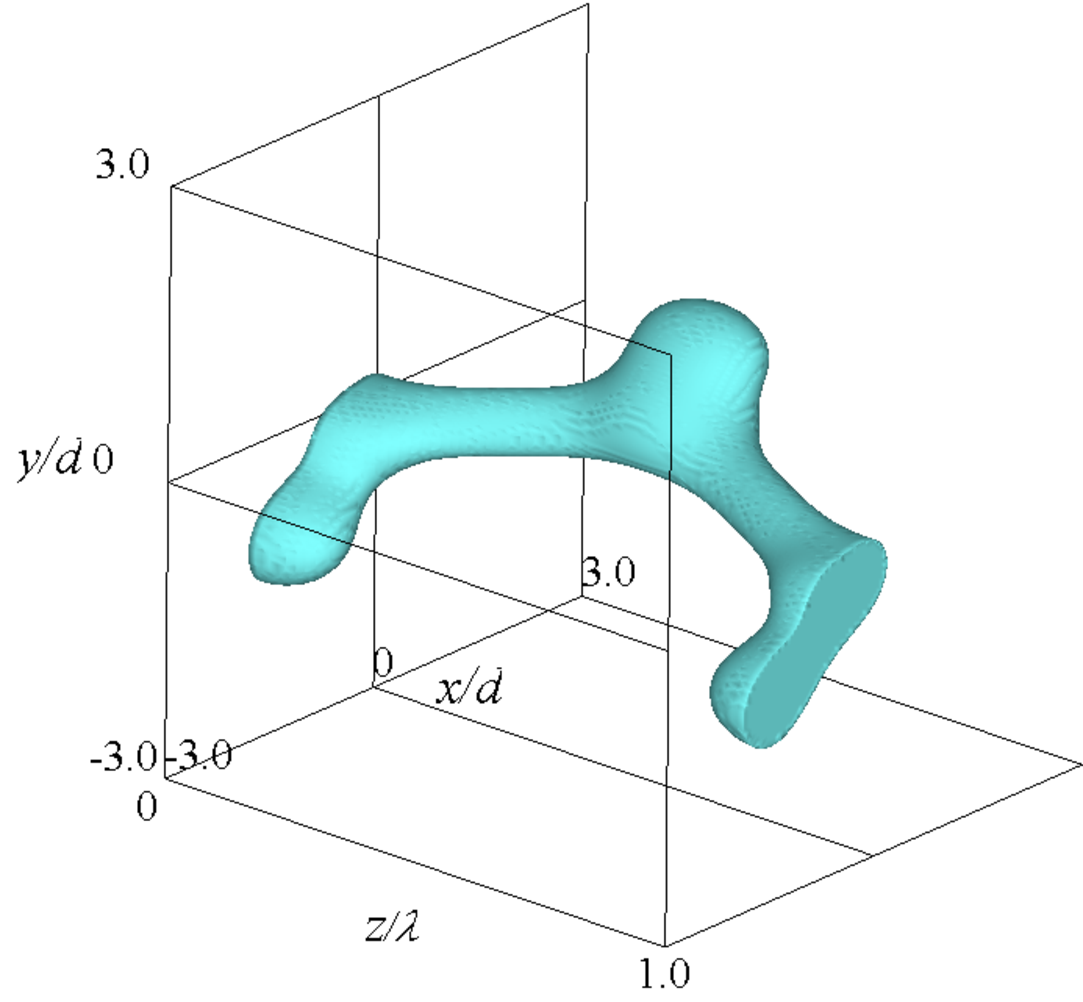} \\
\vspace*{-0.5\baselineskip}
(b)
\end{center}
\end{minipage}

\vspace*{0.5\baselineskip}
\begin{minipage}{0.48\linewidth}
\begin{center}
\includegraphics[trim=0mm 0mm 0mm 0mm, clip, width=75mm]{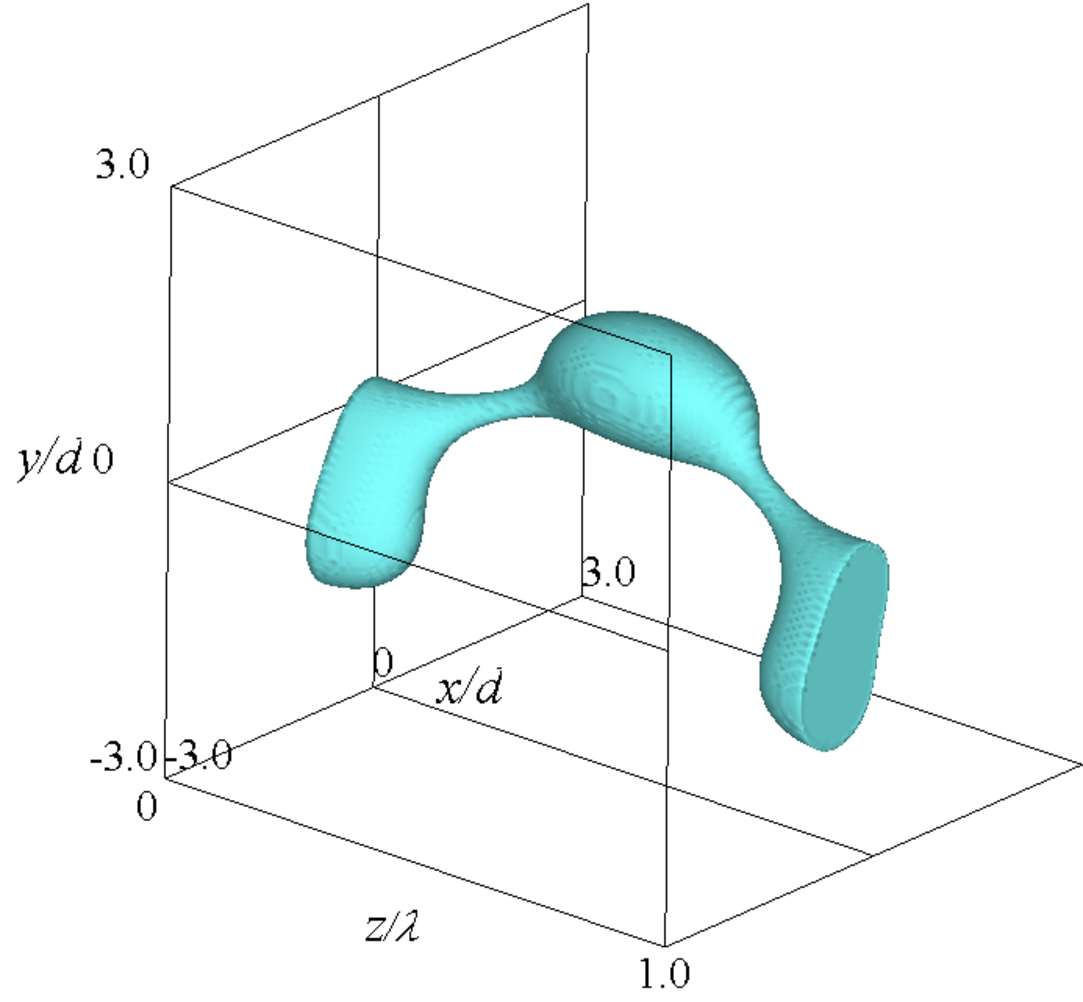} \\
\vspace*{-0.5\baselineskip}
(c)
\end{center}
\end{minipage}
\begin{minipage}{0.48\linewidth}
\begin{center}
\includegraphics[trim=0mm 0mm 0mm 0mm, clip, width=75mm]{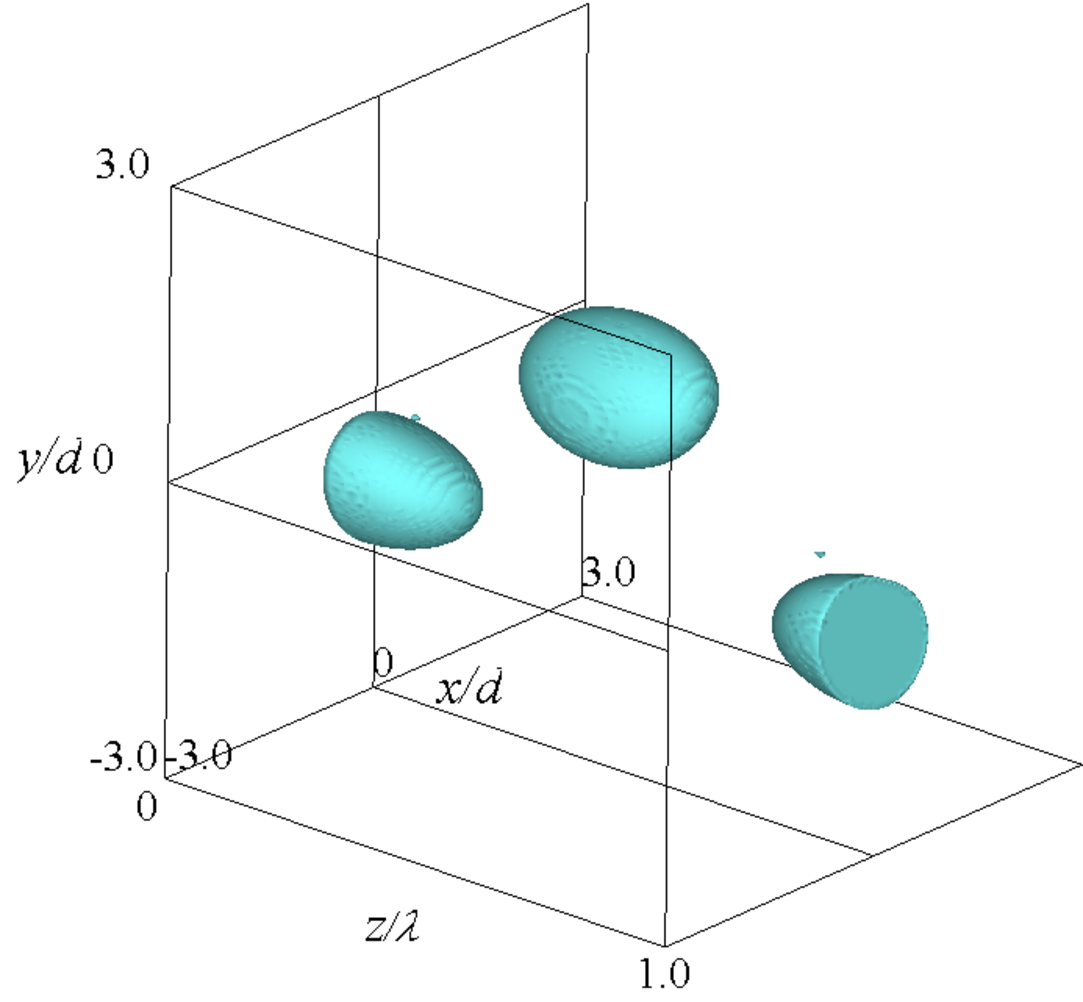} \\
\vspace*{-0.5\baselineskip}
(d)
\end{center}
\end{minipage}
\caption{Time variations of ligament interface: 
(a) $T = 0.6$, (b) $T = 1.0$, (c) $T = 1.6$, and (d) $T = 2.1$: 
$\Delta U/U_\mathrm{ref} = 17.8$, $ka = 0.9$.}
\label{u30k9_time}
\end{figure}

In actual flow fields, gas and liquid flows are turbulent; 
hence, turbulences with various wavenumber components occur at the gas--liquid interface. 
In this analysis, we investigate the effect of such disturbances 
on the deformation and splitting of a liquid ligament. 
First, for the velocity difference $\Delta U/U_\mathrm{ref} = 17.8$, 
where the influence of the shear velocity is the most intensive, 
we vary the dimensionless wavenumber $ka$ as 0.3, 0.5, 0.7, and 0.9. 
We investigate the effect of the wavenumber of the initial disturbance 
at the interface on the deformation and splitting of the liquid ligament.

Figures\ref{u30k3_time} to \ref{u30k9_time} show the time variation of the liquid ligament interface 
at $ka = 0.3$, 0.5, and 0.9 for $\Delta U/U_\mathrm{ref} = 17.8$. 
The result at $ka = 0.7$ was shown in Fig. \ref{u30k7_time}. 
At $T = 0.6$, the liquid ligaments for all $ka$ are deformed into films 
by the shear flow. 
With an increase in $ka$, the portions of $z/\lambda = 0$ and 0.5 in the liquid ligament bulge. 
At $T = 3.0$ and $T = 2.3$ for $ka = 0.3$ and 0.5, respectively, 
turbulence with a higher wavenumber than the initial disturbance occurs 
at the interface of $z/\lambda = 0.1-0.4$ and $0.6-0.9$. 
At $T = 4.3$, 3.0, 2.7, and 2.1 for $ka = 0.3$, 0.5, 0.7, and 0.9, respectively, 
the liquid ligament splits, and the breakup time of the liquid ligament shortens 
as $ka$ increases.

\begin{figure}[!t]
\centering
\begin{minipage}{0.48\linewidth}
\begin{center}
\includegraphics[trim=0mm 0mm 0mm 0mm, clip, width=80mm]{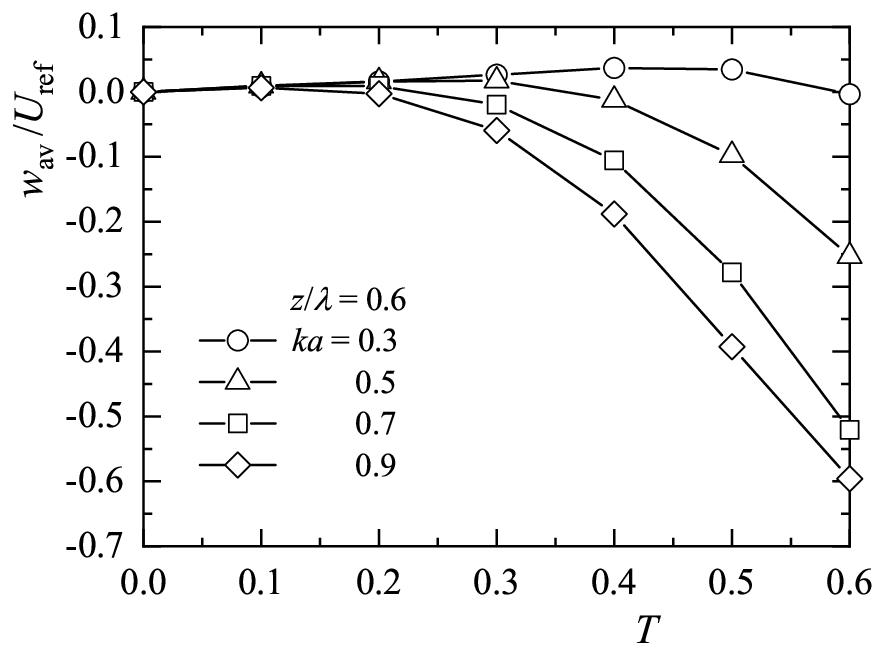} \\
\vspace*{-0.5\baselineskip}
(a)
\end{center}
\end{minipage}
\hspace*{0.02\linewidth}
\begin{minipage}{0.48\linewidth}
\begin{center}
\includegraphics[trim=0mm 0mm 0mm 0mm, clip, width=80mm]{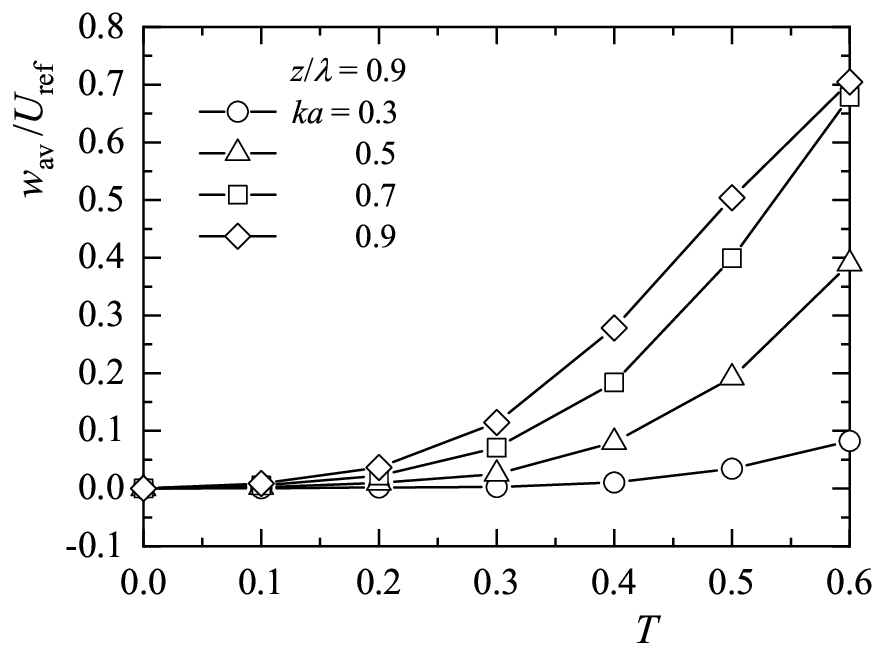} \\
\vspace*{-0.5\baselineskip}
(b)
\end{center}
\end{minipage}
\caption{Time variations of $z$-axial velocity averaged 
in the cross-section at (a) $z/\lambda = 0.6$ and (b) $z/\lambda = 0.9$: 
$\Delta U/U_\mathrm{ref} = 17.8$.}
\label{time_aw}
\end{figure}

Figure \ref{time_aw} shows the time variation of the $z$-axial velocity $w_\mathrm{av}$ 
averaged in the cross-section of the liquid ligament at $z/\lambda = 0.6$ and 0.9. 
At $z/\lambda = 0.6$ and 0.9, 
the liquids flow in the negative and positive directions of the $z$-axis, respectively. 
The absolute value of $w_\mathrm{av}$ increases over time 
for all $ka$ at the two cross-sections. 
Additionally, the magnitude of $w_\mathrm{av}$ increases as $ka$ increases. 
It can be seen from this that the moving liquid velocity 
along the central axis of the liquid ligament differs depending on $ka$, 
and the moving speed increases as $ka$ increases.

\begin{figure}[!t]
\centering
\begin{minipage}{0.48\linewidth}
\begin{center}
\includegraphics[trim=0mm 0mm 0mm 0mm, clip, width=70mm]{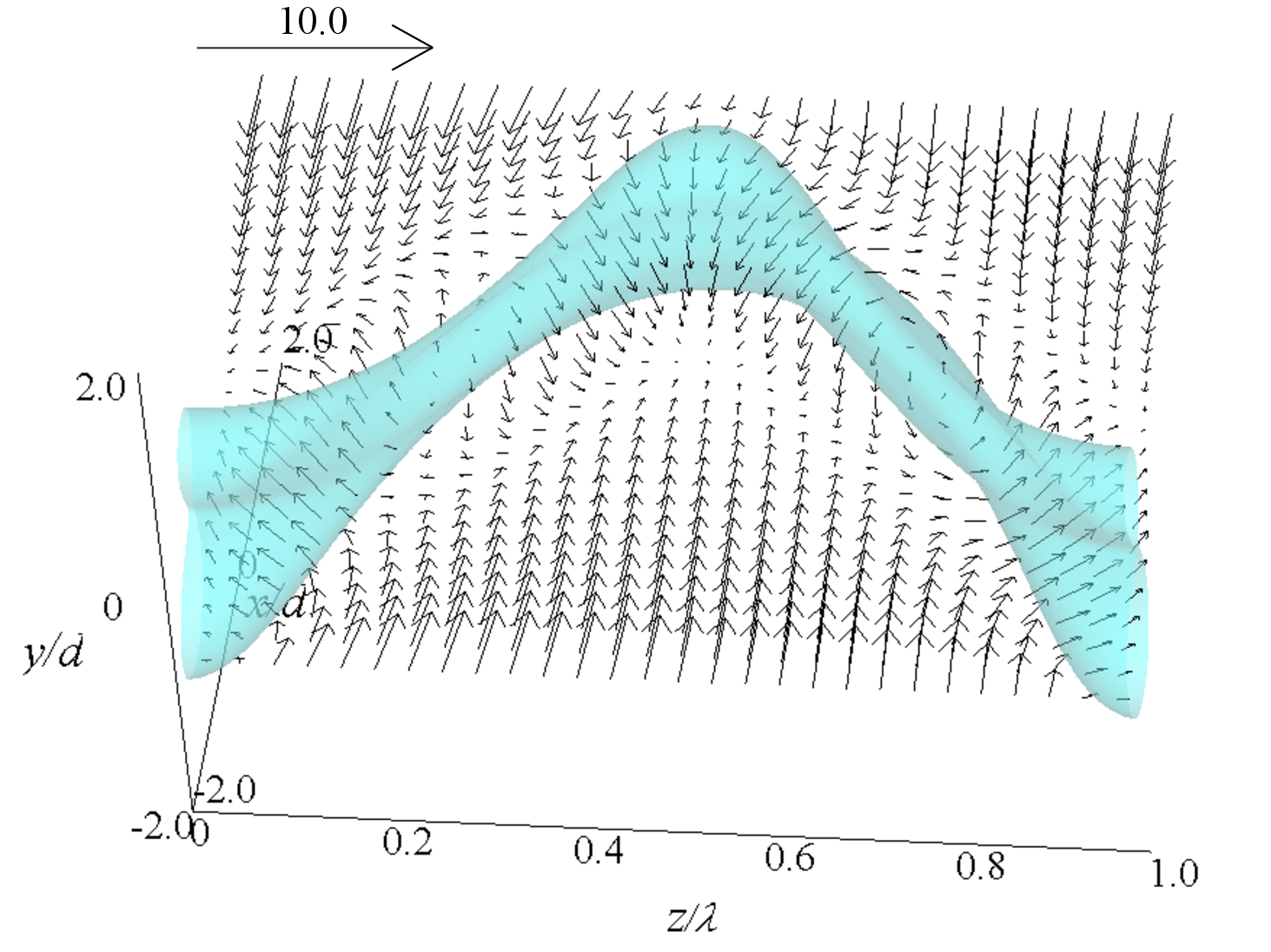} \\
\vspace*{-0.5\baselineskip}
(a)
\end{center}
\end{minipage}
\begin{minipage}{0.48\linewidth}
\begin{center}
\includegraphics[trim=0mm 0mm 0mm 0mm, clip, width=70mm]{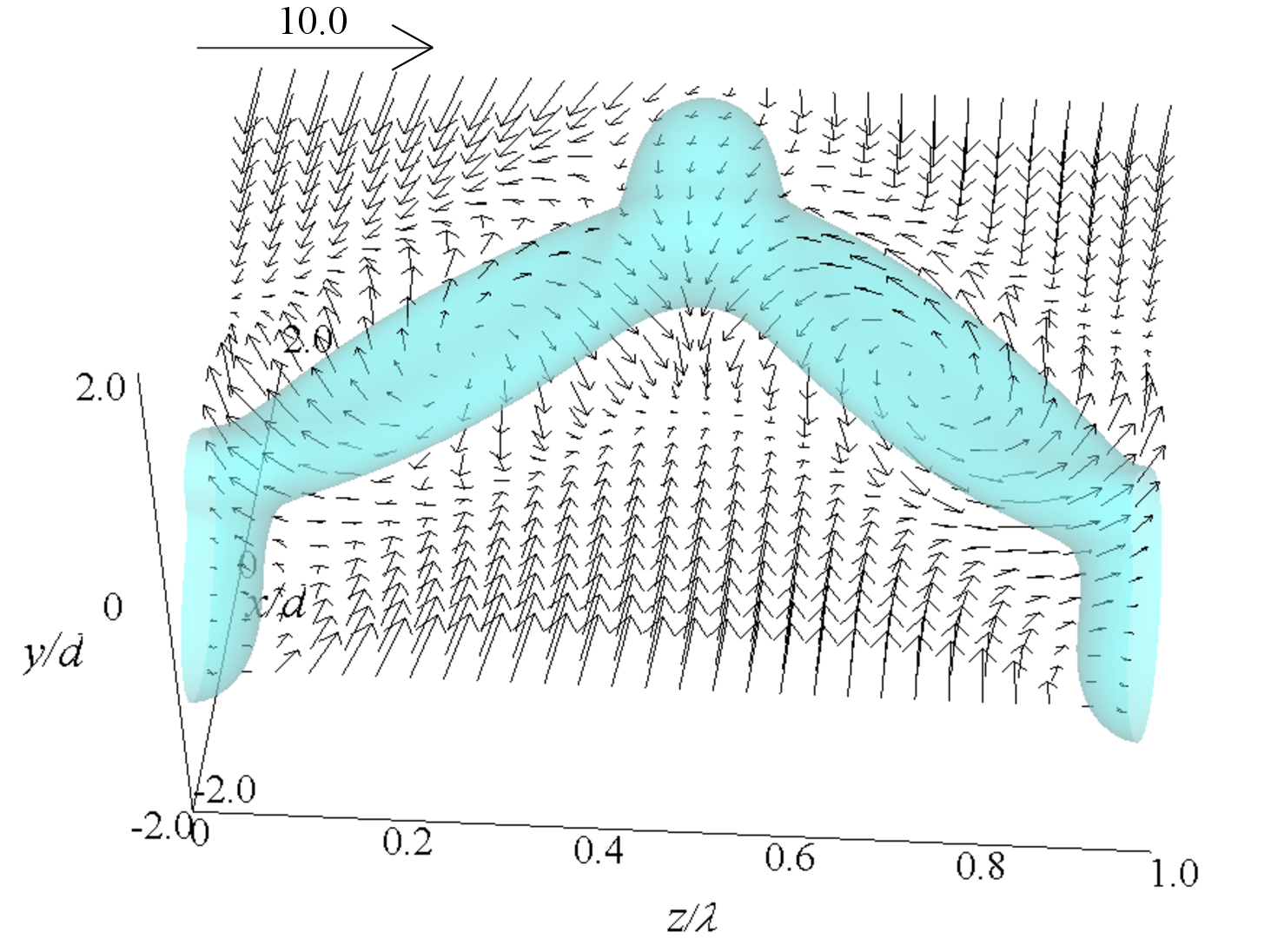} \\
\vspace*{-0.5\baselineskip}
(b)
\end{center}
\end{minipage}
\begin{minipage}{0.48\linewidth}
\begin{center}
\includegraphics[trim=0mm 0mm 0mm 0mm, clip, width=70mm]{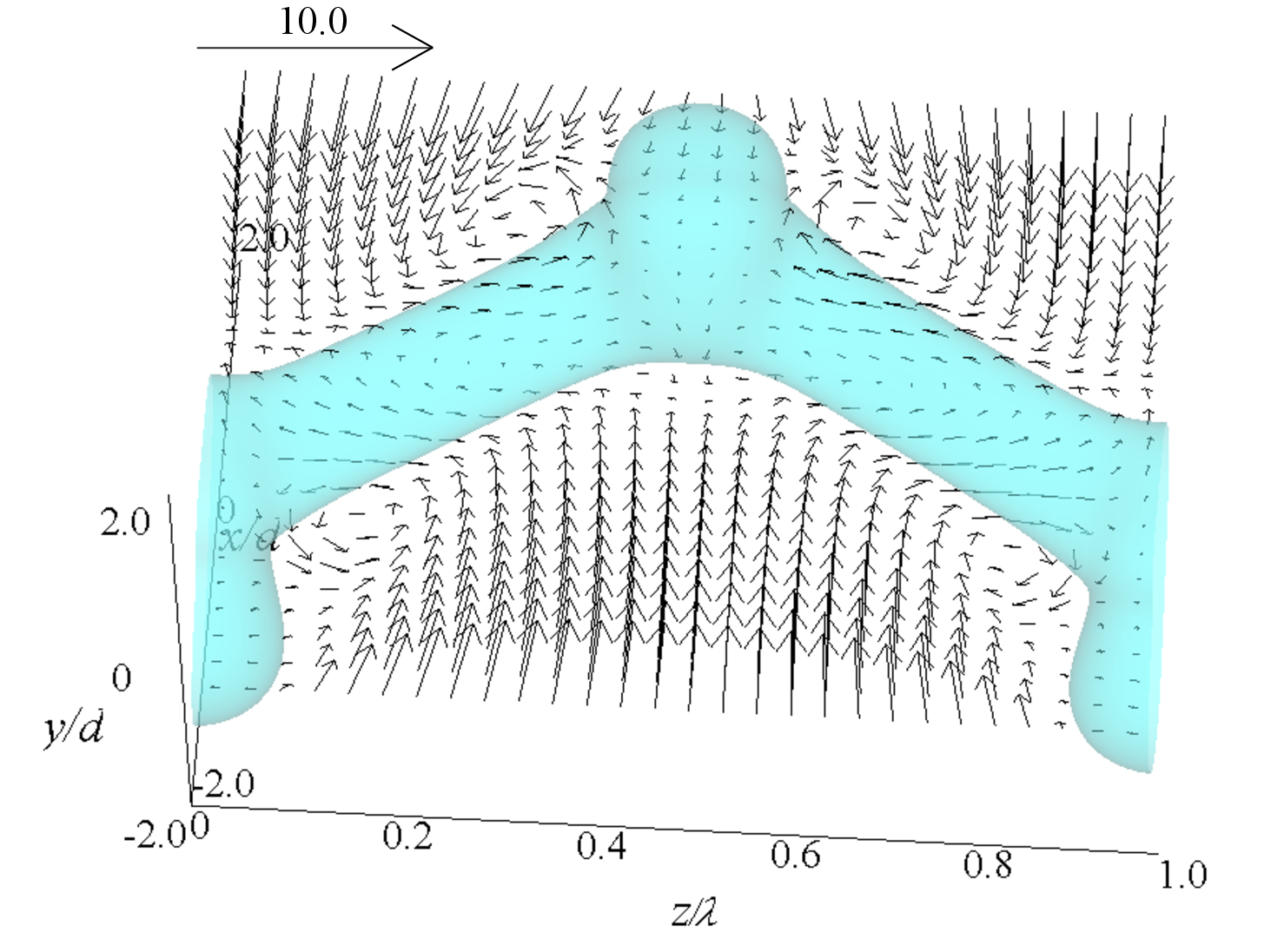} \\
\vspace*{-0.5\baselineskip}
(c)
\end{center}
\end{minipage}
\begin{minipage}{0.48\linewidth}
\begin{center}
\includegraphics[trim=0mm 0mm 0mm 0mm, clip, width=70mm]{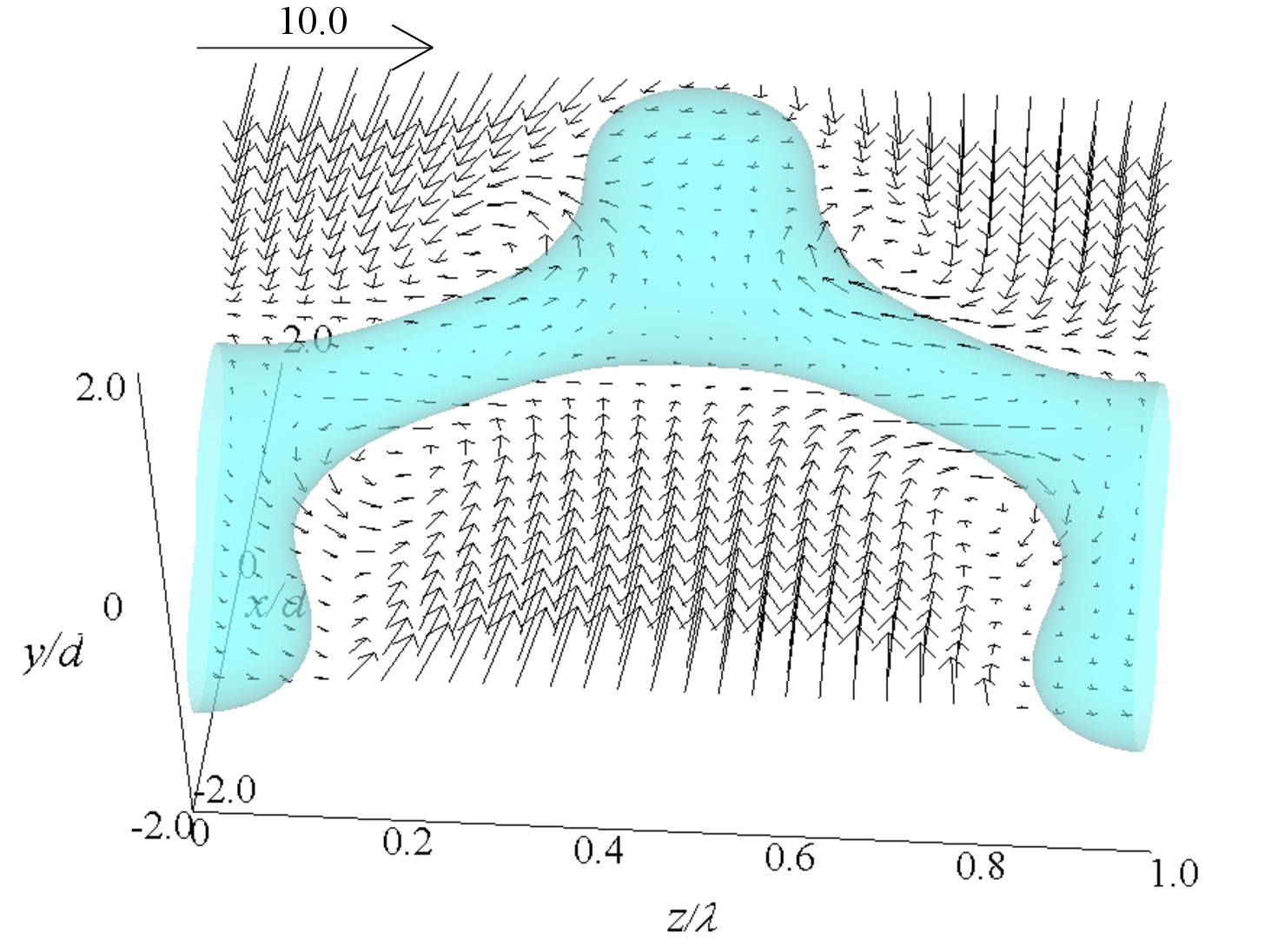} \\
\vspace*{-0.5\baselineskip}
(d)
\end{center}
\end{minipage}
\caption{Ligament interface and velocity vectors: 
(a) $ka = 0.3$, (b) $ka = 0.5$, (c) $ka = 0.7$, and (d) $ka = 0.9$: 
$\Delta U/U_\mathrm{ref} = 17.8$, $T = 1.0$.}
\label{t1vc}
\end{figure}

We investigate the change in the flow field at $T = 1.0$, 
where the difference in the interface shape owing to $ka$ becomes large. 
Figure \ref{t1vc} shows the interface and three-dimensional velocity vectors 
for each $ka$. 
The velocity vector shows the distribution of the cross-section 
across the liquid ligament. 
As $ka$ increases, the bulging parts at $z/\lambda = 0$ and 0.5 
in the liquid ligament grow, and the bulge shape becomes spherical. 
Additionally, the liquid ligament near $z/\lambda = 0.25$ and 0.75 is more constricted. 
At $ka = 0.3$ and 0.5 on the low wavenumber side, 
a three-dimensional vortex is formed at the constriction. 
Near the bulge at $z/\lambda = 0.5$, a flow toward $z/\lambda = 0.5$ occurs. 
Conversely, at $ka = 0.7$ and 0.9 on the high wavenumber side, 
a three-dimensional vortex is formed near the bulge at $z/\lambda = 0$, 0.5, and 1.0. 
Because of these vortices, 
the liquid inside the liquid ligament flows from the constricted part 
to the bulging part, and the bulging part develops.

\begin{figure}[!t]
\begin{center}
\includegraphics[trim=0mm 0mm 0mm 0mm, clip, width=80mm]{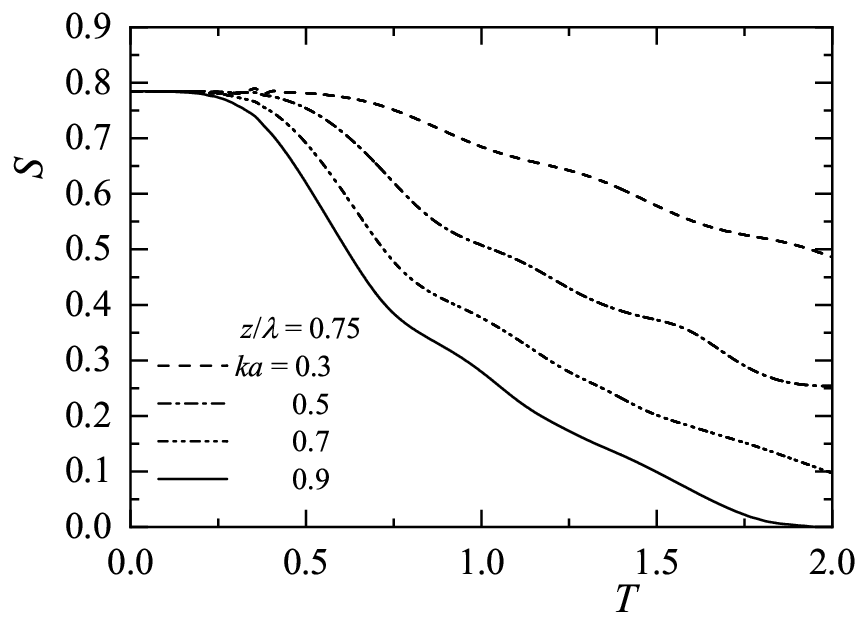}
\end{center}
\vspace*{-1.0\baselineskip}
\caption{Time variations of cross-sectional area at $z/\lambda = 0.75$: 
$\Delta U/U_\mathrm{ref} = 17.8$.}
\label{ka_csarea_time}
\end{figure}

Figure \ref{ka_csarea_time} shows the time variation of the cross-sectional area $S$ 
of the liquid ligament at the cross-section plane $z/\lambda = 0.75$ 
where the liquid ligament constricts. 
The cross-sectional area of the liquid ligament is defined 
using the Heaviside function as follows:
\begin{equation}
   S = \int H(\phi)dxdy.
\end{equation}
For all $ka$, the cross-sectional area of the liquid ligament at $z/\lambda = 0.75$ 
decreases with time. 
As $ka$ increases, the cross-sectional area decreases rapidly. 
As shown in Fig. \ref{time_aw}, 
with an increase in $ka$, the liquid movement along the central axis of the liquid ligament is promoted. 
As shown in Fig. \ref{t1vc}, on the high wavenumber side, 
the vortices that enhance the development of the bulge are formed near the interface. 
An increase in $ka$ shortens the wavelength; 
hence, the liquid inside the ligament shortly moves from the constriction to the bulge. 
Therefore, as $ka$ increases, the cross-sectional area of the liquid ligament 
at $z/\lambda = 0.75$ becomes small in a short time, 
and the breakup time of the liquid ligament becomes short.

\begin{figure}[!t]
\centering
\begin{minipage}{0.48\linewidth}
\begin{center}
\includegraphics[trim=0mm 0mm 0mm 0mm, clip, width=80mm]{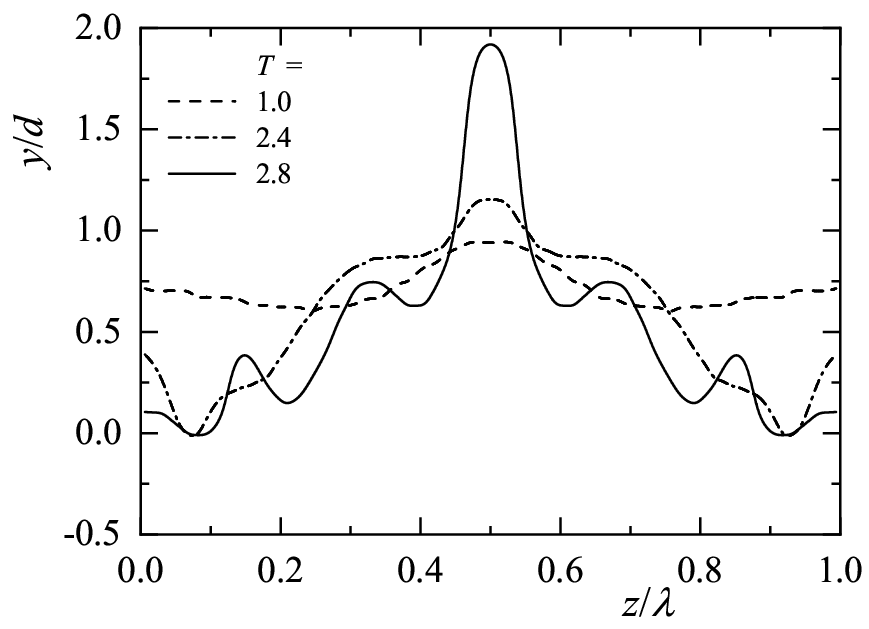} \\
\vspace*{-0.5\baselineskip}
(a)
\end{center}
\end{minipage}
\hspace*{0.02\linewidth}
\begin{minipage}{0.48\linewidth}
\begin{center}
\includegraphics[trim=0mm 0mm 0mm 0mm, clip, width=80mm]{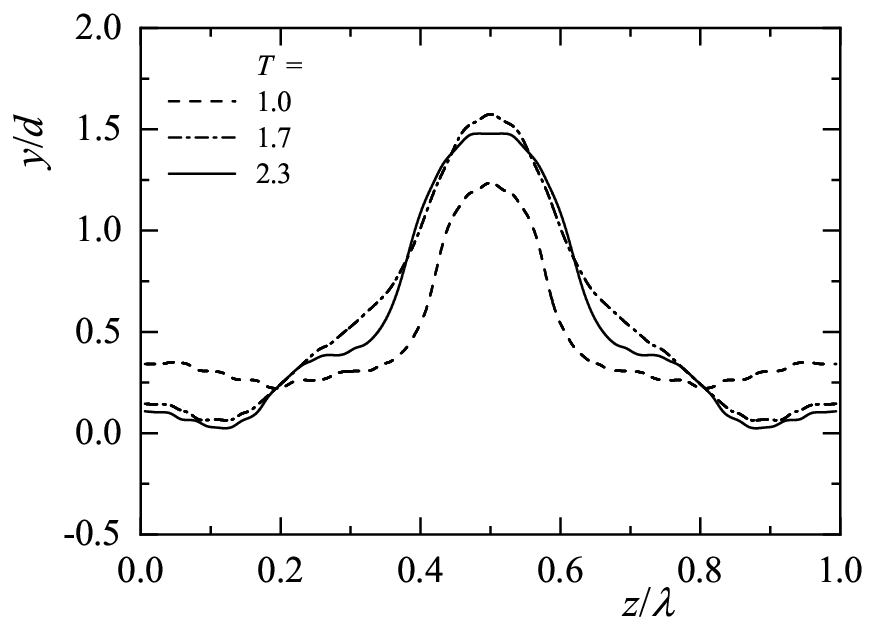} \\
\vspace*{-0.5\baselineskip}
(b)
\end{center}
\end{minipage}
\begin{minipage}{0.48\linewidth}
\begin{center}
\includegraphics[trim=0mm 0mm 0mm 0mm, clip, width=80mm]{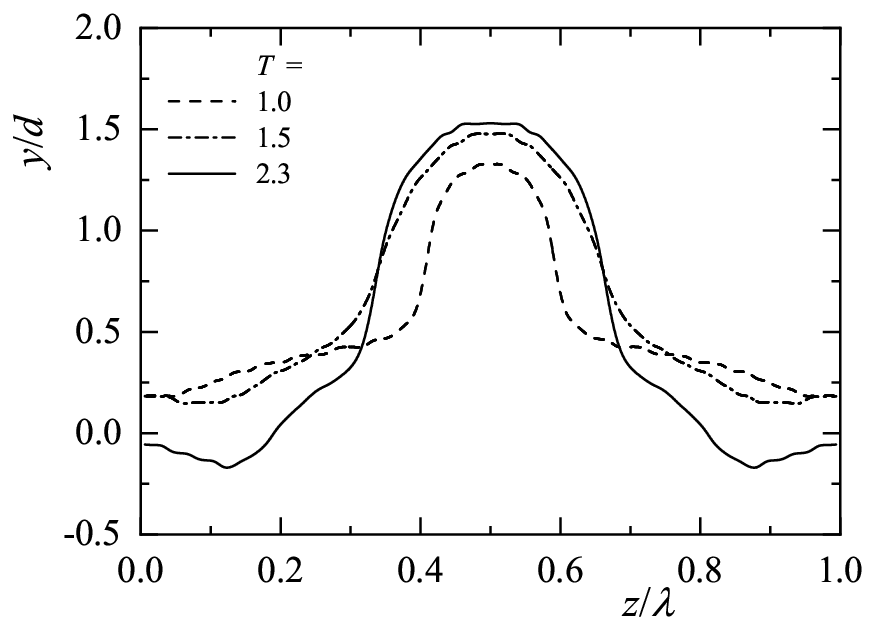} \\
\vspace*{-0.5\baselineskip}
(c)
\end{center}
\end{minipage}
\hspace*{0.02\linewidth}
\begin{minipage}{0.48\linewidth}
\begin{center}
\includegraphics[trim=0mm 0mm 0mm 0mm, clip, width=80mm]{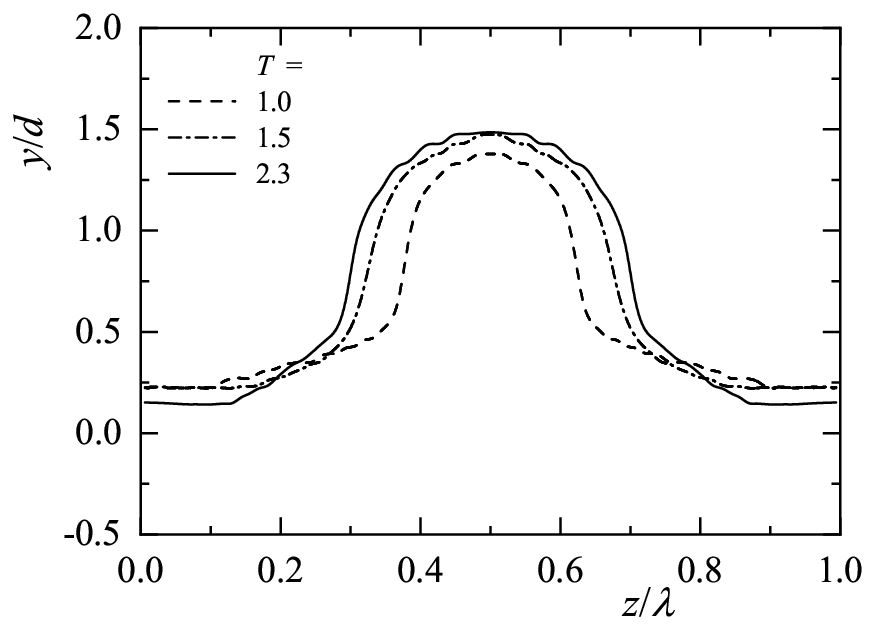} \\
\vspace*{-0.5\baselineskip}
(d)
\end{center}
\end{minipage}
\caption{Time variations of ligament interface for various wavenumbers: 
(a) $ka = 0.3$, (b) $ka = 0.5$, (c) $ka = 0.7$, and (d) $ka = 0.9$: 
$\Delta U/U_\mathrm{ref} = 17.8$.}
\label{timeint}
\end{figure}

As mentioned, 
a liquid ligament breaks up faster in an initial disturbance with a high wavenumber 
than a low wavenumber. 
Conversely, for low wavenumbers, 
as shown in Figs. \ref{u30k3_time}(f) and \ref{u30k5_time}(d), 
new turbulence with a higher wavenumber than the initial disturbance occurs at the interface. 
To quantitatively investigate the difference in the interface shape owing to $ka$, 
Fig. \ref{timeint} shows the time variation in the interface for each $ka$. 
The distribution is the interface at the position $y/d$ 
where the interface is the highest in the $x$--$y$ cross-section. 
At all times for $ka = 0.9$, the interface shape is similar, 
and the initial disturbance grows. 
At $T = 1.0$ and 1.5 for $ka = 0.5$ and 0.7, respectively, 
the parts at $z/\lambda = 0$ and 0.5 of the liquid ligament bulge toward the gas side. 
That is, turbulence with twice the wavenumber of the initial disturbance occurs 
at the interface. 
Moreover, the interface expands with time toward the gas side 
even around $z/\lambda = 0.25$ and 0.75, 
and at $T = 2.3$, turbulence with four times the initial disturbance occurs 
at the interface. 
At $T = 1.0$ for $ka = 0.3$, the parts at $z/\lambda = 0$ and 0.5 of the liquid ligament bulge 
toward the gas side, 
and turbulence with twice the initial disturbance occurs at the interface. 
At $T = 2.4$, the interface bulges slightly toward the gas side 
near $z/\lambda = 0.12$, 0.35, 0.65, and 0.88. 
The raised interface develops over time. 
At $T = 2.8$, near $z/\lambda = 0.15$, 0.35, 0.65, and 0.85, 
the interface rises more toward the gas side, 
and turbulence with a wavenumber six times that of the initial disturbance occurs 
at the interface. 
For $ka = 0.3$, the uplift of the interface toward the gas side is the largest, 
and the nonlinear effect appears most strongly.

It was found from the above results that the liquid ligament deforms more complicatedly 
for the initial disturbance with a low wavenumber than a high wavenumber 
owing to a nonlinear effect. 
Therefore, we considered that the deformation of the liquid ligament would be accelerated 
by adding the high wavenumber disturbance to the low wavenumber disturbance. 
For $\Delta U/U_\mathrm{ref} = 17.8$ and $ka = 0.3$, 
we investigate the effects of low and high wavenumber disturbances 
on the deformation and splitting of the ligament.

\begin{figure}[!t]
\centering
\begin{minipage}{0.32\linewidth}
\begin{center}
\includegraphics[trim=0mm 0mm 0mm 0mm, clip, width=55mm]{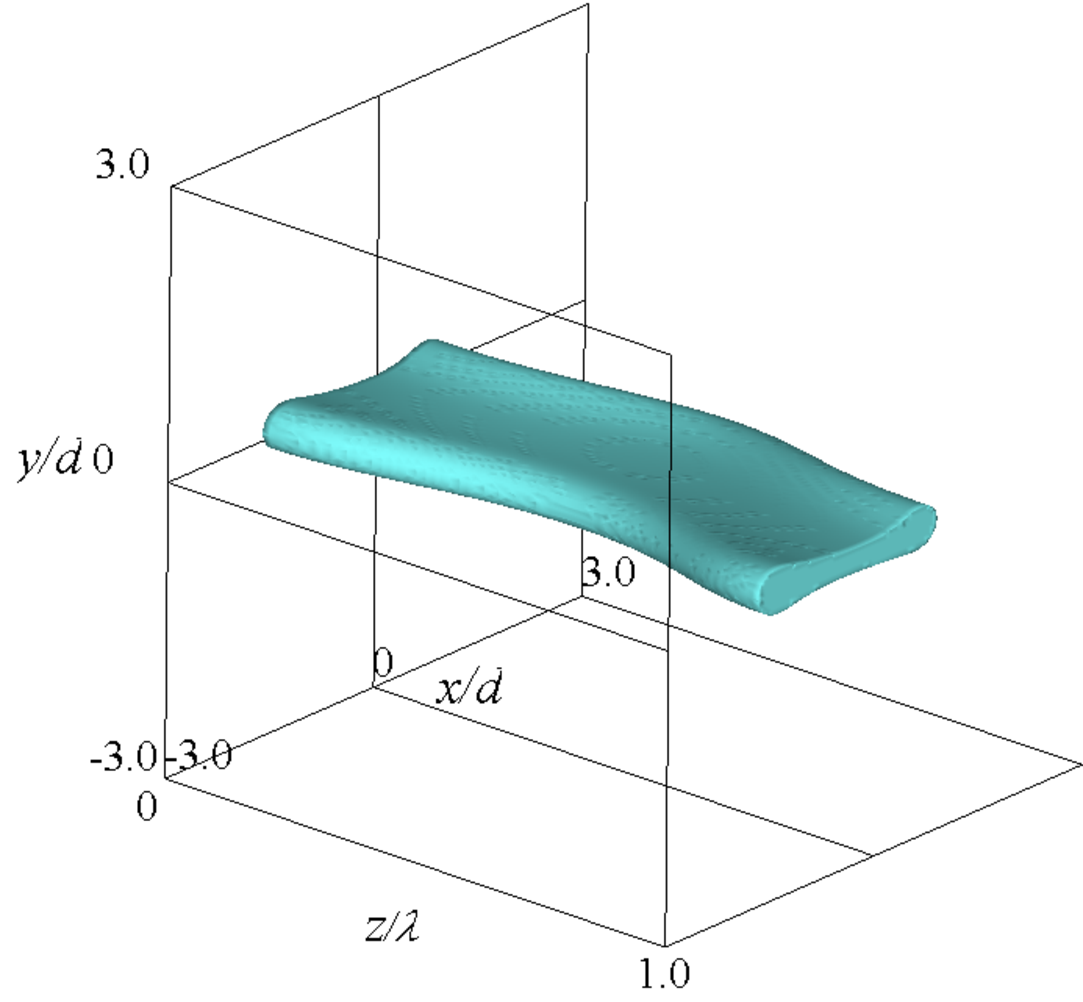} \\
\vspace*{-0.5\baselineskip}
(a)
\end{center}
\end{minipage}
\begin{minipage}{0.32\linewidth}
\begin{center}
\includegraphics[trim=0mm 0mm 0mm 0mm, clip, width=55mm]{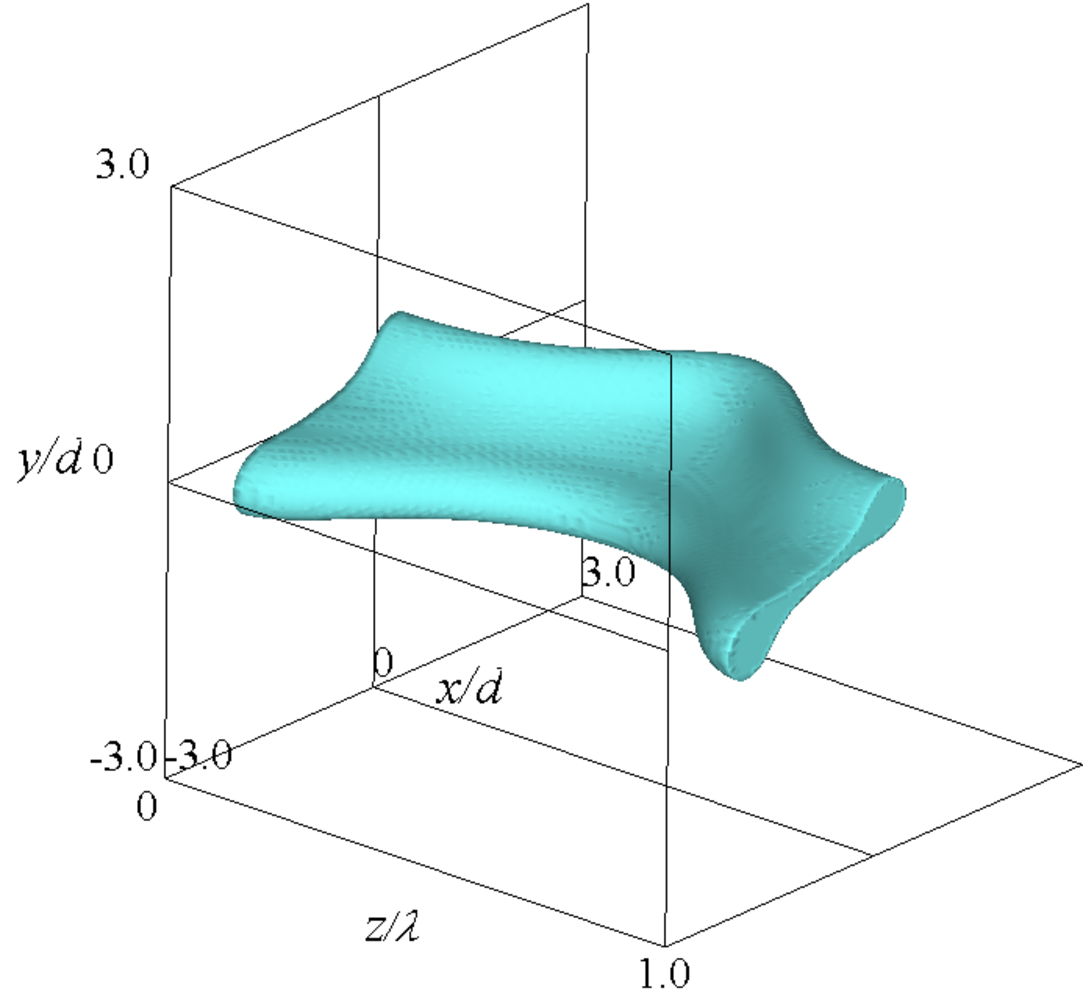} \\
\vspace*{-0.5\baselineskip}
(b)
\end{center}
\end{minipage}
\begin{minipage}{0.32\linewidth}
\begin{center}
\includegraphics[trim=0mm 0mm 0mm 0mm, clip, width=55mm]{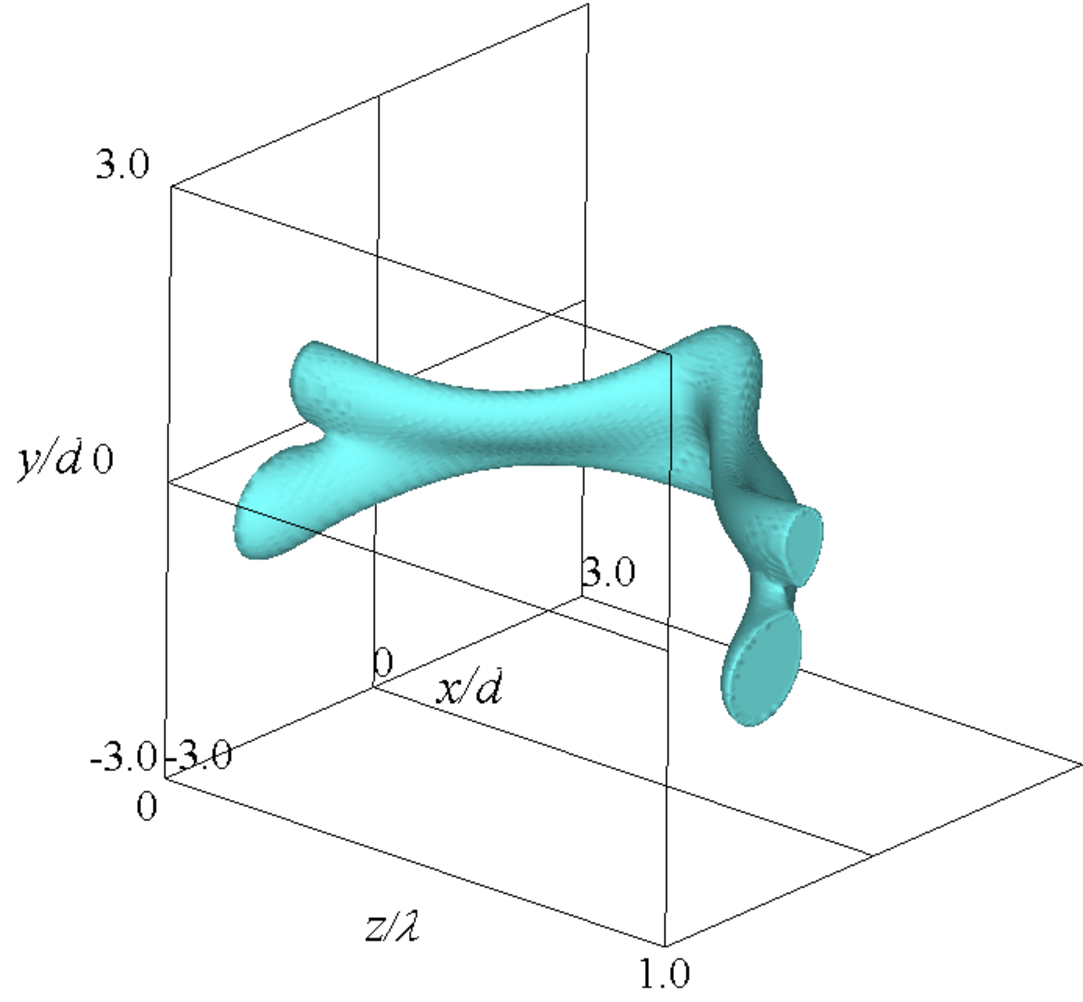} \\
\vspace*{-0.5\baselineskip}
(c)
\end{center}
\end{minipage}

\vspace*{0.5\baselineskip}
\begin{minipage}{0.32\linewidth}
\begin{center}
\includegraphics[trim=0mm 0mm 0mm 0mm, clip, width=55mm]{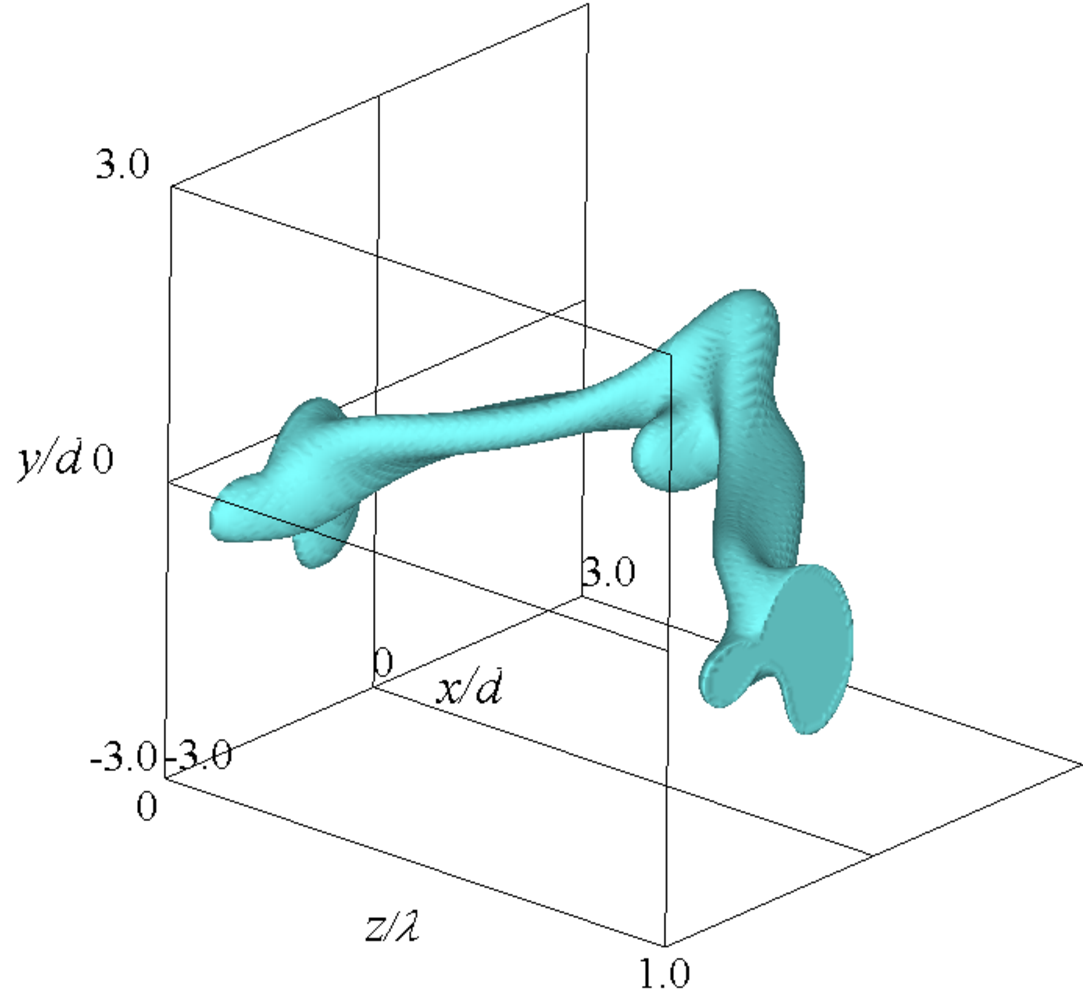} \\
\vspace*{-0.5\baselineskip}
(d)
\end{center}
\end{minipage}
\begin{minipage}{0.32\linewidth}
\begin{center}
\includegraphics[trim=0mm 0mm 0mm 0mm, clip, width=55mm]{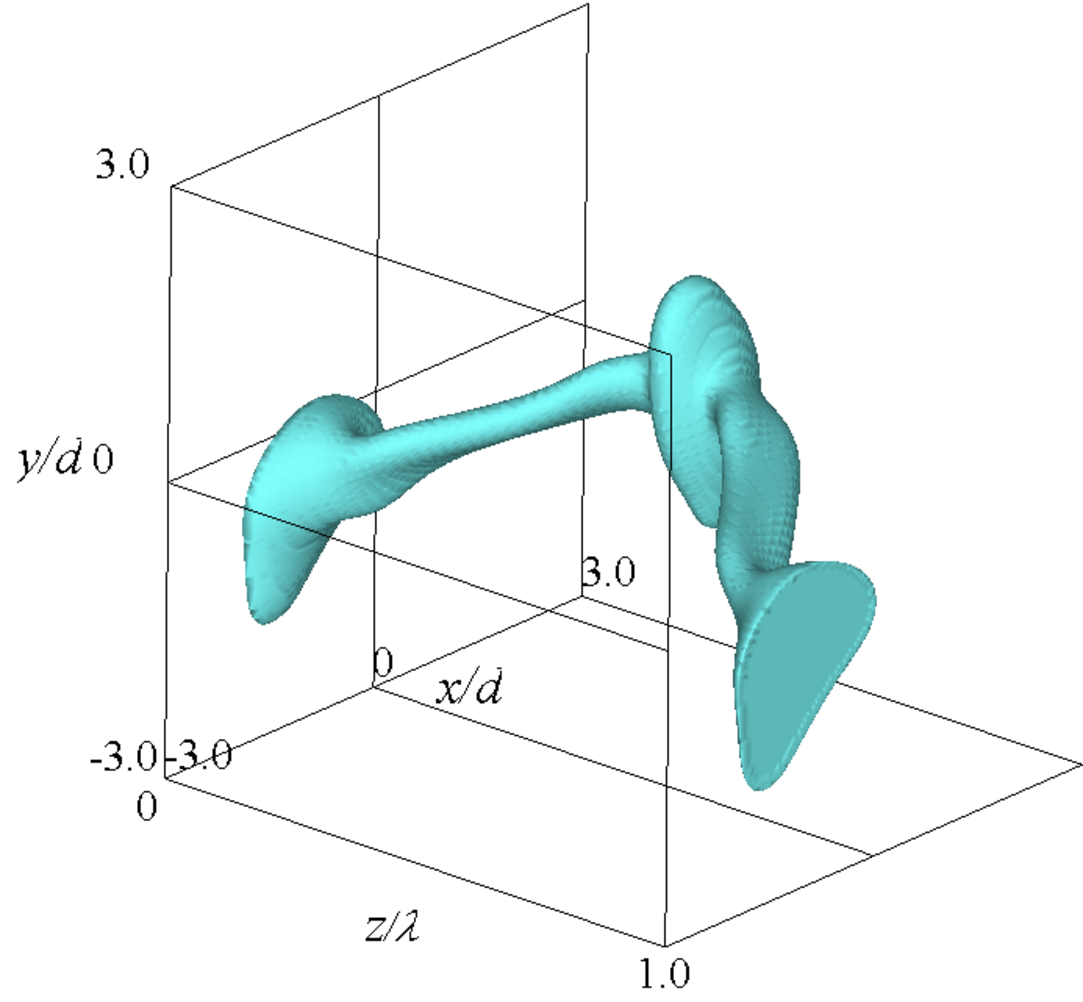} \\
\vspace*{-0.5\baselineskip}
(e)
\end{center}
\end{minipage}
\begin{minipage}{0.32\linewidth}
\begin{center}
\includegraphics[trim=0mm 0mm 0mm 0mm, clip, width=55mm]{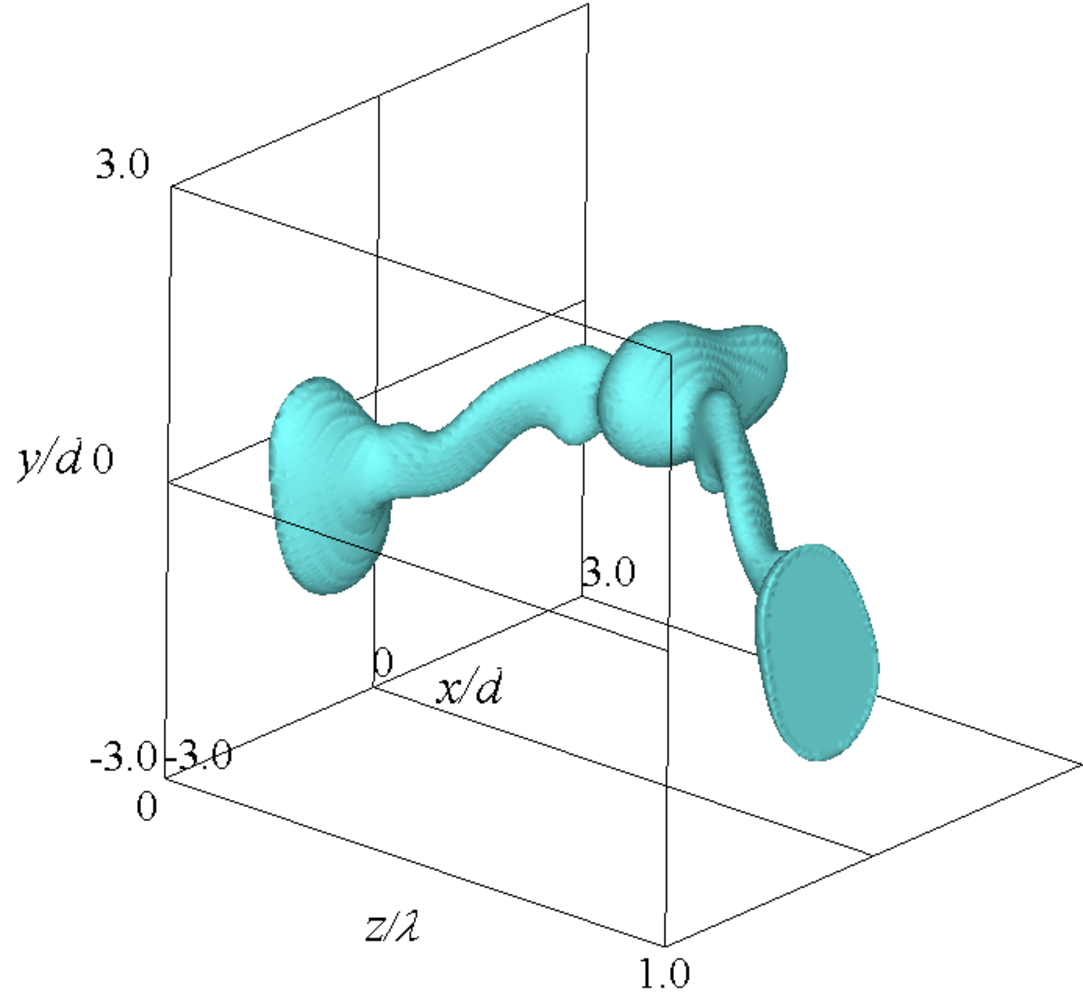} \\
\vspace*{-0.5\baselineskip}
(f)
\end{center}
\end{minipage}

\vspace*{0.5\baselineskip}
\begin{minipage}{0.32\linewidth}
\begin{center}
\includegraphics[trim=0mm 0mm 0mm 0mm, clip, width=55mm]{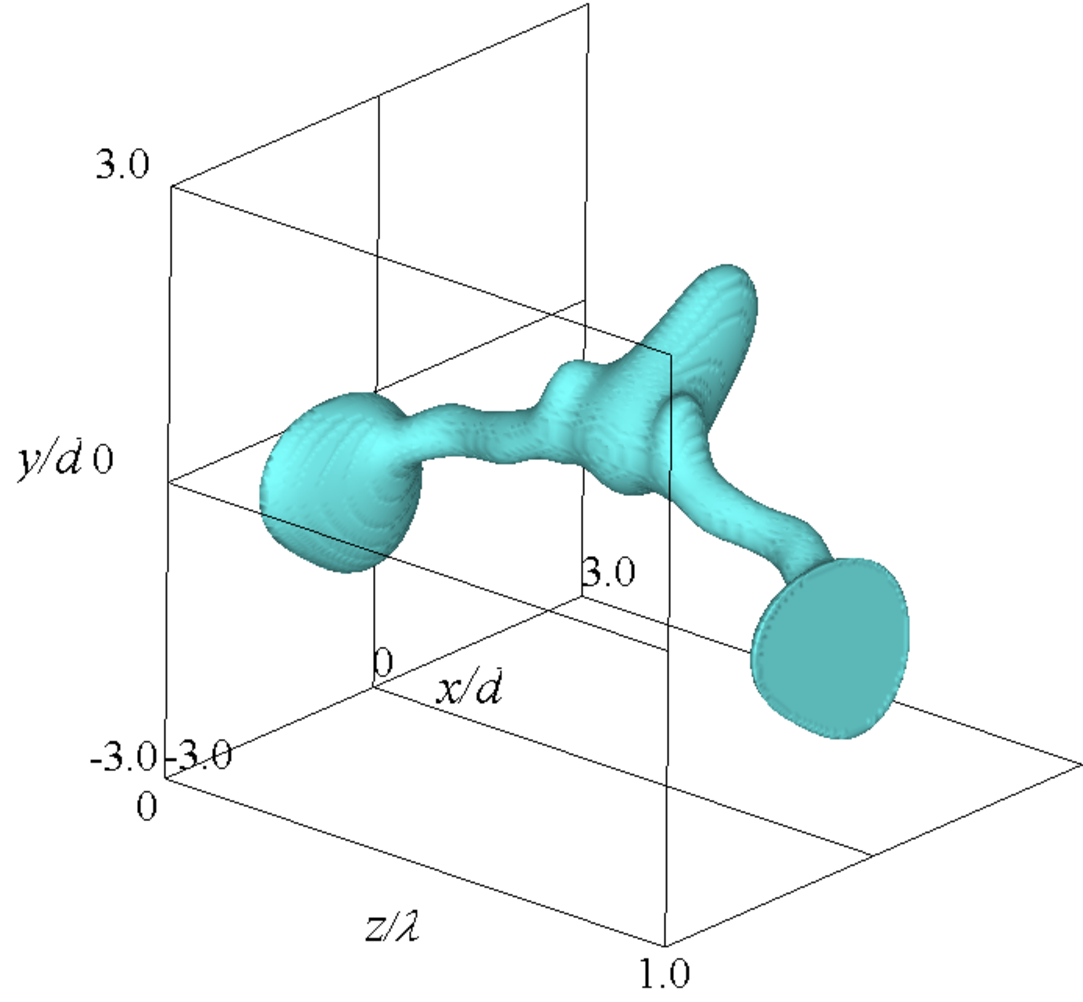} \\
\vspace*{-0.5\baselineskip}
(g)
\end{center}
\end{minipage}
\begin{minipage}{0.32\linewidth}
\begin{center}
\includegraphics[trim=0mm 0mm 0mm 0mm, clip, width=55mm]{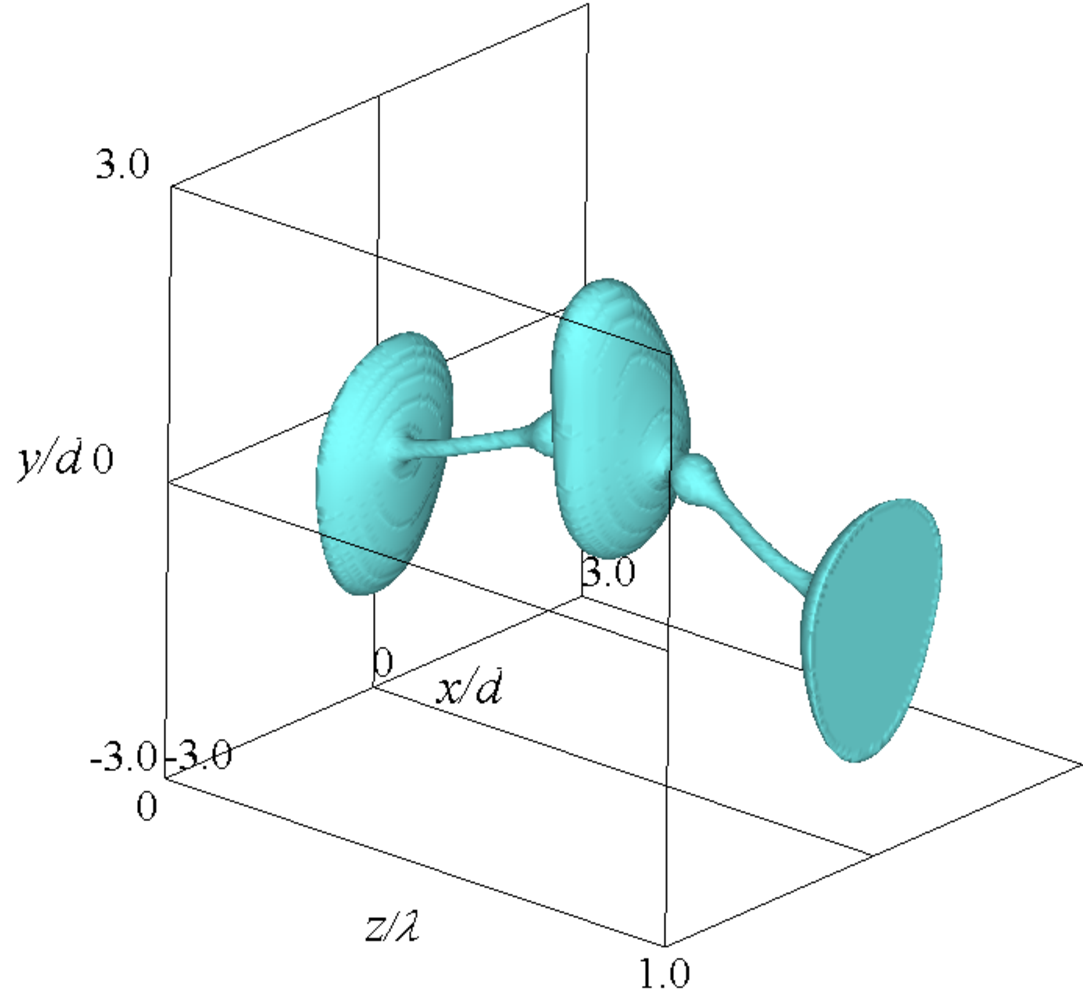} \\
\vspace*{-0.5\baselineskip}
(h)
\end{center}
\end{minipage}
\begin{minipage}{0.32\linewidth}
\begin{center}
\includegraphics[trim=0mm 0mm 0mm 0mm, clip, width=55mm]{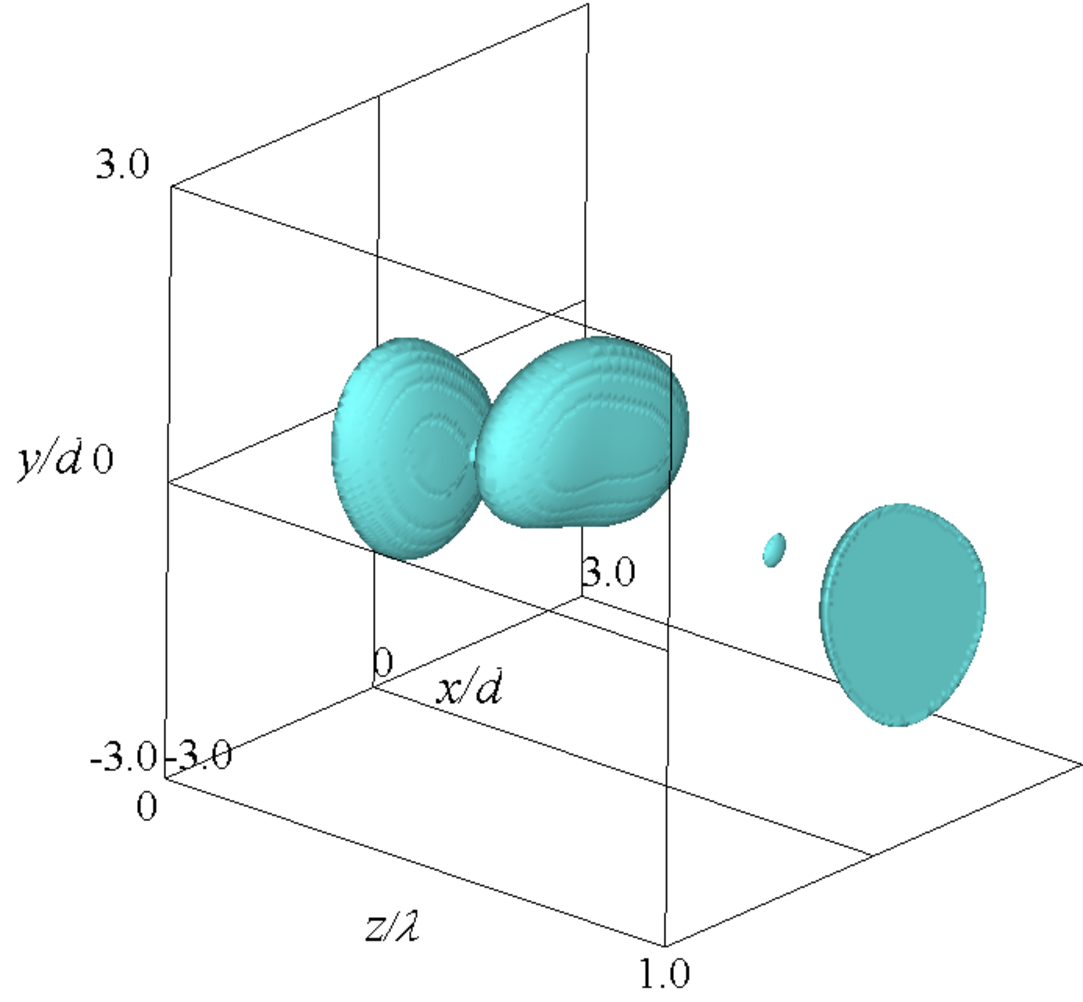} \\
\vspace*{-0.5\baselineskip}
(i)
\end{center}
\end{minipage}
\caption{Time variations of ligament interface: 
(a) $T = 0.3$, (b) $T = 0.6$, (c) $T = 1.0$, (d) $T = 1.6$, 
(e) $T = 2.0$, (f) $T = 2.6$, (g) $T = 3.0$, (h) $T = 3.8$, and (i) $T = 4.6$: 
$\Delta U/U_\mathrm{ref} = 17.8$, $ka = 0.3$, 
$\eta_r = 0.9$, $k_r =2.0$.}
\label{u30k3f2_time}
\end{figure}

\begin{figure}[!t]
\centering
\begin{minipage}{0.32\linewidth}
\begin{center}
\includegraphics[trim=0mm 0mm 0mm 0mm, clip, width=55mm]{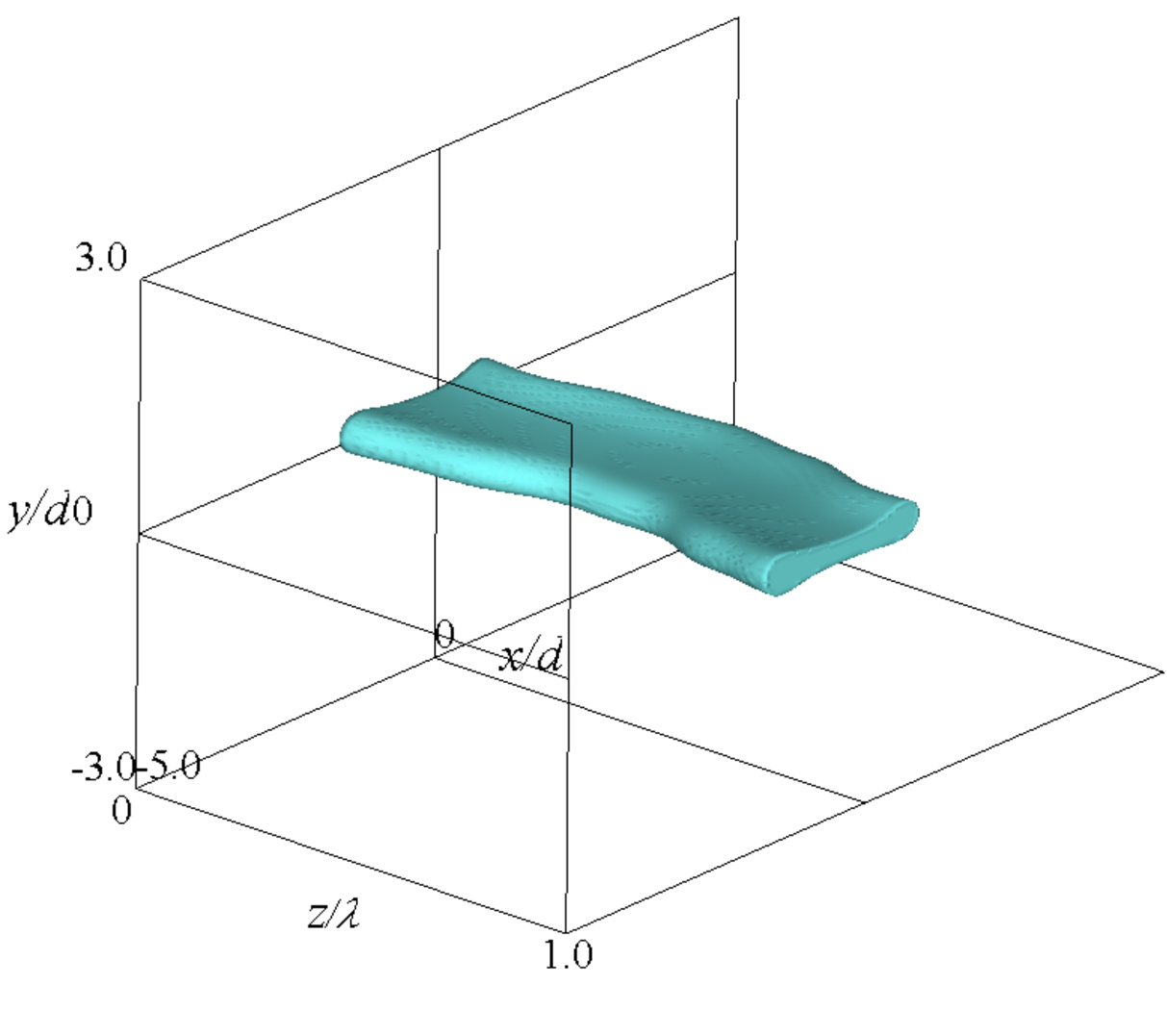} \\
\vspace*{-0.5\baselineskip}
(a)
\end{center}
\end{minipage}
\begin{minipage}{0.32\linewidth}
\begin{center}
\includegraphics[trim=0mm 0mm 0mm 0mm, clip, width=55mm]{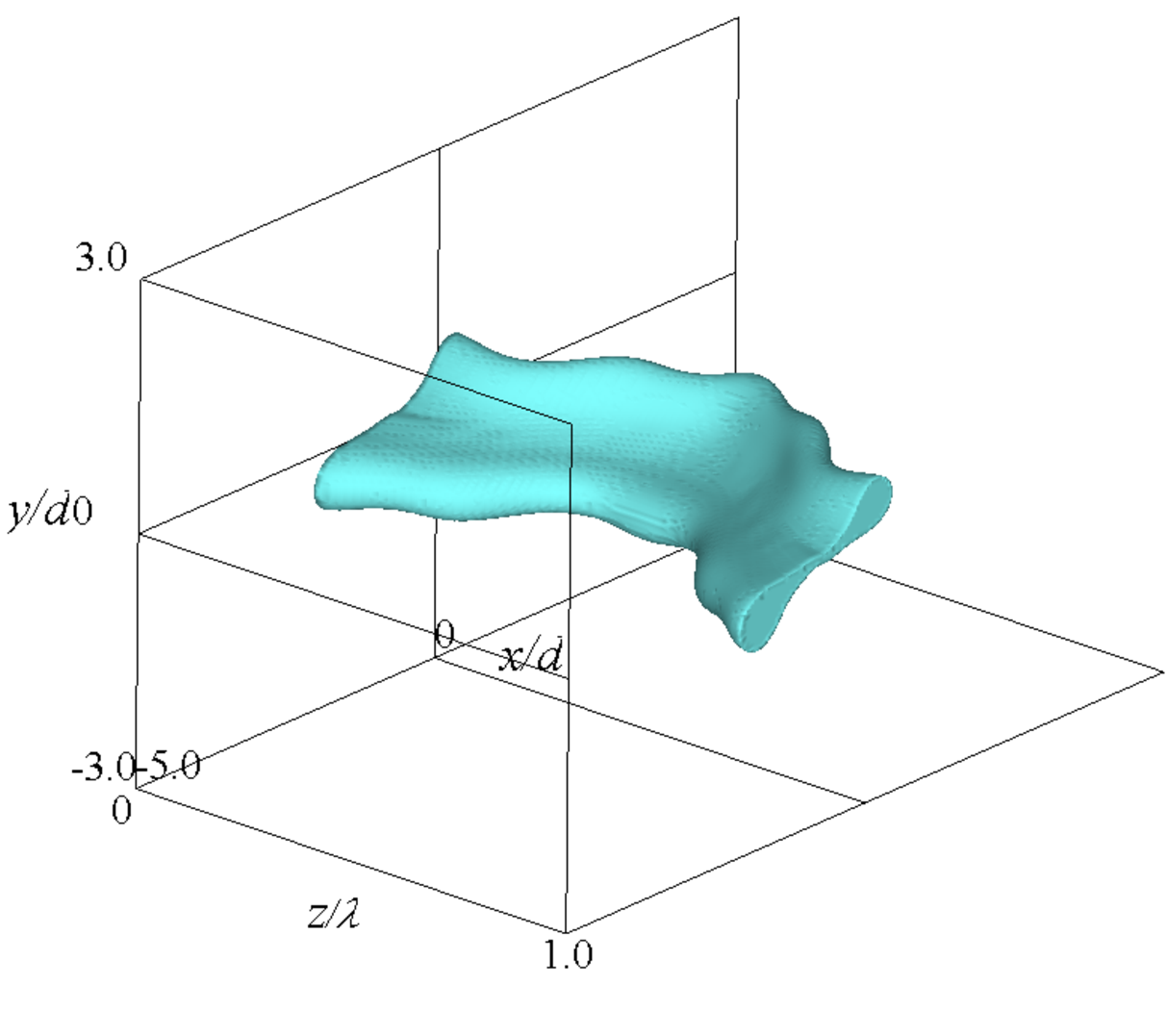} \\
\vspace*{-0.5\baselineskip}
(b)
\end{center}
\end{minipage}
\begin{minipage}{0.32\linewidth}
\begin{center}
\includegraphics[trim=0mm 0mm 0mm 0mm, clip, width=55mm]{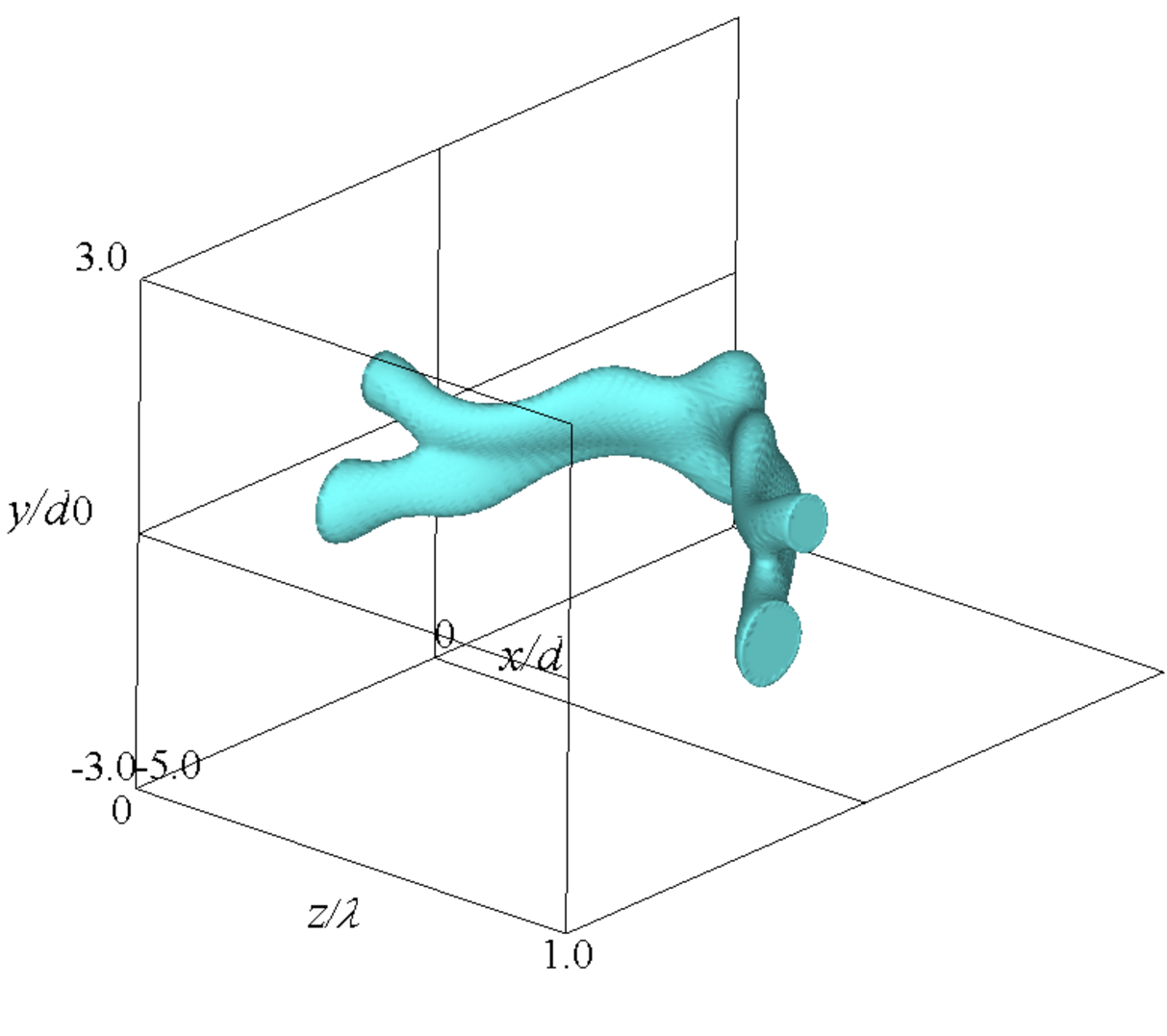} \\
\vspace*{-0.5\baselineskip}
(c)
\end{center}
\end{minipage}

\vspace*{0.5\baselineskip}
\begin{minipage}{0.32\linewidth}
\begin{center}
\includegraphics[trim=0mm 0mm 0mm 0mm, clip, width=55mm]{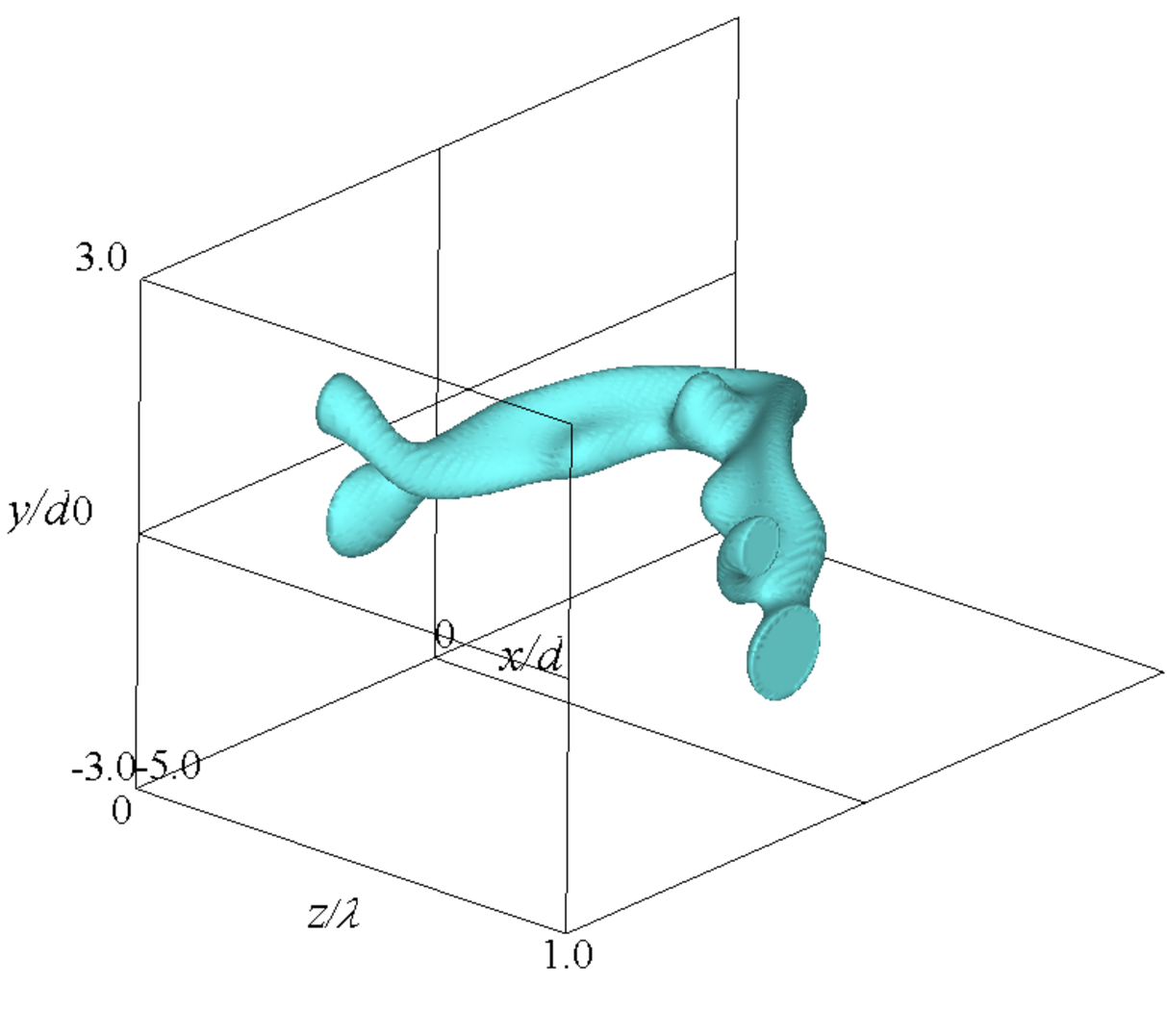} \\
\vspace*{-0.5\baselineskip}
(d)
\end{center}
\end{minipage}
\begin{minipage}{0.32\linewidth}
\begin{center}
\includegraphics[trim=0mm 0mm 0mm 0mm, clip, width=55mm]{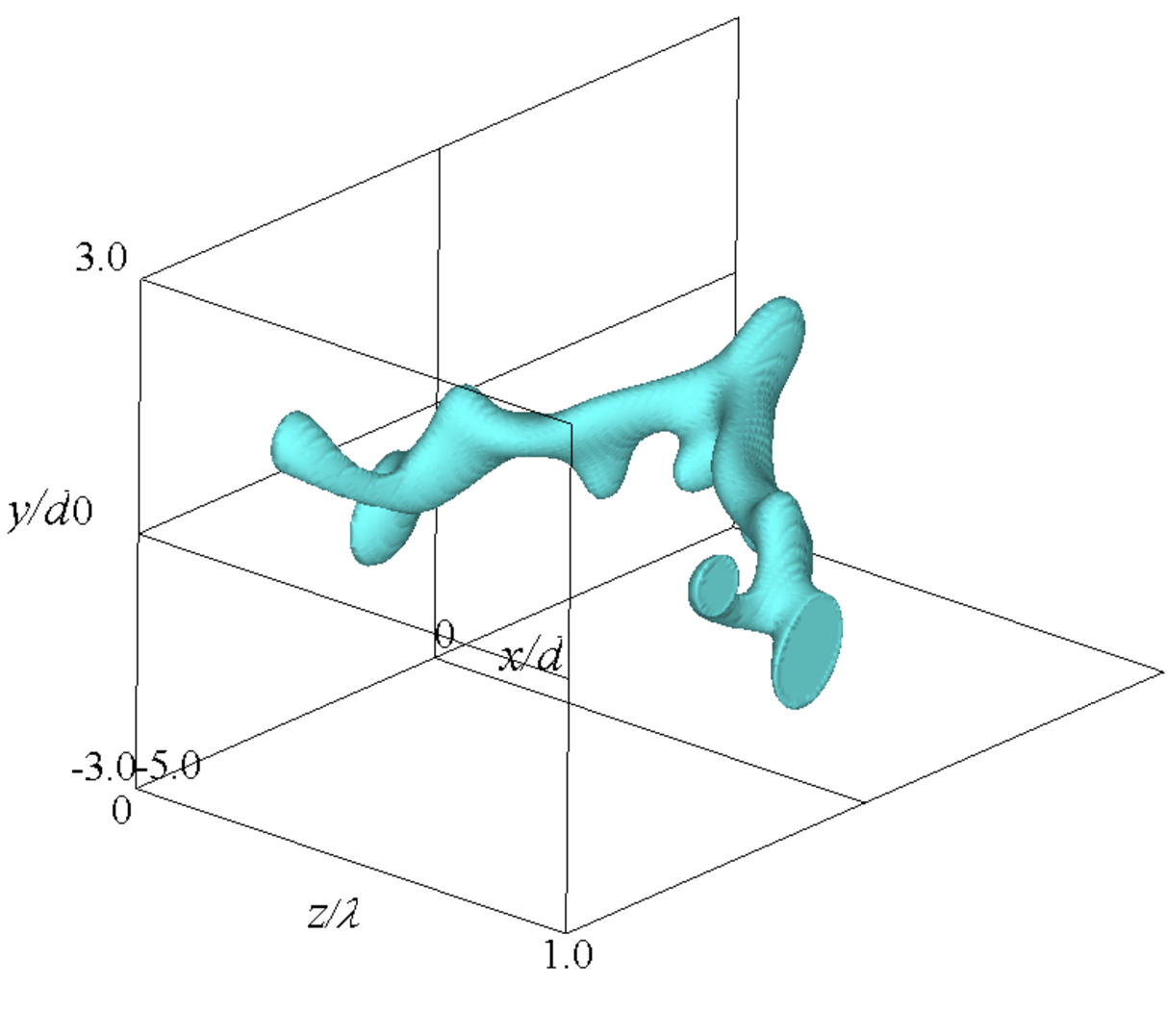} \\
\vspace*{-0.5\baselineskip}
(e)
\end{center}
\end{minipage}
\begin{minipage}{0.32\linewidth}
\begin{center}
\includegraphics[trim=0mm 0mm 0mm 0mm, clip, width=55mm]{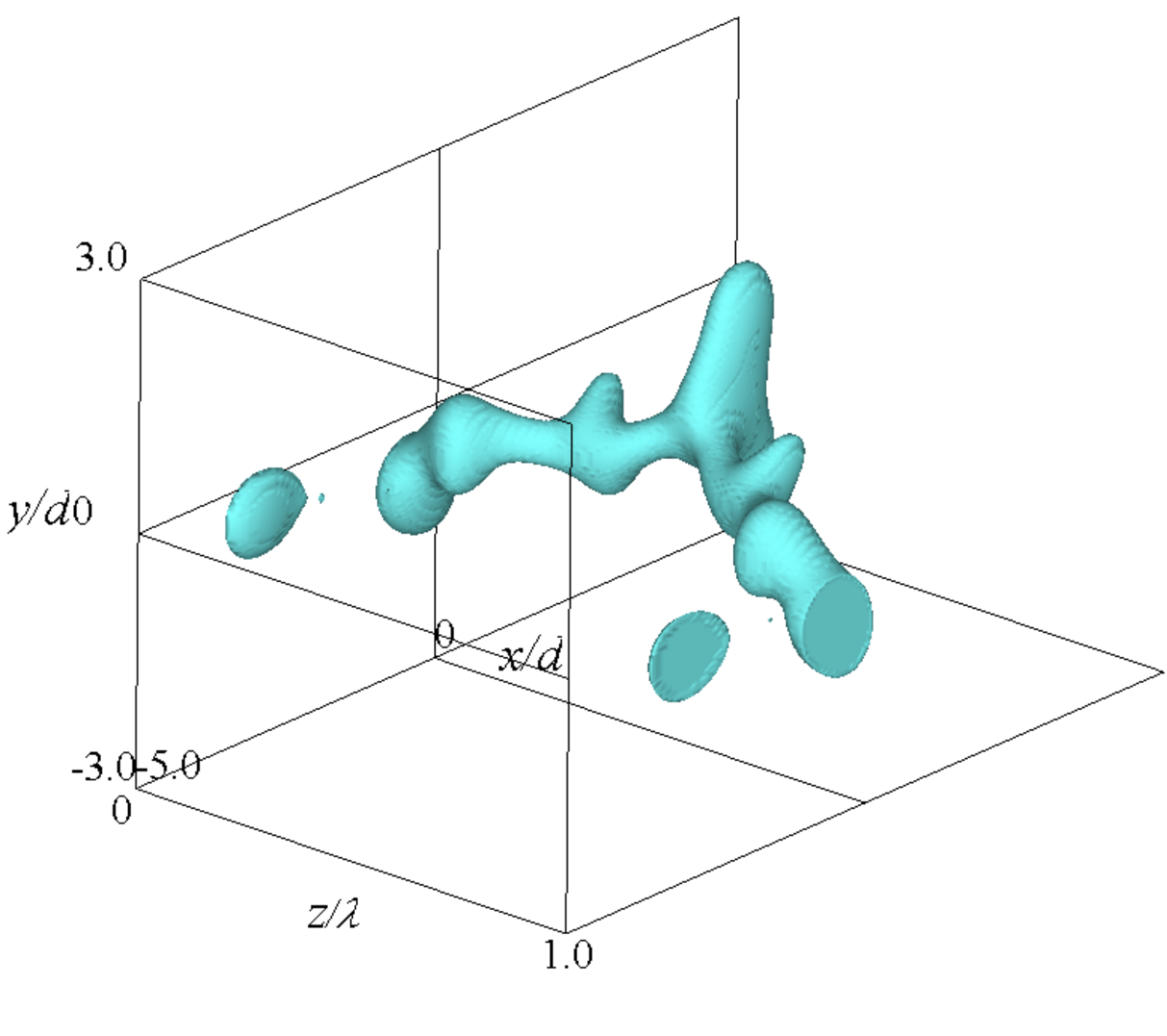} \\
\vspace*{-0.5\baselineskip}
(f)
\end{center}
\end{minipage}

\vspace*{0.5\baselineskip}
\begin{minipage}{0.32\linewidth}
\begin{center}
\includegraphics[trim=0mm 0mm 0mm 0mm, clip, width=55mm]{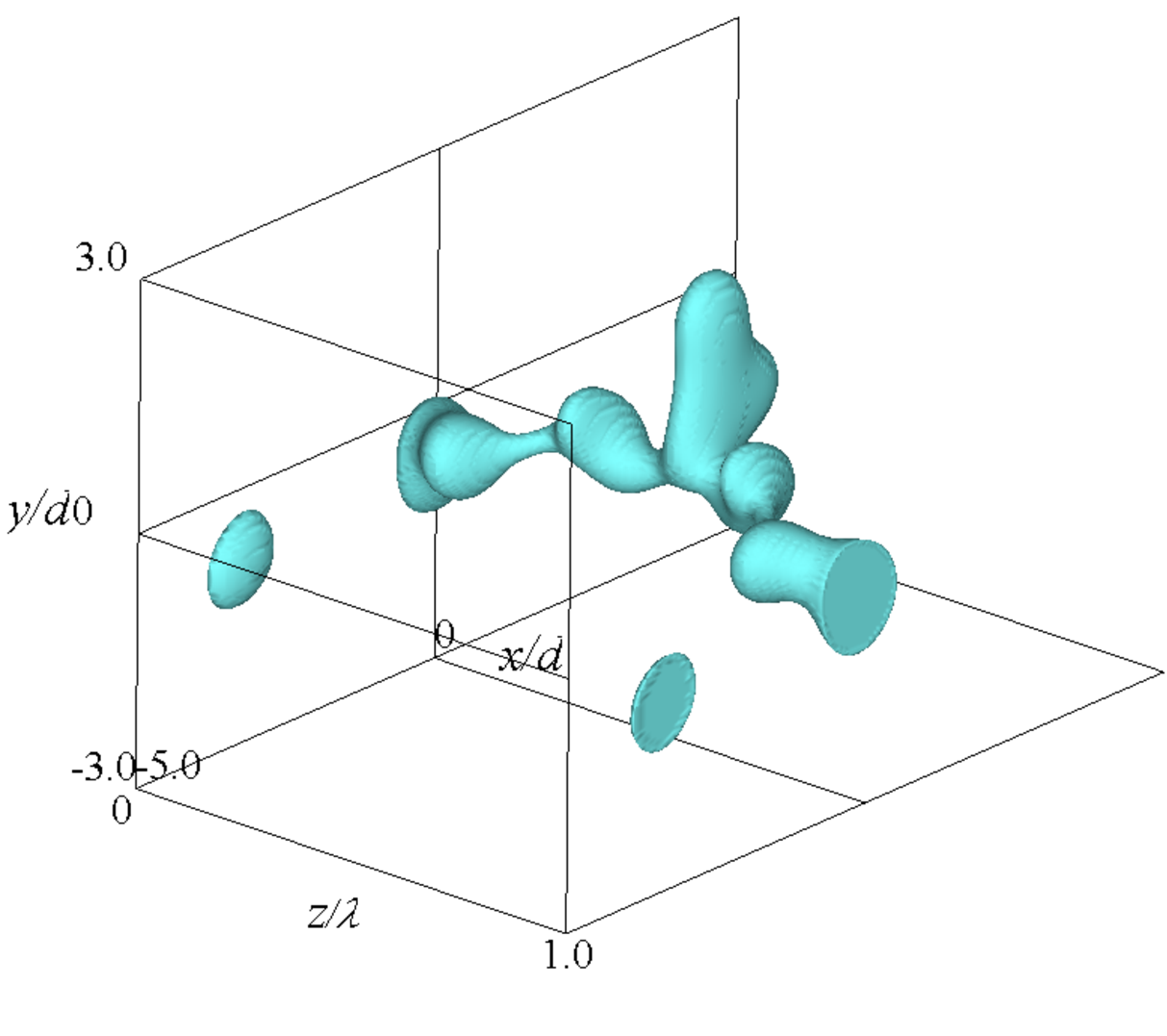} \\
\vspace*{-0.5\baselineskip}
(g)
\end{center}
\end{minipage}
\centering
\begin{minipage}{0.32\linewidth}
\begin{center}
\includegraphics[trim=0mm 0mm 0mm 0mm, clip, width=55mm]{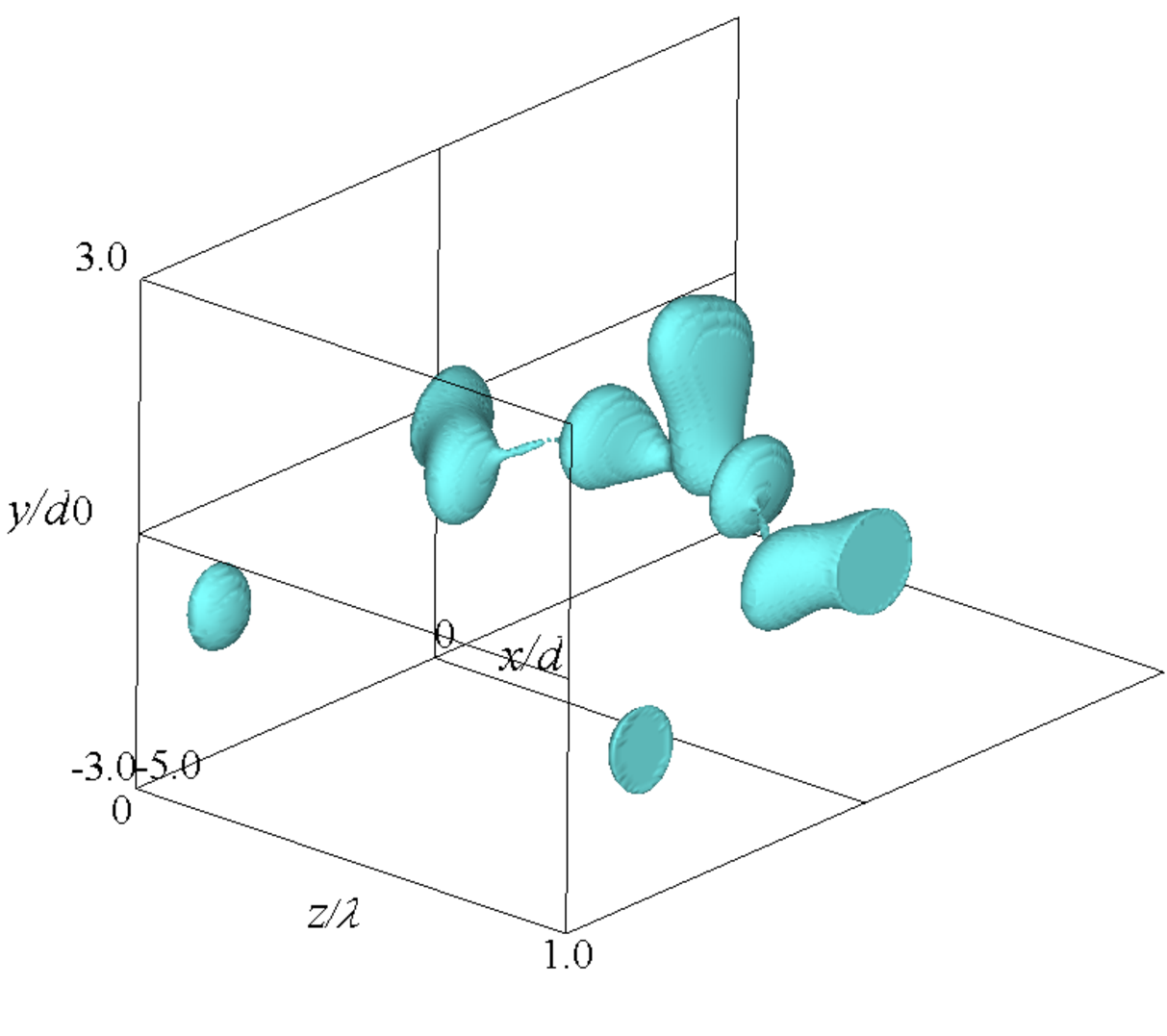} \\
\vspace*{-0.5\baselineskip}
(h)
\end{center}
\end{minipage}
\centering
\begin{minipage}{0.32\linewidth}
\begin{center}
\includegraphics[trim=0mm 0mm 0mm 0mm, clip, width=55mm]{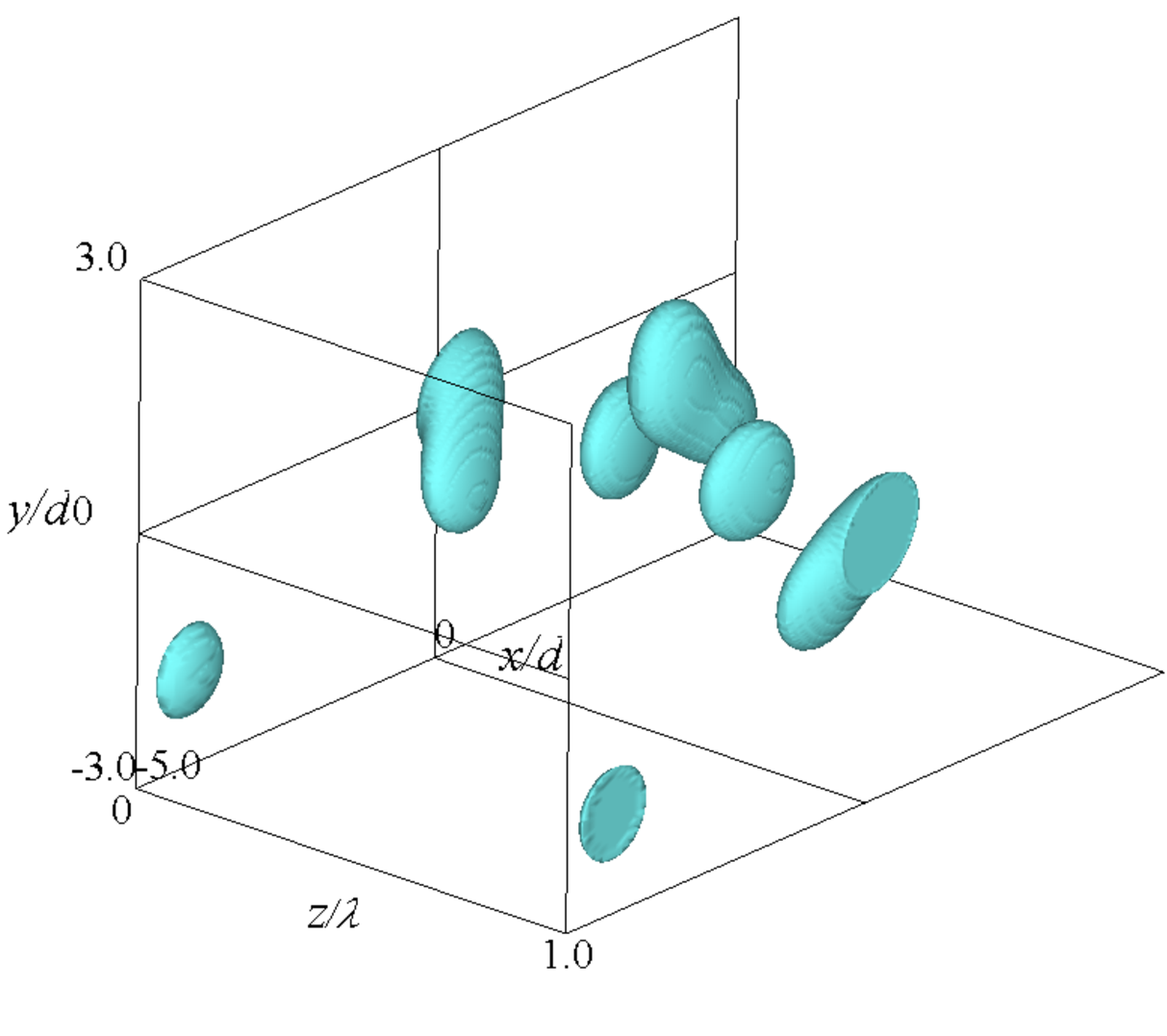} \\
\vspace*{-0.5\baselineskip}
(i)
\end{center}
\end{minipage}
\caption{Time variations of ligament interface: 
(a) $T = 0.3$, (b) $T = 0.6$, (c) $T = 1.0$, (d) $T = 1.3$, 
(e) $T = 1.6$, (f) $T = 2.0$, (g) $T = 2.3$, (h) $T = 2.6$, and (i) $T = 3.0$: 
$\Delta U/U_\mathrm{ref} = 17.8$, $ka = 0.3$, 
$\eta_r = 0.9$, $k_r = 4.0$.}
\label{u30k3f4_time}
\end{figure}

Figures \ref{u30k3f2_time} and \ref{u30k3f4_time} show the time evolution 
of the liquid ligament interface for $\eta_r = 0.9$ at $k_r = 2.0$ and 4.0, respectively. 
At $T = 0.3$ for $k_r = 2.0$, the liquid ligament is deformed into a film 
by the influence of the shear flow. 
At $T = 1.0$, perforation occurs at $z/\lambda = 0$ in the liquid ligament, 
but at $T = 1.6$, the liquid is combined. 
At $T = 2.6$, turbulence with a higher wavenumber than the initial disturbance occurs, 
similarly to the case of $\eta_r = 0$ and $k_r = 0$ shown in Fig. \ref{u30k3_time}. 
When time passes until $T = 4.6$, 
the splitting of the liquid ligament is confirmed, 
and large and small liquid droplets are formed near $z/\lambda = 0$, 0.5 and $z/\lambda = 0.2$, 0.8, respectively. 
For $k_r = 4.0$, the liquid ligament is deformed into a film at $T = 0.3$, 
similarly to the result for $k_r = 2.0$. 
The parts at $z/\lambda = 0$, 0.25, 0.5, and 0.75 of the liquid ligament 
bulge under the initial disturbance and swell further with time. 
At $T = 1.0$, perforation occurs at $z/\lambda = 0$ in the liquid ligament. 
At $T = 2.0$, the part at $z/\lambda = 0$ of the liquid ligament is split by the shear flow. 
The breakup droplets flow in the negative direction of the $x$-axis owing to the airflow. 
At $T = 2.0$, turbulence with a higher wavenumber than the initial disturbance occurs 
at $z/\lambda = 0.1-0.4$ and $0.6-0.9$. 
The bulge formed by this high wavenumber turbulence develops with time, 
and at $T = 3.0$, the liquid ligament breaks up. 
The ligament breakup time for $\eta_r = 0.9$, $k_r = 4.0$ 
becomes shorter than those for $\eta_r = 0$, $k_r = 0$ and $\eta_r = 0.9$, $k_r = 2.0$. 
Additionally, the number of breakup droplets increases.

Herein, we discuss the dependence of the grid on computational results. 
The result obtained using grid2 is compared with that obtained using grid3 
under the condition $\Delta U/U_\mathrm{ref} = 17.8$, $ka = 0.3$, $\eta_r = 0.9$, and $k_r =4.0$. 
As shown at $T = 3.0$ in Fig. \ref{u30k3f4_time}, 
the result with grid2 has two droplets at $z/\lambda = 0$ and one droplet 
each at $z/\lambda = 0.3$, 0.5, and 0.7. 
The result with grid3 was similar to that in Fig. \ref{u30k3f4_time}, 
but the placement of the droplets was slightly different. 
For grid3, there is one droplet at $z/\lambda = 0$, 0.15, and 0.85, 
and one droplet at $z/\lambda = 0.35$, 0.5, and 0.65. 
Here, we compare the droplet volume around $z/\lambda = 0$ 
and the diameters of three droplets around $z/\lambda = 0.3$, 0.5, and 0.7 by using the results of two grids. 
Near $z/\lambda = 0$, the relative error of the droplet volume obtained using grid2 
to the result of grid3 is approximately 1.60\%. 
The relative errors in the droplet diameter near $z/\lambda = 0.3$, 0.5, and 0.7 
are approximately 2.12\%, 0.18\%, and 2.12\%, respectively. 
The difference in the liquid volume and droplet diameter obtained using grid2 and grid3 is kept small. 
Therefore, it can be said that the results obtained using grid2 are valid.

\begin{figure}[!t]
\centering
\begin{minipage}{0.48\linewidth}
\begin{center}
\includegraphics[trim=0mm 0mm 0mm 0mm, clip, width=80mm]{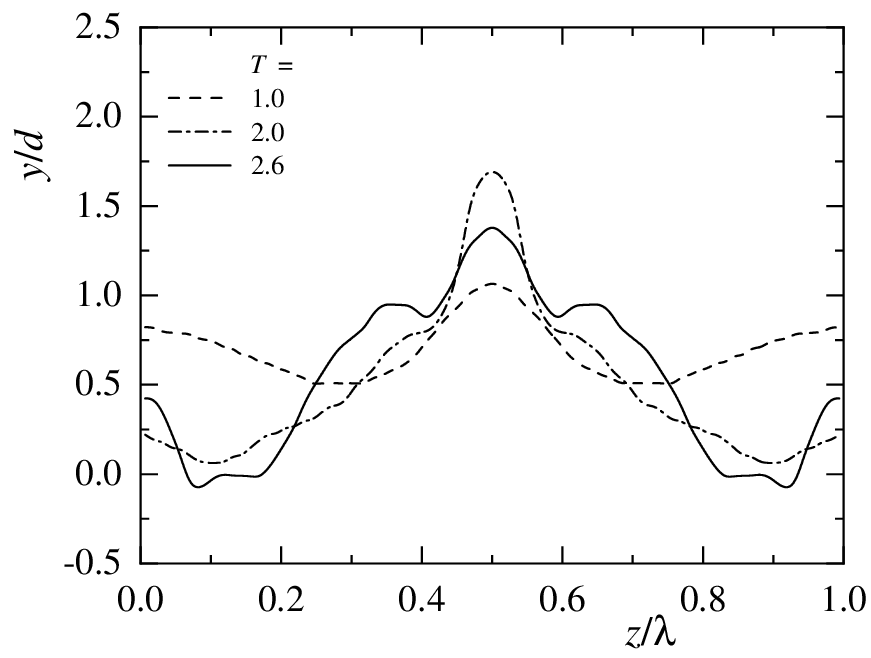} \\
\vspace*{-0.5\baselineskip}
(a)
\end{center}
\end{minipage}
\hspace*{0.02\linewidth}
\begin{minipage}{0.48\linewidth}
\begin{center}
\includegraphics[trim=0mm 0mm 0mm 0mm, clip, width=80mm]{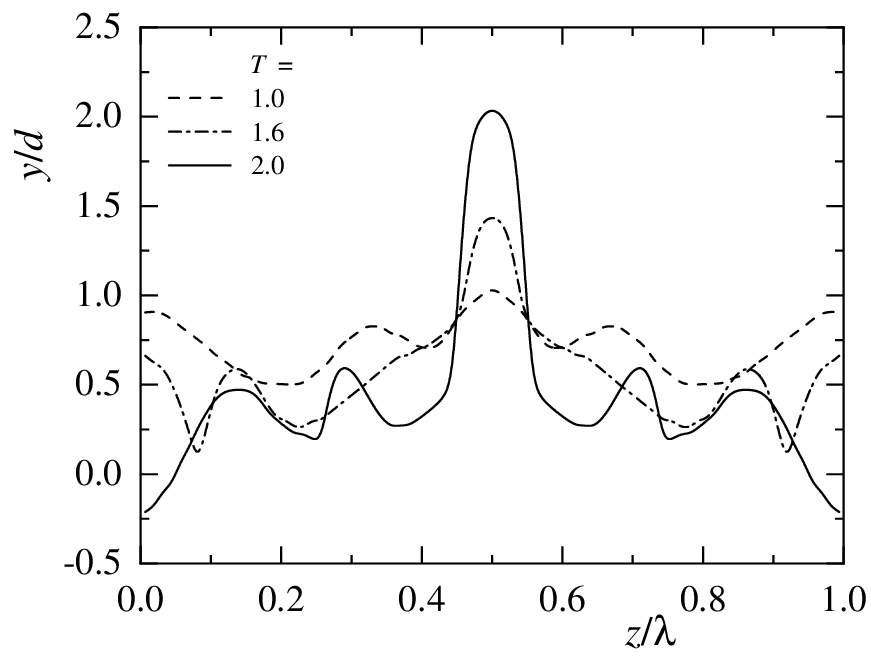} \\
\vspace*{-0.5\baselineskip}
(b)
\end{center}
\end{minipage}
\caption{Time variations of ligament interface for various values of 
$\eta_r$ and $k_r$: 
(a) $\eta_r = 0.9$, $k_r = 2.0$ and 
(b) $\eta_r = 0.9$, $k_r = 4.0$: 
$\Delta U/U_\mathrm{ref} = 17.8$, $ka = 0.3$.}
\label{timeintf}
\end{figure}

Figure \ref{timeintf} shows the time variation of the interface 
for each $\eta_r$ and $k_r$. 
The result for $\eta_r = 0$ and $k_r = 0$ were shown in Fig. \ref{timeint}(a). 
The distribution is the interface at the position $y/d$ 
where the interface is the highest in the $x$--$y$ cross-section. 
At $T = 2.0$ for $\eta_r = 0.9$ and $k_r = 4.0$, 
the position of the droplet interface near $z/\lambda = 0$ is higher than 
that of the liquid ligament interface near $z/\lambda = 0$. 
Hence, except for this droplet, we show the position $y/d$ where the liquid ligament interface is the highest. 
For $k_r = 2.0$, the initial disturbance grows at $T = 1.0$, 
but no new turbulence with high wavenumber occurs. 
At $T = 2.0$, the interface near $z/\lambda = 0.18$, 0.38, 0.68, and 0.82 slightly rises 
toward the gas side. 
At $T = 2.6$, this bulge develops, 
and turbulence with a wavenumber six times that of the initial disturbance is generated 
at the interface. 
This trend is similar to the results for $\eta_r = 0$ and $k_r = 0$. 
Even for $k_r = 4.0$, the initial disturbance grows at $T = 1.0$, 
but no new turbulence with high wavenumber occurs. 
At $T = 1.6$, the interface near $z/\lambda = 0$, 0.15, 0.5, 0.85, and 1.0 rises 
toward the gas side. 
It is also found that the interface near $z/\lambda = 0.38$ and 0.68 slightly bulges 
toward the gas side. 
At $T = 2.0$, two portions of the interface at $z/\lambda = 0.1-0.4$ and $0.6-0.9$ rise toward the gas side, 
and new turbulence with a higher wavenumber than the initial disturbance appears. 
For $\eta_r = 0.9$ and $k_r = 4.0$, 
higher wavenumber turbulence than the initial disturbance occurs on the interface 
at earlier times than the other conditions, 
and the bulge of the interface toward the gas side also increases. 
Therefore, we can see that the nonlinear effect appears stronger 
than the results for $\eta_r = 0$, $k_r = 0$ and $\eta_r = 0.9$, $k_r = 2.0$. 
As shown in Fig. \ref{u30k3f4_time}, 
depending on the strength of the nonlinear effect, 
the deformation process of the liquid ligament, the number of breakup droplets, 
and breakup time are different from the results for $\eta_r = 0$, $k_r = 0$ 
and for $\eta_r = 0.9$, $k_r = 2.0$. 
In the following, we investigate in detail the difference in the deformation of liquid ligaments 
between $\eta_r = 0$, $k_r = 0$ and $\eta_r = 0.9$, $k_r = 4.0$.

\begin{figure}[!t]
\centering
\begin{minipage}{0.48\linewidth}
\begin{center}
\includegraphics[trim=0mm 0mm 0mm 0mm, clip, width=65mm]{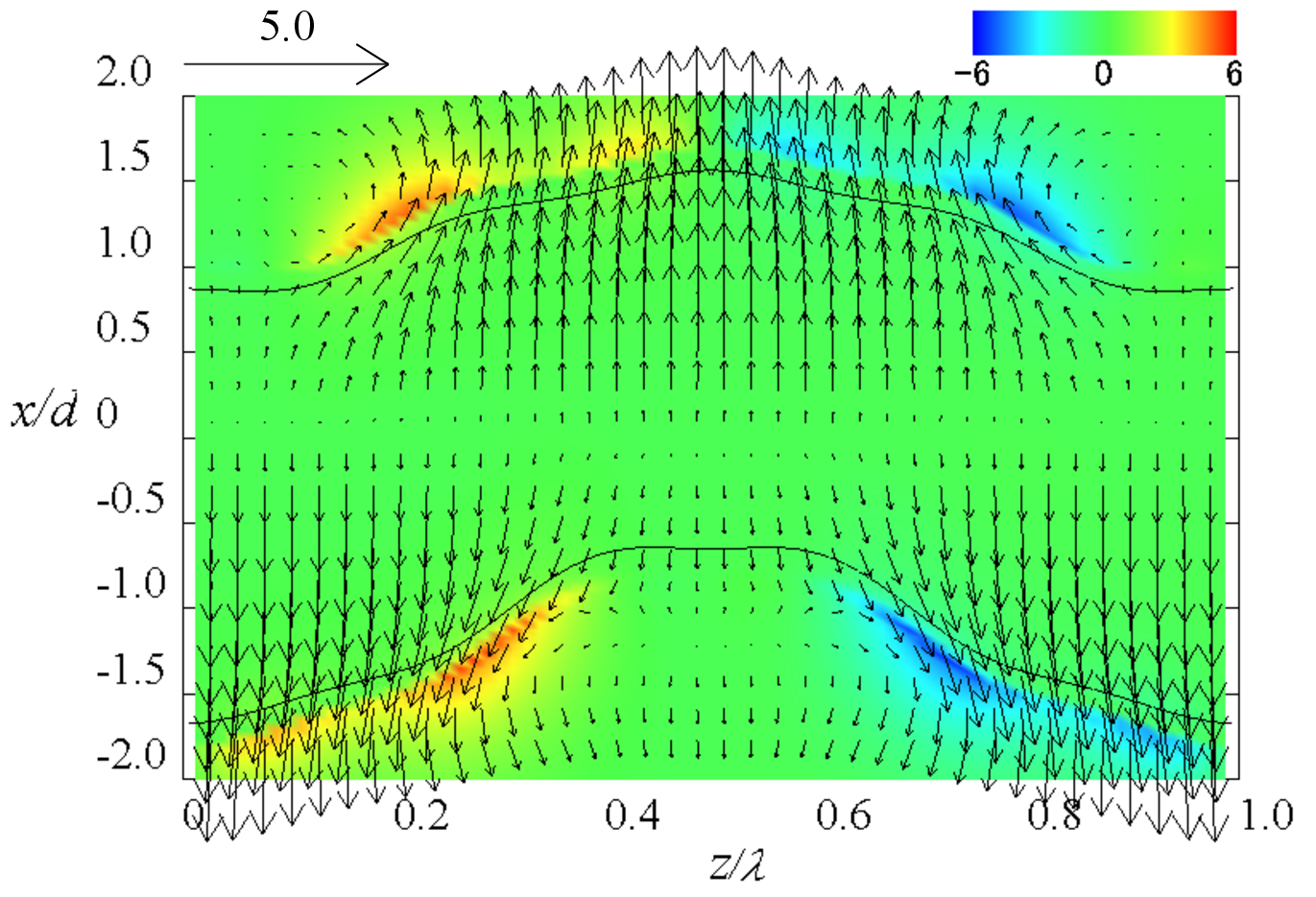} \\
\end{center}
\end{minipage}
\centering
\begin{minipage}{0.48\linewidth}
\begin{center}
\includegraphics[trim=0mm 0mm 0mm 0mm, clip, width=65mm]{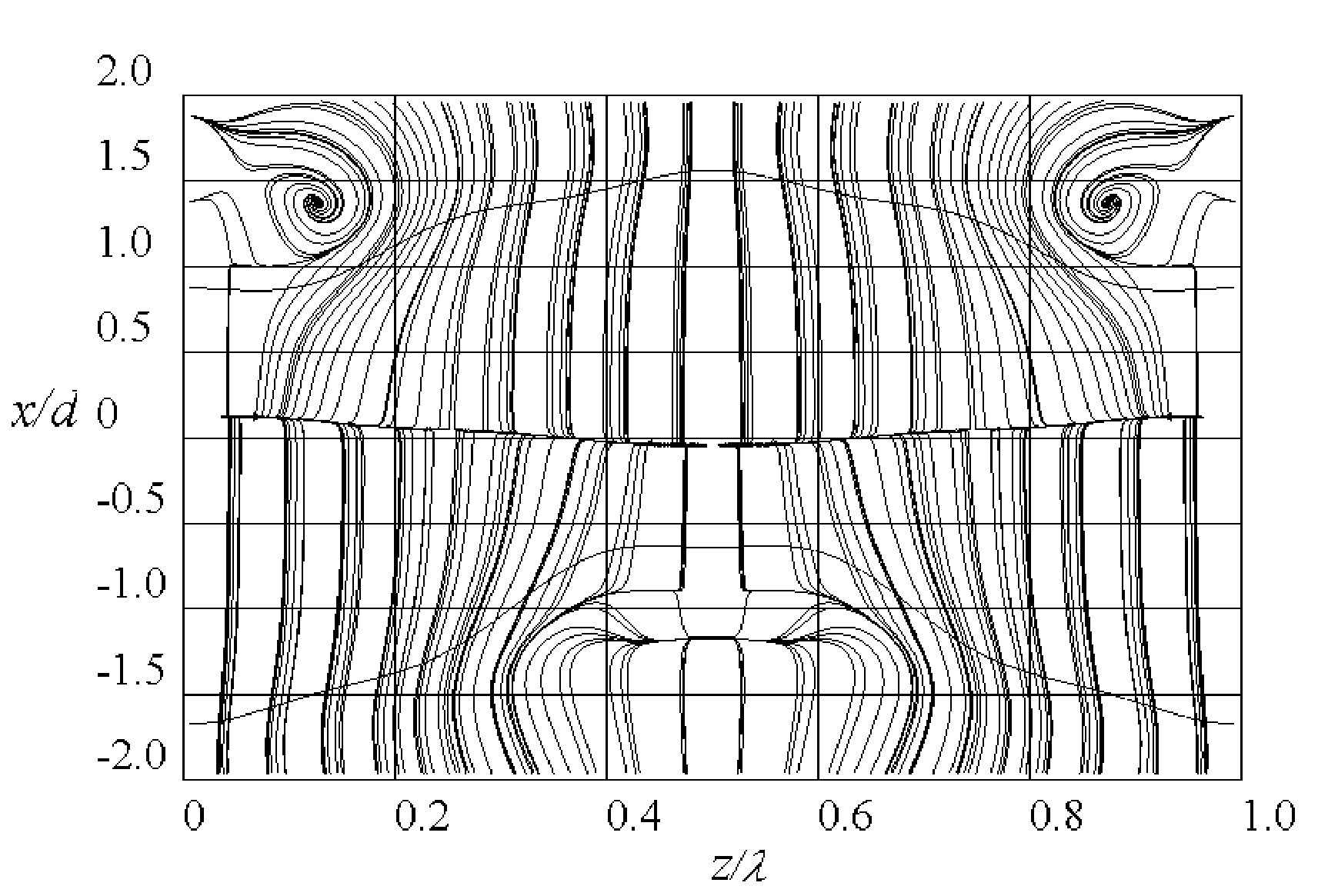} \\
\end{center}
\end{minipage}

(a)

\begin{minipage}{0.48\linewidth}
\begin{center}
\includegraphics[trim=0mm 0mm 0mm 0mm, clip, width=65mm]{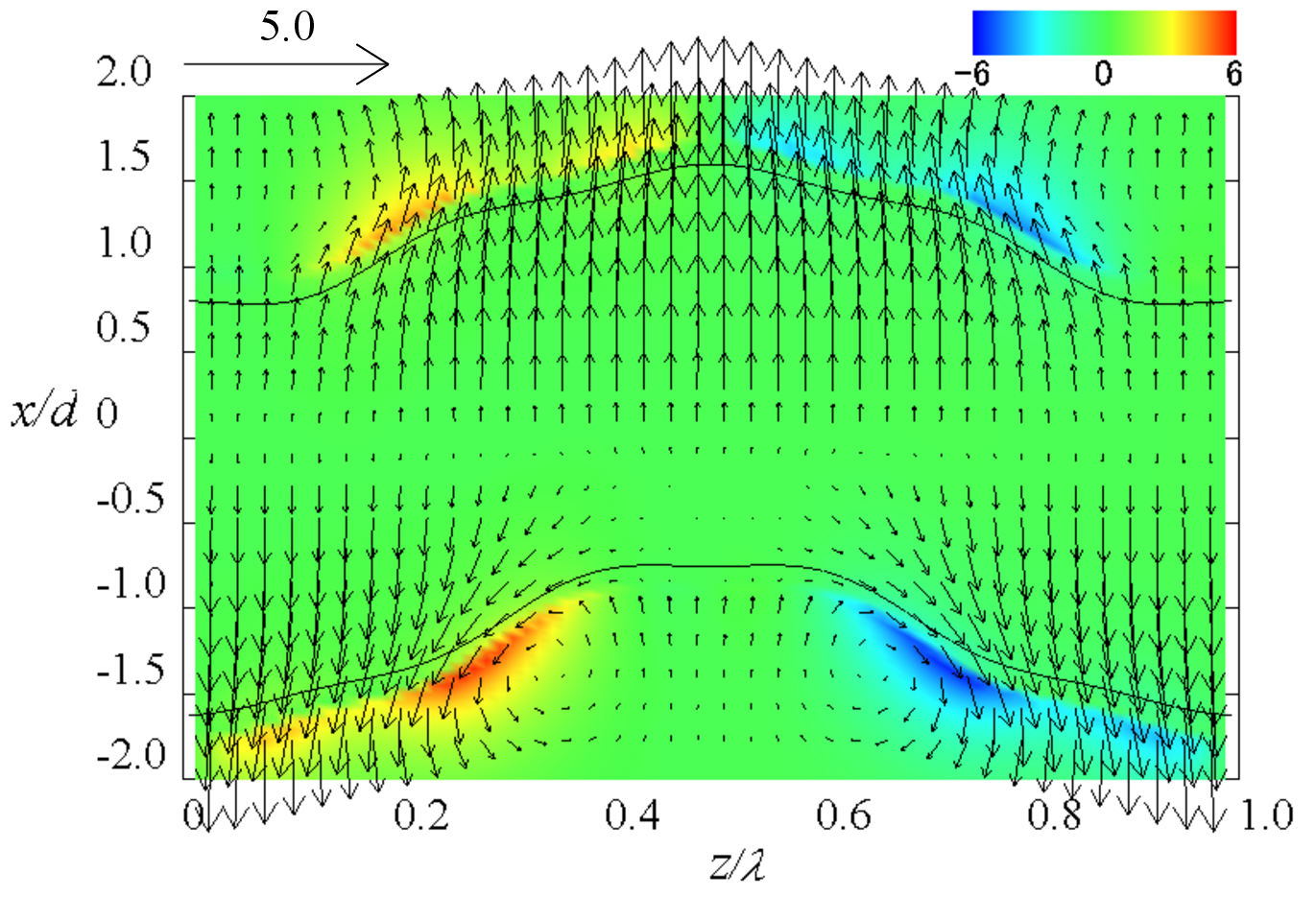} \\
\end{center}
\end{minipage}
\begin{minipage}{0.48\linewidth}
\begin{center}
\includegraphics[trim=0mm 0mm 0mm 0mm, clip, width=65mm]{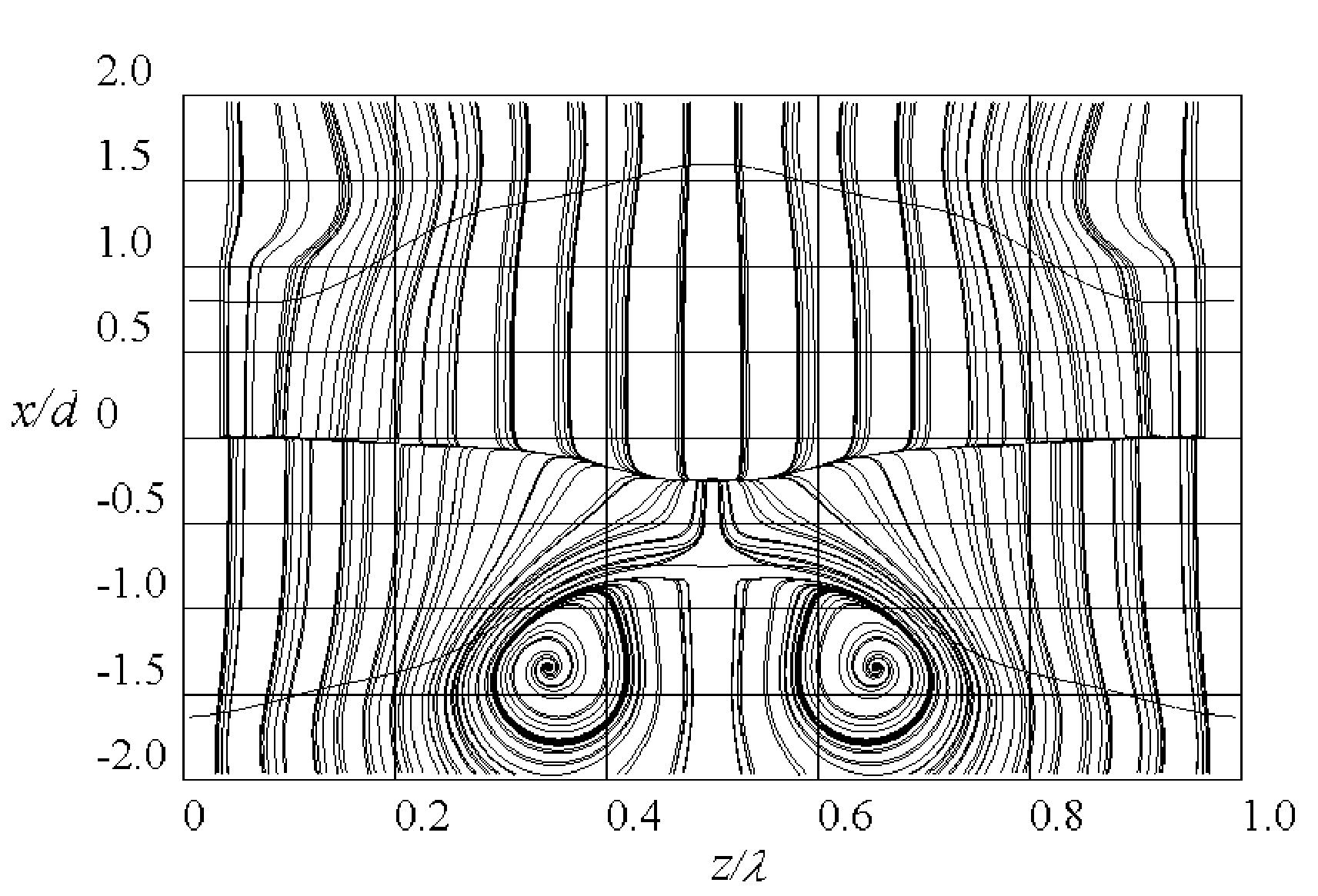} \\
\end{center}
\end{minipage}

(b)

\vspace*{-0.5\baselineskip}
\caption{Ligament interface, velocity vectors, $y$-direction vorticity, 
and streamlines at (a) $y/d = 0.04$ and (b) $y/d = -0.04$: 
$\Delta U/U_\mathrm{ref} = 17.8$, $ka = 0.3$, 
$\eta_r = 0.9$, $k_r = 4.0$, $T = 0.3$.}
\label{f4t03}
\end{figure}

Figure \ref{f4t03} shows the interface, velocity vectors, 
$y$-direction vorticity, and streamlines in the $x$--$z$ cross-section at $y/d = \pm{0.04}$ 
for $\eta_r = 0.9$ and $k_r = 4.0$ at $T = 0.3$. 
At $x/d > 0$ near $z/\lambda = 0.5$, 
the airflow flows in the positive direction of the $x$-axis. 
At $x/d < 0$ near $z/\lambda = 0$ and 1.0, 
the airflow flows in the negative direction of the $x$-axis. 
From the streamlines, we can also see that at $y/d = 0.04$, 
vortices owing to airflow are formed near $(x/d, z/\lambda) = (1.3, 0.1)$ and $(1.3, 0.9)$. 
At $y/d = -0.04$, vortices owing to airflow are formed 
near $(x/d, z/\lambda) = (-1.3, 0.35)$ and $(-1.3, 0.65)$. 
These vortices form positive and negative vorticities in the regions 
around $(x/d, z/\lambda) = (1.5, 0.3)$, $(-1.5, 0.2)$ 
and $(x/d, z/\lambda) = (1.5, 0.7)$, $(-1.5, 0.8)$, respectively. 
The parts around $(x/d,z/\lambda) = (1.25, 0.25)$, $(-1.25, 0.25)$, 
$(1.25, 0.75)$, and $(-1.25, 0.75)$ of the liquid ligament rise. 
It is considered that the vortices move these bulge parts 
along the central axis of the liquid ligament.

\begin{figure}[!t]
\centering
\begin{minipage}{0.48\linewidth}
\begin{center}
\includegraphics[trim=0mm 0mm 0mm 0mm, clip, width=70mm]{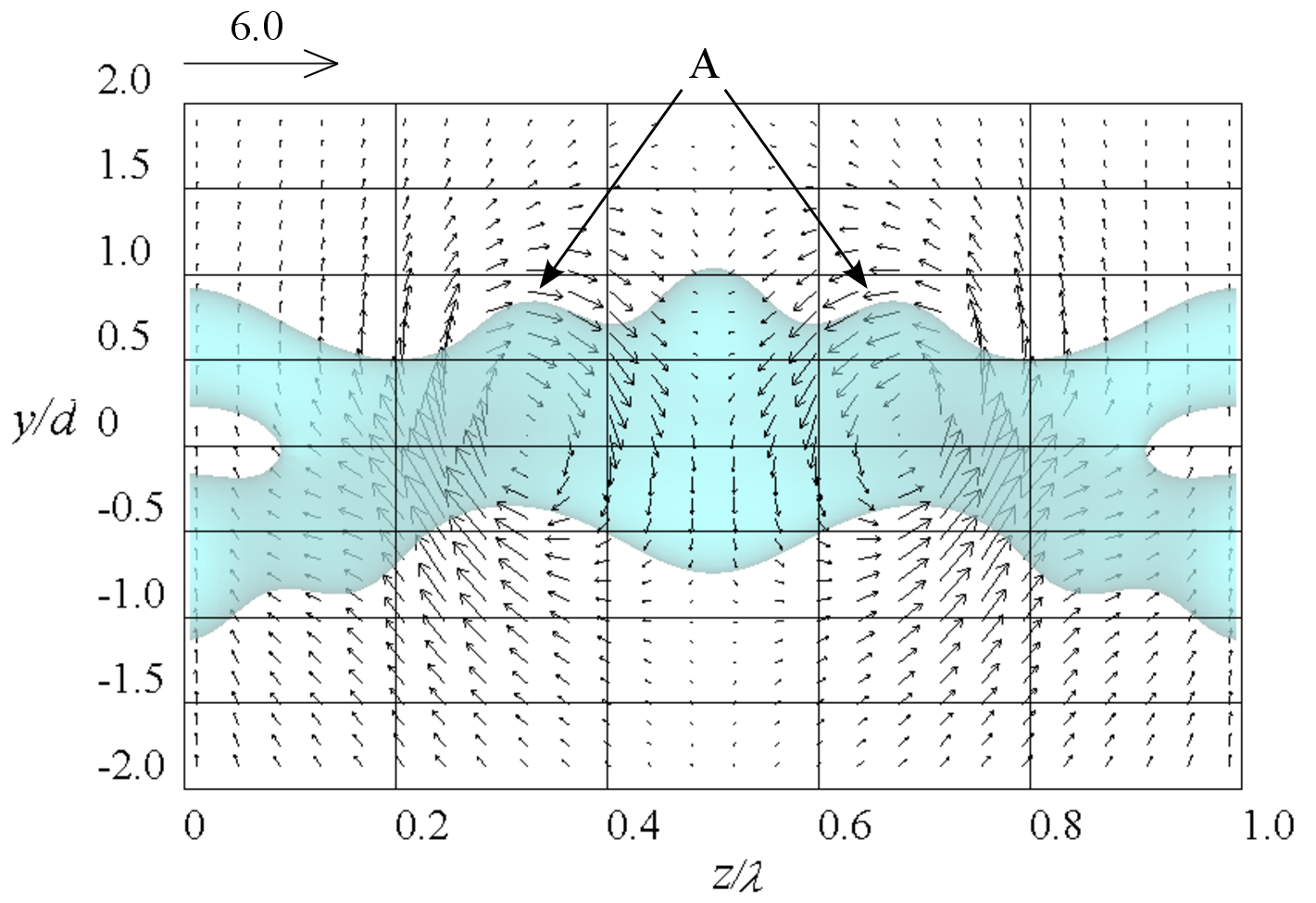}
\end{center}
\end{minipage}
\begin{minipage}{0.48\linewidth}
\begin{center}
\includegraphics[trim=0mm 0mm 0mm 0mm, clip, width=70mm]{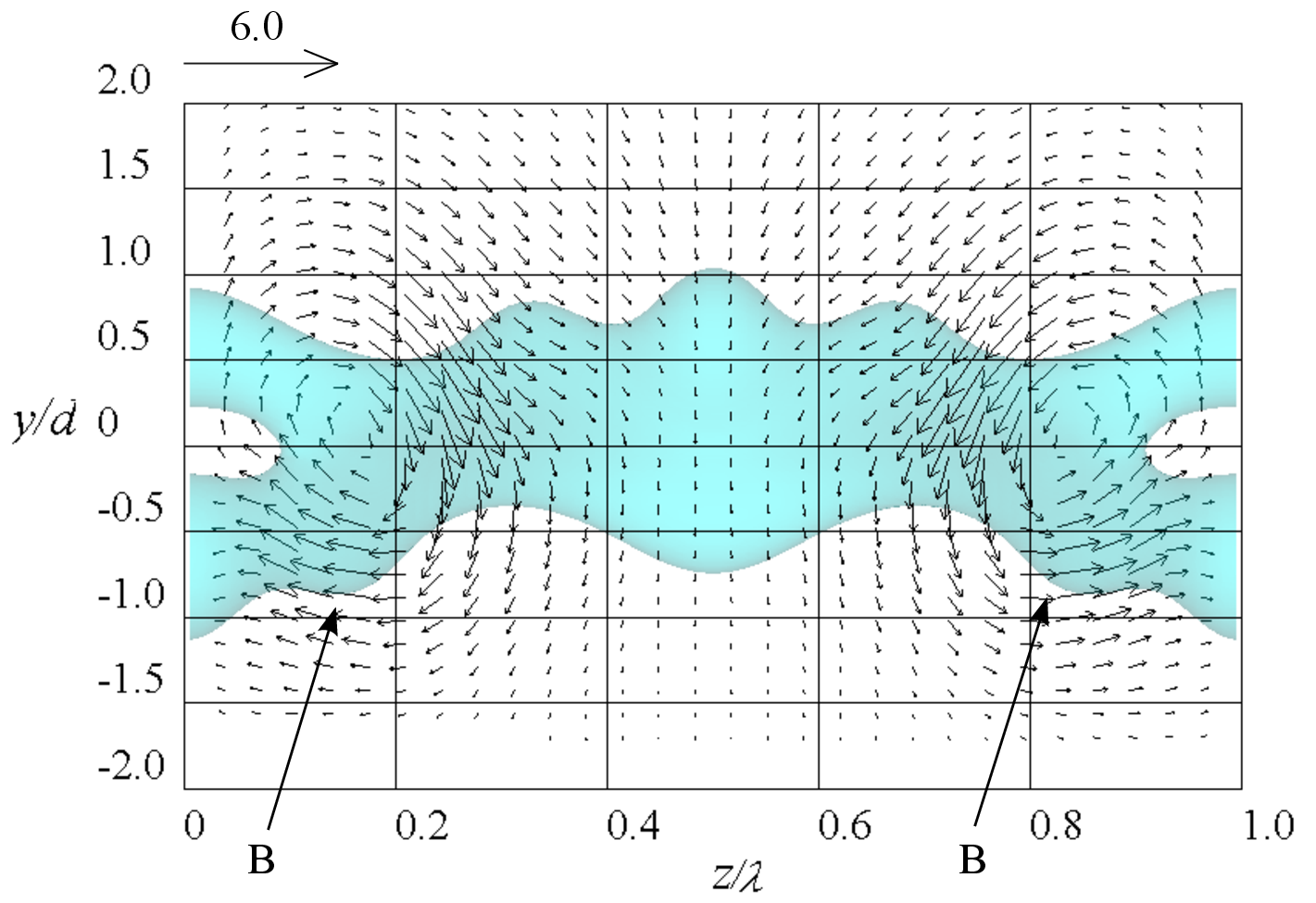}
\end{center}
\end{minipage}

(a)

\begin{minipage}{0.48\linewidth}
\begin{center}
\includegraphics[trim=0mm 0mm 0mm 0mm, clip, width=70mm]{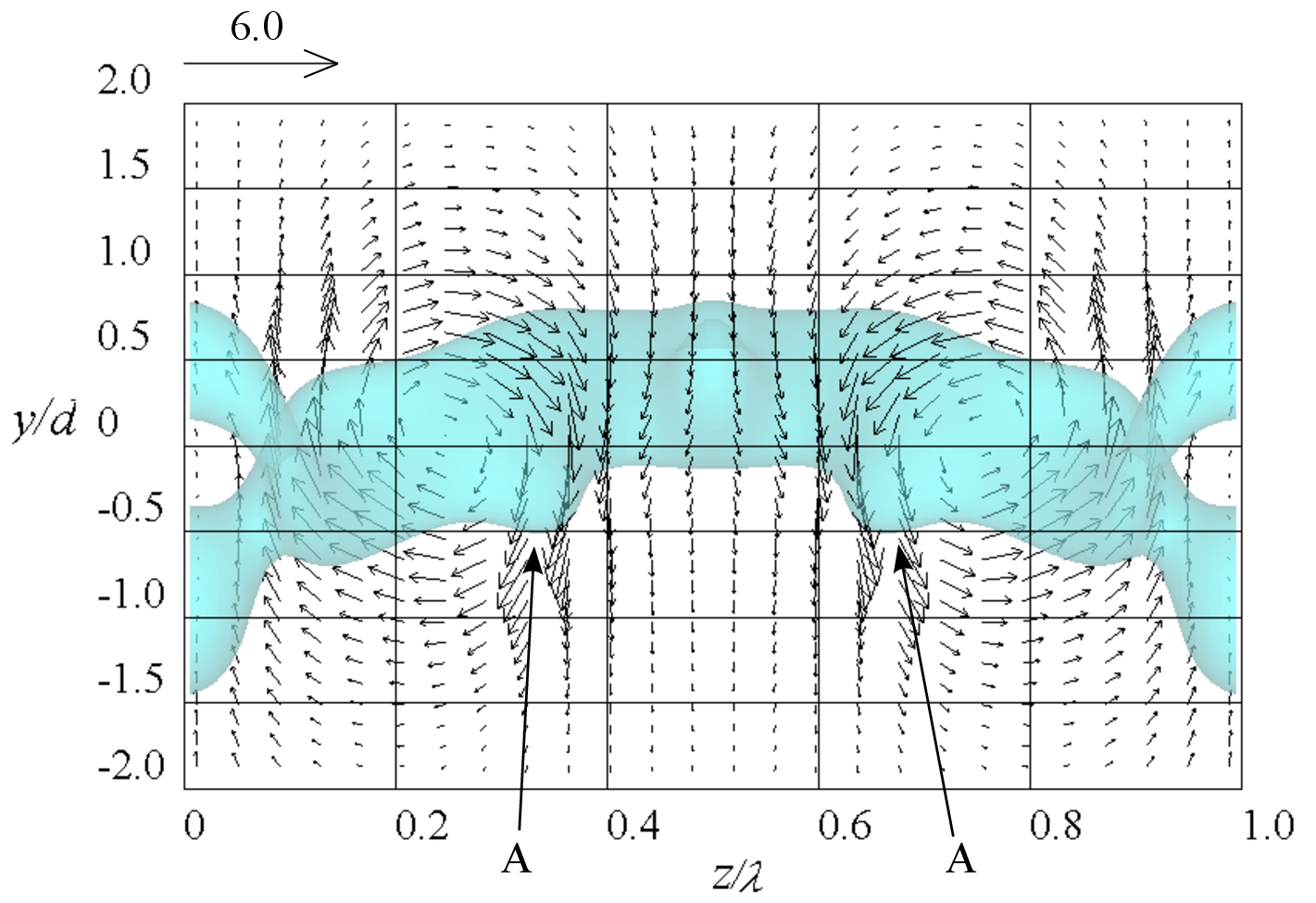}
\end{center}
\end{minipage}
\begin{minipage}{0.48\linewidth}
\begin{center}
\includegraphics[trim=0mm 0mm 0mm 0mm, clip, width=70mm]{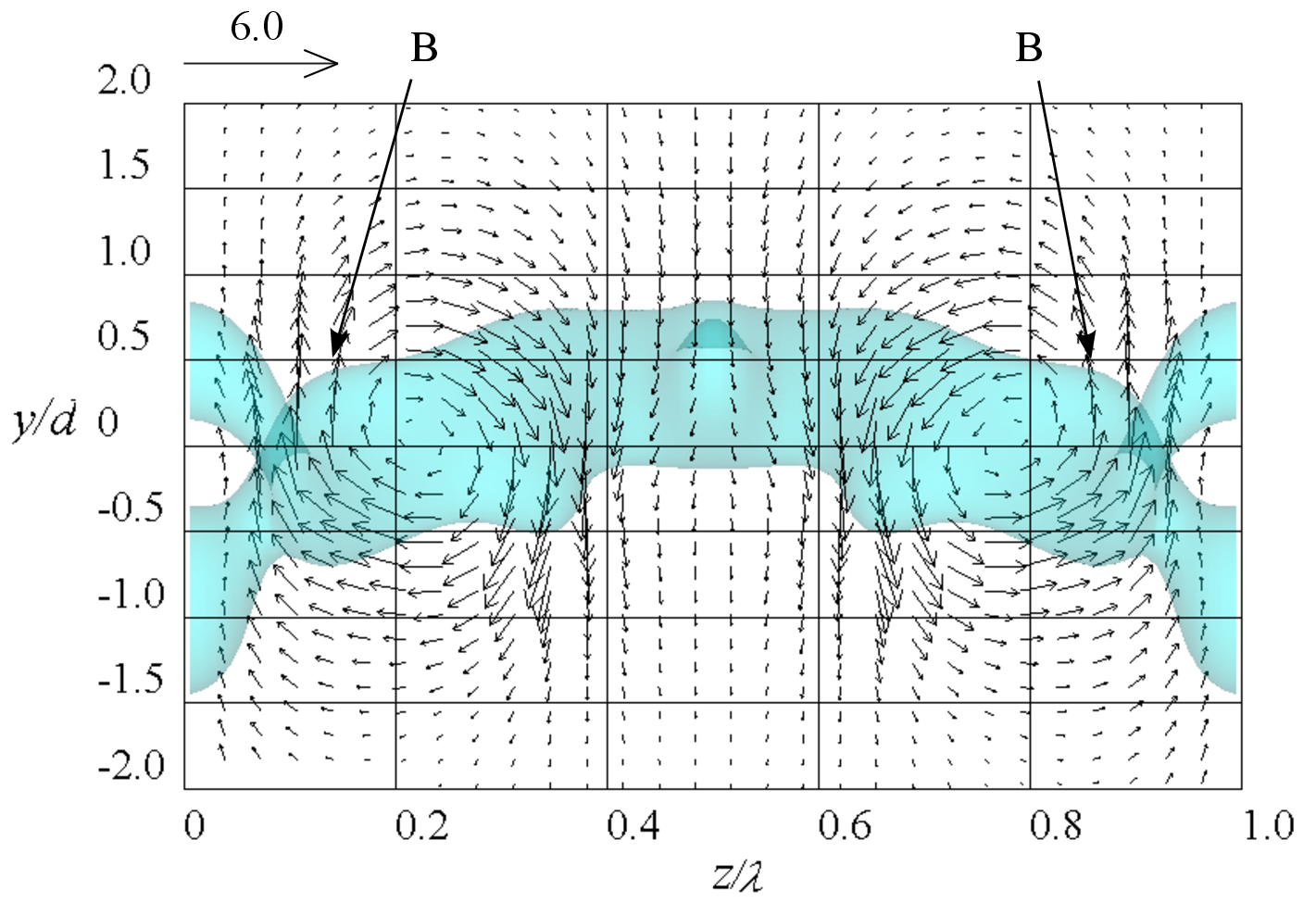}
\end{center}
\end{minipage}

(b)

\begin{minipage}{0.48\linewidth}
\begin{center}
\includegraphics[trim=0mm 0mm 0mm 0mm, clip, width=70mm]{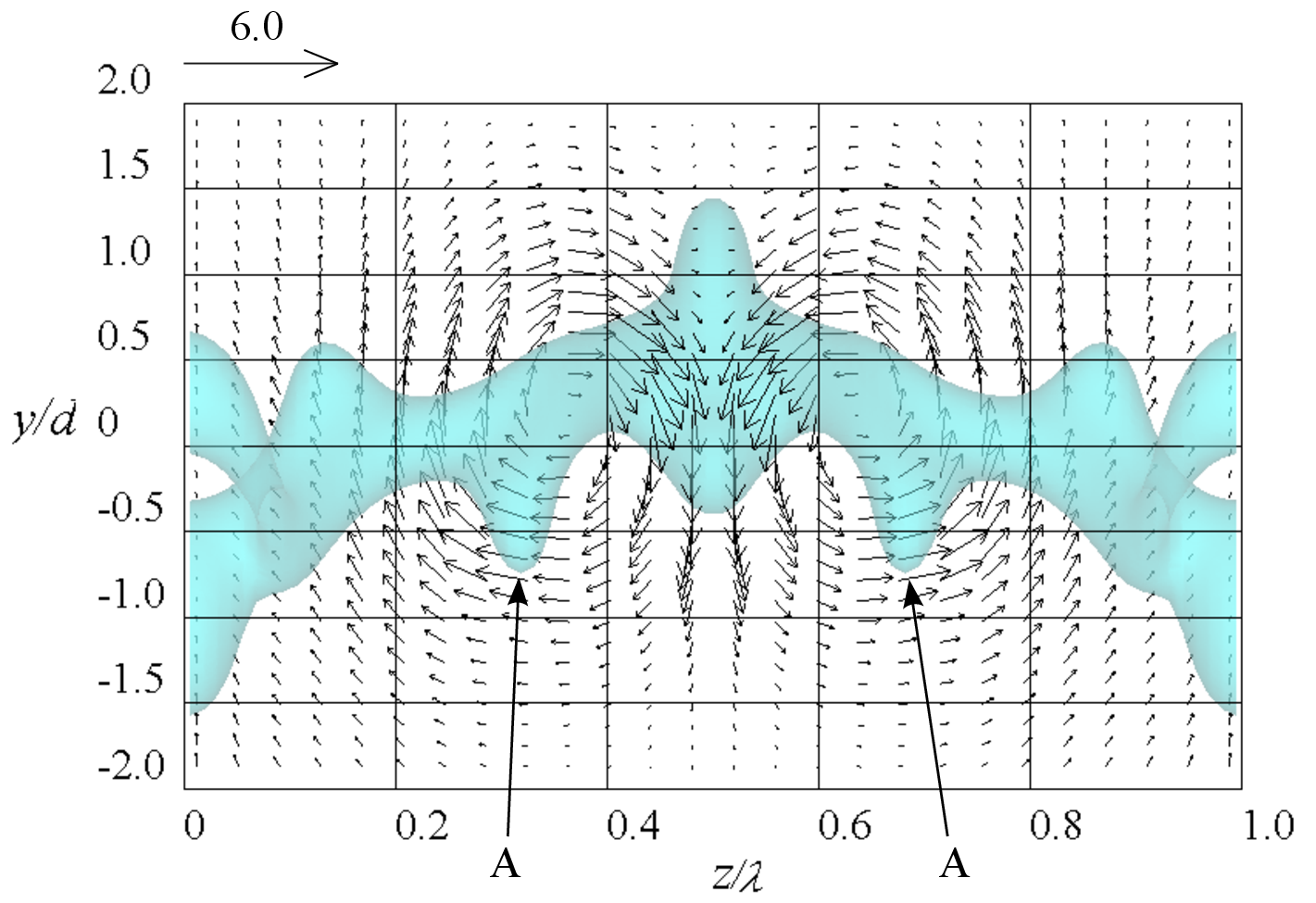}
\end{center}
\end{minipage}
\begin{minipage}{0.48\linewidth}
\begin{center}
\includegraphics[trim=0mm 0mm 0mm 0mm, clip, width=70mm]{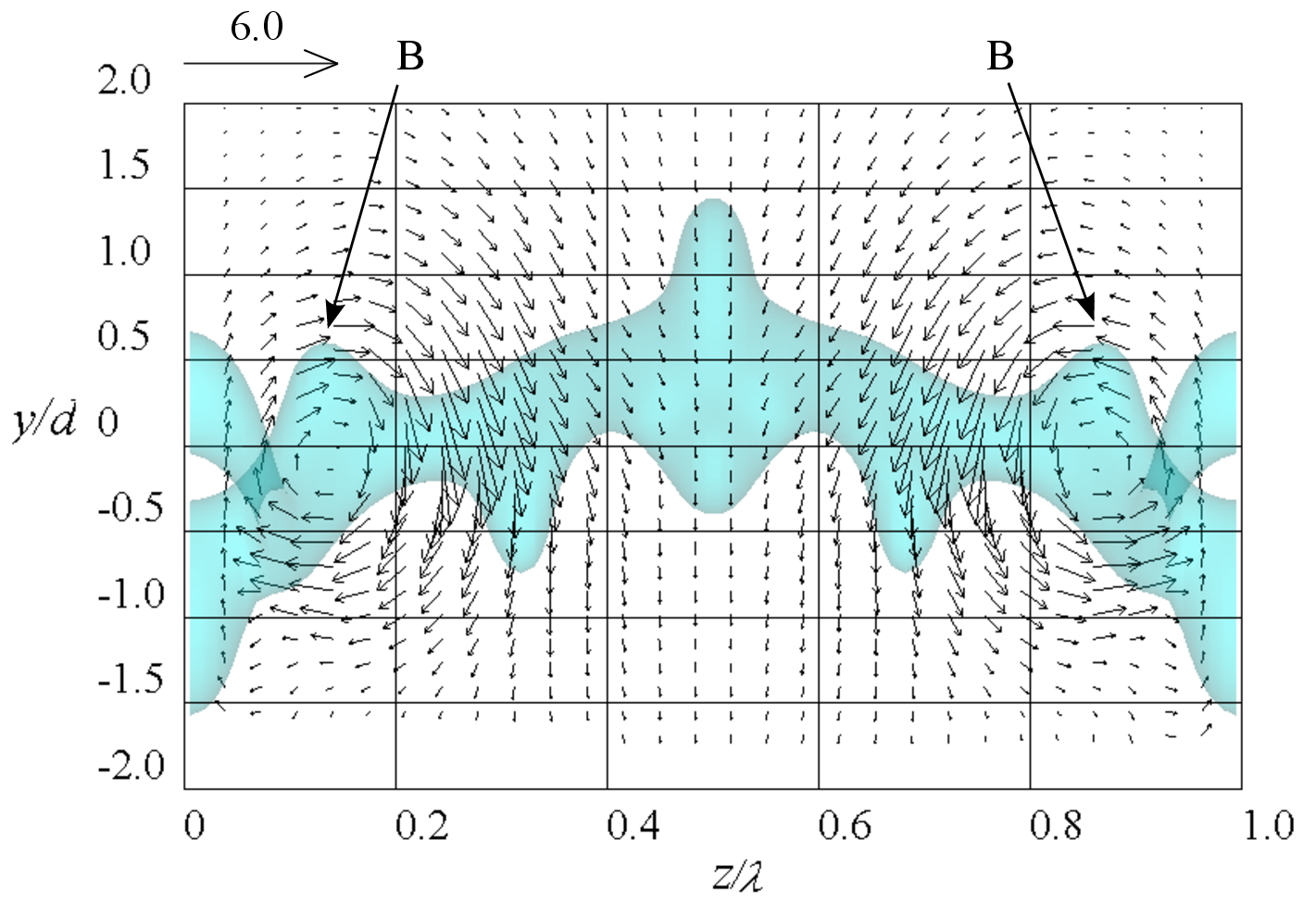}
\end{center}
\end{minipage}

(c)

\begin{minipage}{0.48\linewidth}
\begin{center}
\includegraphics[trim=0mm 0mm 0mm 0mm, clip, width=70mm]{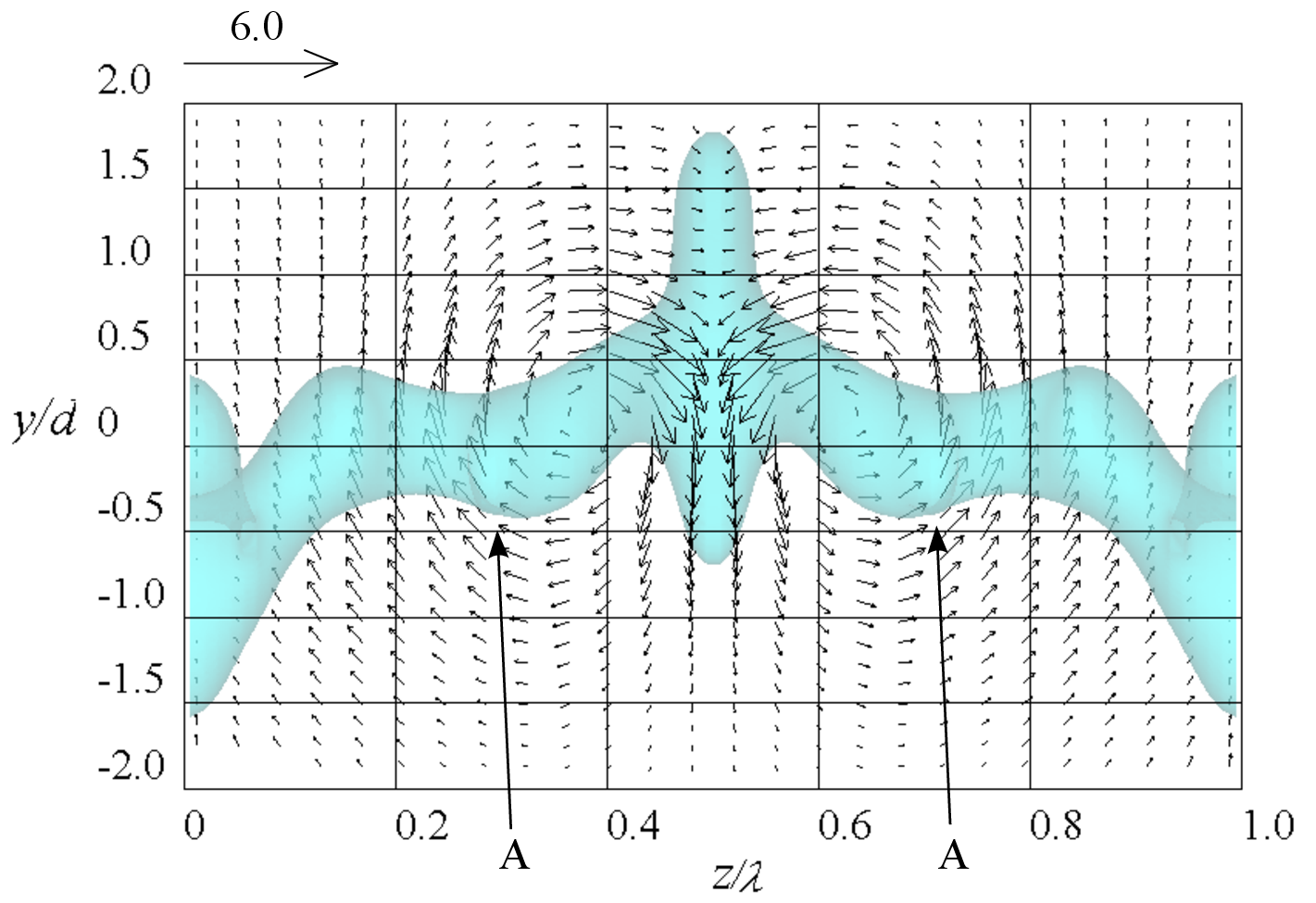}
\end{center}
\end{minipage}
\begin{minipage}{0.48\linewidth}
\begin{center}
\includegraphics[trim=0mm 0mm 0mm 0mm, clip, width=70mm]{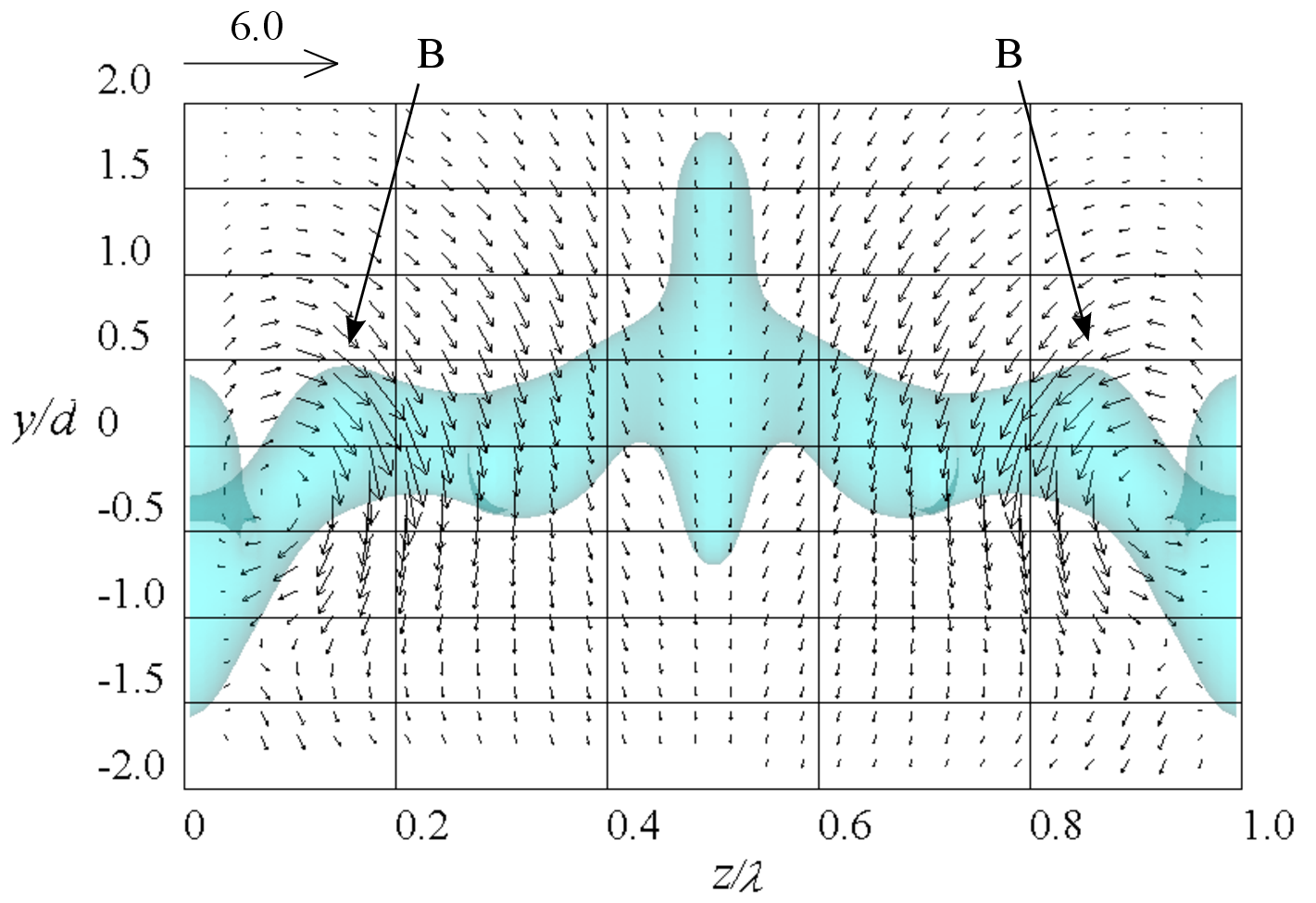}
\end{center}
\end{minipage}

(d)

\vspace*{-0.5\baselineskip}
\caption{Time variations of ligament interface, and velocity vectors: 
(a) $T = 1.0$, (b) $T = 1.4$, (c) $T = 1.6$, and (d) $T = 1.8$: 
$\Delta U/U_\mathrm{ref} = 17.8$, $ka = 0.3$, 
$\eta_r = 0.9$, $k_r = 4.0$.}
\label{f4timevc}
\end{figure}

Figure \ref{f4timevc} shows the time evolution of the interface 
and velocity vector for $\eta_r = 0.9$ and $k_r = 4.0$. 
Part A, shown in the left figure, is the bulge existing at $x/d>0$ 
around $z/\lambda = 0.25$ and 0.75 in Fig. \ref{f4t03}. 
Part B, shown in the right figure, is the bulge existing at $x/d<0$ 
around $z/\lambda = 0.25$ and 0.75 in Fig. \ref{f4t03}. 
Velocity vectors in the $y$--$z$ cross-section where the bulge parts A and B are present are shown. 
At all times, the center of the vortex around the liquid ligament exists inside the ligament. 
Additionally, all bulges develop with time. 
Around the interface of the bulge part A at $T = 1.0$, 
the vortex flows the airflow toward $z/\lambda = 0.5$. 
Near the interface of the bulge part B, 
the vortex flows the airflow toward $z/\lambda = 0$, 1.0. 
Thus, at $T = 1.4$, the bulges A and B move to $z/\lambda = 0.5$ 
and $z/\lambda = 0$, 1.0, respectively. 
At $T = 1.6$, unlike $T = 1.0$, near the interface of the bulge A, 
the airflow flows toward $z/\lambda = 0$, 1.0 owing to the vortex. 
Near the interface of the bulge B, the vortex causes the airflow 
to flow toward $z/\lambda = 0.5$. 
Thus, at $T = 1.8$, the bulges A and B move to $z/\lambda = 0$, 1.0 
and $z/\lambda = 0.5$, respectively. 
In this way, the bulging part of the liquid owing to the initial disturbance is moved 
by the vortex with time along the central axis of the liquid ligament.

\begin{figure}[!t]
\centering
\begin{minipage}{0.48\linewidth}
\begin{center}
\includegraphics[trim=0mm 0mm 0mm 0mm, clip, width=80mm]{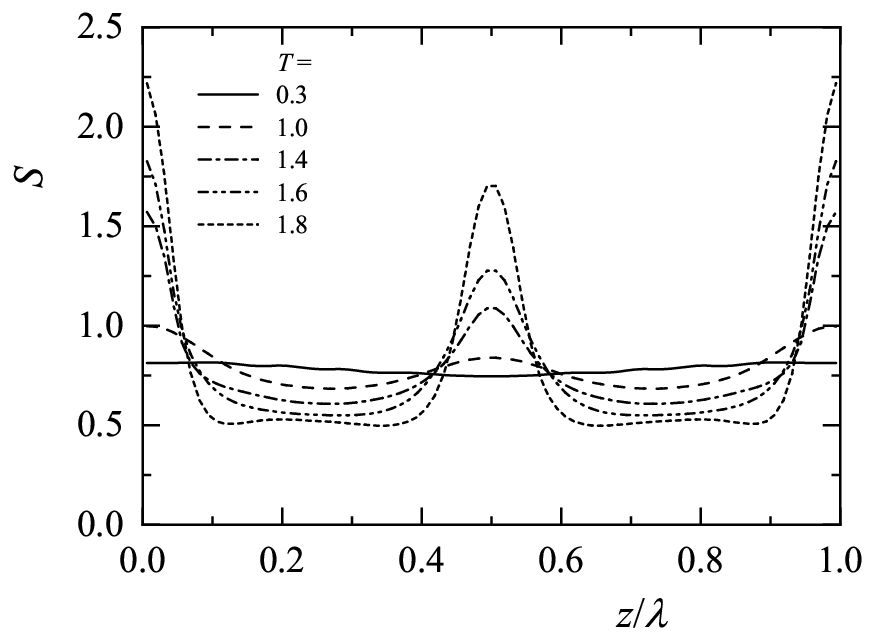} \\
\vspace*{-0.5\baselineskip}
(a)
\end{center}
\end{minipage}
\hspace*{0.02\linewidth}
\begin{minipage}{0.48\linewidth}
\begin{center}
\includegraphics[trim=0mm 0mm 0mm 0mm, clip, width=80mm]{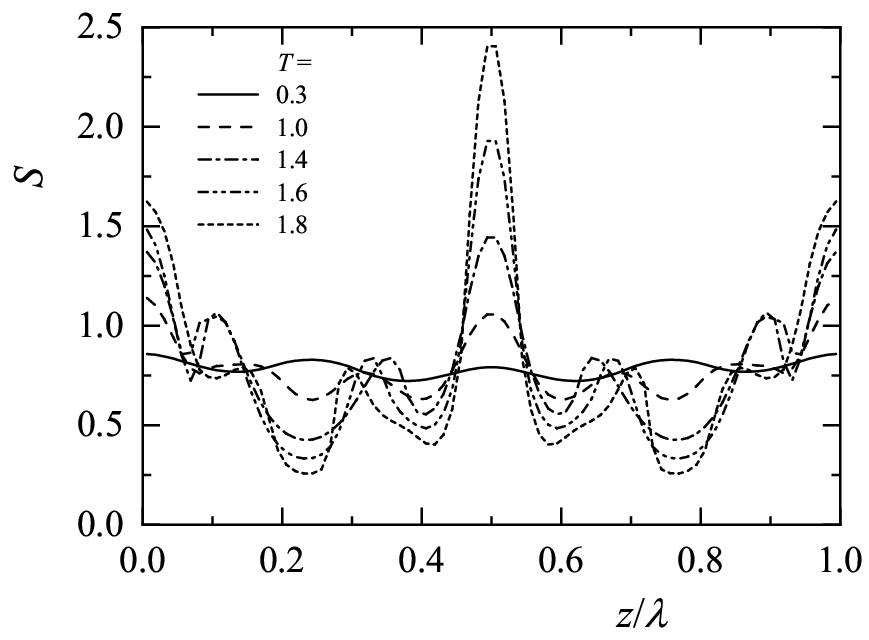} \\
\vspace*{-0.5\baselineskip}
(b)
\end{center}
\end{minipage}
\caption{Time variations of cross-sectional area: 
(a) $\eta_r = 0$, $k_r = 0$ and 
(b) $\eta_r = 0.9$, $k_r = 4.0$: 
$\Delta U/U_\mathrm{ref} = 17.8$, $ka = 0.3$.}
\label{f0f4timecsarea}
\end{figure}

For $\eta_r = 0$, $k_r = 0$ and $\eta_r = 0.9$, $k_r = 4.0$, 
Fig. \ref{f0f4timecsarea} shows the time variation of the cross-sectional area $S$. 
In Fig. \ref{f0f4timecsarea}(a), 
the cross-section areas at $z/\lambda = 0$ and 0.5 increase with time. 
Conversely, the areas at $z/\lambda = 0.25$ and 0.75 decrease with time. 
As shown in Fig. \ref{f4t03}, at $T = 0.3$ for $\eta_r = 0.9$ and $k_r = 4.0$, 
the parts at $z/\lambda = 0.25$ and 0.75 swell owing to the initial disturbance. 
Hence, in Fig. \ref{f0f4timecsarea}(b), the cross sections 
near $z/\lambda = 0.25$ and 0.75 are larger than those near $z/\lambda = 0.1$, 0.4, 0.6, and 0.9. 
The vortices move the bulges at $z/\lambda = 0.1-0.4$ and $0.6-0.9$ 
along the central axis of the liquid ligament. 
Thus, at $T = 1.0$, the cross sections at $z/\lambda = 0.15$, 0.35, 0.65, and 0.85 
are larger than those at $z/\lambda = 0.25$ and 0.75. 
With time, the bulge deforms into a spherical shape owing to surface tension, 
and the liquid ligament around the bulge becomes constricted. 
At $T = 1.8$, the cross sections at $z/\lambda = 0.25$ and 0.75 
for $\eta_r = 0.9$ and $k_r = 4.0$ are smaller than those at $z/\lambda = 0.25$ and 0.75 
for $\eta_r = 0$ and $k_r = 0$. 
It can be seen from this that the deformation of the liquid ligament is enhanced 
for $\eta_r = 0.9$ and $k_r = 4.0$.

\begin{figure}[!t]
\begin{center}
\includegraphics[trim=0mm 0mm 0mm 0mm, clip, width=100mm]{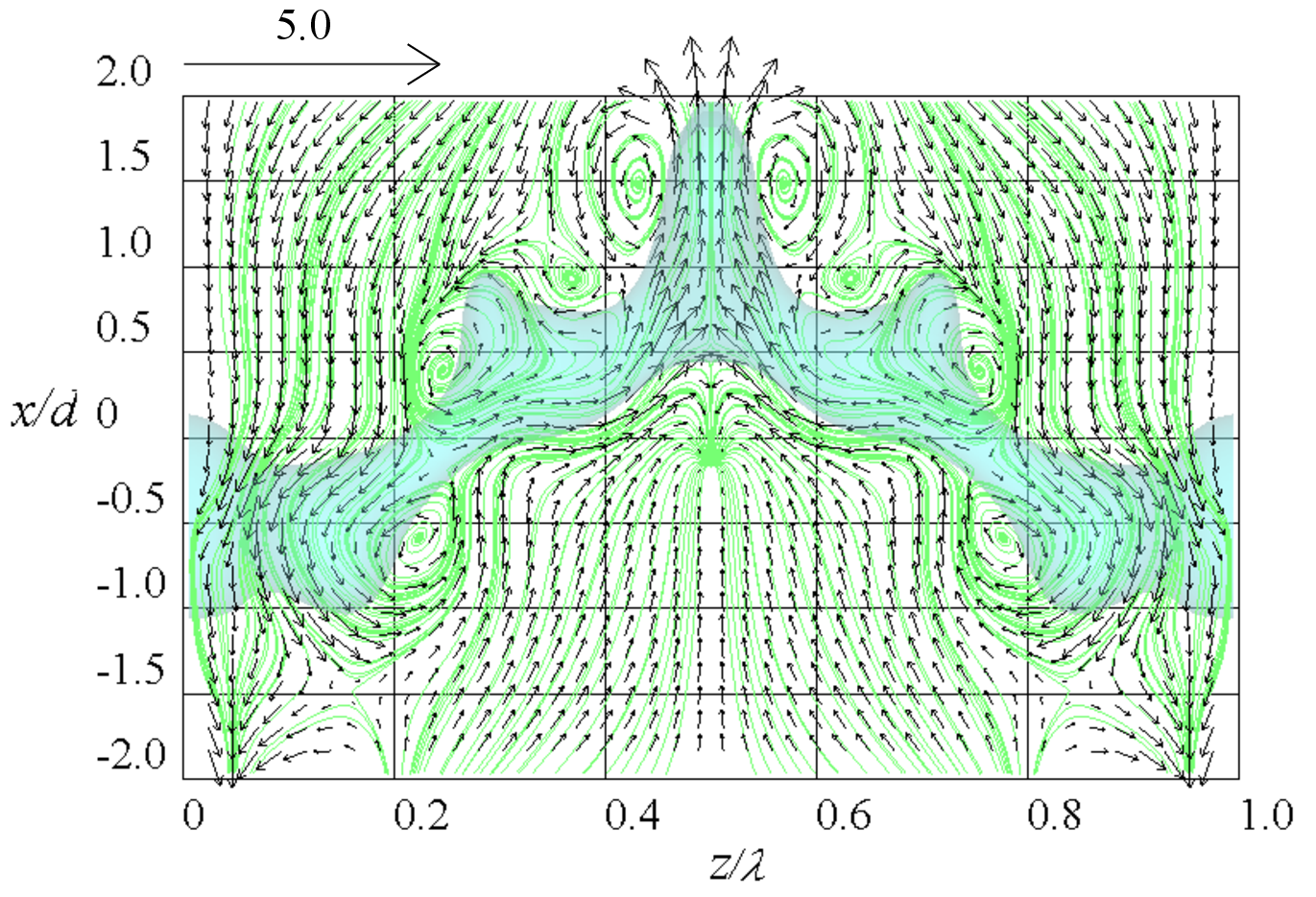}
\end{center}
\vspace*{-1.0\baselineskip}
\caption{Ligament interface, velocity vectors and streamlines 
at $T = 2.0$: $\Delta U/U_\mathrm{ref} = 17.8$, $ka = 0.3$, 
$\eta_r = 0.9$, $k_r = 4.0$.}
\label{f4t2}
\end{figure}

To investigate the effect of the flow field near the interface on the liquid ligament splitting, 
Fig. \ref{f4t2} shows the interface, velocity vectors, and streamline at $T = 2.0$ 
for $\eta_r = 0.9$ and $k_r = 4.0$. 
The velocity vector and streamline are distributions in the $x$--$z$ cross-section at $y/d = 0$. 
As shown in Fig. \ref{timeintf}(b), 
the nonlinear effect causes the liquid ligament to swell at $z/\lambda = 0.15$, 0.3, 0.7, and 0.85. 
A large bulge also occurs at $z/\lambda = 0.5$. 
A vortex owing to the airflow is formed in the wake of each bulge. 
These vortices induce a flow from the constricted part of the liquid ligament 
to the bulging part; 
thus, the bulging part develops further, and the constricted part narrows further. 
Consequently, the deformation of the ligament is accelerated by the vortex, 
and the ligament splits faster than the result of $\eta_r = 0$ and $k_r = 0$.

\begin{figure}[!t]
\begin{center}
\includegraphics[trim=0mm 0mm 0mm 0mm, clip, width=80mm]{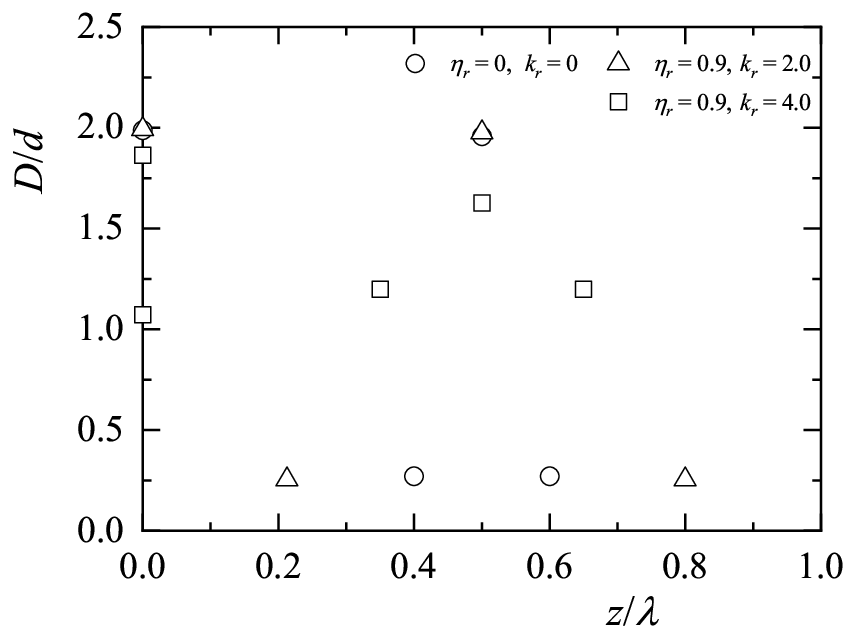}
\end{center}
\vspace*{-1.0\baselineskip}
\caption{Variations of droplet diameter with $\eta_r$ and $k_r$: 
$\Delta U/U_\mathrm{ref} = 17.8$, $ka = 0.3$.}
\label{dsize_f024}
\end{figure}

Figure \ref{dsize_f024} shows the diameter of the breakup droplet. 
Table \ref{dsizearista} shows the droplet diameter, 
which is the arithmetic average of all droplet diameters under each condition, 
and the standard deviation of the droplet diameter. 
In the case of $\eta_r = 0$ and $k_r = 0$, 
the diameters of the droplets existing near $z/\lambda = 0.4$ and 0.6 
are much smaller than those of the droplets formed at $z/\lambda = 0$ and 0.5; 
hence, the average droplet diameter becomes small. 
The standard deviation is high 
owing to the large variation in the diameters of the breakup droplets. 
The authors \citep{Yanaoka&Nakayama_2022} reported the droplet diameters 
of the main and satellite droplets, 
and the average diameter of the two droplets was $1.314d$. 
In this study, as the shear velocity is faster than that in the earlier study, 
the diameter of the droplet is smaller than that of the previous report. 
For $\eta_r = 0.9$ and $k_r = 2.0$, 
the trend is the same as for $\eta_r = 0$ and $k_r = 0$, 
and there is no significant difference in the average droplet size 
and its standard deviation. 
For $\eta_r = 0.9$ and $k_r = 4.0$, more droplets form than other conditions. 
Hence, the total surface area of the liquid is the largest. 
Compared to the results of $\eta_r = 0$, $k_r = 0$ and $\eta_r = 0.9$, $k_r = 2.0$, 
the average droplet size is large, but the standard deviation is low. 
It can be seen from this that the droplet diameters were made uniform 
by adding the disturbance with the high wavenumber component. 
Therefore, for $\eta_r = 0.9$ and $k_r = 4.0$, 
it can be said that the atomization quality of the liquid ligament was improved.

We compare droplet diameters obtained in this study with existing studies 
\citep{Rutland&Jameson_1970,Lafrance_1975,Lakdawala_et_al_2015}. 
A liquid jet generates the main droplet and satellite droplet. 
The diameter $d_s$ of the satellite droplet is smaller than 
the diameter $d_m$ of the main droplet. 
In this study, large and small droplets are also generated, 
similarly to existing results. 
In Rutland and Jameson's experiment \citep{Rutland&Jameson_1970}, 
the diameters are $d_m/d = 2.129$ and $d_s/d = 1.127$ for $ka = 0.34$. 
\citet{Lafrance_1975} reported $d_m/d = 2.402$ and $d_s/d = 1.450$ for $ka = 0.32$. 
In Lakdawala et al.'s calculation \citep{Lakdawala_et_al_2015}, 
the diameters are $d_m/d = 2.350$ and $d_s/d = 1.393$ for $ka = 0.30$. 
In this study, we obtained smaller droplet diameters than existing results. 
In existing research \citep{Rutland&Jameson_1970,Lafrance_1975,Lakdawala_et_al_2015}, 
the direction of shear velocity between gas and liquid is the same as 
the direction of the central axis direction of the liquid column. 
Conversely, the two directions are orthogonal in this study. 
The physical properties of the liquid used also differ. 
It is considered that such a difference causes the difference in droplet diameters.

\section{CONCLUSION}

This study performed a numerical analysis of the deformation and breakup of a liquid ligament 
with various disturbances on the interface in shear flow. 
We investigated the effects of the airflow velocity difference, 
the wavenumber of the initial disturbance on the interface of the liquid ligament, 
and the high wavenumber disturbance on the deformation and splitting of the liquid ligament. 
Consequently, the following findings were obtained.

We analyzed the growth of a liquid ligament interface with an initial disturbance 
in stationary fluid. 
The growth rate of the interface agrees well with the theoretical value, 
and it was found that the present numerical method can capture 
the behavior of the interface.

A three-dimensional flow is generated around the liquid ligament by the shear flow. 
This flow forms vortices near the interface. 
These vortices promote the movement of the liquid inside the liquid ligament. 
When the velocity difference of shear flow increases, 
a nonlinear effect becomes stronger than in the case of the low-velocity difference, 
and turbulence with higher wavenumber components than the initial disturbance occurs 
at the interface. 
This turbulence accelerates the ligament splitting 
and increases the number of breakup droplets. 
Then, the droplet diameters become uniform, and the atomization quality improves.

When an initial disturbance with a low wavenumber is applied to the interface, 
a new turbulence with a higher wavenumber than the initial disturbance is generated 
at the interface, 
and an intensive nonlinear effect appears. 
As the wavenumber of the disturbance applied to the interface increases, 
the liquid moving velocity along the central axis of the liquid ligament increases. 
Additionally, on the high wavenumber side, 
the bulge of the liquid ligament develops earlier owing to the vortex 
near the interface. 
As the wavenumber increases, 
the cross-sectional area of the constricted part of the liquid ligament decreases 
more quickly, and the breakup time of the liquid ligament becomes short.

In addition to the initial reference disturbance with a low wavenumber, 
when turbulence with twice the wavenumber of the reference disturbance is applied to the interface, 
the interface deformation and the splitting of the liquid ligament 
are similar to those with the single reference disturbance. 
Therefore, there is no significant difference 
between the average droplet size and its standard deviation under the two conditions. 
When turbulence with four times the wavenumber is added 
to the reference disturbance with the low wavenumber at the interface, 
the deformation of the liquid ligament is accelerated, 
and the liquid ligament splits faster. 
Additionally, the number of breakup droplets increases; 
hence, the total surface area of the liquid increases. 
The droplet diameters become uniform; 
therefore, the atomization quality of the liquid ligament is improved.

\begin{table}[!t]
\centering
\caption{Arithmetic average of droplet diameters and 
standard deviation of droplet diameter for various values of 
$\eta_r$ and $k_r$: $\Delta U/U_\mathrm{ref} = 17.8$, $ka = 0.3$}
\centering
\begin{tabular}{|c|c|c|}
\hline
  $\eta_r$, $k_r$ & $D/d$ (arithmetic average) & standard deviation \\ \hline
  $\eta_r = 0$, $k_r = 0$  & 1.12 & 0.85 \\ \hline
  $\eta_r = 0.9$, $k_r = 2.0$  & 1.19 & 0.86 \\ \hline
  $\eta_r = 0.9$, $k_r = 4.0$  & 1.39 & 0.30 \\ \hline
\end{tabular}
\label{dsizearista}
\end{table}


\vspace*{1.0\baselineskip}
\noindent
{\bf Acknowledgements.}
The numerical results in this research were obtained 
using supercomputing resources at Cyberscience Center, Tohoku University. 
This research did not receive any specific grant from funding agencies 
in the public, commercial, or not-for-profit sectors. 
We would like to express our gratitude to Associate Professor Yosuke Suenaga 
of Iwate University for his support of our laboratory. 
The authors wish to acknowledge the time and effort of everyone involved in this study.

\vspace*{1.0\baselineskip}
\noindent
{\bf Declaration of interests.}
The authors have no conflicts to disclose.

\vspace*{1.0\baselineskip}
\noindent
{\bf Author ORCID.} \\
H. Yanaoka \url{https://orcid.org/0000-0002-4875-8174}.

\vspace*{1.0\baselineskip}
\noindent
{\bf Author contributions.} \\
{\bf Hideki Yanaoka}: 
Conceptualization (lead); Investigation (equal); 
Methodology (lead); Visualization (equal); Writing - original draft (equal); 
Writing - review \& editing (lead). \\
{\bf Wataru Sakamoto}: 
Formal analysis (lead); Investigation (equal); 
Validation (lead); Visualization (equal); Writing - original draft (equal).


\bibliographystyle{arXiv_elsarticle-harv}
\bibliography{arXiv2024_sakamoto_bibfile}

\end{document}